\documentclass[a4paper,fleqn,usenatbib]{mnras}

\usepackage{letltxmacro}
\usepackage{graphicx}	
\usepackage{amsmath}    
\usepackage{amssymb}   
\usepackage{newtxtext,newtxmath}
\usepackage{multirow}
\usepackage{mathtools}
\usepackage{IEEEtrantools}
\usepackage{float}
\usepackage{rotating}
\usepackage{bm}
\usepackage{footmisc}
\usepackage{booktabs}
\usepackage{threeparttable}
\usepackage{hyperref}
\usepackage{scalefnt}
\LetLtxMacro{\oldtextsc}{\textsc}
\renewcommand{\textsc}[1]{\oldtextsc{\scalefont{1.2}#1}}

\usepackage{relsize}

\usepackage{xcolor}
\usepackage{pdflscape}
\usepackage{array}
\usepackage{threeparttable}
\usepackage{caption}
\usepackage{subcaption}
\usepackage{afterpage}

\newcommand{\cloudy}{\textsc{cloudy}}
\newcommand{\beagle}{\textsc{beagle}}

\newcommand{\starburst}{\textsc{starburst}99}

\newcommand{\CB}{{\small C\&B}}

\newcommand{\JWST}{\textit{JWST}}

\newcommand{\relagn}{\textsc{relagn}}
\newcommand{\relqso}{\textsc{relqso}}

\newcommand{\agnslim}{\textsc{agnslim}}
\newcommand{\GALSEVN}{\textsc{galsevn}}

\newcommand{\Msun}{\hbox{M$_{\rm{\odot}}$}}

\newcommand{\Mstar}{\hbox{$M_{\star}$}}

\newcommand{\hii}{\hbox{H{\sc ii}}}

\newcommand{\zsun}{\hbox{${Z}_{\odot}$}}

\newcommand{\nh}{\hbox{$n_{\mathrm{H}}$}}
\newcommand{\Nh}{\hbox{$N_{\mathrm{H}}$}}

\newcommand{\Us}{\hbox{$U_{\rm{S}}$}}
\newcommand{\xid}{\hbox{$\xi_{\rm{d}}$}}

\newcommand{\CO}{\hbox{C/O}}
\newcommand{\OH}{\hbox{O/H}}

\newcommand{\COsol}{\hbox{(C/O)$_\odot$}}
\newcommand{\NO}{\hbox{N/O}}

\newcommand{\logoh}{\hbox{12 + log(\OH)}}

\newcommand{\Mbh}{\hbox{M$_{\rm{BH}}$}}
\newcommand{\Fedd}{\hbox{F$_{\rm{Edd}}$}}
\newcommand{\Nhstop}{\hbox{N$_{\rm{H}}$}}
\newcommand{\fcovBLR}{\hbox{fcov$_{\rm{BLR}}$}}
\newcommand{\Lbol}{\hbox{L$_{\rm{bol}}$}}
\newcommand{\Lacc}{\hbox{L$_{\rm{acc}}$}}
\newcommand{\vturb}{\hbox{v$_{\rm{turb}}$}}

\newcommand{\nivb}{\hbox{N\,\textsc{iv}]\,$\lambda1485$}}

\newcommand{\loi}{\hbox{O\,\textsc{i}}}

\newcommand{\civ}{\hbox{C\,\textsc{iv}\,$\lambda1549$}}

\newcommand{\heii}{\hbox{He\,\textsc{ii}\,$\lambda1640$}}

\newcommand{\oiii}{\hbox{O\,\textsc{iii}]\,$\lambda1664$}}

\newcommand{\niii}{\hbox{N\,\textsc{iii}]\,$\lambda1750$}}

\newcommand{\ciii}{\hbox{C\,\textsc{iii}]\,$\lambda1908$}}

\newcommand{\nevopt}{\hbox{[Ne\,\textsc{v}]\,$\lambda3426$}}
\newcommand{\lnev}{\hbox{[Ne\,\textsc{v}]}}

\newcommand{\oiid}{\hbox{[O\,{\sc ii}]\,$\lambda\lambda3726,3729$}}
\newcommand{\oiiopt}{\hbox{[O\,\textsc{ii}]$\lambda3727$}}
\newcommand{\neiiiopt}{\hbox{[Ne\,\textsc{iii}]\,$\lambda3869$}}

\newcommand{\heiopta}{\hbox{He\,\textsc{i}\,$\lambda3889$}}
\newcommand{\heioptf}{\hbox{He\,\textsc{i}\,$\lambda4471$}}
\newcommand{\heiiopt}{\hbox{He\,\textsc{ii}$\lambda4686$}}
\newcommand{\hb}{\hbox{H$\beta$}}

\newcommand{\oiiiopt}{\hbox{[O\,\textsc{iii}]$\lambda5007$}}

\newcommand{\oiiioptd}{\hbox{[O\,\textsc{iii}]$\lambda4959,5007$}}
\newcommand{\oiiioptc}{\hbox{[O\,\textsc{iii}]$\lambda4363$}}

\newcommand{\oiopt}{\hbox{[O\,\textsc{i}]\,$\lambda6300$}}
\newcommand{\ha}{\hbox{H$\alpha$}}
\newcommand{\niiopt}{\hbox{[N\,\textsc{ii}]$\lambda6584$}}
\newcommand{\heioptb}{\hbox{He\,\textsc{i}\,$\lambda6678$}}
\newcommand{\heioptc}{\hbox{He\,\textsc{i}\,$\lambda7065$}}

\newcommand{\siiopt}{\hbox{[S\,\textsc{ii}]$\lambda6720$}}

\newcommand{\larivopt}{\hbox{[Ar\,\textsc{iv}]}}
\newcommand{\hg}{\hbox{H$\gamma$}}

\newcommand{\heiopte}{\hbox{He\,\textsc{i}\,$\lambda5875$}}

\newcommand{\neiv}{\hbox{[Ne\,\textsc{iv}]\,$\lambda2423$}}
\newcommand{\nev}{\hbox{[Ne\,\textsc{v}]\,$\lambda3426$}}
\newcommand{\cii}{\hbox{C\,\textsc{ii}]\,$\lambda2326$}}

\newcommand{\lsiliii}{\hbox{Si\,\textsc{iii}]}}

\newcommand{\loiopt}{\hbox{[O\,\textsc{i}]}}
\newcommand{\loii}{\hbox{[O\,\textsc{ii}]}}
\newcommand{\loiii}{\hbox{O\,\textsc{iii}]}}
\newcommand{\lniii}{\hbox{N\,\textsc{iii}]}}

\newcommand{\lniv}{\hbox{N\,\textsc{iv}]}}

\newcommand{\lcii}{\hbox{C\,\textsc{ii}}}
\newcommand{\lciii}{\hbox{C\,\textsc{iii}]}}
\newcommand{\lciv}{\hbox{C\,\textsc{iv}}}

\newcommand{\laliii}{\hbox{Al\,\textsc{iii}}}

\newcommand{\loiiopt}{\hbox{[O\,\textsc{ii}]}}
\newcommand{\loiiiopt}{\hbox{[O\,\textsc{iii}]}}

\newcommand{\lheii}{\hbox{He\,\textsc{ii}}}
\newcommand{\lniiopt}{\hbox{[N\,\textsc{ii}]}}
\newcommand{\lnv}{\hbox{N\,\textsc{v}}}
\newcommand{\lsiiopt}{\hbox{[S\,\textsc{ii}]}}
\newcommand{\lneiiiopt}{\hbox{[Ne\,\textsc{iii}]}}
\newcommand{\lneiv}{\hbox{[Ne\,\textsc{iv}]}}

\newcommand{\loiv}{\hbox{O\,\textsc{iv}]}}

\definecolor{col1}{HTML}{F572B6}
\definecolor{col2}{HTML}{DBAC45}
\definecolor{col3}{HTML}{700002}

\usepackage{xcolor}

\title[High-z spectral diagnostics of massive BHs]{Unified modelling of broad and narrow optical–UV emission lines from massive black holes: interpreting high-redshift JWST observations}

\author[Plat et al.]{
Ad\`ele Plat$^{1}$\thanks{E-mail: adele.plat@epfl.ch, plat@iap.fr},
Michaela Hirschmann$^{1}$,
Emma Curtis-Lake$^{2}$,
Anna Feltre$^{3}$,
Stephane Charlot$^{4}$,
\newauthor
Steven L. Finkelstein$^{5,6}$,
Lorenzo Napolitano$^{7}$,
Fabio Pacucci$^{8}$,
Norbert Pirzkal$^{9}$,
Sandra Raimundo$^{10,11}$,
\newauthor
Anthony Taylor$^{5,6}$,
Alba Vidal-Garcia$^{12}$,
L. Y. Aaron Yung$^{13,14}$\\
$^{1}$Institute of Physics, GalSpec Laboratory, Ecole Polytechnique Federale de Lausanne, Observatoire de Sauverny, Chemin Pegasi 51, \\ 1290 Versoix, Switzerland\\
$^{2}$ Centre for Astrophysics Research, Department of Physics, Astronomy and Mathematics, University of Hertfordshire, Hatfield AL10 9AB, UK \\
$^{3}$INAF-Osservatorio Astrofisico di Arcetri, Largo E. Fermi 5, I-50125, Firenze, Italy\\
$^{4}$Sorbonne Universit\'{e}, CNRS, UMR7095, Institut d'Astrophysique de Paris, F-75014, Paris, France\\
$^{7}$INAF – Osservatorio Astronomico di Roma, via Frascati 33, 00078, Monteporzio Catone, Italy\\
$^{8}$Center for Astrophysics $\vert$ Harvard \& Smithsonian,
Cambridge, MA 02138, USA\\
$^{12}$Observatorio Astronómico Nacional, C/ Alfonso XII 3, 28014 Madrid, Spain\\
$^{5}$Department of Astronomy, The University of Texas at Austin, Austin, TX, USA\\
$^{6}$Cosmic Frontier Center, The University of Texas at Austin, Austin, TX, USA\\
$^{13}$Space Telescope Science Institute, 3700 San Martin Drive, Baltimore, MD 21218, USA\\
$^{14}$Institute of Astronomy and Department of Physics, National Tsing Hua University, Hsinchu 30013, Taiwan\\
$^{9}$ESA/AURA Space Telescope Science Institute, 3700 San Martin Dr, Baltimore MD 21212\\
$^{10}$Physics and Astronomy, University of Southampton, Highfield, Southampton SO17 1BJ, UK\\
$^{11}$DARK, Niels Bohr Institute, University of Copenhagen, Jagtvej 155, Copenhagen N 2200, Denmark
}

\date{Accepted XXX. Received YYY; in original form ZZZ}

\pubyear{--}

\begin{document}
\label{firstpage}
\pagerange{\pageref{firstpage}--\pageref{lastpage}}
\maketitle 

\begin{abstract}
\JWST\ is uncovering a large population of active galactic nuclei (AGN) at high redshift, motivating the extension of models calibrated on local sources to a broader range of physical conditions. We present a consistent suite of broad- and narrow-line photoionisation models based on ionising spectra that vary with black-hole mass and Eddington ratio, spanning sub- to super-Eddington accretion. For the broad-line region (BLR), we investigate the detectability of low-mass accreting black holes. At \(z=6\), broad \ha\ may be detectable down to \(M_{\rm BH}\sim10^{5.7}~\Msun\) for a BH accreting at the Eddington limit under favourable assumptions. We derive line-to-accretion-luminosity corrections, with hydrogen and helium recombination lines providing more robust tracers than metal lines. We also explore which changes in black-hole properties and BLR conditions can contribute to the large \ha\ equivalent widths, elevated Balmer decrements, and weak high-ionisation lines observed in some high-redshift sources, finding that variations in BLR structure can substantially modify the emergent spectrum. For the narrow-line region (NLR), \ha\ and \oiiiopt\ are among the most promising tracers of low-mass accreting black holes, probing \(M_{\rm BH}\sim10^{5.5}~\Msun\) at the Eddington limit. Many classical diagnostic diagrams miss low-metallicity AGN; we identify alternative UV and optical diagnostics that more robustly separate AGN from Pop.~II and Pop.~III stellar photoionisation over the conditions explored. Nevertheless, NLR line ratios are driven primarily by metallicity and ionisation parameter, with only a weak dependence on black-hole mass and Eddington ratio. Looking ahead, these models are intended for Bayesian interpretation of \JWST\ spectra and connection to cosmological simulations through physically motivated priors.
\end{abstract}



\section{Introduction}\label{sec:intro}

Supermassive black holes (SMBHs), with masses of $10^{6}$--$10^{10}\ \Msun$, are believed to reside at the centres of most massive galaxies \citep[e.g.,][]{Magorrian1998,Ferrarese2000,Kormendy2013}. The formation and growth of these SMBHs remain among the major open questions in astrophysics. Several pathways have been suggested for the formation of the first black-hole seeds \citep[e.g.,][]{Rees1984}, including primordial BH seeds, the remnants of Population III stars \citep[light seeds, $10^{2} - 10^{3}$ \Msun, e.g.,][]{Madau2001,Bromm2003}, runaway stellar collisions in dense star clusters \citep[medium seeds, $10^{3} - 10^{5}$ \Msun, e.g.,][]{Portegies2002,Portegies2004,Devecchi2009}, and the direct collapse of low-metallicity gas clouds \citep[heavy seeds, $10^{4} - 10^{6}$ \Msun, e.g.,][]{Loeb1994,Begelman2006,Lodato2006}. Following their formation, these seeds may grow through gas accretion, black-hole mergers, or a combination of both processes \citep[e.g.,][]{Volonteri2003,Pacucci2020}. However, the relative contributions of these seeding and growth mechanisms remain poorly constrained.

Observations of black holes in the early Universe provide a unique opportunity to constrain the formation and growth of SMBHs by probing their masses and accretion rates close to the epoch of seed formation \citep[e.g.,][]{Pacucci2022,Fragione2023}. The advent of the \JWST\ \citep[][]{Gardner2006,Gardner2023} has revolutionized this field by uncovering a large population of candidate type~I and type~II AGN at high redshift \citep[$4 \lesssim z \lesssim 12$, e.g.,][]{Castellano2026,Chavez2025,Napolitano}, extending AGN studies to lower luminosities and black-hole masses than previously possible \citep[e.g.,][]{Goulding2023,Onoue2023,Kokorev2023,Kocevski2023,Harikane2023,Maiolino2024,Furtak2024,Fujimoto2024,Ren2025,Mazzolari2025,Scholtz2025,Taylor2025b,Juodvabalis2026}. These discoveries now enable the characterization of black-hole masses and accretion rates over a much wider range of the high-redshift AGN population than was previously accessible.
While some studies have reported black holes that appear overmassive relative to the stellar masses of their host galaxies compared to the local $M_{\rm BH}$--$M_\star$ relation \citep[e.g.,][]{Pacucci2023,Furtak2024,Maiolino2024,Gupta2026}, potentially favouring massive seeds or rapid, possibly super-Eddington, growth, other studies have identified AGN consistent with the local relation \citep[e.g.,][]{Ren2025}. These contrasting results highlight both the potential of high-redshift AGN to constrain SMBH formation and the need for robust methods to identify AGN and accurately infer their physical properties.

These faint high-redshift AGN candidates exhibit spectral properties that differ from those of their local counterparts. While the difficulty of classical optical diagnostic diagrams in distinguishing AGN from star-forming (SF) galaxies at high redshift, owing to their low metallicities, had been recognized well before \JWST\ \citep[e.g.,][]{Feltre2016,Hirschmann2019}, and has since been confirmed by the location of the narrow emission lines of high-redshift type~I AGN \citep[e.g.,][]{Maiolino2024,Juodvabalis2026}, these sources also display several unexpected properties. These include weak X-ray emission \citep[e.g.,][]{Ananna2024,Madau2024,Pacucci2024,Yue2024,Maiolino2025,Mazzolari2025}, unusually large equivalent widths of the broad \ha\ emission \citep[e.g.,][]{Maiolino2025}, and weak broad and/or narrow high-ionisation emission lines \citep[e.g.,][]{Zucchi2025,Wang2025,Brazzini2026,Juodvabalis2026}.

In addition, some of these candidate type~I AGN belong to a newly identified class of compact red sources, known as Little Red Dots (LRDs) \citep[e.g.,][]{Labbe2023,Furtak2023,Furtak2024,Kocevski2023,Matthee2024,Greene2024,Kocevski2025,Barro2026}. LRDs are characterized by their distinctive V-shaped spectral energy distributions and exhibit several unusual properties, 
extremely strong Balmer breaks and Balmer absorption lines in some sources, weak X-ray emission, and weak high-ionisation emission lines \citep[e.g.,][]{Juodvabalis2024,Ji2025,Inayoshi2025,Maiolino2025,Naidu2025,deGraaff2025b,deGraaff2025a,Taylor2025,Wang2025,Rusakov2026,Brazzini2026,PerezGonzalez2026}. 

These observations suggest that the physical conditions in high-redshift AGN may differ substantially from those in their low-redshift counterparts. Robustly inferring the physical properties of high-redshift AGN therefore requires photoionisation models that not only account for the dependence of the ionising spectrum on black-hole mass and Eddington ratio, including the super-Eddington accretion regime, but also incorporate the physical conditions responsible for the distinctive spectral properties revealed by \JWST.

Photoionisation models of AGN have been extensively developed over the past decades for both the narrow-line region \citep[NLR, e.g.,][]{Dopita2002,Groves2004,Feltre2016,Thomas2016,Nakajima2022b,McKaig2024,Zhang2026} and the broad-line region \citep[BLR, e.g.,][]{Radovich1994,Baldwin1995,Hamann2002,Nagao2006,Floris2024,Zhang2026}. However, many of these models adopt ionising spectra that do not explicitly depend on black-hole mass and Eddington ratio, focus on black holes more massive than $10^{6}\ \Msun$ \citep[e.g.,][]{Panda2019,Zhang2025}, or assume solar metallicity \citep[e.g.,][]{Temple2020,Sarkar2021,Temple2021}. Such assumptions may not be appropriate for the low-mass, rapidly accreting AGN population now being uncovered by \JWST.
Importantly, despite the extensive modelling of the NLR and BLR individually, models that simultaneously and consistently predict the emission from both the NLR and BLR remain comparatively rare. Existing approaches often rely on simplified assumptions, such as describing both regions with a single photoionisation component \citep[e.g.,][]{Cleri2023} or representing the BLR with a single cloud rather than a distribution of clouds spanning a range of physical conditions \citep[e.g.,][]{Zhang2026}. Such simplifications may limit their ability to reproduce the diversity of emission-line properties observed in high-redshift AGN.

More recently, several models have been proposed to explain the nature of LRDs. These include scenarios in which the emission arises from dense gas surrounding an accreting black hole \citep[e.g.,][]{Inayoshi2025,Ji2025,Naidu2025,deGraaff2025b,Sneppen2026,Torralba2026,Pacucci2026}, quasi-star models \citep[e.g.,][]{Begelman2008,Gentile2026}, supermassive stars \citep[e.g.,][]{Chisholm2026}, and models in which the observed continuum is produced by thermalized blackbody emission from stellar-like atmospheres \citep[e.g.,][]{Liu2026}.
Additionally, super-Eddington accretion has been invoked to explain several of the unusual spectral properties of high-redshift AGN, including their weak X-ray emission \citep[e.g.,][]{Madau2024,Pacucci2024,Lambrides2024,Madau2025,Madau2026}.

In this paper, we develop a new suite of unified photoionisation models for both the NLR and the BLR. The models self-consistently account for the dependence of the ionising spectrum on black-hole mass and Eddington ratio, including the super-Eddington accretion regime. Covering a wide range of physical parameters, this extensive grid is designed to reproduce the emission-line properties of AGN across cosmic time, from classical local type~I and type~II AGN to high redshift faint AGN and the newly discovered LRD population.

Using these models, we investigate the detectability of black-hole accretion with \JWST\ down to the intermediate-mass black-hole regime, $10^3 \Msun$. We explore spectroscopic diagnostics capable of distinguishing AGN from stellar ionising sources, including Population~II and Population~III stellar populations, derive new bolometric corrections for both type~I and type~II AGN, and assess the extent to which black-hole masses and Eddington ratios can be constrained for type~II AGN in the absence of broad emission lines. Finally, we interpret the physical properties of recently observed high-redshift type~I and type~II AGN within the framework of our models.

In Section~\ref{sec:method}, we describe our modelling approach for the BLR, the NLR, and \hii\ regions ionized by Population~I, Population~II, and pristine first-generation Population~III stars. In Section~\ref{sec:validation}, we validate our models against observations of low-redshift AGN.
In Section~\ref{sec:TypeI}, we investigate the detectability of broad AGN emission lines with \JWST, derive new bolometric corrections for type~I AGN, and interpret the spectral properties of faint high-redshift broad-line sources. In Section~\ref{sec:TypeII}, we investigate the detectability of narrow AGN emission lines with \JWST, derive new bolometric corrections for type~II AGN, and explore diagnostics for distinguishing black-hole accretion from other sources of ionising radiation.
Finally, in Section~\ref{sec:discussion}, we discuss the uncertainties of our modelling approach and examine how alternative assumptions may account for some of the intriguing spectral properties of high-redshift AGN, such as their weak high-ionisation emission lines.

\section{Photoionisation models}\label{sec:method}

We employ the photoionisation code \cloudy\ (version c23.01; \citealt{CLOUDY2023}) to construct a grid of models describing the optical and ultraviolet (UV) nebular line and continuum emission of active galactic nuclei (AGN). In Sec.~\ref{sec:method_SED}, we present the spectral energy distribution (SED) adopted to model emission from the accretion disc surrounding the central BH. Sections~\ref{sec:method_BLR} and \ref{sec:method_NLR} detail the parameterisation of the broad-line region (BLR) and the narrow-line region (NLR), respectively. To facilitate emission-line diagnostics that distinguish between star-forming (SF) and AGN-dominated galaxies, Sec.~\ref{sec:method_stellar} briefly summarises the grid of photoionisation models for \hii\ regions, including contributions from Population~III (Pop~III) stars.

\subsection{The incident ionising spectra of AGN}\label{sec:method_SED}
   
For the incident ionising spectra of AGN, we adopt SEDs which describe the emission from BH accretion discs across both sub- and super-Eddington regimes, from $\log(\dot{\rm{M}}/\dot{\rm{M}}_{\rm{Edd}}) = -1.5$ to 2. In the sub-Eddington case, the accretion-disc emission is modelled using the \relqso\ model \citep{Hagen2023}. The total emission arises from three components: (i) an outer, standard geometrically thin and optically thick disc, (ii) a warm Comptonization region, and (iii) a hot Comptonization region. The \relqso\ model is a simplified version of the more general \relagn\ model \citep{Hagen2023}, in which several adjustable parameters are fixed following the prescriptions of \citet{Kubota2018}. In particular, the energy dissipated in the hot corona is set to $0.02~L_{\rm Edd}$, thereby directly linking the X-ray emission to the Eddington ratio. Although this relation was calibrated using fits to nearby AGN and is expected to be broadly representative of typical AGN, it may not entirely capture the diversity of properties observed in individual sources. Nevertheless, it significantly reduces the number of free parameters in the model, which is advantageous given the uncertainties in modelling accretion-disc emission.  We fix the BH spin to zero. 
The spin depends on a range of physical processes \citep[e.g.][]{Ricarte2023,Ricarte2025}, while its distribution remains poorly constrained observationally. We therefore discuss the impact of this assumption further in Section~\ref{sec:discumodeluncertainties}. The remaining free parameters of the model are then the BH mass, \Mbh, and the Eddington ratio, $\Fedd = \dot{\rm{M}}/\dot{\rm{M}}_{\rm{Edd}}$. For simplicity, we adopt a single average inclination angle of $45^{\circ}$ between the accretion disc and the narrow-line region. Varying the inclination angle modifies the shape of the incident radiation; however, we assume that the NLR is not predominantly dominated by clouds viewing the accretion disc at extreme polar angles (i.e. $0^{\circ}$ or $90^{\circ}$). As a result, the impact of inclination is expected to be small compared to variations driven by other model parameters (see Sec.~\ref{sec:discumodeluncertainties}). 
Note that we adopt a different inclination for the broad-line region, as described in Section~\ref{sec:method_BLR}.  

In the super-Eddington regime, we adopt the \agnslim\ model from \citet{Kubota2019}. As in the sub-Eddington case, the emission is composed of three components; however, the accretion flow is  modelled  as a slim disc rather than a geometrically thin disc. As a result, the radial dependence of the emitted flux scales approximately as $1/r^2$, rather than $1/r^3$ as in the sub-Eddington regime. 
The \agnslim\ model includes a large number of parameters describing the properties of each emission component. To focus on the dependence on BH mass and Eddington ratio, we fix all other parameters to their fiducial values recommended for high \Fedd\ AGN \citep[spectral index of the warm Comptonisation component, $\Gamma_{\rm{warm}}=2.7$, electron temperature for the warm Comptonisation component, $\rm{kTe}_{\rm{warm}}=0.2\rm{keV}$, outer radius of the warm Comptonisation component, $\rm{r}_{\rm{warm}}=20 \rm{R_{g}}$, spectral index of the hot Comptonisation component, $\Gamma_{\rm{hot}}=2.4$, electron temperature for the hot Comptonisation component, $\rm{kTe}_{\rm{hot}}=100\rm{keV}$, outer radius of the hot Comptonisation component, $\rm{r}_{\rm{hot}}=10 \rm{R_{g}}$, see][]{Kubota2019}. We discuss models powered by alternative super-Eddington SEDs, including X-ray-weak accretion-disc SEDs, in Sec.~\ref{sec:discumodeluncertainties}.

\subsection{Broad-line region AGN models}\label{sec:method_BLR}

To model emission from the BLR in AGN, we construct two sets of models. The first assumes a multi-cloud scenario, and is calibrated to reproduce local type~I AGN, as described in Section~\ref{sec:method_BLR_fid}. The underlying model grid, however, spans a broad range of physical conditions, allowing us to explore variations around these fiducial assumptions and investigate how potential differences in the BLR properties of high-redshift type~I AGN may affect their emission-line spectra (see Section~\ref{sec:discuhighz}). The second set extends to large column densities of dense gas and is designed to capture the extreme conditions observed in some high-redshift LRDs; this grid is presented in Sec.~\ref{sec:method_BLR_bb}.

\subsubsection{Fiducial BLR multi-cloud models}\label{sec:method_BLR_fid}

\begin{figure}
\includegraphics[width=1\linewidth]{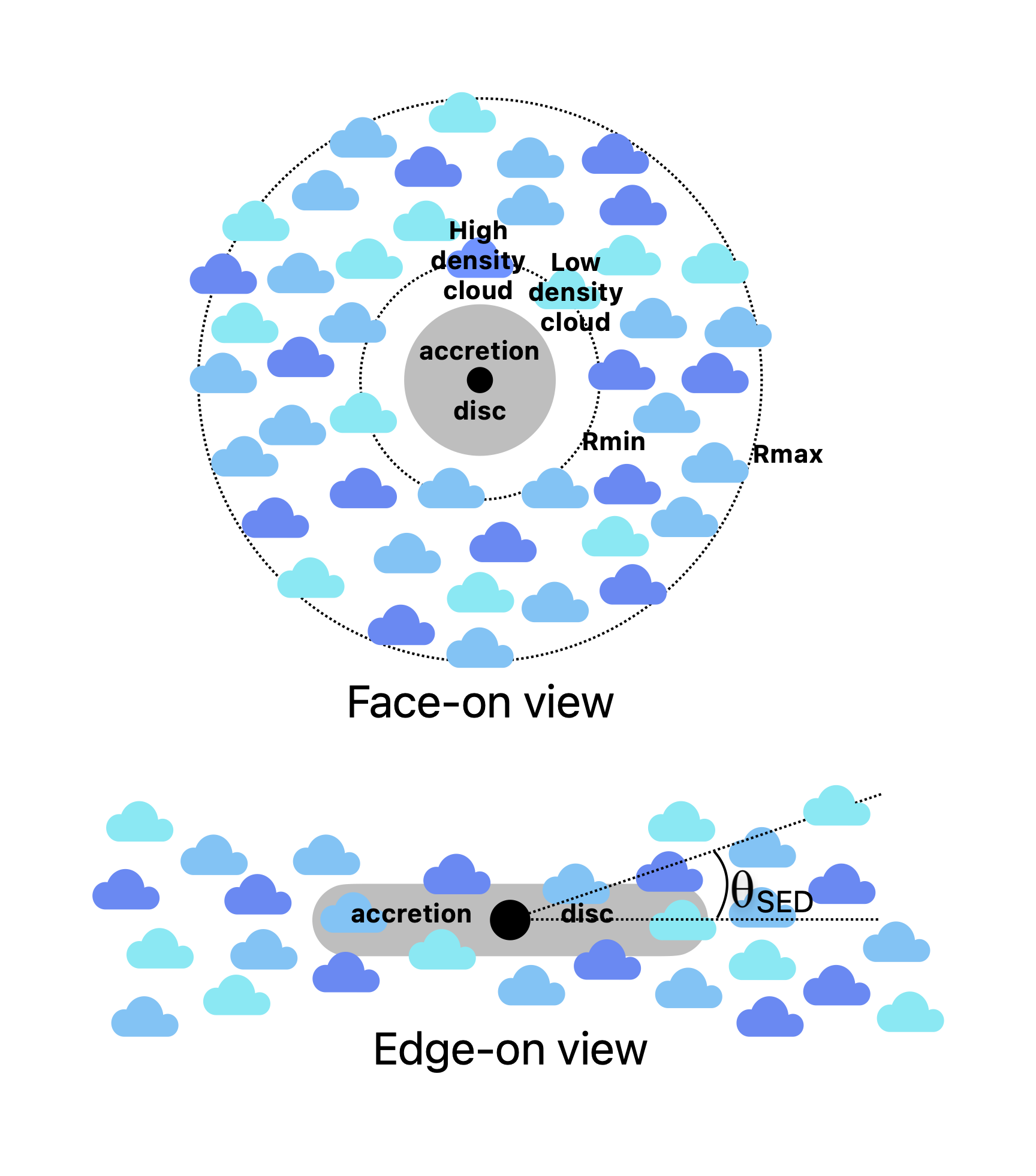}
\caption{Schematic view of the multi-cloud modelling approach for the BLR. The BLR is represented as a collection of clouds spanning a range of gas densities, from $\nh = 10^{8}\ \mathrm{cm^{-3}}$ to $10^{14}\ \mathrm{cm^{-3}}$, illustrated by different shades of blue. The clouds are distributed over distances between $\rm R_{\rm min}$ and $\rm R_{\rm max}$ from the accretion disc, defined by the level of incident ionising flux ($16 \leq \log (\Phi/\mathrm{photons}\ \mathrm{s}^{-1} \mathrm{cm}^{-2}) \leq 24$). 
The incident radiation originates from the accretion-disc emission as seen by the BLR clouds at an angle $\theta_{\rm SED}$.}
\label{fig:SchemaBLR}
\end{figure}

To model the emission from the BLR in AGN, we adopt the locally optimally emitting cloud (LOC) framework introduced by \citet{Baldwin1995} (see also \citealt{Radovich1994}). In this approach, the BLR is represented as an ensemble of individual clouds spanning a range of gas densities and distances from the BH accretion disc, and the total line luminosity is computed as the sum of the emission coming from all these clouds. A schematic illustration of this configuration is shown in Fig.~\ref{fig:SchemaBLR}. Such multi-cloud models were developed to better reproduce the observed quasar emission-line spectra, as single-cloud models generally fail to account for all emission lines simultaneously \citep[e.g.,][]{Radovich1994}. Below, we first describe the modelling of individual BLR clouds, and then outline the construction of the multi-cloud model.

The line and continuum emission from single BLR clouds are computed using \cloudy\ by making the following assumptions.
The incident radiation illuminating the BLR clouds is given by the accretion-disc SED described in Section~\ref{sec:method_SED}. Since the radiation emitted by the accretion disc depends on viewing angle, the SED seen by the BLR clouds depends on their location relative to the disc. Here, we assume that the BLR is located along the edge of the accretion disc (see Fig.~\ref{fig:SchemaBLR}), as suggested by reverberation-mapping studies \citep[e.g.,][]{Grier2017b,Williams2018}, and therefore adopt an average angle of $\theta_{\rm{SED}} = 22.5^{\circ}$ between the accretion disc and the BLR clouds. This simplification is appropriate when the covering factor of the BLR, \fcovBLR, is below unity, as is typically inferred for local AGN \citep[$\fcovBLR \approx 0.1 - 0.4$, e.g.,][]{Peterson2006,Ruff2012}. If instead the BLR covering factor approaches unity, the angle between the accretion disc and the clouds should vary with the cloud position. The impact of the angle between the BLR clouds with respect to the disc is further discussed in Section~\ref{sec:discumodeluncertainties}.
For the gas in the BLR, we further adopt the metal abundance patterns of \citet{Gutkin2016} (\zsun=0.01524). But we extend the metallicity range, Z, from 0.0001 up to 0.200, to account for the potentially highly super-solar metallicity seen in quasars \citep[e.g.,][]{Hamann2002,Nagao2006,Wang2022}. And we discuss in Sec.\ref{sec:discumodeluncertainties} the impact of different N/O and C/O abundances. Dust grains are not included in the models, as dust is expected to be sublimated within the BLR \citep[e.g.,][]{Netzer1993}. The Cloudy calculations are stopped when either the electron fraction drops to 0.01 (we include only emission from ionised gas) or the total hydrogen column density reaches $\log (\Nhstop/\mathrm{cm^{-2}}) = 23$ (whichever occurs first). We adopt this column density for our fiducial grid, as we find that it reproduces well the observed properties of quasars (see Section~\ref{sec:validation_BLR}). 
We explore the impact of varying this parameter over the range $19 \leq \log (\Nhstop/\rm{cm}^{-2}) \leq 25$ in Section \ref{sec:discussion}.
We also include microturbulence with velocities, \vturb, of 0 and $250\ \mathrm{km\ s^{-1}}$ in the fiducial models. The impact of dissipative heating is discussed in Section~\ref{sec:discussion}.

\begin{figure}
\includegraphics[width=1\linewidth]{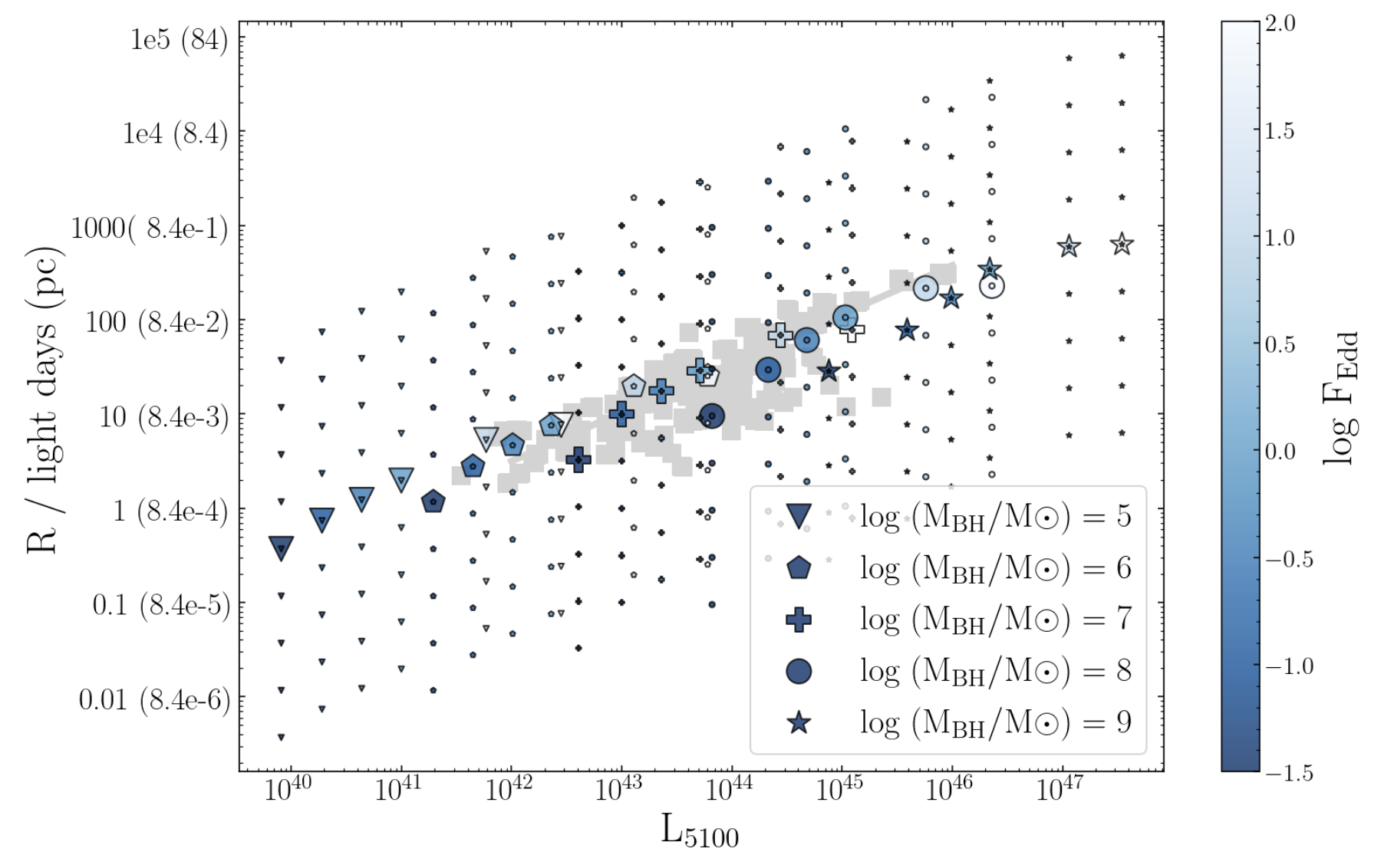}
\caption{Radius–luminosity relation. Grey symbols show observational measurements for local AGN from reverberation-mapping studies using \hb\ line \citep{Bentz2009,Bentz2013,Grier2017}. The models consist of multiple BLR clouds distributed over a range of distances from the accretion disc. Small blue symbols indicate the radial distribution of individual BLR clouds defined such that $\log (\Phi/\mathrm{photons}\ \mathrm{s}^{-1} \mathrm{cm}^{-2}) = 16$ to 24, while the larger blue symbols show the BLR radius corresponding to $\log (\Phi/\mathrm{photons}\ \mathrm{s}^{-1} \mathrm{cm}^{-2}) = 20$ for each model. Symbol shapes correspond to different black-hole masses: $\log (\Mbh/\Msun) = 5$ (downward triangles), 6 (pentagons), 7 (crosses), 8 (circles), and 9 (stars). The models are colour-coded by Eddington ratio, from dark blue in the sub-Eddington regime to light blue in the super-Eddington regime.}
\label{fig:RLrel}
\end{figure}

As an observational reference for the radial scales explored by our BLR models, we show in Fig.~\ref{fig:RLrel} the \hb\ radius–luminosity relation inferred from reverberation-mapping studies. The grey symbols represent observational constraints from reverberation-mapping studies based on the \hb\ line \citep{Bentz2009,Bentz2013,Grier2017}. Observationally, the radius corresponds to a characteristic distance of the broad \hb--emitting region from the accretion disc, as inferred from reverberation mapping, while the luminosity is the AGN continuum luminosity at $5100$\AA. In reality, the BLR is unlikely to be confined to a single radius but is instead spatially extended, as suggested by reverberation-mapping studies comparing delays between high- and low-ionisation emission lines \citep[e.g.,][]{Clavel1991,Horne2021,Shen2024}. The LOC approach, which assumes a distribution of cloud distances rather than a single characteristic radius, is therefore consistent with the observational evidence for an extended BLR. 

Given this spatial extent, it is convenient to adopt a parameterisation that naturally captures the radial distribution of the BLR clouds. Rather than describing the BLR in terms of an inner radius of the BLR clouds through a relation of the form $R \propto L_{5100}^{\alpha}$, we instead parameterize the clouds using the flux of ionising-photon at the incident radius of the BLR cloud, $\log \Phi$. This choice facilitates direct comparison with previous studies in the literature \citep[e.g.,][]{Baldwin1995,Korista1997,Hamann2002}. This parameterisation is further motivated by the fact that the observed radius--luminosity relation follows approximately $R \propto L_{5100}^{0.519}$ \citep[e.g.,][]{Bentz2009}, while the hydrogen-ionising photon production rate $Q_{\rm H}$ scales nearly linearly with $L_{5100}$. Since $\Phi \propto Q_{\rm H}/R^{2}$, fixing $\log \Phi$ to a constant value implies $R \propto L_{5100}^{0.5}$, in close agreement with the observed relation. We find that adopting $\log (\Phi/\mathrm{photons}\ \mathrm{s}^{-1} \mathrm{cm}^{-2}) = 20$ reproduces the observed radius--luminosity relation obtained using \hb. This is shown in Fig.~\ref{fig:RLrel}, where the blue symbols indicate the inner radii of the  modelled  BLR clouds; and the larger symbols denote the radius corresponding to $\log (\Phi/\mathrm{photons}\ \mathrm{s}^{-1} \mathrm{cm}^{-2}) = 20$.
Models with higher BH masses and Eddington ratios produce larger $L_{5100}$ luminosities, and therefore place the BLR clouds at larger distances from the accretion disc in order to reproduce the observed radius–luminosity relation.
In addition, we allow $\log \Phi$ to vary between 16 and 24 for the distribution of individual clouds, consistent with previous BLR-modelling efforts \citep[e.g.,][]{Korista1997,Hamann2002,Nagao2006}. This range corresponds to BLR-cloud distances spanning approximately $10^{-7}$ to $100$ pc from the accretion disc across the full range of adopted BH masses and Eddington ratios \footnote{We do not require the multi-cloud model to recover the observed \hb\ radius-luminosity relation, since the reverberation-mapping radius traces the responsivity-weighted radial distribution of the line-emitting gas rather than its luminosity-weighted radial distribution \citep[e.g.,][]{Korista2004}.}. In Fig.~\ref{fig:RLrel}, the smaller symbols show the full radial distribution of the modelled BLR clouds.

The BLR is classically thought to be truncated at large radii when the gas temperature drops below the dust sublimation temperature, since dust then forms and absorbs a significant fraction of the ionising photons \citep{Netzer1993}. 
\citet{Korista2000} and \citet{Nagao2006} found that this outer BLR radius corresponds to an ionising-photon flux of $\log (\Phi/\mathrm{photons}\  \mathrm{s}^{-1} \mathrm{cm}^{-2}) \simeq 18$. We therefore restrict the cloud distribution in our fiducial model to $\log (\Phi/\mathrm{photons}\ \mathrm{s}^{-1} \mathrm{cm}^{-2}) \geq 18$, which we find reproduces well the observed properties of quasars, as shown in Section~\ref{sec:validation_BLR}.

In addition to this radial distribution, we also consider a range of gas densities for the BLR clouds, spanning $\log (\nh/\mathrm{cm^{-3}}) = 8$ to $14$, with the density assumed to be constant within each cloud. In the fiducial model, we adopt cloud distributions proportional to $1/\nh$ and $1/R$ for the density and radius, respectively\footnote{This corresponds to clouds being equally weighted with a grid spacing of 1 dex in \nh\ and $\Phi$.}. This assumption follows the work of \citet{Baldwin1997}, who found that such distributions reproduce observed quasar emission-line spectra well. 
A more detailed discussion of the validity and performance of the multi-cloud BLR modelling is presented in Sec.~\ref{sec:validation_BLR_multi}. 
In addition, we explore more distributions for the radius and gas density in Secs. \ref{sec:discumodeluncertainties} and \ref{sec:discuhighz}.

The density and radial cloud distributions are normalised to a BLR covering factor of unity. A lower covering factor can then be applied by scaling the integrated multi-cloud line luminosities by a factor \fcovBLR.
While the broad-line luminosities for different covering factors can be directly obtained by scaling them, for the computation of line equivalent widths (EW) we have to make additional assumptions, depending on the BLR covering factor. For a covering factor of unity, we define the continuum luminosity as the sum of the attenuated incident continuum transmitted through the BLR clouds and the nebular continuum emission \footnote{This implies that the continuum is effectively observed along directions close to the plane of the accretion disc, since the clouds are illuminated by a low-inclination SED. However, this assumption is unlikely to significantly affect our conclusions, as the BLR covering factor is not expected to be unity in realistic systems. For lower covering factors, we instead adopt a different prescription in which the continuum is viewed at a higher inclination, closer to face-on.}. 
For covering factors below unity, we assume that the observer has a nearly face-on view of the accretion disc, which we fix to an average angle of $67.5^{\circ}$ between the disc plane and the line of sight. In this case, the observed continuum is taken to be the sum of the direct accretion-disc emission (evaluated at $67.5^{\circ}$) and \fcovBLR\ times the nebular continuum (diffuse component). 
This geometry corresponds to the standard configuration for local type~I AGN, in which the BLR lies near the outer edge of the disc while the observer views the accretion-disc emission close to face-on (see Fig.~\ref{fig:SchemaBLR}). For the same reason, unless otherwise specified, the bolometric luminosity also refers to the luminosity of the accretion disc evaluated at $67.5^{\circ}$.
Instead, when a covering factor of 1 is assumed, we use the net transmitted continuum, which is the sum of the attenuated incident radiation with the nebular continuum emission.

We further assume that each cloud is directly illuminated by the intrinsic accretion-disc radiation. In reality, multiple clouds may lie along the same line of sight from the accretion disc. Since clouds at small radii are likely density-bounded \footnote{The cloud does not extend sufficiently far to reach the ionisation front.}, ionising radiation can escape and subsequently illuminate clouds at larger radii, but with a partially attenuated incident spectrum. One consequence is that distant clouds may experience a lower ionising-photon flux than expected from a simple $1/R^{2}$ geometric dilution, which primarily affects the inferred radius associated with a given value of $\Phi$. In addition, because the attenuation is wavelength-dependent, the shape of the incident spectrum may also be modified at large radii. Finally, this implies that the total covering factor of the BLR cloud population could in principle exceed unity if a fraction of the clouds are density-bounded. While accounting for these effects is beyond the scope of this work, the simplified modelling adopted here provides a first-order approximation of the BLR, which is already more realistic than the often-assumed single-cloud BLR models. Nevertheless, these caveats should be kept in mind when interpreting the model predictions and comparing them with observations.

\subsubsection{Single-cloud, Balmer-break BLR models}\label{sec:method_BLR_bb}

In addition to the multi-cloud LOC models described in Section~\ref{sec:method_BLR_fid}, we also investigate single-cloud models selected to produce the pronounced Balmer breaks observed in some LRDs.
Many LRDs exhibit strong Balmer breaks and/or Balmer absorption features, indicating the presence of a large column density of hydrogen in the $n = 2$ state along the line of sight \citep[e.g.,][]{Juodvabalis2024,Inayoshi2025,Ji2025,deGraaff2025a,deGraaff2025b,Naidu2025,Taylor2025,Tang2025,Kokorev2025,Rusakov2026,Matthee2026,DEugenio2026,Sneppen2026,Asada2026,Torralba2026,Pacucci2026}. While the origin of the ionising radiation exciting this dense hydrogen gas remains uncertain, there is also ongoing debate over the origin of the optical continuum itself, namely whether it arises from a fully thermalized, optically thick atmosphere with a temperature of $\sim5000$ K \citep[e.g.,][]{Liu2026} or from nebular/recombination emission \citep[e.g.,][]{Sneppen2026b,Rusakov2026}, with recent observations also challenging models in which the BH is fully enclosed by an optically thick envelope \citep[e.g.,][]{Geris2026}. As the geometry and degree of thermalization of the gas remain uncertain, here we focus on the non-fully-thermalized regime, considering dense gas irradiated by an accreting BH. 
Not all LRDs exhibit Balmer breaks \citep[e.g.,][]{PerezGonzalez2026}; however, we focus on the most extreme cases to investigate the physical properties of clouds capable of producing these features. Our aim is not to provide a comprehensive model of LRDs, but rather to explore the emission-line properties of BLR clouds that can produce strong Balmer breaks.
In our model, we assume that the gas responsible for the Balmer break also produces emission lines, motivated by the observed correlation between Balmer-break strength and the total \ha/\hb\ ratio in LRDs \citep[e.g.,][]{deGraaff2025a}. This contrasts with alternative scenarios in which the broad emission lines are instead produced by young star clusters surrounding the BH \citep{Inayoshi2026}.
We also note that the broad Balmer-line profiles observed in some LRDs may not arise solely from virial motions around the BH. Several studies have suggested that electron scattering in dense ionized gas contributes to the observed line shapes \citep[e.g.,][]{Rusakov2026,Torralba2026,Nikopoulos2026,Matthee2026}, while others argue that they are instead produced by virial motions within a stratified BLR \citep[e.g.,][]{Madau2026b}, or virial motions of stars \citep[e.g.,][]{Baggen2024,Nandal2026}. We do not attempt to model the line profiles in this work, focusing instead on the predicted line strengths.

To construct these models, we adopt the same parameterisation as for the individual clouds composing the fiducial multi-cloud BLR models. These models therefore correspond to a specific subset of BLR clouds drawn from the same underlying model grid. We apply a selection to retain only clouds that satisfy the following criteria: a Balmer break strength \footnote{The Balmer break strength is computed by evaluating the continuum, $f_{\nu}$, between 3630-3638~\AA\ and 3654-3665~\AA.} between 1.5 and 10 (the maximum observed value is $\sim$7.7; \citealt{Naidu2025}), and an \ha\ EW greater than 100~\AA, consistent with the EW of the broad \ha\ emission commonly observed in LRDs \citep[e.g.,][]{Rusakov2026}.
This selection preferentially includes models that produce sufficiently large column densities of hydrogen in the $n = 2$ state to generate a Balmer break through bound--free absorption. These conditions correspond to clouds with densities of $\sim 10^{9}~\mathrm{cm^{-3}}$ and ionising-photon fluxes of $\sim 10^{18}~\rm{photons}\ \mathrm{s^{-1}\ cm^{-2}}$.
Since the Balmer break is highly sensitive to the amount of partially ionized gas surrounding the BH, we also consider models with larger total hydrogen column densities, up to $\Nh = 10^{25}~\mathrm{cm^{-2}}$ (compared to $\Nh = 10^{23}~\mathrm{cm^{-2}}$ in the fiducial models). These models generally produce stronger Balmer breaks \citep[see also][]{Pacucci2026}, allowing a broader range of physical conditions (in density and ionising flux) to satisfy our selection criteria.

In the scenario considered here, producing a Balmer break requires the clouds to lie along the line of sight between the accretion disc and the observer. 
Therefore, unlike in the fiducial models -- where the accretion disc emission is assumed to be seen directly and the clouds are located primarily along the edge of the disc -- we instead consider a configuration in which the observer views the accretion-disc radiation after it has passed through the clouds. In this case, the observed continuum is composed of the attenuated incident radiation transmitted through the cloud, together with the nebular emission scaled by the covering factor.

In addition to this grid of Balmer-break models, we highlight a representative model similar to models previously used to reproduce LRD spectra. This model, which we refer to as the typical LRD-like model, has a hydrogen density of $\nh = 10^{10}~\mathrm{cm^{-3}}$, an ionising-photon flux of $\Phi = 10^{19}~\mathrm{photons\ s^{-1}cm^{-2}}$, a hydrogen column density of $\Nh = 10^{24}~\mathrm{cm^{-2}}$, a metallicity of $Z = 0.001$, and a turbulent velocity of 120~km~s$^{-1}$. It is computed assuming a BH mass of $10^{7}~\Msun$ and an Eddington ratio of 0.1. This set of parameters is similar to the fiducial model of \citet{Ji2025} used to reproduce observed LRD properties. We fix the covering factor of this model to 0.4, which produces a Balmer-break strength of $\sim 9.7$.

In both the BLR models used to construct the fiducial multi-cloud models and the single-cloud Balmer-break models, we exclude clouds that are optically thick to electron scattering, as this regime is not handled by \cloudy.

\subsection{Narrow-line region AGN models}\label{sec:method_NLR}

To construct the model grids for the processed AGN spectrum in the NLR, we follow the general methodology of \citet{Feltre2016}, with the key difference that we adopt accretion-disc SEDs dependent on BH mass and Eddington ratio (described in Sec.~\ref{sec:method_SED}) in place of a fixed broken power-law ionising spectrum.
We assume a constant density of $\log\ (\nh/\rm{cm}^{-3})=3$ in the fiducial grid and explore the impact of a higher density of $\log\ (\nh/\rm{cm}^{-3})=6$ in Section~\ref{sec:TypeIIdiag}. We stop the calculations when the electron fraction drops below 0.01. 

We use the abundances as defined in \citet{Gutkin2016}, varying the metallicity from 0.0001 to 0.060, and adopting carbon-to-oxygen ratio of (C/O) =\COsol and 0.52\COsol. We fix the dust-to-metal mass ratio, $\xid$, to 0.3. We include microturbulence and dissipative heating as described in \citet{Mignoli2019}, with a turbulence velocity of 100 km/s.  This addition was made to better reproduce observed line ratios including resonant lines, \lciv\ and \lnv. The ionisation parameter is varied from $\log \Us = -5$ to $-1$.
We discuss in Sec.~\ref{sec:discumodeluncertainties} the impact of some of the model choices ($\xid=0.3$, N/O, C/O and \nh).

\subsection{Stellar \hii\ region models}\label{sec:method_stellar}

In addition to AGN emission, in this work, we also consider stellar continuum and line emission from the photoionisation models of \citet{Gutkin2016}. These models are computed with \cloudy, adopting Charlot $\&$ Bruzual (hereafter, \CB) stellar population synthesis models \citep[updated version of][]{Bruzual2003} as the input stellar radiation field. They further assume a constant hydrogen density of $\nh = 10^{2}~\mathrm{cm^{-3}}$ and are stopped when the electron fraction falls below 0.01.
We explore the impact of a higher density of $\log\ (\nh/\rm{cm}^{-3})=6$ in Section~\ref{sec:TypeIIdiag}.
We explore metallicities in the range $Z = 0.0001$–0.030, with $\CO = \COsol$ and $0.52~\COsol$, and a fixed dust-to-metal mass ratio of $\xi_d$ = 0.3. The ionisation parameter spans $\log \Us = -4$ to $-1$. Since we focus on high-redshift galaxies, we adopt an upper stellar-mass cutoff of 300~\Msun, which produces a harder ionising spectrum than a cutoff at 100~\Msun\ \citep[e.g.,][]{Gutkin2016,Plat2019}. We assume a constant star-formation rate with ages of 3 and 10~Myr; the younger population similarly yields a harder ionising spectrum. Both choices therefore provide a conservative setup when assessing diagnostics to distinguish AGN from star-forming emission.

In addition to the Pop~II and I \hii\ region models described above, we also consider stellar ionising radiation from first, metal-free Population~III stars \citep[see][for a recent review]{Venditti2026}. These stars are expected to produce extremely hard ionising spectra, which may be degenerate with the ionising radiation from AGN \citep[e.g.,][]{Nakajima2022b}. Here, we use the models of \citet{Lecroq2025}, based on the \GALSEVN\ code. We consider two cases for the ISM metallicity ionized by Pop~III stars. In the first scenario, Pop~III stars ionize their pristine birth environment, corresponding to a dust-free medium with $\mathrm{Z}_{\mathrm{ISM}} = 10^{-11}$. In the second scenario, we assume that some of the first Pop~III stars have already exploded as pair-instability supernovae (PISNe), slightly enriching the surrounding ISM \citep[e.g.,][]{Wiklind2013}. In this case, we adopt an ISM metallicity $\mathrm{Z} = 10^{-4}$ with $\CO = 0.52~\COsol$, consistent with a ISM enriched by the first PISN explosions \citep{Vanni2024}.

\section{Validation of the photoionisation models}\label{sec:validation}

In this section, we validate our photoionisation models against key observational constraints from low-redshift galaxies ($z \lesssim 5$). While our model grids span a broad range of physical conditions, designed to encompass both local AGN and the physical properties expected at high redshift, it is important to first establish that they can reproduce the well-characterised spectral properties of nearby AGN. We therefore compare our predictions with observations of local sources, considering separately diagnostics associated with the BLR (Section~\ref{sec:validation_BLR}) and the NLR (Section~\ref{sec:validation_NLR}). These comparisons allow us to assess whether the adopted parameterisations of the BH accretion-disc SED and gas properties provide a reliable description of observed AGN line ratios and luminosities in the local Universe. Having established this baseline, we then extend the models to interpret and predict the emission-line properties of high-redshift type~I AGN (Section~\ref{sec:TypeI}) and type~II AGN (Section~\ref{sec:TypeII}).

\subsection{BLR validation}\label{sec:validation_BLR}

We start with the validation of the BLR models. We first confront the predictions from single-cloud BLR models to the ones from the combined multi-cloud models obtained using the LOC approach in Section~\ref{sec:validation_BLR_multi}. We then compare the model predictions with observations of local broad-line AGN in Section~\ref{sec:validation_BLR_z0}. 

\subsubsection{Line ratios from multi-cloud models versus single-cloud models}\label{sec:validation_BLR_multi}

\begin{figure*}
\includegraphics[width=1\linewidth]{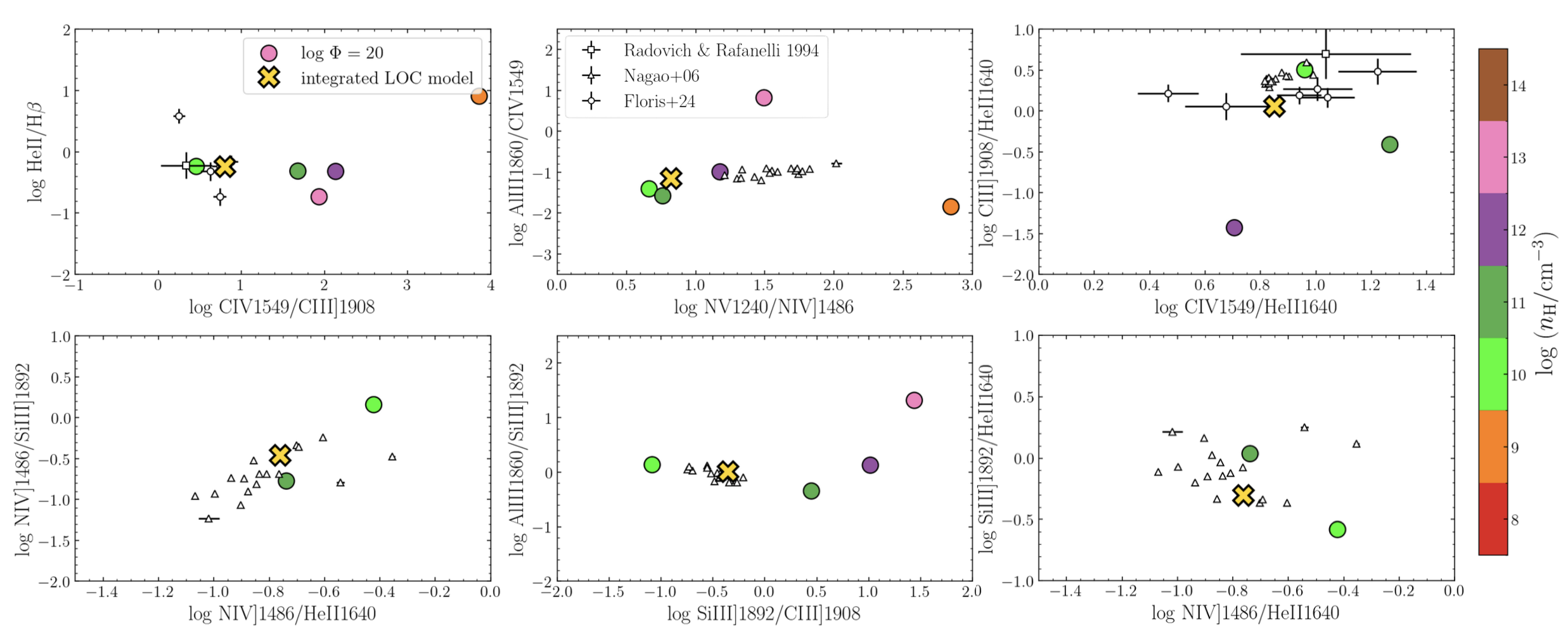}
\caption{Comparison of BLR models computed using single clouds and the integrated multi-cloud approach. The figure shows six emission-line ratio diagrams. The single-cloud models with $\log (\Phi/\rm{photons}\ \rm{s}^{-1}\ \rm{cm}^{-2}) = 20$ (which reproduce the observed radius - luminosity relation, see text for details) are highlighted with black outlines and are colour-coded according to gas density, as indicated by the colour bar. The integrated multi-cloud model prediction is shown as a yellow cross. All models are shown for $\log \Mbh/\Msun = 9$, $\log \Fedd = -0.5$, Z=0.030, and $\vturb = 250\ \mathrm{km\ s^{-1}}$. Observational data for broad-line AGN from \citet{Radovich1994,Nagao2006,Floris2024} are shown as black squares.}
\label{fig:BLRsinglemulti}
\end{figure*}

The adoption of a radial distribution of BLR clouds is motivated by the stratified structure of the BLR inferred from reverberation-mapping studies. In addition to this physical motivation, the multi-cloud (or LOC) model was originally introduced because it provides a better fit to quasar emission-line spectra than single-cloud photoionisation models. Emission lines are sensitive both to the ionisation parameter, which determines the ionisation state of the gas, and to the local gas conditions, such as temperature and density. As a result, each emission line is most efficiently produced for a specific combination of gas density and distance from the accretion disc. A distribution of cloud densities and radii therefore naturally accounts for the range of emission-line ratios observed in broad-line AGN, whereas single-cloud models assuming a single ionisation parameter and gas density generally fail to reproduce all lines simultaneously \citep[e.g.,][]{Baldwin1995,Baldwin1997,Korista1997}. In this section, we illustrate this point by comparing the predictions of the integrated BLR models with those obtained from single-cloud calculations.

To understand the differences between the single- and multi-cloud predictions, we first discuss how the physical conditions of individual clouds shape their emission-line spectra.
The ionisation parameter is proportional to $\Phi/\nh$, such that, at fixed density, models with higher $\Phi$ have higher ionisation parameters. This leads to increased ratios of high- to low-ionisation potential lines \citep[e.g.,][]{Baldwin1995}, such as \lciv/\lciii, \lheii/\hb, \lciv/\laliii\ and \lnv/\lniv. In addition, the column density required to reach the hydrogen-ionisation front scales with the ionisation parameter. Since we stop the models at a fixed column density of $\Nh = 10^{23}\ \mathrm{cm^{-2}}$, or when the electron fraction falls below 0.01, models with very large ionisation parameters can be density-bounded. This could further enhance the ratios of high- to low-ionisation potential lines \citep[e.g.,][but see also Sec.~\ref{sec:discumodeluncertainties}]{Korista1997}.
Higher ionisation fluxes can also lead to larger line optical depths. In particular, this can affect lines such as \ha, which can become optically thick at high ionisation parameter and density \citep[e.g.,][]{Korista1997}.

The gas density also strongly affects the emission-line spectrum, both through its connection to the ionisation parameter, column density and through collisional processes. In the high-density environment of the BLR, collisionally excited lines are influenced by both collisional excitation and collisional de-excitation, depending on their critical densities and the local gas density. For example, \citet{Marziani2015} suggested that \lsiliii/\lciii\ and \laliii/\lsiliii\ can be used as BLR density diagnostics because of the different critical densities of these transitions. 
Finally, in extreme cases, when a line becomes optically thick in combination with high densities, the line emission can approach thermalization, especially for \ha, \lciv\ and \lheii\ \citep[e.g.,][]{Korista1997,Hamann2002,Popovic2003,Ilic2012}. 

Having described how variations in the physical conditions of individual clouds can affect the predicted line ratios, we now assess whether integrating over this range of conditions can reproduce the observed BLR emission. To this end, we compile a literature sample of type~I AGN, including composite spectra of quasars and Seyfert~1 galaxies from Netzer (1990; see the compilation in \citealt{Radovich1994}), 
the composite spectra of quasars at $2 \leq z \leq 4.5$ from \citet{Nagao2006}, and local quasars from \citet{Floris2024}.
We find that multi-cloud models composed of clouds with hydrogen column densities $\log (\Nhstop/\mathrm{cm}^{-2}) \geq 23$ provide in general a better match to the observations than models with lower column densities. We also compare models without dissipative heating with those including dissipative heating following the prescription of \citet{Bottorff2002}, assuming a dissipation length scale equal to the BLR cloud size. We find that models without dissipative heating provide a overall better match to the observations (see also Section~\ref{sec:discussion}).

In Fig.~\ref{fig:BLRsinglemulti}, we compare these observational data (black symbols) with the predictions of a multi-cloud BLR model for a BH mass of $10^{9}\ \Msun$, an Eddington ratio of $\log (\Fedd) = -0.5$, a metallicity of 0.030, and $\vturb = 250\ \mathrm{km\ s^{-1}}$ (gold cross). 
The figure includes 6 emission-line ratio diagnostic diagrams: \heii/\hb\ versus \lciv/\lciii, \laliii/\lciv\ versus \lnv/\lniv, \lciii/\heii\ versus \lciv/\heii, \lniv/\lsiliii\ versus \lniv/\heii, \laliii/\lsiliii\ versus \lsiliii/\lciii, and \lsiliii/\heii\ versus \lniv/\heii.
In addition, we highlight the single-cloud models with $\log (\Phi/ \rm{photons}\ \rm{s}^{-1}\ \rm{cm}^{-2} )= 20$, which correspond to clouds following the observed \hb-derived radius–luminosity relation (see Section~\ref{sec:method_BLR}). These models are represented by filled circles, colour-coded according to their gas density. 

Overall, models with low $\Phi$ values (not shown in Fig~\ref{fig:BLRsinglemulti}) tend to lie closer to the observed line ratios than models with the highest $\Phi$ values, with the exception of \lnv/\lniv. Focusing on models with $\log (\Phi/\rm{photons}\ \rm{s}^{-1}\ \rm{cm}^{-2}) = 20$, we find that \lciv/\lciii, \lciv/\heii, \lniv/\lciii\ and \lciii/\heii\ are well reproduced by clouds with $\log (\nh/\rm{cm}^{-3}) = 10$  (light green circles). 
However, this density overestimates \lniv/\lsiliii\ and \lniv/\heii\ and underestimates \lsiliii/\lciii, \lsiliii/\heii\ and \lnv/\lniv\ which are better reproduced by models with densities between $\log (\nh/\rm{cm}^{-3}) = 11$ and 12 (dark green and purple circles).
In each of these diagrams, individual cloud models can reproduce the observed line ratios. However, the same single-cloud model does not simultaneously reproduce all line ratios, and moreover, the model that best matches a given line ratio does not necessarily satisfy the observed radius–luminosity relation.
On the other hand, the multi-cloud model provides a more consistent overall description of the different emission-line ratios.

We caution that this comparison has several limitations. The line ratios shown in these diagrams are drawn from different observational samples, while the radius--luminosity relation is calibrated using independent datasets.
In addition, the integrated multi-cloud model does not perfectly reproduce all of the observed line ratios. 
This is the case, for example, for the \lnv/\lniv\ ratio, which is not well reproduced by the multi-cloud models shown in this figure.

This is consistent with previous studies \citep[e.g.,][]{Hamann1993,Hamann1996,Bottorff2000,Shemmer2002,Nagao2006}, which have found that some quasars exhibit unusually strong \lnv\ emission that is difficult to reproduce with standard photoionisation models. Several explanations have been proposed, including high metallicity and/or enhanced nitrogen abundance (although this does not necessarily enhance \lnv\ relative to \lniv), variations in the ionisation and physical conditions of the BLR gas, microturbulence and associated dissipative heating (but see Section~\ref{sec:discumodeluncertainties}), and continuum pumping. In addition, because \lnv\ is a resonance line, its emission may be enhanced by resonant scattering of Ly$\alpha$ and continuum photons in outflowing gas. In the following, we therefore focus on other line ratios that do not involve \lnv. A comprehensive exploration of the parameter space, including the distribution functions of cloud radius and density, is beyond the scope of this work.
Similar caution is required for ratios involving \lciv, which is also a resonance line and whose emission may be affected by resonant scattering in the presence of outflowing gas \citep[see also][]{Temple2023}. Despite these limitations, the multi-cloud model provides a more realistic overall representation of the observations than a single-cloud model.

\subsubsection{Comparison with observations at low redshift}\label{sec:validation_BLR_z0}

\begin{figure*}
\includegraphics[width=1\linewidth]{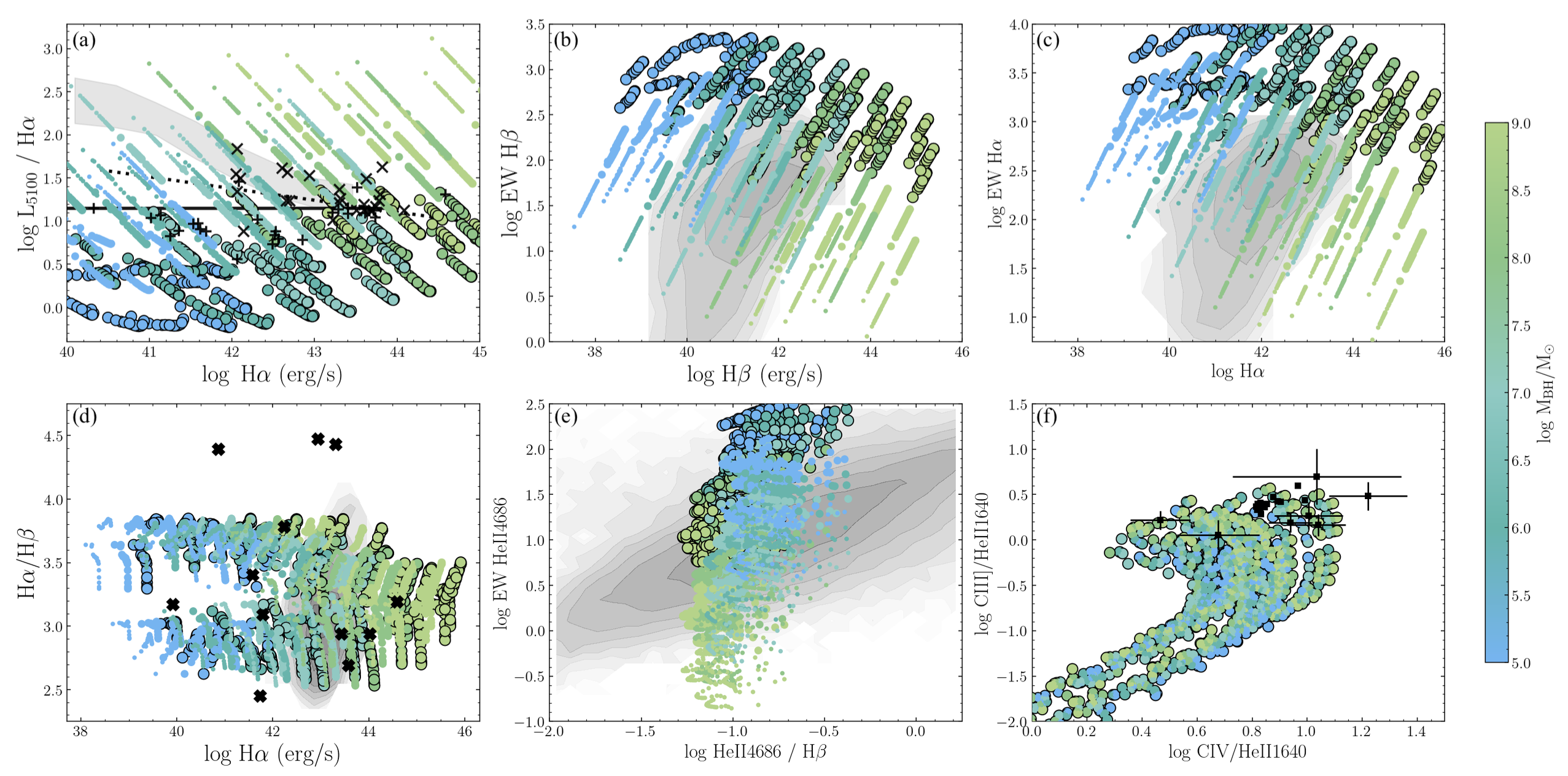}
\caption{Comparison of the BLR models with observations of local broad-line AGN. The parameters of the BLR models are fixed to their fiducial values (see Sec.~\ref{sec:method_BLR_fid}). In all panels, BLR model predictions are shown as circles, colour-coded by black-hole mass from $\log(\Mbh/\Msun)=5$ (blue) to $\log(\Mbh/\Msun)=9$ (green). Big circles with black outlines correspond to a BLR covering factor of 1, while smaller circles without outlines correspond to a covering factor of 0.4 and 0.1. (a) Continuum luminosity at 5100\AA/\ha\ versus \ha\ luminosity. The grey region represents the quasar distribution from \citet{Stern2012}; the solid line shows the relation derived by \citet{Stern2012}, and the dotted line shows the relation from \citet{Greene2005}. Black crosses ($\times$) correspond to individual quasars from the Bright Quasar Survey \citep[][]{Sanders1989}, using measurements reported in \citet{Stern2012}. The plus symbols ($+$) show individual sources from \citet{DallaBonta2025}.
(b) and (c) \hb\ equivalent width (\ha\ equivalent width) as a function of \hb\ (\ha) luminosity. Grey contours indicate the observational distributions from \citet{Liu2019}.
(d) \ha/\hb\ ratio as a function of \ha\ luminosity. Grey contours show observational data from \citet{Dong2008}, while black crosses indicate individual sources from \citet{Popovic2003}.
(e) \heiiopt-EW as a function of \heiiopt/\hb. The Grey contours show the observational data from \citet{Wu2022}. 
(f) \ciii/\heii\ ratio as a function of \civ/\heii. Observational data for broad-line AGN from \citet{Radovich1994,Nagao2006,Floris2024} are shown as black squares.}
\label{fig:valBLRz0}
\end{figure*}

In this subsection, we primarily focus on the ability of the fiducial multi-cloud models to reproduce observations of low-redshift AGN.  The fiducial model grid corresponds to the multi-cloud LOC integration of BLR clouds distributed as $1/\nh$ and $1/R$, restricted to $\log \Phi \geq 18$ photons cm$^{-2}$ s$^{-1}$, and adopting a total hydrogen column density of $\Nh = 10^{23}\ \mathrm{cm^{-2}}$ (Sec.~\ref{sec:method_BLR_fid}). Fig.~\ref{fig:valBLRz0} compares the predicted multi-cloud emission-line properties with observations of low-redshift AGN \citep[][]{Sanders1989,Radovich1994,Popovic2003,Nagao2006,Dong2008,Stern2012,Liu2019,Wu2022,DallaBonta2025,Floris2024}. In all panels, the multi-cloud models are colour-coded by BH mass. Larger circles with black outlines correspond to a BLR covering factor of unity, while smaller circles without outline indicate a covering factor of 0.4 and 0.1.  For comparison, we also show the predictions of the single-cloud, LRD-like BLR models (Sec.~\ref{sec:method_BLR_bb}), assuming covering factors of unity (dark brown shaded region) and 0.1 (light brown shaded region).

In subplot (a), we show the L$_{5100}$/\ha\ ratio as a function of \ha\ luminosity. The models are compared with observations from \citet{Stern2012} (grey shaded region). Their sample consists of 3579 Type~I AGN from the SDSS Seventh Data Release \citep[][]{Schneider2010} with detected broad \ha\ emission at $z < 0.31$. The models reproduce well the observed relation at the high-\ha-luminosity end for a BLR covering factor of $\approx 0.4$, but tend to deviate at low \ha\ luminosities (or would require much lower covering factors). This behaviour might be interpreted as an increasing contribution from the host galaxy at low \ha\ luminosities, where the continuum L$_{5100}$ emission is no longer dominated by the AGN continuum \citep[see][]{Stern2012}.
The $\times$ symbols show the position of 26 individual quasars from \citet{Stern2012} that are also part of the Bright Quasar Survey \citep[][]{Sanders1989}. These objects were selected to be point sources with blue optical colours, and therefore have a reduced host-galaxy contribution to the continuum \citep[][]{Stern2012}. The $+$ symbols show the sample of \citet{DallaBonta2025}, which consists of reverberation-mapped AGN at $z < 0.2$, for which the stellar host contribution has been removed from the continuum. Both sets of objects are well reproduced by the fiducial grid of models across the \ha\ luminosity range.

Finally, we also show commonly used empirical relations to infer the optical luminosity of AGN from \ha\ emission. These include the relation derived by \citet{Stern2012} using the sample described above, and the relation derived by \citet{Greene2005} from a sample of SDSS AGN selected to have weak stellar contributions to the optical continuum. While these relations are useful for estimating continuum luminosities from \ha\ measurements, they do not capture the large scatter present in the observations. The models naturally reproduce this large scatter, spanning approximately 1 dex in L$_{5100}$/\ha\ at fixed covering factor. In particular, the models predict a slight increase of L$_{5100}$/\ha\ with BH mass. This trend arises because lower-mass BHs produce a higher ratio of ionising photons to optical continuum luminosity, i.e. a larger $Q_{\rm H}$/L$_{5100}$. At fixed Eddington ratio, accretion-disc models predict higher disc temperatures for lower-mass BHs, approximately following $T \propto \left[L/(L_{\rm Edd}\times \Mbh)\right]^{1/4}$ \citep[e.g.,][]{Shakura1973,Novikov1973}. This shifts the disc emission peak to higher energies and increases the number of hydrogen-ionising photons emitted per unit optical continuum luminosity, leading to stronger \ha\ emission at fixed optical continuum luminosity. This result highlights the need for caution when applying bolometric corrections at high redshift, where AGN are often powered by lower-mass BHs \citep[but see also][]{Davis2007}.
In Section \ref{sec:TypeILAGNcal}, we explore in more detail uncertainties in bolometric corrections and their dependence on various physical parameters.

Subplots (b) and (c) in Fig.~\ref{fig:valBLRz0} show the \hb\ and \ha\ equivalent widths (EWs) as a function of \hb\ and \ha\ luminosity, respectively. The models predict a trend resembling the Baldwin effect \citep[][]{Baldwin1977}, with decreasing Balmer-line EWs associated with increasing BH mass. This reflects the lower ratio of ionising photons to optical continuum luminosity in higher-mass BHs, i.e. a smaller $Q_{\rm H}$/L$_{\rm cont}$ ratio, which results in weaker line emission relative to the optical continuum.
The grey contours show the distribution of the \citet{Liu2019} sample of local broad-line AGN. We note that their EWs are computed by subtracting a flat continuum, rather than a stellar continuum, which may affect the measured values. Overall, we find that the bulk of the observed distribution for both EWs is well reproduced by models with a BLR covering factor of 0.4. The lowest EW values can be reproduced either by models with lower covering factors or by a significant stellar contribution to the observed continuum, which could reduce the measured EWs.

Subplot (d) in Fig.~\ref{fig:valBLRz0} shows the \ha/\hb\ ratio as a function of $\log \ha$ luminosity. The fiducial models predict Balmer decrements spanning $\ha/\hb = 2.53$ to $3.80$, with a median value of 3.03. 
The scatter in the predicted Balmer decrement arises from the high densities and large line optical depths characteristic of BLR gas, where collisional processes, partial thermalization of the excited hydrogen levels, and radiative-transfer effects modify the relative Balmer-line emissivities \citep[e.g.][]{Korista1997,Korista2004}.
The grey contours show observational samples from \citet{Dong2008} and \citet{Liu2019}, while the black crosses correspond to individual quasars from \citet{Popovic2003}. The \citet{Dong2008} sample consists of 446 Seyfert~1 galaxies and quasi-stellar objects from the SDSS Fourth Data Release at $z \leq 0.35$, selected to have blue optical continua in order to minimize the impact of dust reddening. The observed \ha/\hb\ ratios span $\sim$2.3 to 4.2, with a mean value of 3.1. Overall, our models reproduce the range covered by these observations quite well. As noted by \citet{Dong2008}, the relatively narrow distribution of \ha/\hb\ is naturally reproduced by multi-cloud models, since integrating over a wide range of cloud densities and ionisation parameters produces an effective average Balmer decrement.
The sample of \citet{Liu2019} (not shown) consists of 14,584 broad-line AGN from the SDSS Seventh Data Release at $z < 0.35$. The \ha/\hb\ ratio exhibits a significantly larger scatter than in \citet{Dong2008}, primarily because the observed ratios are affected by dust attenuation. The median \ha/\hb\ value is $\sim 4$.
The \citet{Popovic2003} sample consists of 14 AGN at $z < 0.2$. The Balmer decrement ranges from 2.45 to 6.91, with a median value of 3.295. The observed scatter is interpreted by \citet{Popovic2003} as arising from partial local thermodynamic equilibrium effects and/or dust reddening. 
Additional uncertainty in the measured \ha/\hb\ ratio may arise from the line-fitting procedure, particularly from the blending of \ha\ with the neighboring \lniiopt\ lines and the resulting difficulty in decomposing the broad and narrow emission-line components.
Overall, the fiducial models reproduce the bulk of the observed distributions, if we account for the potential effect of dust attenuation. High \ha/\hb\ may also be present without dust reddening, and we further explore in Section~\ref{sec:discussion} the physical conditions that lead to large \ha/\hb\ ratios.

Subplot (e) in Fig.~\ref{fig:valBLRz0}, illustrates \heiiopt\ EW as a function of \heiiopt/\hb. The grey contours represent the observational sample of \citet{Wu2022}, consisting of 750,414 broad-line quasars from SDSS DR16Q at redshifts $0.1 \leq z \leq 6$. The fiducial models reproduce the bulk of the observed distribution, although the data extend to a wider range of \heiiopt/\hb\ ratios. Adopting different distribution functions for the BLR cloud density and radius would allow the models to cover a larger region of the diagram. Here, however, we restrict the analysis to the fiducial cloud distribution, which already provides a good match to the main locus of the observations (see also Section~\ref{sec:discuhighz}).

Finally, in subplot (f), we show \lciii/\heii\ as a function of \lciv/\heii. The models reproduce well the observed ratios of the \citet{Nagao2006} sample of 21 composite quasar spectra. However, the quasar sample of \citet{Floris2024} and the composite spectrum of \citet{Radovich1994} extend to larger \lciv/\heii\ values than predicted by the fiducial models. As in the case of \heiiopt/\hb, exploring alternative distributions of BLR cloud densities and radii would likely broaden the region spanned by the models in this diagram. In addition, since \lciv\ is a resonant line, its observed flux may be affected by resonant scattering in outflowing material, potentially complicating direct comparisons with BLR photoionisation models \citep[e.g.,][]{Temple2023}.

\subsection{NLR validation}\label{sec:validation_NLR}

\begin{figure*}
\includegraphics[width=1\linewidth]{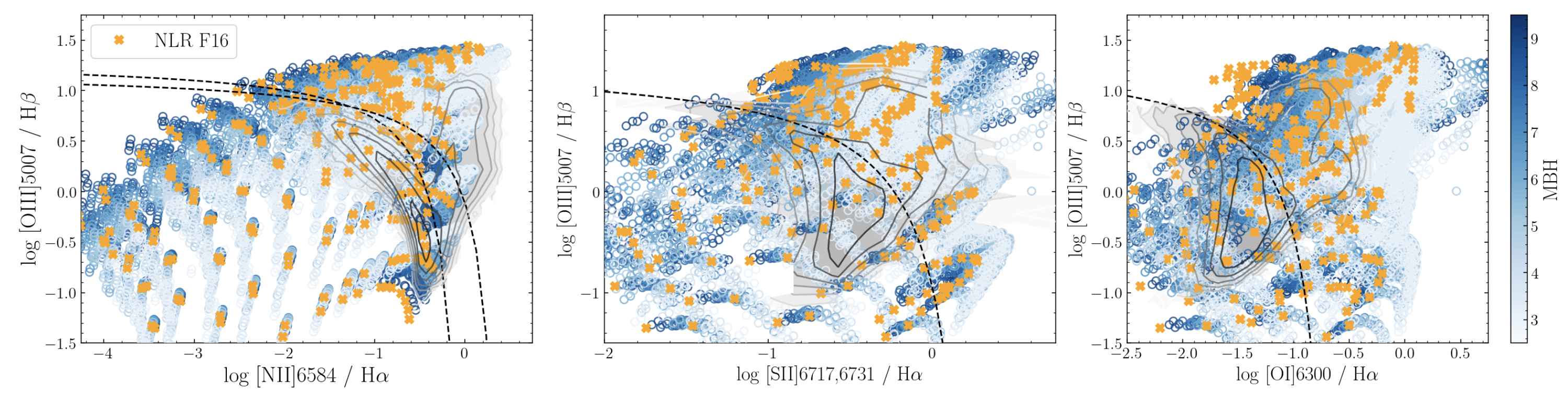}
\caption{Validation of the new NLR models. Line-ratio diagnostic diagrams comparing the new grid of NLR photoionisation models from this work (blue circles) with SDSS galaxies (grey contours) and the NLR models of \citet{Feltre2016} (orange crosses). The models are colour-coded by black-hole mass, from dark to light shades corresponding to high to low \Mbh. The dashed lines show the demarcation criteria from \citet{Kewley2001} and \citet{Kauffmann2003}.}
\label{fig:NLRBPT}
\end{figure*}

To further validate the new NLR models, we compare their predictions with observations in the three classical optical line-ratio diagrams \citep[][]{Baldwin81,Veilleux1987}: $\loiiiopt/\hb$ versus $\lniiopt/\ha$, $\lsiiopt/\ha$, and $\loiopt/\ha$, shown in the left, middle, and right panels of Fig.~\ref{fig:NLRBPT}, respectively. The figure displays the grid of NLR photoionisation models from this work (blue circles, colour-coded by $\Mbh$), computed using a fiducial set of parameters: a nitrogen-to-oxygen abundance ratio following the relation of \citet{Gutkin2016}, a dust-to-metal ratio of $\xid = 0.3$, a turbulent velocity of $100~\mathrm{km~s^{-1}}$, and a gas density of $\log (\nh/\mathrm{cm}^{3}) = 3$. These model predictions successfully reproduce the location of SDSS DR7 \citep[][]{Abazajian2009} AGN and star-forming galaxies (grey contours) in the optical line-ratio diagrams.

For comparison, we also show the grid of NLR photoionisation models from \citet[][orange crosses]{Feltre2016}, updated to include microturbulence and dissipative heating \citep[as described in][]{Mignoli2019}. These models adopt the same gas parameters as our fiducial grid (N/O following \citealt{Gutkin2016}, \xid = 0.3, turbulent velocity of 100~km~s$^{-1}$, and $\log (\nh/\mathrm{cm}^{3}) = 3$), but, in contrast to our models, they assume power-law ionising spectra with indices of $-2$ and $-1.2$. Our new grid spans a region of parameter space similar to that covered by the models of \citet{Feltre2016}, but extends to slightly higher values of \lniiopt/\ha, \lsiiopt/\ha, and \loiopt/\ha. This extension arises from the harder ionising radiation produced for certain combinations of BH mass and Eddington ratio, particularly at low \Mbh. Yet, the absence of strong systematic differences between the two model grids suggests that the primary parameters governing the position of photoionisation models in these optical line-ratio diagrams are ionisation parameter and metallicity, rather than the detailed shape of the adopted ionising SED.

\section{Detectability and diagnostics of high-redshift type~I AGN}\label{sec:TypeI}

In this section, we employ our BLR model grid to investigate the properties of {\it high-redshift} type~I AGN. We first assess the detectability of broad emission lines at high redshift with JWST, focusing on the dependence on BH mass and Eddington ratio (Sec.~\ref{sec:TypeIfluxlimit}). We then explore theoretical bolometric corrections for broad emission-line luminosities and examine their sensitivity to key physical parameters (Sec.~\ref{sec:TypeILAGNcal}). Finally, we compare our model predictions with recent high-redshift observations to interpret the spectral properties of type~I AGN and LRDs revealed by JWST.

\subsection{Detectability of broad emission lines of type~I AGN with JWST/NIRSpec}\label{sec:TypeIfluxlimit}

\begin{figure*}
\begin{centering}
\includegraphics[width=1\linewidth]{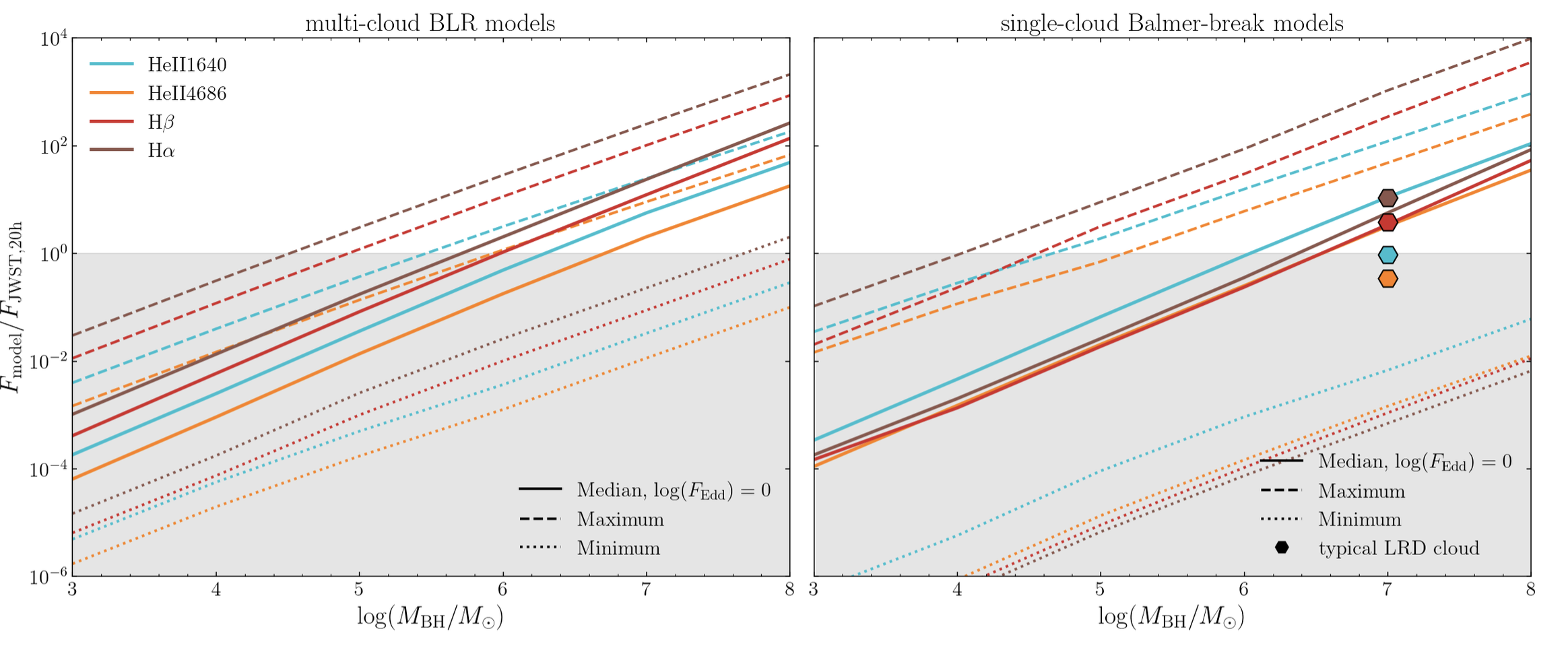}
\par\end{centering}
\caption{Detectability of broad emission lines with JWST at $z = 6$. The plot shows the ratio of modelled BLR line flux to the JWST flux limit required to achieve S/N = 5 using the high resolution gratings with an integration time of 20 h, for a subset of emission lines, as a function of black-hole mass over the range $\log (\Mbh/\Msun) = 3$–8. The models assume an BLR covering factor of 0.4. Solid lines show the median line flux for models with $\log (\Fedd) = 0$. Dashed lines show the maximum line flux across all values of \Fedd\ and metallicity $Z$. Dotted lines show the minimum line flux across the same parameter space. Each colour correspond to a different line, as indicated in the legend. The grey shaded region highlights fluxes below the detection limit. The left panel shows the fiducial BLR models, while the right panel shows single-cloud Balmer-break models. In this panel, the hexagons highlight the typical LRD cloud model defined in Section~\ref{sec:method_BLR_bb}.}
\label{fig:BLRdetecJWST_sum}
\end{figure*}

\begin{figure*}
\begin{centering}
\includegraphics[width=1\linewidth]{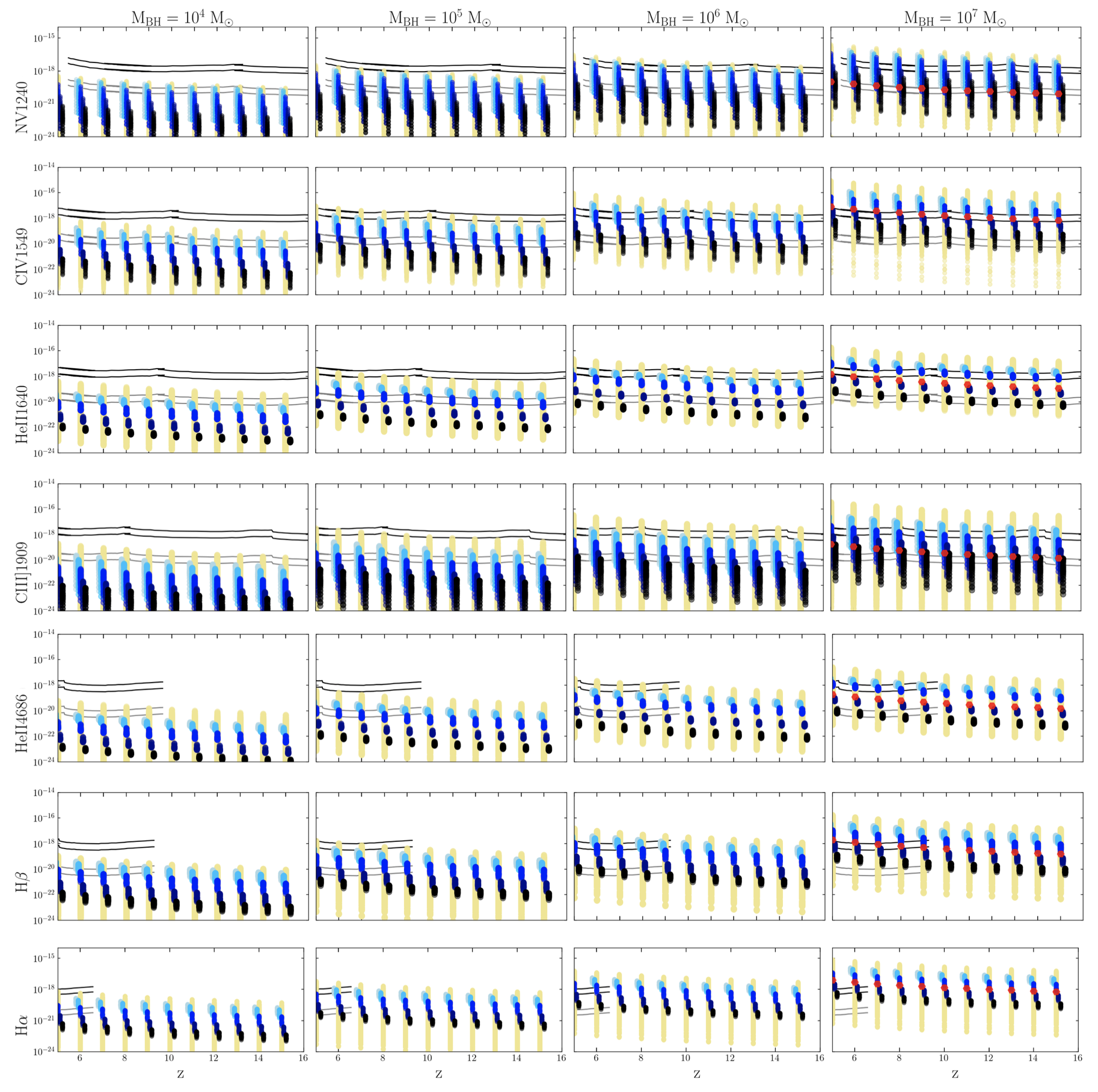}
\includegraphics[width=0.8\linewidth]{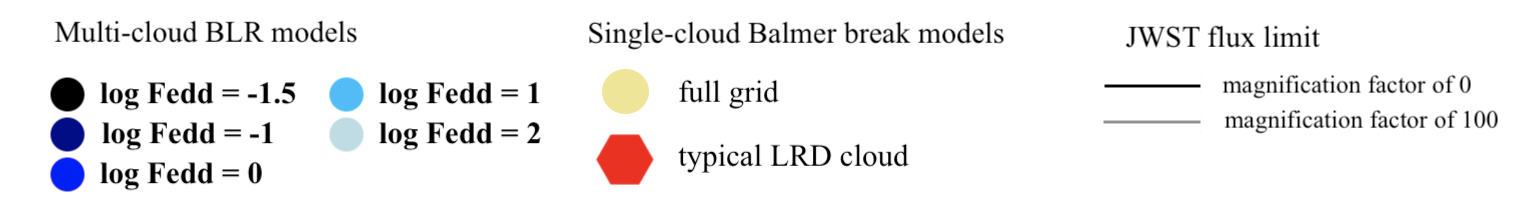}
\par\end{centering}
\caption{Detectability of broad emission lines with JWST. The panels show the predicted line fluxes as a function of redshift, assuming a BLR covering factor of \fcovBLR = 0.4. The models are colour-coded by Eddington ratio (see legend), and each column corresponds to a different black-hole mass, spanning $\log (\Mbh/\Msun) = 4$ to 7. single-cloud Balmer-break models are shown in pale yellow, for all Eddington ratios from $\log \Fedd = -1.5$ to 2, and the typical LRD model (see Sec.~\ref{sec:method_BLR_bb}) is highlighted with a red hexagon in the $\log (\Mbh/\Msun) = 7$ panel. Each row corresponds to a different emission line. The predicted fluxes are compared with the JWST/NIRSpec high resolution gratings detection limits required to achieve S/N = 5, assuming a line width fixed to 1500 km s$^{-1}$, for integration times ranging from 2 to 20 hours (black lines). We also show how the detection limit changes assuming lensing magnification factors of 100 (grey lines).}
\label{fig:BLRJWSTdec}
\end{figure*}

The primary goal of this section is to determine under which conditions, namely, BH masses and Eddington ratios, broad lines of type~I AGN can be detected at high redshift with JWST/NIRSpec. To this end, we assess the detectability of broad emission lines from massive BHs by comparing model predictions with JWST/NIRSpec sensitivity limits for a range of exposure times. 

We compute the JWST/NIRSpec detection limits using \texttt{Pandeia} \citep{Pontoppidan2016} for the high-resolution grating/filter combinations (G140H/F070LP, G235H/F170LP, G395H/F290LP), evaluating total integration times between 2 and 20~h. We assume an unresolved point source and model each broad emission line as a Gaussian with a fixed rest-frame velocity width of $1500~\mathrm{km\ s^{-1}}$, with no underlying continuum. The simulations adopt the NIRSpec MOS configuration with 19 groups per integration, two integrations per exposure, and the NRSIRS2 readout pattern; the number of exposures is varied to sample the total exposure time, with the resulting signal-to-noise ratio interpolated between the pre-computed exposure-time grid points. We define the detection limit as the line flux corresponding to an integrated line signal-to-noise ratio of $\mathrm{S/N}=5$. These limits should therefore be regarded as a best-case estimate of the sensitivity to an isolated broad emission line. In particular, they do not account for the additional uncertainty associated with decomposing a broad component from superposed narrow-line emission, nor for the loss of sensitivity introduced by fitting the line profile. Tests in which a Gaussian line is recovered through explicit profile fitting (e.g. using MCMC) indicate that the limiting line flux can be higher by a factor of $\sim2.5$, depending on the adopted fitting procedure. The limits shown here should consequently be interpreted as optimistic instrumental sensitivity estimates rather than as the detection threshold expected from a full broad- and narrow-line decomposition.

In Figure~\ref{fig:BLRdetecJWST_sum}, we show the ratio of the modelled emission-line fluxes to the \JWST\ detection limits for an integration time of 20~h at $z=6$, as a function of BH mass, from $\log (\Mbh/\Msun)=3$ to 8. We consider a subset of emission lines: \heii, \heiiopt, \hb, and \ha. The left panel shows the median ratios predicted by the fiducial multi-cloud models at $\log \Fedd = 0$, together with the minimum and maximum values across the full model grid. The right panel shows the corresponding predictions for the single-cloud Balmer-break models and the typical-LRD model. All models are scaled assuming a covering factor of 0.4.
We first focus on the fiducial multi-cloud BLR models. Assuming $\log \Fedd = 0$, the brightest lines, \ha\ and \hb, reach S/N $>5$ for BH masses of $\log (\Mbh/\Msun) \geq 5.7$, while the \lheii\ lines require BH masses above $\log (\Mbh/\Msun) \geq6.2$.
For highly super-Eddington accretion ($\log \Fedd = 2$), \ha\ becomes detectable down to $\log (\Mbh/\Msun) \geq 4.5$. Conversely, at the low Eddington ratio of $\log \Fedd = -1.5$, the emission lines become detectable only for BH masses of $\log (\Mbh/\Msun) \approx 8$.
The behaviour is similar for the single-cloud Balmer-break models, although the predicted line fluxes exhibit a larger scatter. In the super-Eddington case, \ha\ reaches S/N $=5$ at $\log (\Mbh/\Msun) \approx 4$, while for $\log \Fedd = 0$, this threshold is reached at $\log (\Mbh/\Msun) \approx 6.2$. For the typical LRD cloud with $\log (\Mbh/\Msun) = 7$, both \ha\ and \hb\ lie above the S/N $=5$ threshold, whereas \heii\ and \heiiopt\ remain below it, consistent with the non-detection of \heii\ in many LRD spectra.

More detailed predictions are shown in Figure~\ref{fig:BLRJWSTdec}, which presents the broad-line fluxes predicted by the fiducial multi-cloud BLR models (blue circles, colour-coded by Eddington ratio) as a function of redshift for a wider set of emission lines. The different panels correspond to BH masses spanning $\log (\Mbh/\Msun) = 4$--$7$ from left to right. 
The predicted line fluxes span several orders of magnitude, such that the difference in sensitivity between exposure times of 2 and 20~h appears relatively small in comparison.

The fiducial models assume a BLR covering factor of $\fcovBLR = 0.4$. Increasing the covering factor enhances the line fluxes, but only moderately: adopting $\fcovBLR = 1$ increases fluxes by $\sim 0.4$ dex. Consequently, models in which \ha\ is detectable only at the Eddington limit could instead be detected at slightly lower accretion rates, around $\log \Fedd \sim -0.6$, reflecting the approximate scaling of \ha\ luminosity with Eddington ratio.

We first discuss the predictions for the multi-cloud BLR models. Fig.~\ref{fig:BLRJWSTdec} illustrates that most emission-line fluxes fall below the detection limits for BHs with masses below $\log (\Mbh/\Msun) = 5$, with the exception of \ha\ (and to some extend \hb) in the case of super-Eddington accretion. For $\log (\Mbh/\Msun) = 6$, several additional lines become detectable, including \lnv, \lciv, \heii, \lciii, and \hb, but primarily for super-Eddington accretion.
In contrast, \ha\ remains detectable down to $\log \Fedd = 0$ and above. 

A promising way to probe lower-mass BHs is through gravitational lensing \citep[e.g.,][]{Bogdan2024}. We illustrate this by showing detection limits assuming magnification factors of 100 (as can be achieved, for example, in the GLIMPSE \JWST\ programs \citealt{Atek2025}). With a magnification factor of 10, most emission lines become detectable for $\log \Fedd \geq 0$ at $\log (\Mbh/\Msun) = 6$. For a magnification factor of 100, several lines from sub-Eddington accreting BHs also become detectable at the same mass. In addition, \ha\ could be detected down to $\log (\Mbh/\Msun) = 3$ in the case of super-Eddington accretion. At $\log (\Mbh/\Msun) = 4$, \lnv, \lciv, \hb, and \ha\ all lie above the detection limit for super-Eddington accretion, while at $\log (\Mbh/\Msun) = 5$, \ha\ remains detectable even in the sub-Eddington regime.
While H and He lines primarily depend on the Eddington ratio and BH mass, metal lines are also strongly sensitive to metallicity and therefore exhibit a larger scatter in flux (the models are shown for the full range of metallicity of the grid, see Section~\ref{sec:method_BLR}). In particular, at very low metallicities, metal lines are expected to be significantly more difficult to detect.

We emphasize that these limits correspond to a signal-to-noise ratio of S/N = 5 for line detection, and do not account for the additional difficulty of decomposing broad and narrow component,
 which would likely make detections more challenging. We also do not include the effects of dust reddening, which would further reduce the observed fluxes.

We further show in Fig.~\ref{fig:BLRJWSTdec} the distribution of single-cloud, Balmer-break models (yellow filled dots). These models exhibit a much larger scatter in emission-line fluxes compared to the fiducial multi-cloud BLR models, reflecting the wide range of densities and ionising-photon fluxes they sample. For a given BH mass and accretion rate, they extend to both higher and lower fluxes.
The highest fluxes, which enhance line detectability, arise from two main effects. First, the inclusion of models with larger gas column densities ($\log (\Nh/\mathrm{cm^{-2}})= 25$) allows the line-emitting regions to extend further compared to models with $\log (\Nh/\mathrm{cm^{-2}})= 23$, particularly for lines that are produced throughout the cloud, such as \ha. Second, higher line fluxes can be obtained for specific combinations of gas density and ionising-photon flux that maximize the reprocessing of ionising photons. For example, models with lower ionising-photon flux $\Phi$ can produce stronger \ha\ emission, when normalised to the same total rate of ionising photons, compared to models with higher $\Phi$, which often have larger optical depths.
The red hexagons in Fig.~\ref{fig:BLRJWSTdec} highlight the fluxes of the typical LRD-like model, computed for a BH mass of $10^{7}\ \Msun$ and an Eddington ratio of $\log \Fedd = -1$. This model tends to produce stronger emission-line fluxes than the fiducial multi-cloud model with the same BH mass and Eddington ratio, particularly for \ha\ and \hb, bringing them above the detection threshold. This suggests that, for a given \Mbh\ and \Fedd, such physical conditions could lead to more easily detectable broad emission lines of lower-mass BHs at early cosmic times.

\subsection{Bolometric corrections for broad line luminosities of type~I AGN}\label{sec:TypeILAGNcal}

\begin{figure*}
\includegraphics[width=1\linewidth]{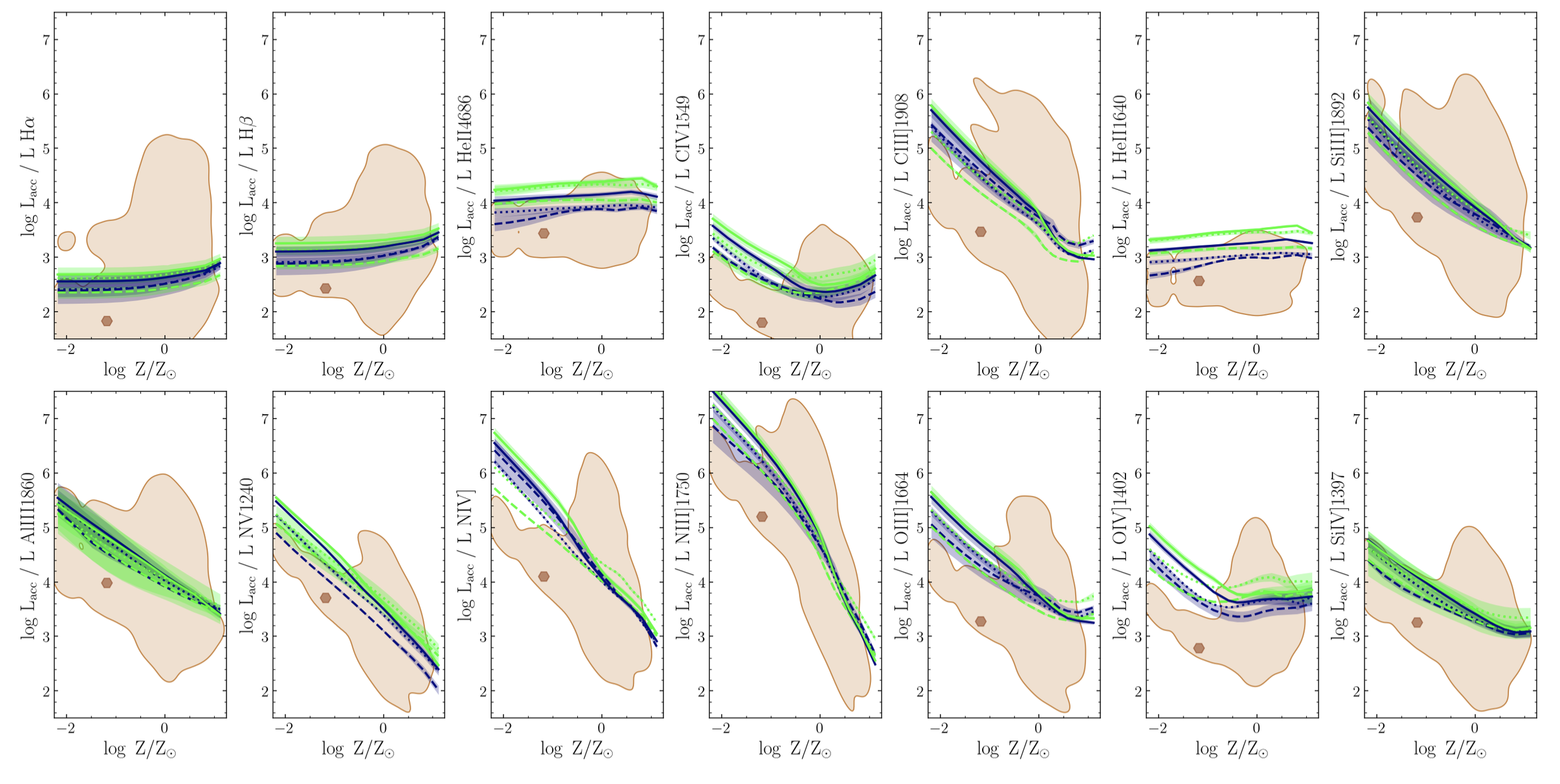}\\
\includegraphics[width=1.\linewidth]{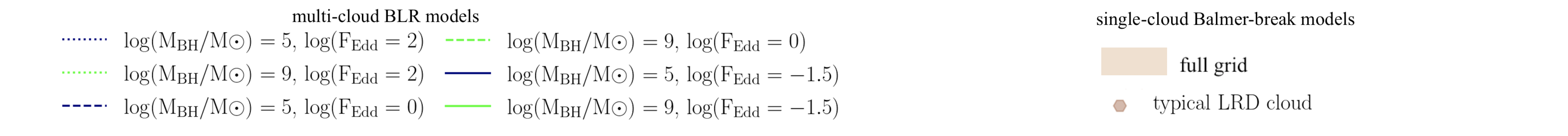}
\caption{Bolometric correction as a function of metallicity for various emission lines. Models are shown for a BLR covering factor of 0.4. Green curves correspond to a black-hole mass of $10^{9}~\Msun$, while blue curves correspond to $10^{5}~\Msun$. Solid, dashed, and dotted curves indicate $\log \Fedd = -1.5$, 0, and 2, respectively. The coloured shaded region around the lines show the dispersion between microturbulent velocity of 0 and 250~km/s. The brown region show the distribution of the single-cloud Balmer-break models, for all \Mbh\ and \Fedd, and the brown hexagon the position of the typical LRD-like model (see Sec.~\ref{sec:method_BLR_bb}).}
\label{fig:BLRLbollinesfid}
\end{figure*}

In this section, we further explore predicted bolometric corrections for broad emission lines of type~I AGN, to estimate the gas accretion rate (modulo the assumed radiative efficiency) onto the BH and thus, to understand how efficiently BHs grow at early cosmic epochs. Different types of bolometric corrections exist in literature, based either on the observed continuum luminosity or on emission-line luminosities, to infer the bolometric accretion luminosity. These corrections are often calibrated using observations of local type~I AGN \citep[e.g.,][]{Greene2005,Richards2006,Stern2012,Duras2020}. However, such calibrations may introduce significant uncertainties when applied to high-redshift sources, since high-redshift AGN likely have physical properties that differ from those of their low-redshift counterparts. Some efforts have been made to account for these differences; for example, \citet{Greene2026} derived new bolometric calibrations specifically for LRDs using observed LRD spectra.
In parallel, theoretical predictions have been developed for bolometric corrections based on continuum luminosities, i.e. $\Lbol/L_{\rm cont}$, to account for the impact of \Mbh\ and \Fedd\ \citep[e.g.,][]{Netzer2019,Temple2023,Azadi2025}, and to explore the effects of super-Eddington accretion \citep[e.g.,][]{Pacucci2024}. Theoretical predictions have also been proposed for bolometric corrections based on \ha\ emission, i.e. $\Lbol/L_{\rm H\alpha}$, where the \ha\ luminosity is estimated using analytical prescriptions \citep[e.g.,][]{Wilkins2025}.
In our study, we explore the ratio $\Lacc/L_{\rm line}$ for the most luminous broad emission lines, where \Lacc\ is the luminosity of the accretion disc estimated for an average angle of $45^{\circ}$.
We investigate the impact of the main free parameters of our models — \Mbh, \Fedd, and metallicity $Z$ — using the fiducial BLR multi-cloud grid. These predictions are particularly important for interpreting high-redshift sources, where typical \Mbh, \Fedd, and $Z$ are likely to differ from those in local AGN.

These bolometric corrections $\Lacc/L_{\rm line}$ are shown in Fig.~\ref{fig:BLRLbollinesfid} as a function of metallicity, for $\log(\Mbh/\Msun)=5$ and 9, and for $\log \Fedd = -1.5$, 0, and 2 (different lines). We assume a covering factor of the BLR of 0.4. As expected, metal lines show a strong dependence on metallicity, in contrast to H and He lines, which display a much smaller range of $\Lacc/L_{\rm line}$ ratios.
The strongest metallicity dependence is found for nitrogen lines, with a decrease in $\Lacc/L_{\rm line}$ with $Z$ of approximately 3, 3.5, and 5 dex for \lnv, \lniv, and \lniii, respectively. This strong dependence arises from the adopted nitrogen abundance prescription of \citet{Gutkin2016}, which includes secondary nitrogen enrichment \citep[see also][]{Hamann2002}. Other metal lines also strengthen with increasing metallicity (which decreases $\Lacc/L_{\rm line}$) due to the higher elemental abundances, but with a shallower trend, showing variations of $\sim$2.5 dex for \lciii, \lsiliii, and \laliii.
For \lciv\ and \loiv, the line strength increases with metallicity and then flattens, or even decreases, at the highest metallicities. Several effects likely contribute to this behaviour. First, the decrease in electron temperature with increasing metallicity reduces the emissivity of collisionally excited lines \citep[e.g.,][]{Hamann2002,Nagao2006}. This is combined with a reduction of the C$^{3+}$ zone at high metallicity, due to stronger absorption of the high-energy photons required to ionize C$^{3+}$ \citep[][]{Huang2023}. In addition, the temperature dependence of recombination rates favours C$^{2+}$ over C$^{3+}$ at high metallicity. Finally, \lciv\ may become optically thick at very high metallicities, further decreasing the emergent line flux. As a result, the variations in $\Lacc/L_{\rm line}$ with $Z$ are smaller for \lciv\ and \loiv, of approximately 1.5 dex.
For all of these lines, variations in the SED shape due to \Mbh\ and \Fedd\ have a much smaller impact on $\Lacc/L_{\rm line}$ than metallicity, producing a scatter of approximately 0.5 to 1 dex at fixed metallicity.
A consequence of the strong metallicity dependence of metal lines is that the bright lines commonly observed in local AGN may become more difficult to detect at high redshift, if BLRs at high redshift have systematically lower metallicities. 
Studies of luminous quasars find little evidence for an evolution in BLR metallicity out to $(z\sim6-7)$, with typically supersolar inferred metallicities \citep[e.g.,][]{Nagao2006,Wang2022}. However, these studies predominantly probe luminous quasars, and the metallicities and physical conditions of the BLRs in the much fainter AGN population at high redshift remain considerably less well constrained.

The dependence of H and He lines on metallicity is weak, and the $\Lacc/L_{\rm line}$ ratios show a total scatter of only $\sim$0.7 dex. At fixed metallicity, the scatter decreases to $\sim$0.4 dex for \ha\ and \hb, and $\sim$0.6 dex for \heii. The variations of $\Lacc/\heii$ primarily follow the variations of $Q_{\rm HeII}/\Lacc$. Decreasing the BH mass shifts the accretion-disc emission towards higher energies, and increasing the Eddington ratio has a similar effect ($T \propto \left[L/(L_{\rm Edd}~\Mbh)\right]^{1/4}$). At the highest accretion rates, however, the SED becomes softer again in the super-Eddington regime.
There is a slight increase of $\Lacc/\ha$ and $\Lacc/\hb$ with metallicity. This weak metallicity dependence relates to the fraction of the incident ionising luminosity converted into escaping Balmer-line emission, which depends jointly on the thermal and ionisation structure, collisional excitation, and radiative transfer.
In addition, the variation of these ratios with \Mbh\ and \Fedd\ does not strictly follow $Q_{\rm H}/\Lacc$, because it also depends on the detailed shape of the ionising SED, and its impact on the ionisation state and temperature of the gas. 
In dense BLR gas, radiative-transfer effects associated with large line optical depths cause the \ha\ and \hb\ emissivities to depend on the local gas conditions and need not respond linearly to $Q_{\rm H}$ \citep[e.g.,][]{Korista1997,Korista2004,Wang2025}. Collisional excitation, particularly for \ha, introduces an additional contribution to the Balmer-line emission.

We also show the distribution of bolometric corrections for the single-cloud, Balmer-break models in Fig.~\ref{fig:BLRLbollinesfid}, which spans a very wide range of values (brown shaded region). For example, these models exhibit a total scatter of $\sim 5.2$ dex for \ha, $\sim 4.7$ dex for \hb, and $\sim 2.9$ dex for \heiiopt. This suggests that using inappropriate bolometric corrections may introduce large systematic uncertainties when inferring the accretion luminosity of sources with such physical properties.
The typical LRD-like model is shown as a brown cross. This model tends to lie near the lower edge of the bolometric-correction predictions spanned by the full grid of single-cloud, Balmer-break models. Its predicted $\Lacc/L_{\rm line}$ ratios are lower than those of the fiducial multi-cloud BLR model with the same \Mbh, \Fedd, and $Z$, by about $\sim$0.5 dex for \ha, \hb\ and \heiiopt\ (see also Sec.~\ref{sec:TypeIfluxlimit}). This implies that using the standard bolometric corrections would lead to an overestimate of the accretion luminosity by a similar amount.

Another bolometric correction commonly used in the literature is the relation between the bolometric luminosity and the continuum luminosity at 5100~\AA, $L_{5100}$, such as $\Lbol = 9\times\ L_{5100}$ \citep{Marconi2004} or $\Lbol = 10.33\times\ L_{5100}$ \citep{Richards2006}. This relation is often used directly, or combined with an empirical calibration between \ha\ luminosity and $L_{5100}$ \citep[e.g.,][]{Greene2005,Stern2012}. However, adopting a fixed bolometric correction does not capture the variations in SED shape predicted by the theoretical \relqso\ models of \citet{Kubota2018,Kubota2019,Hagen2023}. In particular, the predicted ratio $\Lbol/L_{5100}$ varies by up to $\sim1$ dex, reaching its largest values at low BH masses \citep[see also][]{Netzer2019,Temple2023,Azadi2025,Greene2026}.

\subsection{Interpretation of broad emission lines of high-redshift type~I AGN}\label{sec:TypeIhighz}

\begin{figure*}
\begin{centering}
\includegraphics[width=1\linewidth]{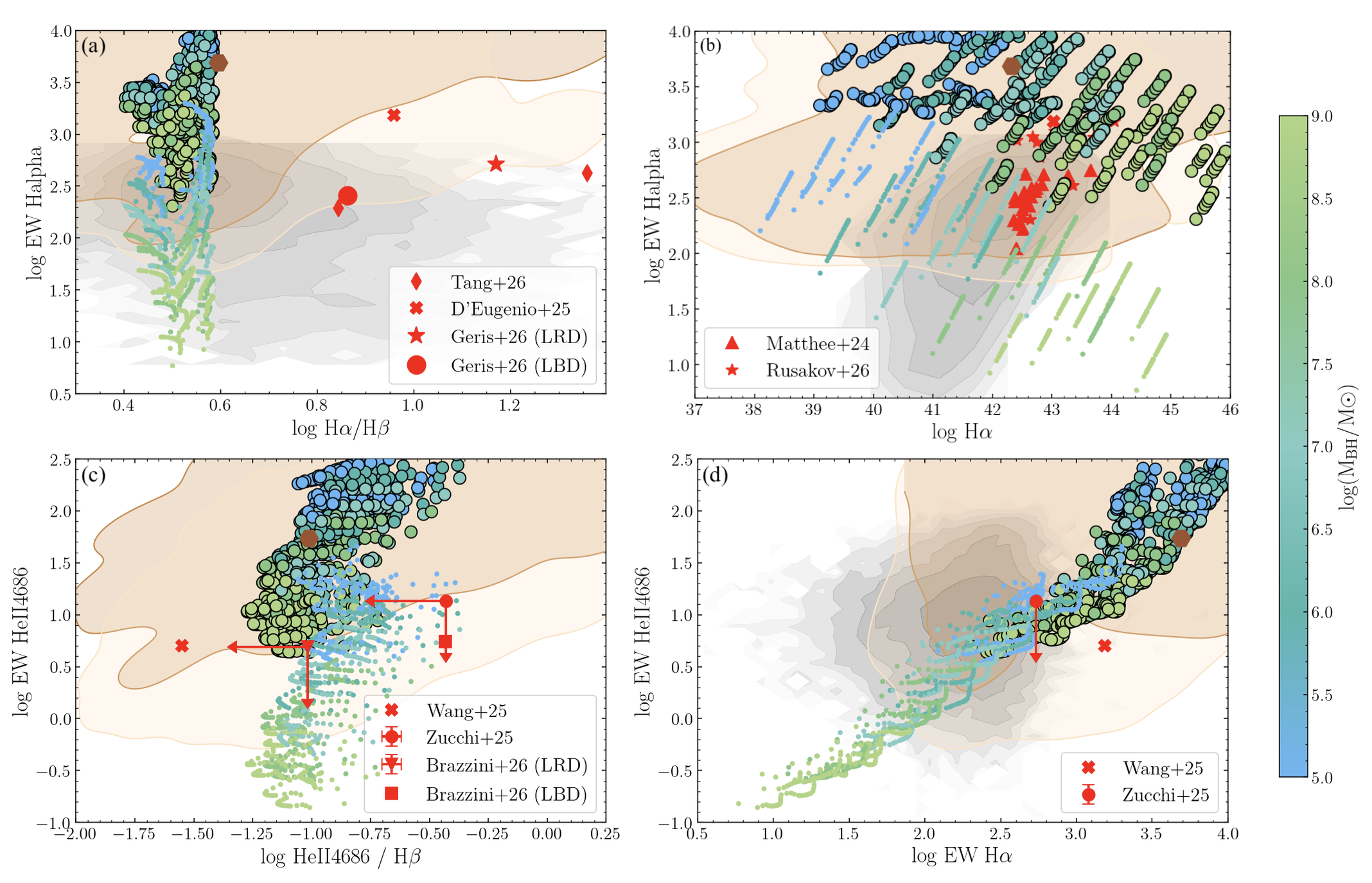}\\
\includegraphics[width=0.4 \linewidth]{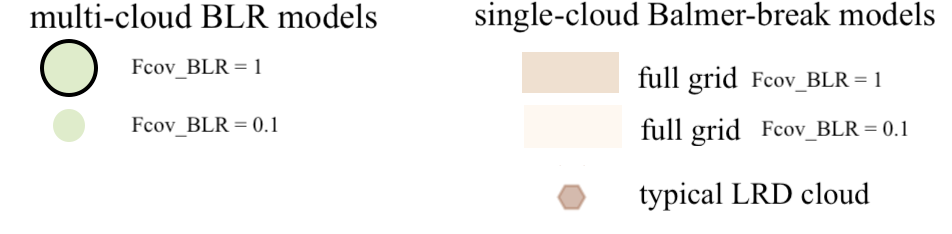}
\par\end{centering}
\caption{Comparison of BLR model predictions with high-redshift observations of broad-line AGN and LRDs. (a) EW(\ha) as a function of log \ha/\hb. (b) EW(\ha) as a function of \ha\ luminosity. (c) EW(\heiiopt) as a function of \heiiopt/\hb. (d) EW(\heiiopt) as a function of EW(\ha).
The multi-cloud fiducial BLR models are shown as circles, colour-coded by black-hole mass, from blue ($\log \Mbh/\Msun = 5$) to green ($\log \Mbh/\Msun = 9$). Models with a BLR covering factor of 1 are shown as larger circles with black edges, while models with a covering factor of 0.1 are shown as smaller circles without edges.
The brown shaded regions show the distribution of single-cloud, Balmer-break models, for covering factors of 1 (dark brown) and 0.1 (light brown). The dark brown hexagon indicates the “typical LRD-like” single-cloud model, shown here with a covering factor of 0.4. 
Dust is not included in the models, but the black arrow shows the direction of attenuation.
Observations of high redshift Type I AGN and LRDs \citep[][]{Matthee2024,Wang2025,Zucchi2025,Brazzini2026,DEugenio2025,Rusakov2026,Tang2026,Geris2026} are shown with red symbols.
For comparison, we also show the position of local AGN from \citet{Liu2019} (a,b) and \citet{Wu2022} grey shaded contours (c,d).}
\label{fig:EWlrBLRfid}
\end{figure*}

\JWST\ has revealed a population of faint AGN (with lower luminosities than pre-\JWST\ high-redshift AGN) and LRDs, which appear to exhibit spectral properties that differ from those of local type~I AGN. These include, for example, strong \ha\ equivalent widths and low X-ray emission \citep{Maiolino2025}, as well as weak high-ionisation emission lines in some sources \citep[e.g.,][]{Lambrides2024,kokorev2024,Ji2025,Wang2025,Zucchi2025,Brazzini2026,Geris2026,Chavez2025}. In this section, we explore how some of these features can be interpreted in the context of our models.

In Fig.~\ref{fig:EWlrBLRfid}, we compare our model predictions with a compilation of broad emission-line measurements for high-redshift type~I AGN and LRDs. This compilation includes samples of broad \ha\ emitters from \citet{Matthee2024} at $z = 4.2$--5.5 and from \citet{Rusakov2026} at $z = 3.4$--6.7, as well as a stacked spectrum of type~I AGN at $z > 5$ from \citet{Zucchi2025} and an LRD and LBD stacks from \citet{Geris2026}.
We also include a few individual sources: GN-28074, an LRD at $z = 2.26$ \citep{Juodvabalis2024,Brazzini2026}; GS-3073, a “Little Blue Dot” (LBD) at $z = 5.5$ \citep{Brazzini2026}; RUBIES-EGS-49140, an LRD at $z = 6.68$ \citep{Wang2025,DEugenio2025}; two LRDs, Abell2744-QSO1 at $z = 7.04$ and UNCOVER-2476 at $z = 4.02$, from \citet{Tang2026}; and a stack of LRD and LBD spectra from \citet{Geris2026}.
We also show the samples of local type~I AGN from \citet{Liu2019} and \citet{Wu2022}. 

In all panels of Fig.~\ref{fig:EWlrBLRfid}, the fiducial grid of the multi-cloud BLR models are colour-coded by BH mass. Larger circles with black outlines correspond to a BLR covering factor of unity, while smaller circles without outline indicate a covering factor of 0.1. The distribution of single-cloud, Balmer-break, BLR models is shown with dark brown shaded region for a covering factors of unity and light brown shaded region for a covering factor of 0.1. The typical LRD-like model, with a covering factor of 0.4, is plotted as a brown cross.

We note that, in all the diagrams of Fig.~\ref{fig:EWlrBLRfid}, the single-cloud, Balmer-break models span a wide region of parameter space, encompassing the fiducial models and extending to more extreme values, often beyond those probed by the observational samples. This behaviour primarily reflects the fact that these models sample a broad range of BLR conditions in density and radius, rather than being combined into a single integrated model.

We first focus on the Balmer decrement: panel (a) of Fig.~\ref{fig:EWlrBLRfid} illustrates \ha\ EW as a function of \ha/\hb. The mean \ha/\hb\ ratio of the fiducial multi-cloud BLR models, $\sim 3.4$, is consistent with the median value of $\sim 3.9$ observed in the local SDSS sample of \citet{Liu2019} (see also Section~\ref{sec:validation_BLR_z0}), and is close to the expectation from case-B recombination, indicating that recombination processes dominate the emission of these Balmer lines.
However, the SDSS distribution shows a tail toward higher \ha/\hb\ ratios, and a similar trend is observed in the LRDs of \citet{DEugenio2025} and \citet{Tang2026} and the LRD stack from \citet{Geris2026}, which exhibit Balmer decrements in the range $\sim 7$–23.
Such high ratios can be explained by the presence of dust \citep[see e.g.,][]{Liu2019}, or significant collisional excitation of \ha\ emission and partial local thermodynamic equilibrium effects \citep[e.g.,][]{Popovic2003}. In the case of LRDs, the presence of dense, large column densities of partially ionized hydrogen further suggests that collisional contributions, radiative-transfer effects and resonance scattering may play an important role \citep[e.g.,][]{deGraaff2025a,DEugenio2025,Chang2026,Nikopoulos2026}.
This interpretation is consistent with the distribution of the single-cloud, Balmer-break models, for which the predicted \ha/\hb\ ratios extend up to $\sim 25$, while the typical LRD-like model yields a moderately elevated value of $\sim 4$.
In addition, the grid of single-cloud Balmer-break models also extends to \ha/\hb\ ratios below the Case B value. These low decrements arise in regions of parameter space where the Balmer lines, particularly \ha, become optically thick, such that radiative-transfer effects modify the relative escape of \ha\ and \hb\ photons \citep[e.g.,][]{Scarlata2024,Pirzkal2024}.

\citet{Maiolino2025} found that, in a sample of faint high-redshift broad-line AGN, the observed \ha\ EWs tend to be larger than those measured in lower-redshift SDSS type~I AGN. To investigate this, we show in panel (b) \ha\ EW as a function of \ha\ luminosity. While the \ha\ EWs of the \citet{Matthee2024} sample lie within the main distribution of the low-redshift sample from \citet{Liu2019} (grey shaded area), those from \citet{Rusakov2026} tend to be systematically higher.
As discussed in Section~\ref{sec:validation_BLR_z0}, the fiducial multi-cloud BLR models, assuming a covering factor of 0.4, reproduce well the region where most local type~I AGN are observed. At fixed covering factor, decreasing the BH mass shifts the models toward higher \ha\ EWs and lower bolometric luminosities, which may partly explain the large EWs observed at high redshift. This trend is primarily driven by variations in the ratio of optical continuum emission to the number of ionising photons produced by the accretion disc (see Section~\ref{sec:validation_BLR}), resulting in higher EWs at lower BH masses. In addition, the large observed EWs may also reflect higher BLR covering factors in these high-redshift AGN.
The single-cloud, Balmer-break models extend to larger \ha\ EWs due to the wide range of physical conditions they encompass, which generally leads to enhanced \ha\ emission (see Sections~\ref{sec:TypeIfluxlimit} and \ref{sec:TypeILAGNcal}). In particular, this enhancement may be driven by increased collisional excitation of \ha\ in clouds with large column densities of partially ionized gas. These conditions could therefore contribute to the higher EWs observed in LRDs compared to typical AGN.

Another characteristic of newly discovered faint high-redshift AGN and LRDs is the presence of weak broad high-ionisation emission lines in some sources \citep[e.g.,][]{Lambrides2024,Ji2025,Wang2025,Zucchi2025,Brazzini2026}. To investigate this, we show in panel (c) \heiiopt\ EW as a function of \heiiopt/\hb, and in panel (d) \heiiopt\ EW as a function of \ha\ EW.
First, we examine which fiducial multi-cloud models produce the lowest \heiiopt\ emission relative to other lines. We find a slight decrease of \heiiopt/\hb\ of $\approx 0.5$ dex with increasing BH mass, which follows the decrease of $Q_{\rm HeII} / Q_{\rm H}$ with \Mbh. In addition, lower metallicities shift the predicted \heiiopt/\hb\ ratios downward by $\approx 0.5$~dex.
The upper limits on the \heiiopt\ flux and EW for the $z > 5$ stacked sample of \citet{Zucchi2025} are consistent with the predictions of the multi-cloud BLR models. However, this is not the case for the LRD source RUBIES-EGS-49140 \citep{Wang2025}, for which none of our fiducial multi-cloud models reproduce the observed strength of the Balmer lines (\ha\ and \hb) relative to \heiiopt.

By contrast, the extreme edge of the single-cloud, Balmer break model distribution reaches the observed ratios of this source. This behaviour is primarily driven by the ionisation parameter (proportional to $\Phi/\nh$), with models at lower ionisation parameter producing lower \heiiopt/\hb.
Alternatively, several explanations have been proposed to account for the low \heii/\hb\ ratios. In particular, these low ratios may indicate a softer ionising radiation field produced by the accretion disc \citep[e.g.,][]{Wang2025,Zucchi2025}. Such a softening could be linked to super-Eddington accretion \citep{Lambrides2024,Pacucci2024,Ji2025}, anisotropic emission from geometrically thick super-Eddington accretion discs \citep[e.g.,][]{Madau2025,Madau2026}, or the presence of dense gas attenuating the ionising radiation \citep[e.g.,][]{Brazzini2026}. 
Another possibility is a contribution from stellar radiation \citep[e.g.,][]{Wang2025}, in particular if the broad line region is ionized by young star clusters rather than the BH accretion disc \citep[e.g.,][]{Inayoshi2026}.
Here, we note that, under our adopted assumptions for super-Eddington accretion discs (see Section~\ref{sec:method_SED}), the models do not naturally predict a decrease in \heiiopt/\hb\ (see also Section~\ref{sec:discumodeluncertainties}). In Section~\ref{sec:discuhighz}, we discuss how modifying some of our model assumptions may help reproduce the weak \heiiopt\ emission observed in some of these high-redshift sources.

\section{Detectability and diagnostics of high-redshift  type~II AGN}\label{sec:TypeII}

In this section, we focus on Type II AGN, restricting our analysis to emission from the narrow-line region and neglecting any AGN continuum component. We assess their detectability with JWST (Sec. \ref{sec:TypeIIfluxlimit}), derive bolometric corrections for narrow-line luminosities (Sec. \ref{sec:TypeIILAGNcal}), and re-visit existing as well as develop novel diagnostics to distinguish Type II AGN powered by intermediate-mass and supermassive BHs from star-forming galaxies and Pop III stellar populations (Sec. \ref{sec:TypeIIdiag}).

\subsection{Detectability of narrow emission lines of type~II AGN with JWST/NIRSpec}\label{sec:TypeIIfluxlimit}

\begin{figure}
\begin{centering}
\includegraphics[width=1\linewidth]{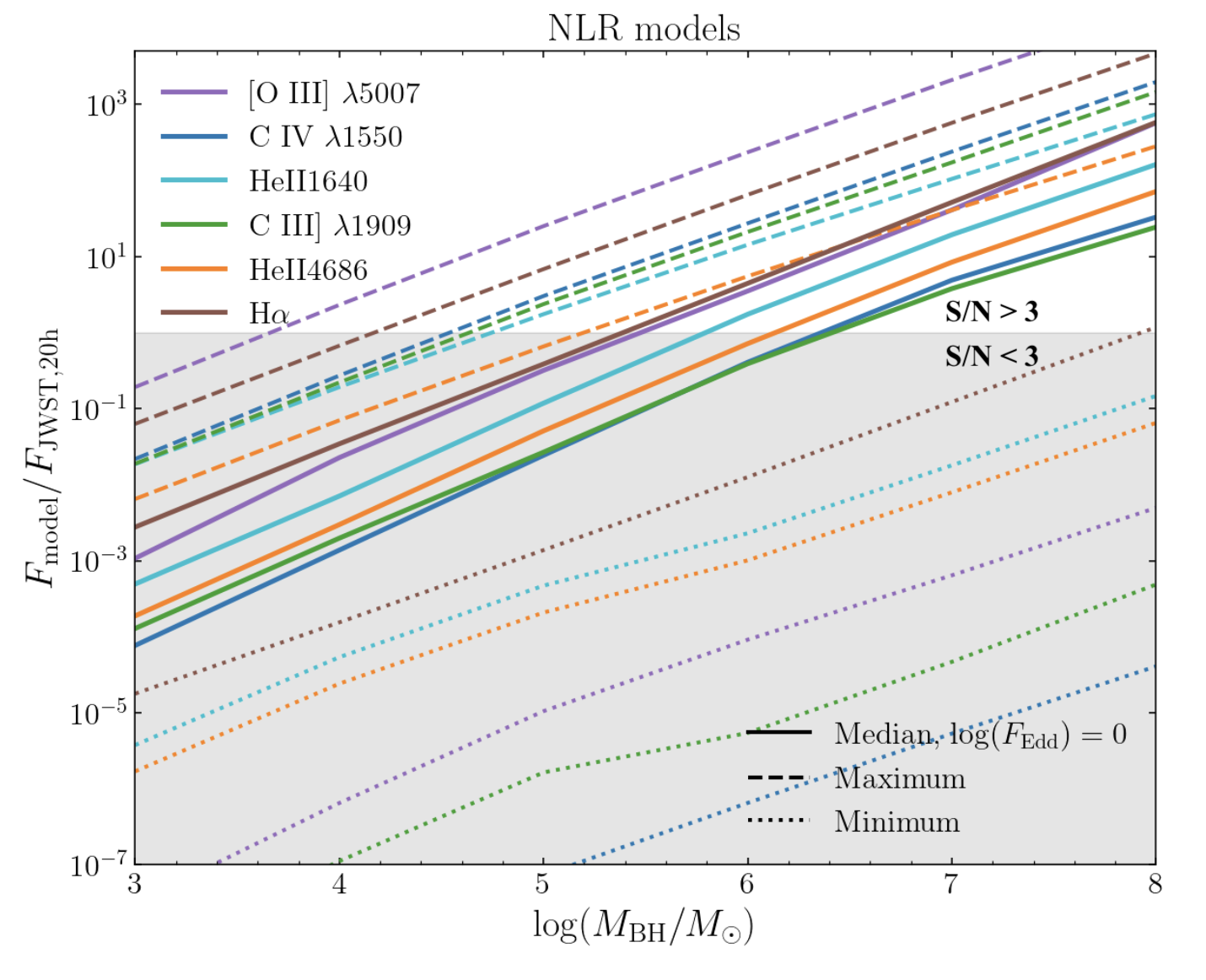}
\par\end{centering}
\caption{Detectability of narrow emission lines with JWST at $z = 6$. The plot shows the ratio of modelled NLR line flux to the JWST flux limit required to achieve S/N = 3 with an integration time of 20 h, for several of the main AGN emission lines, as a function of black-hole mass over the range $\log (\Mbh/\Msun) = 3$–8. The models assume an NLR covering factor of 0.1. Solid lines show the median line flux for models with $\log (\Fedd) = 0$. Dashed lines show the maximum line flux across all values of \Fedd, ionisation parameter, and metallicity $Z$. Dotted lines show the minimum line flux across the same parameter space. Each colour correspond to a different line, as indicated in the legend. The grey shaded region highlights fluxes below the detection limit.}
\label{fig:NLRdetecJWST_sum}
\end{figure}

\begin{figure*}
\begin{centering}
\includegraphics[width=1\linewidth]{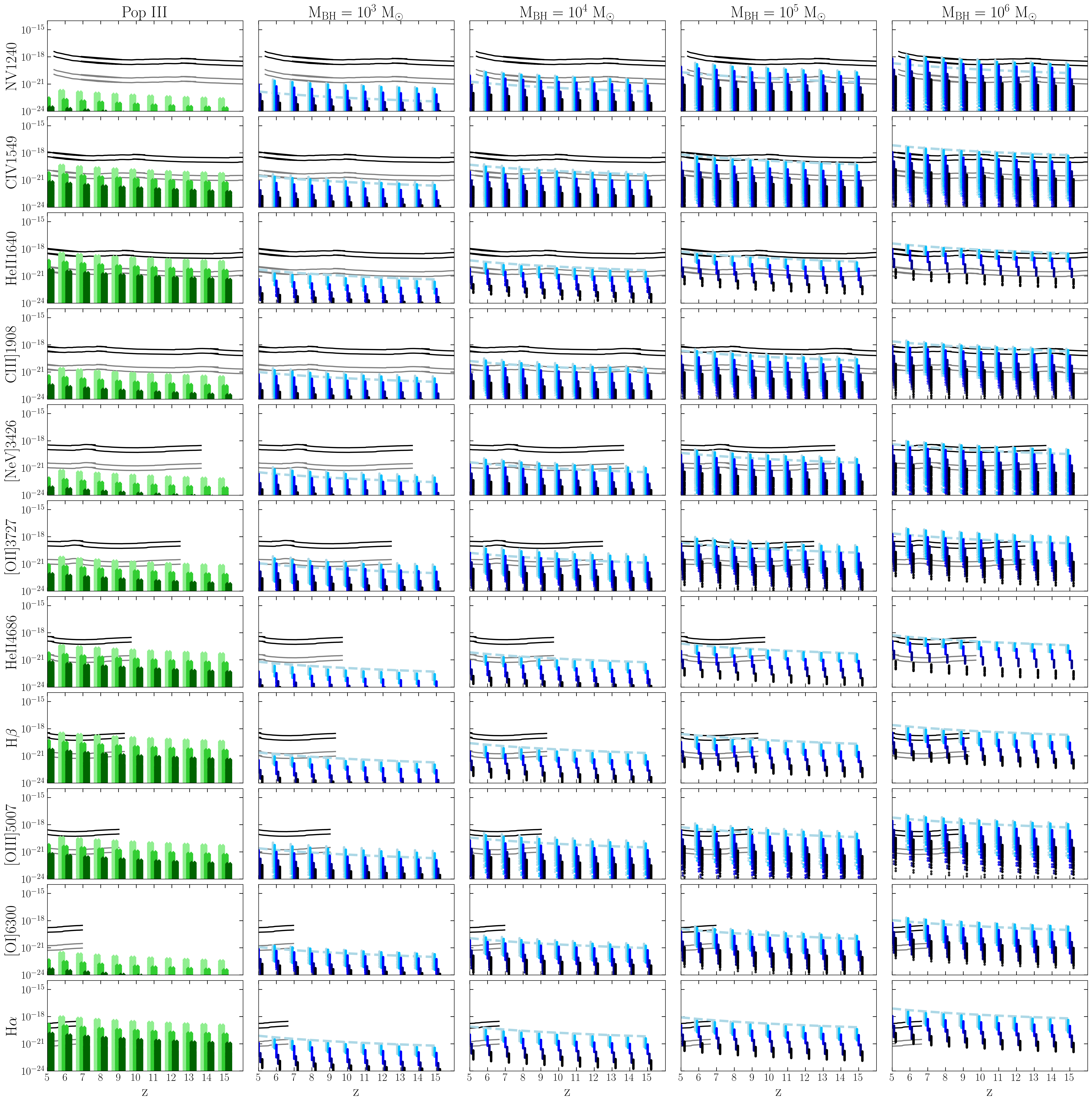}
\includegraphics[width=0.9\linewidth]{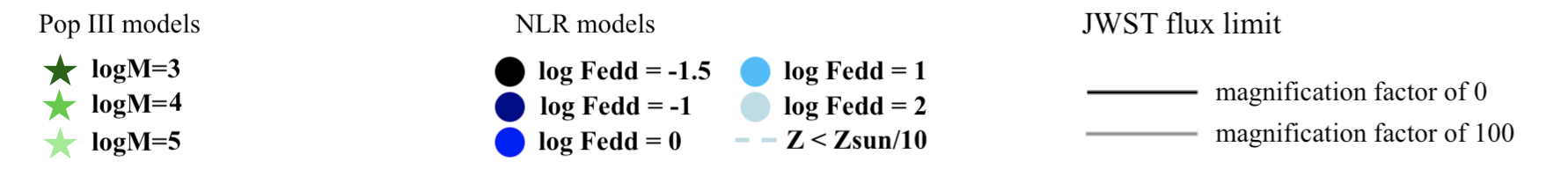}
\par\end{centering}
\caption{Detectability of narrow emission lines with JWST. The panels show the predicted line fluxes as a function of redshift. The first column shows pop III star models with masses of $\log (\rm M_{*}/\Msun) = 3$ (dark green) to 5 (light green). The other columns show the NLR models, assuming a NLR covering factor of 0.1. The models are colour-coded by Eddington ratio, and each column corresponds to a different black-hole mass, spanning $\log (\Mbh/\Msun) = 3$ to 6. The dashed line shows the maximal flux for models with $Z<0.1\zsun$. Each row corresponds to a different emission line. The predicted fluxes are compared with the JWST/NIRSpec high resolution gratings detection limits required to achieve S/N = 3, assuming a line width fixed to 50 km s$^{-1}$, for integration times ranging from 2 to 20 hours (black lines). We also show how the detection limit changes assuming lensing magnification factors of 100 (grey). 
}
\label{fig:NLRdetecJWST}
\end{figure*}

We now investigate the detectability of narrow emission lines in high-redshift AGN with JWST, focusing on intermediate-mass BHs, to establish the minimum BH mass at which narrow emission lines become observable, which may help to constrain BH seeding models and early growth. Fig.~\ref{fig:NLRdetecJWST_sum} shows the ratio of the modelled narrow line fluxes to the \JWST\ flux limit required to achieve a signal-to-noise ratio of S/N = 3, as a function of BH mass over the range $\log (\Mbh/\Msun) = 3$–8 at redshift $z=6$. The \JWST\ flux limits are computed using Pandeia, using the same set-up as described in Section~\ref{sec:TypeIfluxlimit}. The predictions are shown for NIRSpec high-resolution grating spectroscopy and an integration time of 20~h. 
We present predicted ratios of  modelled  line fluxes to S/N flux limit for the main narrow-line emission lines: \ha, \loiiiopt, \heiiopt, \heii, \lciii, and \lciv\ (different coloured lines). We assume an NLR covering factor of 0.1, which is consistent with the range of estimated values from observations by \citet{Baskin2005} and \citet{Netzer2009}; higher covering factors would increase detectability accordingly. We show the median line flux ratios for models with $\log (\Fedd) = 0$ (solid lines), as well as the maximum (dashed lines) and minimum (dotted lines) predicted fluxes across the full range of ionisation parameters ($\log \rm U  = -5$ to -1), metallicities ($Z = 0.0001$ to 0.060), and Eddington ratios ($\log \Fedd = -1.5$ to 2), at a given BH mass. 

H- and He-line fluxes (orange, blue and brown lines) span a narrower range of predicted flux at fixed BH mass compared to metal-line ratios (green, red and lilac lines), because they are primarily sensitive to the number of ionising photons, which is mainly controlled by \Fedd\ and \Mbh. For a BH accreting at the Eddington limit, \ha\ becomes detectable for masses $\log (\Mbh/\Msun) \approx 5.5$, and down to lower masses $\log (\Mbh/\Msun) \approx 4$ in the case of strongly super-Eddington accretion ($\log \Fedd =2$). The \lheii\ lines are generally weaker than \ha\ and cross the detection threshold only for masses around $\log (\Mbh/\Msun) = 6$, unless the BH is accreting in the super-Eddington regime. In the lowest-flux models, corresponding to low Eddington ratios $\log \Fedd = -1.5$ and low ionisation parameters $U$, the \lheii\ lines lie approximately 3 dex below the detection limit even at $\log (\Mbh/\Msun) = 6$.
The predictions are qualitatively similar for \lciv\ and \lciii\ to those of \lheii. They reach a S/N of 3 at $\log (\Mbh/\Msun) \approx 6.2$, for a BH accreting at the Eddington limit. However, because these lines are also strongly sensitive to metallicity, \CO\ abundance and ionisation parameter, the lowest-flux models fall about 7 dex and 6 dex below the detection limit, respectively, at $\log (\Mbh/\Msun) = 6$.
As for \ha, \loiiiopt\ becomes detectable for BH masses of $\log (\Mbh/\Msun) \approx 5.5$ when accreting at the Eddington limit, and down to $\log (\Mbh/\Msun) = 3.8$ in the most extreme cases, corresponding to models with $\log \Fedd = 2$, high ionisation parameter, and metallicities sufficiently high to provide abundant oxygen ions, but not so high as to significantly decrease the electron temperature.
This assumes an NLR covering factor of 0.1. For a covering factor of unity, \ha\ and \loiiiopt\ would be detectable down to a $\log (\Mbh/\Msun) \approx 4.5$ instead when accreting at $\log \Fedd = 0$.

In Fig.~\ref{fig:NLRdetecJWST}, we extend our detectability analysis to a larger set of emission lines, also including \lnv, \lnev, \loii, \hb\ and \loi\ as well as a wider redshift range, spanning $5 \leq z \leq 15$. 
The \JWST\ flux limit required to reach a S/N of 3 is shown with black lines, for integration times of 2 and 20~h, for the high resolution gratings.
We show the redshift evolution of predicted line fluxes from our models for BH masses in the range $\log (\Mbh/\Msun) = 3$–6, colour-coded according to the Eddington ratio. We also highlight the maximum predicted flux for models with $Z < 0.1\zsun$ (light blue dashed line), since high-redshift galaxies are expected to have lower metallicities.
For comparison, the first column of Fig.~\ref{fig:NLRdetecJWST} illustrates Population~III stellar models described in Section \ref{sec:method_stellar}, for stellar masses in the range $\log (\Mstar/\Msun) = 3$–5. For the highest assumed mass, \ha\ and \hb\ are above the S/N threshold for some models, and \heii\ barely reaches this limit, only in the most extreme cases and for an integration time of 20~h. These maximum values are typically reached at young stellar ages and assuming a top-heavy IMF. This highlights the difficulty of detecting Pop~III star emission with \JWST.
To extend this analysis to higher BH masses, a similar figure is presented in Annex~\ref{sec:JWSTlimhighBHmass}, covering the range $\log(\Mbh/\Msun) = 7$–9 and showing that most emission lines become detectable for such supermassive BHs, at least when accreting close to or above Eddington.

Regarding the NLR models, Fig.~\ref{fig:NLRdetecJWST} highlights the scatter in the predicted line fluxes at fixed \Fedd\ and \Mbh. As anticipated, metal lines exhibit a much larger scatter than H and He lines. In particular, the predicted fluxes for different values of \Fedd\ significantly overlap for metal lines, and only specific combinations of metallicity and ionisation parameter yield the highest fluxes.

Turning to metallicities of $Z < 0.1~\zsun$, for a BH with $\log (\Mbh/\Msun) = 3$, none of the emission-line fluxes lie above the detection limit. 
At $\log (\Mbh/\Msun) = 5$, \ha, \loiiiopt, and \hb\ become detectable, but mostly in the super-Eddington regime. 
At $\log (\Mbh/\Msun) = 6$, additional lines become accessible at high Eddington ratios, including \loiopt, \lheii, \loiiopt, \lciii\ and \lciv. 
It is only at $\log (\Mbh/\Msun) = 7$ that some lines become detectable in the sub-Eddington regime, namely \ha, \loiopt, \loiiiopt, \hb, and \loiiopt. 
Instead, the high-ionisation lines \lheii, \lnev, and \lnv, which are particularly useful for diagnosing AGN activity, are detectable only for accretion at the Eddington limit and above or for higher BH masses.
Thus, detecting these high-ionisation lines in an unlensed source would mostly favour either a high black-hole mass or accretion near or above the Eddington limit. Conversely, their non-detection does not rule out AGN activity, and selecting AGN solely through these lines may miss lower-mass or more weakly accreting black holes.

We also illustrate in Fig.~\ref{fig:NLRdetecJWST} the impact of gravitational lensing magnification, highlighting in grey the detection limits corresponding to magnification factor of 100. Such strong magnification enables the detection of the brightest lines, such as \ha\ and \loiiiopt, even down to BH masses of $\log(\Mbh/\Msun) = 3$.
In addition, several fainter high-ionisation lines, including \lheii\ and \lciv, may become detectable in the sub-Eddington regime for BH masses of $\log(\Mbh/\Msun) = 6$ under similarly strong magnification. 
This demonstrates that JWST lensing programs (e.g., GLIMPSE) are crucial for probing the high-redshift BH population down to lower masses.

\subsection{Bolometric calibrations for narrow line luminosities of type~II AGN}\label{sec:TypeIILAGNcal}

\begin{figure*}
\begin{centering}
\includegraphics[width=1\linewidth]{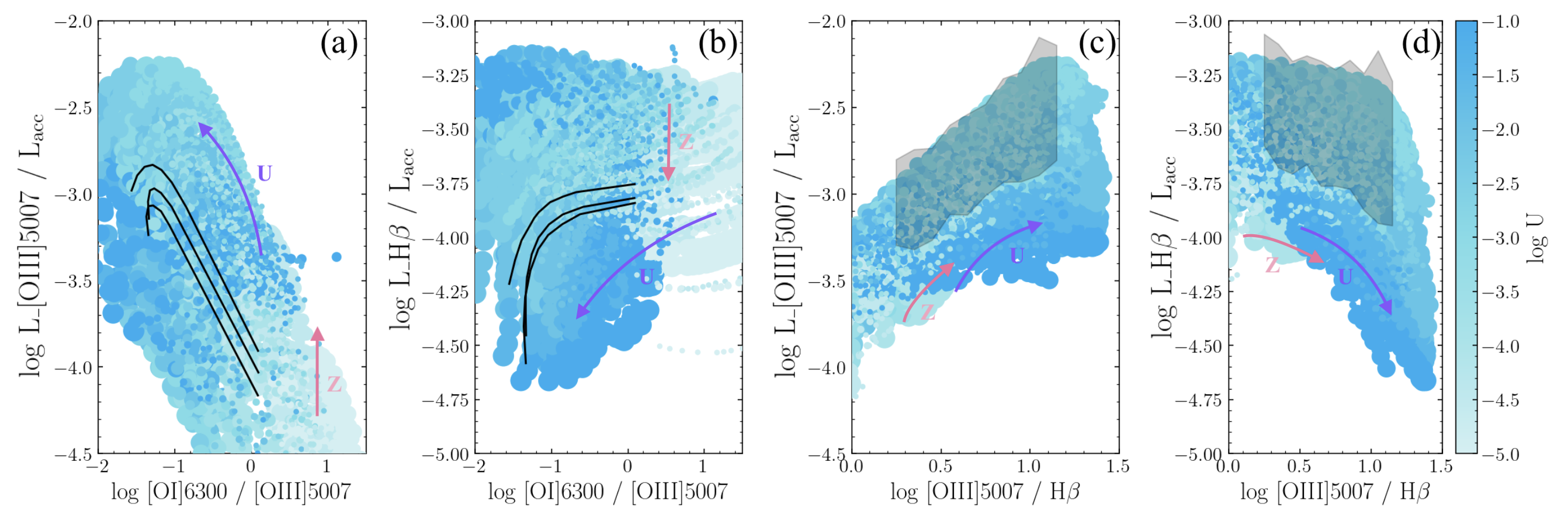}
\par\end{centering}
\caption{NLR bolometric correction for \loiiiopt\ (panels a, c) and \hb\ (b, d) as a function of the line ratios \loiiiopt/\hb\ (c, d) and \loiopt/\loiiiopt\ (a, b). NLR models are represented by blue circles, colour-coded by the ionisation parameter from log U = -5 (light) to -1 (dark). Marker size scales with BH mass, ranging from 10$^3$ \Msun\ (small) to 10$^9$ \Msun\ (large). All models assume an NLR covering factor of 0.1. The solid black lines show the theoretical predictions from \citet{Netzer2009} (for \zsun,  $3\times \zsun$ and $3\times \zsun$ with constant pressure), shifted down by 1~dex to have a covering factor of 0.1, while the grey shaded regions indicate the standard deviation of the bolometric corrections for the local sample of Type I AGN from \citet{Netzer2007}, as derived by \citet{Netzer2009}, shifted up by 0.7~dex to account for dust correction.}
\label{fig:NLRbolN09}
\end{figure*}

\begin{figure*}
\begin{centering}
\includegraphics[width=1\linewidth]{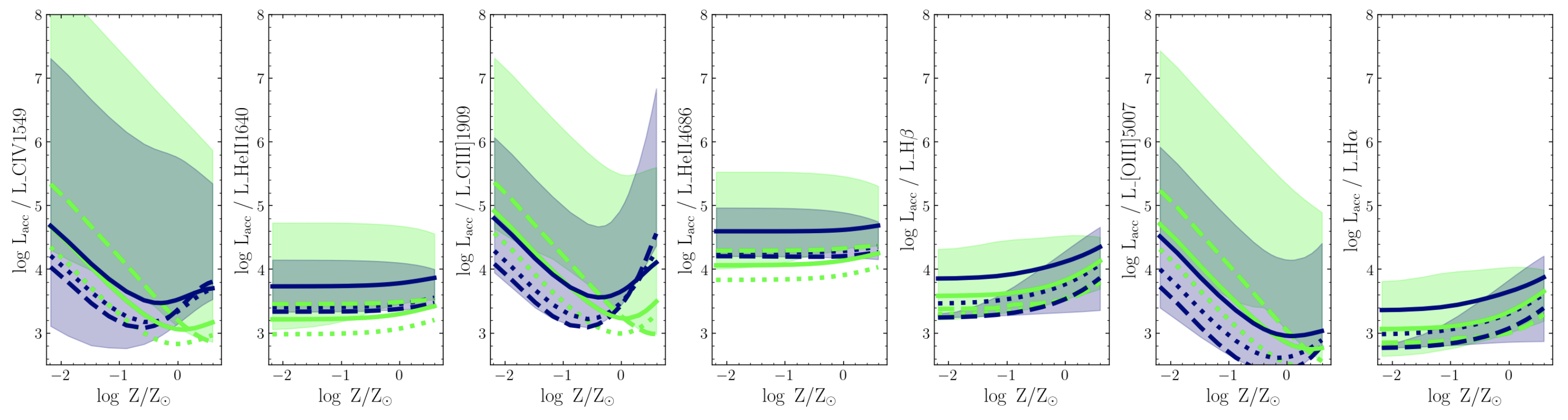}\\
\includegraphics[width=0.8\linewidth]{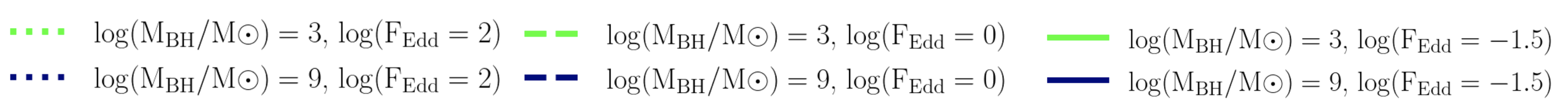}
\par\end{centering}
\caption{Bolometric luminosity correction for various emission lines as function of the metallicity. Models are shown for a covering factor of the NLR of 0.1. Green lines are for a black-hole mass of $10^{3}\rm{M}_{\odot}$ and the blue lines for $10^{9}\rm{M}_{\odot}$. Solid, dashed and dotted lines correspond to Eddington ratio of -1.5, 0 and 2 in log10 scale, with a fixed ionisation parameter of -2.5. The shaded region shows the range of values for an Eddington ratio of log \Fedd\ = 0 for a black-hole mass of $10^{3}\rm{M}_{\odot}$ (green) and $10^{9}\rm{M}_{\odot}$ (blue), the range correspond to models with different ionisation parameter.}
\label{fig:NLRbol}
\end{figure*}

As for the broad-line analysis in Sec.~\ref{sec:TypeILAGNcal}, here, we investigate bolometric corrections but for type~II AGN to estimate their BH accretion rates and, thus, assess how efficiently BHs grow at early cosmic epochs. In Fig.~\ref{fig:NLRbolN09}, we compare our model predictions with the commonly used bolometric corrections of \citet{Netzer2009}.
In this work, they derived bolometric corrections for \loiiiopt\ and \hb\ using photoionisation models, expressing the corrections as a function of \loiopt/\loiiiopt. This formulation accounts for the dependence of \loiiiopt\ emission on the ionisation parameter, while \hb\ is found to be less sensitive to such variations. They also report only a weak dependence on metallicity and on the shape of the ionising SED.

In subplots (a) and (b) of Fig.~\ref{fig:NLRbolN09}, we present the bolometric correction ratios \loiiiopt/\Lacc\ and \hb/\Lacc\ as a function of \loiopt/\loiiiopt. Our NLR model predictions are shown for a covering factor of 0.1, using blue circles colour-coded by ionisation parameter, from $\log U = -5$ (light) to $-1$ (dark), with marker size increasing with BH mass from $10^{3} \Msun$ to $10^{9} \Msun$. The bolometric corrections from \citet{Netzer2009} (solid black lines), shifted downward by 1 dex to match the assumed covering factor,
are consistent with the predictions from our models. However, our models exhibit a significantly larger scatter at fixed \loiopt/\loiiiopt, of $\approx 1$–2 dex for \loiiiopt\ and $\approx 1.5$ dex for \hb.
This scatter is primarily driven by variations in metallicity. For \loiiiopt, changing the metallicity leads to variations of $\approx 1.5$ dex: increasing metallicity both enhances the abundance of O$^{2+}$, boosting the line emission, and reduces the electron temperature, which lowers the emissivity, resulting in a complex, non-monotonic behaviour. For \hb, increasing metallicity tends to decrease the line flux due to the absorption of hydrogen-ionising photons by dust, especially at high $\log U$, when the dust column densities are large, leading to a scatter of $\approx 0.5$ dex in \hb/\Lacc. Also, variations in the SED shape driven by \Mbh\ and \Fedd\ introduce an additional scatter of $\approx 0.75$ dex in both \loiiiopt/\Lacc\ and \hb/\Lacc.

In addition to theoretical estimates, \citet{Netzer2009} provide empirical bolometric calibrations based on type~I and type~II AGN samples. Panels (c) and (d) of Fig.~\ref{fig:NLRbolN09} show \loiiiopt/\Lacc\ and \hb/\Lacc\ as a function of \oiiiopt/\hb, contrasting our model predictions with the dispersion reported for the type~I sample of \citet{Netzer2007}. The latter are shifted upward by 0.7 dex to account for dust attenuation (Section~2.2.2 of \citealt{Netzer2009}).

While we again find an overall good agreement between the models and the observations, we caution that several uncertainties affect the observational estimates. In particular, while the bolometric luminosity is known for the photoionisation model predictions, \citet{Netzer2009} estimate \Lacc\ using a simple relation between $L_{5100}$ and \Lacc, which does not capture the full complexity of variations in the accretion-disc SED shape (see Sec.~\ref{sec:TypeILAGNcal}).
Additional sources of uncertainty include the dust correction and the contribution of star formation to the narrow emission lines.
\citet{Silcock2025} compared the predictions of the \citet{Feltre2016} photoionisation models with these observations, finding good agreement at high metallicity but significant discrepancies at low metallicity. They concluded that relying solely on the dependence of the bolometric corrections (\loiiiopt/\Lacc\ and \hb/\Lacc) on \oiiiopt/\hb\ is insufficient, given the strong dependence on both metallicity and ionisation parameter.
We find similar results, with an additional source of scatter arising from variations in the SED shape driven by \Mbh\ and \Fedd. Rather than relying on simple empirical relations, a more robust approach to infer the bolometric luminosity of type~II AGN would be to use a Bayesian fitting framework, such as \beagle\ \citep[][]{Chevallard2016,Vidal2024} or \textsc{CIGALE} \citep[][]{Zhang2026}, allowing all relevant physical parameters to be constrained simultaneously.

We present in Fig.~\ref{fig:NLRbol} our predicted bolometric corrections, $\Lacc/\rm{L}_{\rm line}$, for several of the brightest narrow emission lines, including \lciv, \heii, \lciii, \heiiopt, \hb, \loiiiopt, and \ha, as a function of metallicity. To also assess the dependence on \Mbh\ and \Fedd, the models are shown for BH masses of $\log (\Mbh/\Msun) = 3$ and 9 (green and blue lines, respectively), and Eddington ratios of $\log \Fedd = -1.5$, 0, and 2 (solid, dashed and dotted lines, respectively), at a fixed ionisation parameter of $\log U = -2.5$. The additional scatter introduced by variations in the ionisation parameter is illustrated by showing the full range of $\log U$ for models with $\log \Fedd = 0$, using green ($\log (\Mbh/\Msun) = 3$) and blue ($\log (\Mbh/\Msun) = 9$) shaded regions.
As in the BLR case (Section~\ref{sec:TypeILAGNcal}), H and He lines are the most promising tracers, as they show a weaker dependence on metallicity than metal lines and therefore exhibit a smaller scatter in $\Lacc/\rm{L}_{\rm line}$. Variations in the SED shape driven by \Mbh\ and \Fedd\ introduce a scatter of $\sim 0.75$ dex for \heii, \heiiopt, \ha, and \hb. In contrast, the ionisation parameter leads to a larger dispersion, of $\sim 1.5$ dex for \lheii\ lines and $\sim 1$ dex for \ha\ and \hb. The stronger dependence of \lheii\ on $U$ is driven by the increasing fraction of He$^{2+}$ at high ionisation parameter.
We note, however, that the dependence on $\log U$ is not straightforward and also depends on the SED shape through \Mbh\ and \Fedd.

Metal-line bolometric corrections exhibit a similar dependence on \Mbh\ and \Fedd\ as those based on H and He lines. However, their sensitivity to the ionisation parameter is significantly stronger. In particular, the dependence of \lciv, \lciii, and \loiiiopt\ on $U$ reaches $\sim 4$ dex, $\sim 2$ dex, and $\sim 2.5$ dex, respectively. This reflects the strong sensitivity of these lines to the abundance of the corresponding ion species, which varies significantly with $\log U$.
This behaviour is further combined with a strong dependence on metallicity. In general, $\Lacc/\rm{L}_{\rm line}$ decreases with increasing metallicity due to the higher abundance of heavy elements, but can increase again at the highest metallicities as the electron temperature decreases, reducing the emissivity of these collisionally excited lines.
Overall, the strong, multi-parameter dependence of these bolometric corrections implies that simple emission-line calibrations are subject to substantial uncertainties, motivating the need for more sophisticated, physically informed estimates (metallicity, ionisation parameter etc...).

\subsection{Diagnostics for BH seeds}\label{sec:TypeIIdiag}

We now examine line-ratio diagnostic diagrams, with a particular focus on identifying BH seeds among different sources of ionising radiation. For this purpose, we define BH seed like models as a subset of the NLR model grid presented in Section~\ref{sec:method_NLR}, with BH masses in the range $10^{3}$–$10^{5}~\Msun$. 
We further restrict these models to low metallicities, $Z \leq 0.001 \sim 0.1~\zsun$, motivated by the expectation that BH seeds preferentially form in metal-poor environments \citep[see e.g.,][]{Volonteri2021}.
We also define SMBH models as NLR models with BH masses $\log (\Mbh/\Msun) \geq 6$, without imposing any constraint on metallicity.
In general, we find that BH mass appears to have only a minor impact on the line ratios predicted by the NLR models, which are instead primarily driven by metallicity and ionisation parameter.

The fiducial models are computed with gas densities of $\nh = 10^{3}\ \rm cm^{-3}$ for the NLR models and $\nh = 10^{2}\ \rm cm^{-3}$ for the \hii\ region models. However, recent \JWST\ observations have revealed some high-redshift galaxies with inferred gas densities as high as $\gtrsim10^{5}$--$10^{6}\ \rm cm^{-3}$ \citep[e.g.,][]{Senchyna2023,Topping2024,Topping2025}. To investigate the impact of such extreme conditions, we compute additional NLR and \hii\ region models with $\nh = 10^{6}~\rm cm^{-3}$.
Such densities exceed the critical densities of many collisionally excited lines. Assuming an electron temperature of $10^4$ K, the critical densities of the O$^{2+}$ levels $^1D_{2}$ (\oiiioptc) and $^1S_{0}$ (\oiiiopt) are $6.8\times10^5$ and $2.6\times10^7\rm cm^{-3}$, respectively \citep{Osterbrock2006,Telles2014}. As the density increases above the critical density of the $^1D_{2}$ level, \oiiiopt\ emission is suppressed by collisional de-excitation. At the same time, collisional excitation increasingly populates the $^1S_{0}$ level, enhancing \oiiioptc\ \citep[e.g.,][]{Osterbrock2006}. Consequently, the ratio \oiiioptc/\oiiiopt\ increases above densities of $\sim10^{5}\rm cm^{-3}$.
Similarly, the critical densities of O$^+$ $^2D_{5/2}^0$ ([OII]3727), O$^+$ $^2D_{3/2}^0$ ([OII]3729), and N$^+$ $^1D_{2}$ (\niiopt) are $1.5\times10^{4}$, $3.4\times10^{3}$, and $6.6\times10^{4}\rm cm^{-3}$, respectively \citep{Osterbrock2006}. These lines therefore experience significant collisional de-excitation at high density, particularly \loiiopt, significantly decreasing \loiiopt/\loiiiopt\ and \loiiopt/\hg\ ratios (by almost 2~dex from $\nh=10^{3}$ to $10^{6}\rm cm^{-3}$).
The collisional de-excitation of some of these forbidden optical lines also moves the NLR models into the SF region in the P2 versus P1 diagram.
In contrast, the critical densities of most UV lines (\lciv, \lciii, \loiii, \lniii, and \lniv) are of the order of $10^{9}$--$10^{10}\rm cm^{-3}$ \citep{Hamann2002}, so they are only weakly affected by collisional de-excitation. As a result, ratios involving pairs of these lines remain largely unchanged at densities around $10^{6}\rm cm^{-3}$.
Another set of lines strongly affected by the gas density are the optical He~I recombination lines \heiopta, \heioptf, \heiopte, \heioptb, and \heioptc. These transitions can be enhanced by collisional excitation, particularly at high electron temperatures (and thus low metallicity) \citep{Clegg1987}, and are also subject to fluorescence processes, which increase the strength of all these lines except \heiopta\ \citep[e.g.,][]{Izotov1997}.
Overall, increasing the density mainly suppresses low-ionisation optical forbidden lines, while leaving most UV line ratios largely unaffected.
In addition, the two-photon continuum, which dominates the nebular UV continuum, is strongly suppressed at these high densities \citep[e.g.,][]{Osterbrock2006,Raiter2010,Topping2024}. As a result, the UV continuum is reduced in sources with high ionisation parameters, where it is otherwise dominated by nebular emission, leading to larger UV emission-line equivalent widths. In particular, this effect enhances the predicted \heii-EW\ of Pop~III regions \citep[see, e.g.,][]{Rusta2026,Maiolino2026}.

Our goal is to assess whether widely used line-ratio diagnostic diagrams, originally developed to distinguish AGN from SF emission at low redshift, remain applicable to the identification of BH seeds, and to propose alternative diagnostics that provide a better separation. We first discuss diagrams that fail to distinguish BH-seed models from other ionising sources (Sec.~\ref{sec:TypeIIdiagfail}). We then present diagnostics that are degenerate between AGN and Pop.~III emission, as well as diagrams that can distinguish AGN emission at low hydrogen density (Sec.~\ref{sec:TypeIIdiagdeg}). Finally, we investigate diagnostics that provide a better separation between AGN emission and Pop.~II and Pop.~III SF emission (Sec.~\ref{sec:TypeIIdiaggood}). We caution, however, that none of these diagrams can clearly distinguish BH-seed models from other low-metallicity NLR models.

\subsubsection{Diagnostics degenerate between emission from accreting BH seeds and stars}\label{sec:TypeIIdiagfail}

\begin{figure*}
\begin{centering}
\includegraphics[width=1\linewidth]{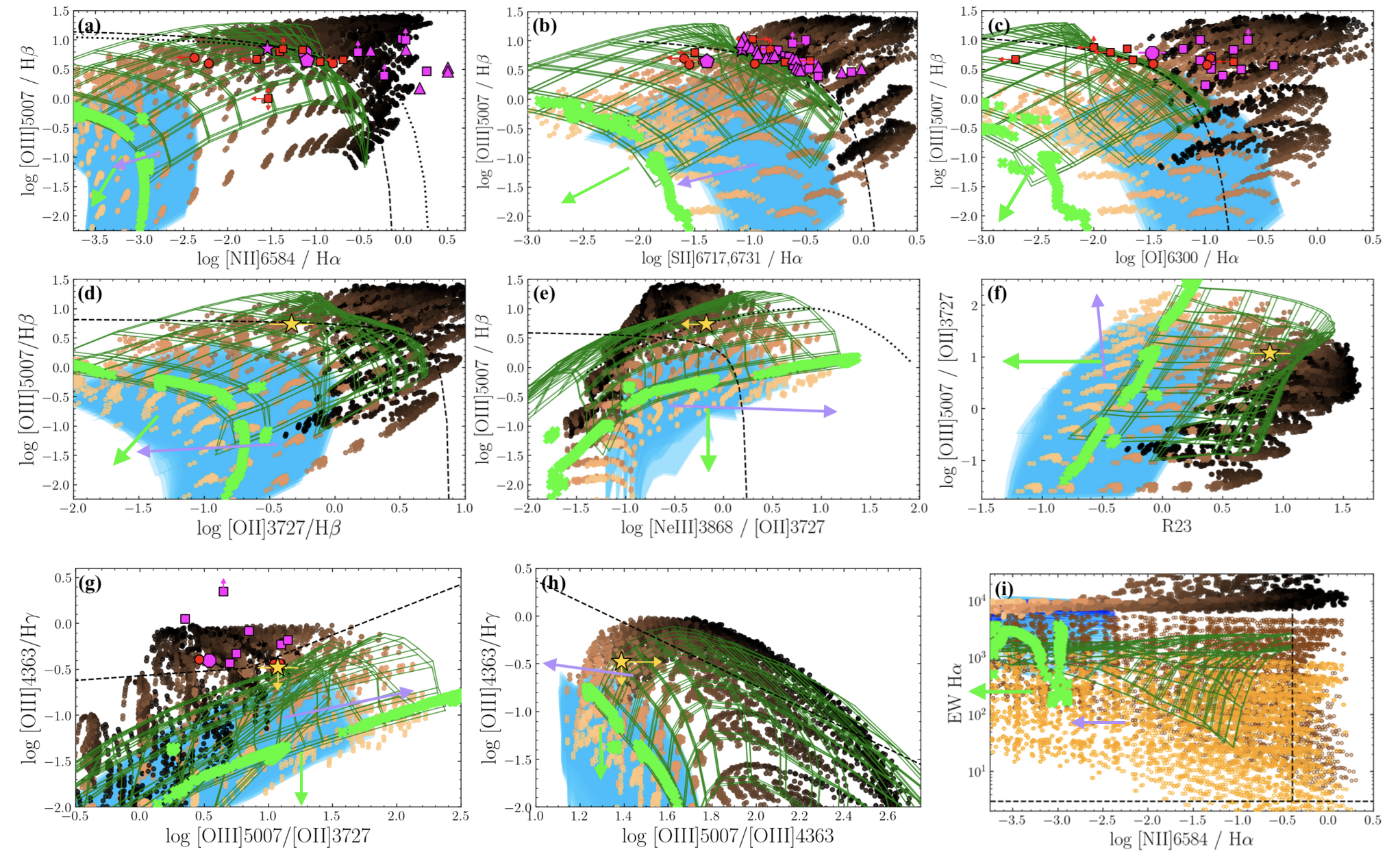}
\includegraphics[width=0.3\linewidth]{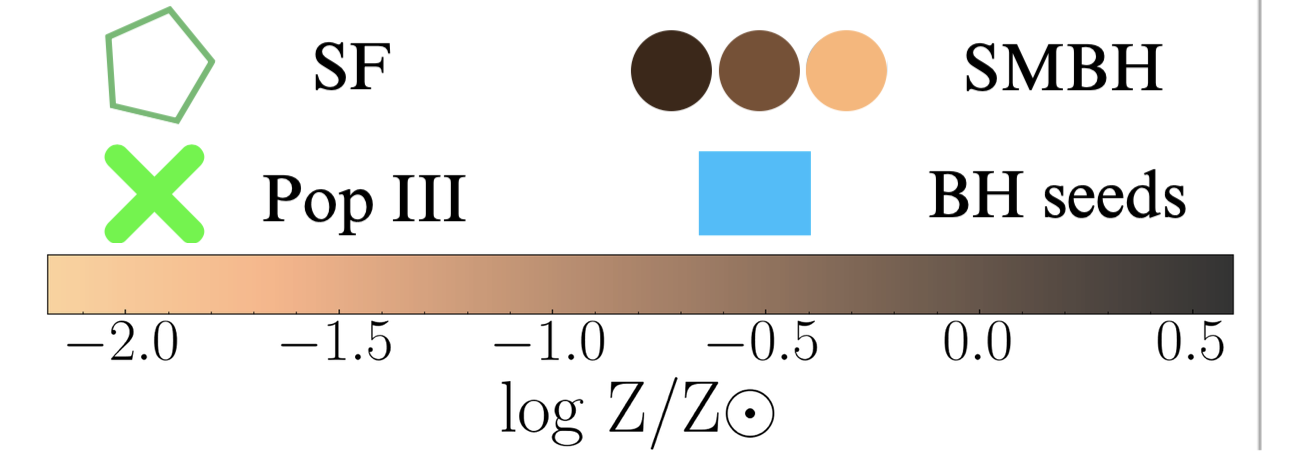}
\includegraphics[width=0.6\linewidth]{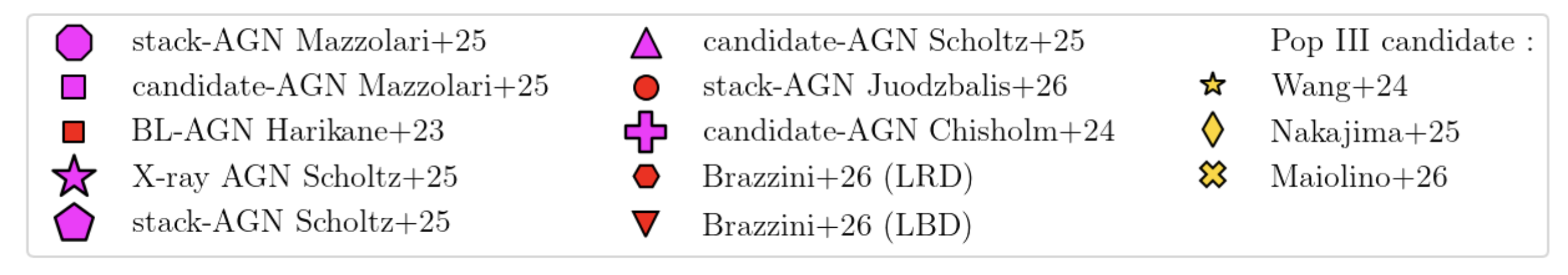}
\par\end{centering}
\caption{Line-ratio diagrams that fail to identify BH seeds. The dark green grid shows the location of star-forming \hii\ region models, while the light green crosses indicate Pop.~III-ionized \hii\ regions. Pop.~III models in a pristine ISM produce extremely weak metal lines and therefore do not appear directly in some of the diagrams; their direction is instead indicated by arrows. In this case, the full range of Pop.~III models thus extends from the visible points (corresponding to slightly enriched ISM) toward the arrow directions.
The orange circles show the predictions for supermassive BH NLR models (i.e. $\log (\Mbh/\Msun) \geq 6$), colour-coded by metallicity from light ($Z = 0.0001$) to dark ($Z = 0.060$). The blue shaded region shows the predictions for BH seed models, corresponding to low-metallicity ($Z \leq 0.001$) and low BH-mass ($\log (\Mbh/\Msun) \leq 5$) NLR models. EWs are computed assuming either pure nebular emission from the NLR (filled symbols) or including the accretion-disc continuum, for NLR covering factors of 0.1 (empty orange circles and light blue line) and 0.9 (empty brown circles and dark blue line).
The pink and red symbols show \JWST\ observations of candidate AGN, selected via narrow lines (pink) and broad lines (red). These include the stacked AGN from \citet{Mazzolari2025} (pink octagon), individual AGN candidates from \citet{Mazzolari2025} (pink squares), broad-line AGN from \citet{Harikane2023} (red squares), X-ray AGN from \citet{Scholtz2025} (pink star), stacked AGN from \citet{Scholtz2025} (pink pentagon), individual AGN candidates from \citet{Scholtz2025} (pink triangles), stacks of BL AGN from \citet{Juodvabalis2026} (red circles) and an AGN candidate from \citet{Chisholm2024} (pink cross). The yellow symbols show candidates Pop~III sources from \citet{Wang2024} (star), \citet{Nakajima2025} (diamond) and \citet{Maiolino2026} (cross).
(a) \oiiiopt/\hb\ versus \niiopt/\ha. 
(b) \oiiiopt/\hb\ versus \siiopt/\ha.
(c) \oiiiopt/\hb\ versus \oiopt/\ha. 
(d) \oiiiopt/\hb\ versus \oiiopt/\hb.
(e) \oiiiopt/\hb\ versus \neiiiopt/\oiiopt.
(f) \oiiiopt/\oiiopt\ versus R23=(\oiiioptd+\oiiopt)/\hb.
(g) \oiiioptc/\hg\ versus \oiiiopt/\oiiopt.
(h) \oiiioptc/\hg\ versus \oiiiopt/\oiiioptc.
(i) \ha-EW versus \niiopt/\ha. 
In each panels, the lines indicate classification criteria separating AGN-dominated from star-forming galaxies: \citet{Kewley2001} (dotted) and \citet{Kauffmann2003} (dashed) in (a), \citet{Kewley2001} in (b) and (c), \citet{Lamareille2010} in (d), \citet{Backhaus2022} (dashed) and \citet{Arevalo2025} (dotted) in (e), \citet{Mazzolari2024} in (g) and (h), and \citet{Stevenson2026} in (i).
The purple arrow shows the direction of high density models.}
\label{fig:NLRdiagbad}
\end{figure*}

We first show in Fig.~\ref{fig:NLRdiagbad} a set of commonly used diagnostic diagrams that fail to distinguish BH seeds from other ionising sources. The BH seed models are shown as a blue shaded region, while the SMBH models are represented by circles, colour-coded by metallicity from $Z = 0.0001$ (light orange) to $Z = 0.060$ (dark brown).
We compute the EWs of the NLR models under two different assumptions for the underlying continuum. First, we assume that the continuum is entirely dominated by the nebular emission from the NLR, corresponding to a type~II AGN in which the UV and optical continuum from the accretion disc is strongly attenuated by the dusty torus. These EWs are shown as filled circles.
We also consider the case in which the accretion-disc continuum is directly visible. In this case, we assume NLR covering factors of 0.9 and 0.1, shown as empty brown and orange circles, respectively.
In addition to these two sets of NLR models, we also include the emission from star-forming \hii\ regions (dark green grid) and Pop~III \hii\ regions (light green crosses).
The effect of increasing density (from $10^{2} - 10^{3}$ to $10^{6}\rm cm^{-3}$) on the predicted emission-line ratios and EWs is illustrated by the purple arrows in the diagnostic diagrams where the density has the largest impact. The length of each arrow indicates approximately by how much the ratios change.
We also show observational samples of high-redshift AGN candidates from \JWST\ studies in the literature, selected via narrow lines (pink symbols) and broad lines (red symbols). These include the stacked AGN from \citet{Mazzolari2025}, individual AGN candidates from \citet{Mazzolari2025}, broad-line AGN from \citet{Harikane2023}, an X-ray AGN source from \citet{Scholtz2025}, stacked AGN from \citet{Scholtz2025}, individual AGN candidates from \citet{Scholtz2025}, stacks of broad-line AGN from \citet{Juodvabalis2026}, broad line sources (an LRD and an LBD) from \citet{Brazzini2026} and an AGN candidate from \citet{Chisholm2024}.
The yellow symbols show candidates Pop~III sources from \citet{Wang2024}, \citet{Nakajima2025} and \citet{Maiolino2026}.

The first diagnostic diagram we explore is the BPT diagram (panel a), \loiiiopt/\hb\ versus \lniiopt/\ha, introduced by \citet{Baldwin81} to distinguish AGN from star-forming (SF) emission. We show the demarcation criteria of \citet{Kewley2001} and \citet{Kauffmann2003}, which separate AGN-dominated from SF-dominated regions. We find that the BH seed models lie deep within the SF-dominated region of the diagram.
This behaviour is driven by the low metallicity of these models. Decreasing the metallicity reduces both \loiiiopt/\hb\ and \lniiopt/\ha\ in NLR and SF models, as fewer metals are available in the gas phase. The effect is even stronger for \lniiopt/\ha, due to the dependence of N/O on metallicity through secondary nitrogen enrichment. We will discuss the impact of elevated \NO\ abundances in Sec.~\ref{sec:discumodeluncertainties}.
This limitation of the BPT diagram in identifying low-metallicity AGN has been previously noted \citep[e.g.,][]{Feltre2016}, implying that a significant fraction of AGN could be missed at high redshift if relying solely on this diagnostic.
Quantitatively, we find that $\sim 80\%$ of the models at $Z = 0.006 \approx 0.4\zsun$, fall outside the AGN region defined by \citet{Kauffmann2003}, and at $Z = 0.004$ when using the criterion of \citet{Kewley2001}.
In addition, very high density ($\nh = 10^6 \ \rm{cm}^{-3}$) reduces \lniiopt/\ha, moving the NLR model closer to the SF region.
Many high-redshift AGN candidates lie in the AGN region, as expected given that this diagnostic was used for their selection. However, a fraction of sources—particularly those identified via broad lines—fall within the SF region of the diagram. This may indicate either a low-metallicity NLR or a significant contribution from star formation. A few sources extend beyond the range covered by the predictions in the AGN region, which could point to enhanced nitrogen abundances compared to those assumed in the models.

In panels (b) and (d), we show \loiiiopt/\hb\ versus \lsiiopt/\ha\ and \loiiiopt/\hb\ versus \loiiopt/\hb, respectively. These diagrams exhibit similar behaviour to the BPT diagram, but are not affected by secondary nitrogen enrichment. As a result, the BH seed models do not lie as deeply within the SF region, since the metallicity dependence is a bit less pronounced. However, they would still be missed due to their low metallicity.
Quantitatively, we find that $\sim 80\%$ of the models at $Z = 0.002$ ($\approx 0.15\sun$) fall outside the AGN region defined by \citet{Kewley2001} in the \loiiiopt/\hb\ versus \lsiiopt/\ha\ diagram, and similarly for the criterion of \citet{Lamareille2010} in the \loiiiopt/\hb\ versus \loiiopt/\hb\ diagram.
In addition, very high density ($\nh = 10^6 \ \rm{cm}^{-3}$) reduces \lsiiopt/\ha, and strongly suppress \loiiopt, moving the NLR model even deeper within the SF region.

The \loiiiopt/\hb\ versus \loiopt/\ha\ diagnostic diagram shown in panel (c) performs somewhat better, with models down to $Z = 0.0005$ lying predominantly within the AGN-dominated region defined by \citet{Kewley2001}. Only the lowest-metallicity BH seed models fail this criterion, while even a modest metal enrichment significantly boosts the \loiopt\ emission, making it much stronger relative to SF emission.
This behaviour arises because X-rays produced by the accretion disc penetrate deeper into the gas, heating an extended partially ionized region and enhancing the collisional excitation of \loiopt. We caution, however, that if the NLR is density-bounded rather than ionisation-bounded, as assumed in these models, the \loiopt\ emission would be reduced, shifting the models back toward the SF region of the diagram.
We also explore X-ray weak accretion-disc SEDs (see Sec.~\ref{sec:discumodeluncertainties}) and find that such ionising spectra result in lower \loiopt/\ha\ ratios that nevertheless remain higher than those of the SF models.
In addition, this diagnostic is degenerate with emission from fast radiative shocks, which also produce strong X-ray emission and therefore enhanced \loiopt/\ha\ ratios. We further discuss this in Sec.~\ref{sec:discumodeluncertainties}.

In panel (e), we show \loiiiopt/\hb\ versus \lneiiiopt/\loiiopt. Most of the parameter space covered by the AGN and SF models overlaps, except for a subset of AGN models with the highest \loiiiopt/\hb\ ratios at moderate \lneiiiopt/\loiiopt. These correspond to models with $Z \gtrsim 0.008$ and $\log U \gtrsim -3.5$, where \loiiiopt\ and \loiiopt\ are boosted in the NLR models relative to the SF models due to the higher electronic temperature and the extended partially ionized region.
The criteria of \citet{Backhaus2022} (dashed line) and \citet{Arevalo2025} (dotted line) are less restrictive, such that part of the SF model grid lies within their AGN regions. However, despite this broader selection, low-metallicity AGN models are still largely missed. More than $80\%$ of the low-metallicity AGN models remain classified as SF, for $Z < 0.001$ or $\log U < -3.5$ using the criterion of \citet{Backhaus2022}, and for $Z < 0.004$ using that of \citet{Arevalo2025}.
In addition, at very high density, the increased \lneiiiopt/\loiiopt\ ratio moves the SF models into the AGN region defined by \citet{Backhaus2022}.

In panel (f), we show the \loiiiopt/\loiiopt\ versus $\rm{R23} = \log( (\oiiioptd+\oiiopt) / \hb)$ diagram. Similar to the previous diagram, as expected from the harder ionising radiation produced by AGN accretion discs, the NLR models generally exhibit larger R23 values at fixed \loiiiopt/\loiiopt\ than the SF models. At high metallicity ($Z > 0.008$), the combination of high metal abundances and elevated electronic temperatures in the NLR models boosts the R23 ratio, producing a region of the diagram not populated by the SF models.
However, this separation disappears for the BH seed models. Due to their low metallicity, they overlap with the SF models at low R23 value.

Panels (g) and (h) show the diagnostic diagrams proposed by \citet{Mazzolari2024}: \oiiioptc/\hg\ versus \oiiiopt/\oiiopt\ and \oiiioptc/\hg\ versus \oiiiopt/\oiiioptc, respectively. Only models with $Z \geq 0.004$ and $\log U \geq -3$ satisfy the AGN selection criteria, again missing the low-metallicity AGN and BH seed models.
The demarcation in panel (g) at least excludes the SF models. This is because the harder AGN SED produces higher electron temperatures at high metallicity, boosting the \oiiioptc\ emission, while the extended partially ionized region decreases \oiiiopt/\oiiopt\ at high ionisation parameter. 
An increase in density moves the NLR models towards higher \oiiioptc/\hg\ and \oiiiopt/\oiiopt, as well as lower \oiiiopt/\oiiioptc, along the demarcation criteria from \citet{Mazzolari2024}. 
This is also the case of the SF models. However a combination of low and high density \hii\ regions can shift the SF models in the AGN only region of the diagram (see Section~\ref{sec:discumodeluncertainties}). 
In contrast, the diagnostic shown in panel (h) does not cleanly separate AGN from SF models, as the AGN models are shifted approximately diagonally toward higher \oiiioptc, broadly following the demarcation line itself.

In panel (i), we show \ha-EW as a function of \lniiopt/\ha, together with the criterion of \citet{Stevenson2026} used to distinguish AGN from star-forming galaxies. Most AGN models fall within the SF region defined by this diagnostic, with only models at relatively high metallicity ($Z > 0.020$) lying within the AGN region. Lower-metallicity AGN, including the BH seed models, remain classified as SF.

In conclusion, all of the diagnostic diagrams discussed above select only a subset of AGN, primarily those with relatively high metallicity. As a result, they would miss a significant population of low-metallicity AGN, which is particularly problematic when attempting to identify BH seeds at high redshift.

\subsubsection{Diagnostics degenerate with Population~III emission and gas density}\label{sec:TypeIIdiagdeg}

\begin{figure*}
\begin{centering}
\includegraphics[width=1\linewidth]{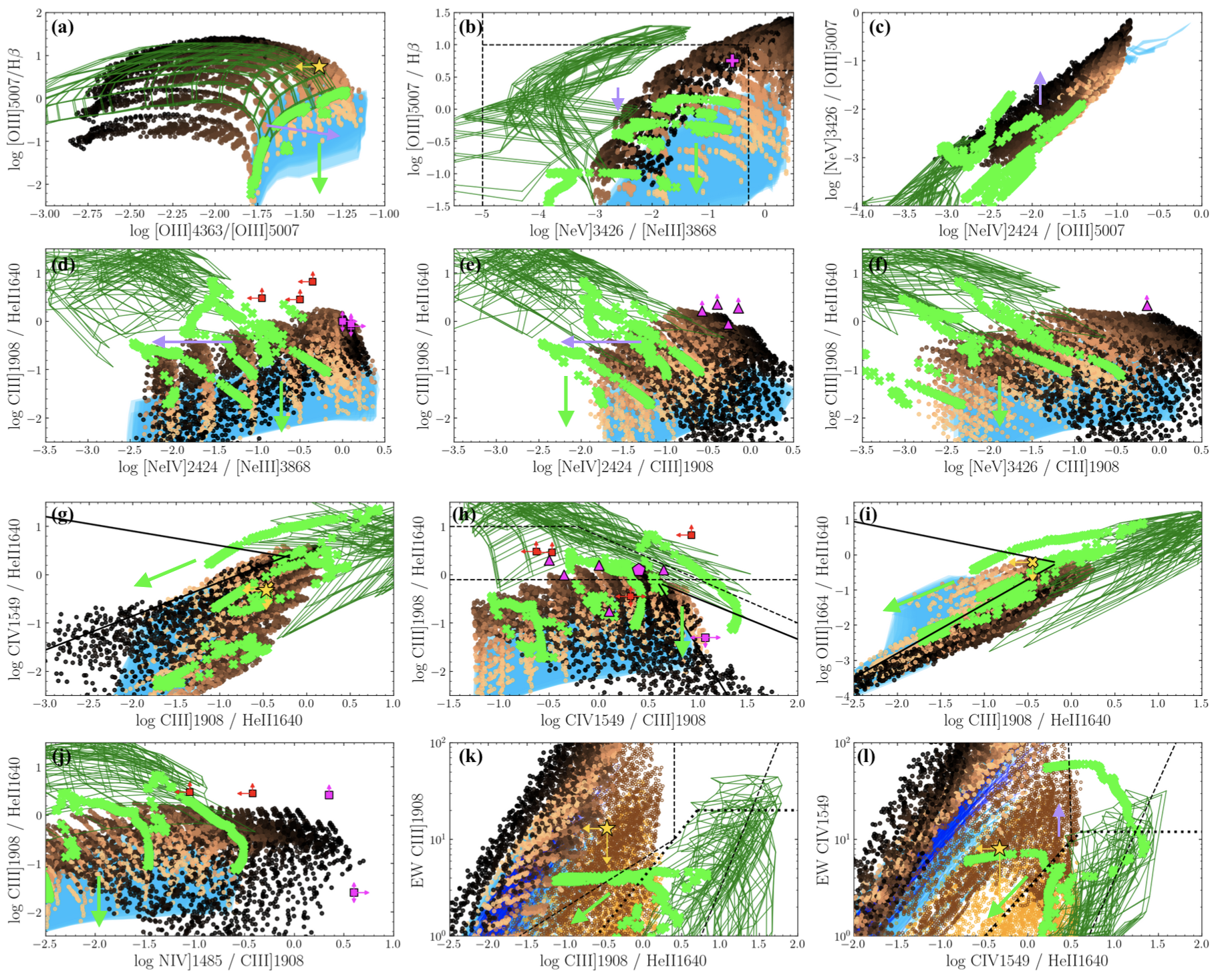}
\includegraphics[width=0.3\linewidth]{fig/NLR_diag_leg_mod.png}
\includegraphics[width=0.6\linewidth]{fig/NLR_diag_leg_obs.png}
\par\end{centering}
\caption{Line-ratio diagnostic diagrams in which AGN are partially degenerate with Pop-III \hii\ regions. Symbols and colors as in Fig.~\ref{fig:NLRdiagbad}.
(a) \oiiiopt/\hb\ versus \oiiioptc/\oiiiopt.
(b) \oiiiopt/\hb\ versus \nev/\neiiiopt.
(c) \nev/\oiiiopt\ versus \neiv/\oiiiopt.
(d) \ciii/\heii\ versus \neiv/\neiiiopt.
(e) \ciii/\heii\ versus \neiv/\ciii.
(f) \ciii/\heii\ versus \nev/\ciii.
(g) \ciii/\heii\ versus \civ/\heii.
(h) \ciii/\heii\ versus \civ/\ciii.
(i) \oiii/\heii\ versus \ciii/\heii.
(j) \ciii/\heii\ versus \nivb/\ciii.
(k) \ciii-EW versus \ciii/\heii.
(l) \civ-EW versus \civ/\heii.
In panel (b), the lines show the demarcation criteria of \citet{Cleri2023}, separating star-forming galaxies, AGN, composite sources, and emission from Pop~III or intermediate-mass BHs. In panels (h), (k), and (l), the dashed lines indicate the criteria of \citet{Hirschmann2019}, separating star-forming galaxies, AGN, and composite regions, while the dotted lines in (k) and (l) show the criteria of \citet{Nakajima2018}, distinguishing star-forming galaxies from AGN. In panel (g), (i) and (j), the solid lines show the demarcation criteria of \citet{Rusta2025} to identify Pop~III stars. The purple arrow shows the direction of high density models.}
\label{fig:NLRdiagdegpopiii}
\end{figure*}

\begin{figure*}
\begin{centering}
\includegraphics[width=1\linewidth]{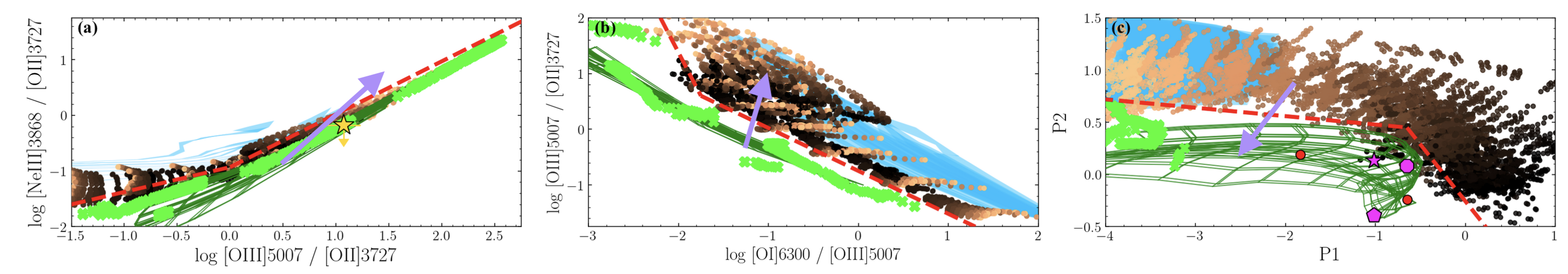}
\includegraphics[width=0.3\linewidth]{fig/NLR_diag_leg_mod.png}
\includegraphics[width=0.5\linewidth]{fig/NLR_diag_leg_obs.png}
\par\end{centering}
\caption{Diagrams in which most NLR models are separated from star-forming (SF) regions, including Pop.~III, for gas densities of $\nh = 10^{2}$--$10^{3}~\rm cm^{-3}$. Symbols and colours are the same as in Fig.~\ref{fig:NLRdiagbad}. The red lines indicate the proposed demarcation between AGN-dominated regions and regions degenerate with emission from star-forming galaxies (Pop.~I, II and III). At higher densities ($\nh = 10^{6}~\rm cm^{-3}$), these diagnostics become strongly degenerate, reducing their ability to distinguish AGN from star formation.
(a) \neiiiopt/\oiiopt\ versus \oiiiopt/\oiiopt.
(b) \oiiiopt/\oiiopt\ versus \oiopt/\oiiiopt.
(c) P2 = -0.63*log(\lniiopt/\ha) + 0.78*log(\lsiiopt/\ha) versus P1  = 0.63*log(\niiopt/\ha) + 0.51*log(\lsiiopt/\ha) + 0.59*log(\loiiiopt/\hb), as defined in \citet{Ji2020}.
}
\label{fig:NLRdiagdegnh}
\end{figure*}

We now turn to diagnostic diagrams that are more successful at separating AGN from typical star-forming galaxies, but still show a strong degeneracy with Pop~III stellar populations. These diagrams are shown in Fig.~\ref{fig:NLRdiagdegpopiii}.

In panel (a), we show \oiiiopt/\hb\ versus \oiiioptc/\oiiiopt. The BH seed models occupy the lower-right region of the diagram, together with the Pop~III models. This is driven by their low metallicity and hard ionising radiation field, which decreases \oiiiopt/\hb\ by favouring higher ionisation species, while increasing \oiiioptc/\oiiiopt\ through the elevated electron temperatures reached at low metallicity.
However, this diagnostic is also sensitive to gas density. At densities above $10^{5}\ \rm{cm}^{-3}$, both the SF and NLR models are shifted toward higher \oiiioptc/\oiiiopt\ ratio (see discussion above), further preventing the identification of BH seed in this diagram.

In panel (b), we show the diagnostic diagram proposed by \citet{Cleri2023}, originally designed to distinguish between SF galaxies (left), composite sources (middle), AGN (upper right), and Pop~III/IMBH sources (lower right). We find that many SF models extend into both the composite and AGN regions, while a substantial fraction of the AGN models also falls within the composite region.
Nevertheless, the AGN and Pop~III models are generally offset from the bulk of the SF models toward higher \lnev/\lneiiiopt\ ratios at fixed \loiiiopt/\hb, reflecting their harder ionising radiation fields. The AGN region defined by \citet{Cleri2023} mainly selects NLR models with the highest ionisation parameters.
The BH seed and Pop~III models tend to exhibit low \loiiiopt/\hb\ ratios due to their low metallicities, but span a broad range of \lnev/\lneiiiopt\ depending on ionisation parameter, causing them to lie partly within the composite region.
The narrow-line IMBH candidate of \citet{Chisholm2024} lies in the region occupied by the AGN models.

Panel (c) shows \lnev/\loiiiopt\ versus \lneiv/\loiiiopt. Owing to their harder ionising radiation fields, both the AGN and Pop~III models are shifted toward the upper-right region of the diagram relative to the bulk of the SF models. The BH seed models extend to even higher values of these ratios, since low-mass BH accretion discs generally produce harder ionising spectra than higher-mass BHs.
This diagram therefore shows a partial degeneracy between Pop~III and AGN models, although a fraction of the AGN models still occupies a distinct region of the parameter space.

Panels (d), (e), (f), (h) and (j) show \lciii/\lheii\ as a function of \lneiv/\lneiiiopt, \lneiv/\lciii, \lnev/\lciii, \lciv/\lciii\ and \lniv/\lciii\ respectively. Compared to standard SF models, both AGN and Pop~III models produce stronger \lheii\ emission, as well as enhanced \lneiv\ and \lnev\ emission, due to their harder ionising spectra.
The narrow-line-selected AGN candidates predominantly lie within the AGN region of these diagrams, whereas the broad-line-selected AGN tend to be more consistent with SF or composite AGN/SF emission. Some sources lie at the edge of the \lniv/\lciii\ produced by the models, which could point to elevated nitrogen abundances.
At high density, \lneiv/\lneiiiopt\ and \lneiv/\lciii\ significantly decrease.

Panels (g) and (i) show \lciv/\lheii\ and \loiii/\lheii\ as a function of \lciii/\lheii, respectively. In these diagrams, the AGN and Pop~III models separate clearly from the SF models. However, they remain highly degenerate with each other, since both produce strong \lheii\ emission, and since the metal-to-\lheii\ line ratios in the Pop~III models approach zero in the pristine ISM case.

Lastly, panels (k) and (l) show EW-\lciii\ as a function of \lciii/\lheii\ and EW-\lciv\ as a function of \lciv/\lheii, respectively. These diagrams also exhibit a partial degeneracy between AGN and Pop~III emission. Although the AGN models can reach larger EWs, they also extend to lower values, particularly depending on the assumed NLR covering factor. For the lowest covering factors, some AGN models even fall within the SF regions defined by \citet{Nakajima2018} and \citet{Hirschmann2019,Hirschmann2023}.

In Fig.~\ref{fig:NLRdiagdegnh}, we present diagnostic diagrams that provide a good separation between AGN and star-forming (SF) emission, including Pop.~III, for gas densities of $\nh = 10^{2}$--$10^{3}\ \rm cm^{-3}$. However, at higher densities ($\nh = 10^{6}\ \rm cm^{-3}$), either the SF models shift into the AGN region or the AGN models move into the SF region, significantly increasing the overlap between the two populations. Consequently, these diagnostics can only be applied reliably if the gas density is known independently.
New demarcation lines defining regions dominated by AGN emission in the low density regime are presented in Table~\ref{tab:diagNLRgood}.

In panels (a), we show \lneiiiopt/\loiiopt\ as a function of \loiiiopt/\loiiopt. The AGN models are shifted toward higher \lneiiiopt/\loiiopt\ ratios at fixed \loiiiopt/\loiiopt, especially at low ionisation parameter and for some low-BH-mass models, reflecting their more extended Ne$^{2+}$ zones compared to the SF models. However, at high density, the suppression of \loiiopt\ moves the SF models in the AGN region.

We show \loiiiopt/\loiiopt\ as a function of \loiopt/\loiiiopt\ in panel (b). This diagram provides a good separation between AGN and SF emission, due to the stronger \loiopt\ emission produced in AGN. However, a decrease in \loiopt\ emission in the case of a density-bounded NLR would move the AGN models into the SF region (see Section~\ref{sec:discumodeluncertainties}). In addition, increasing the density moves the SF models in the AGN region, due to the reduction of \loiiopt\ emission.

Finally, in panel (c), we show the reprojection diagram introduced by \citet{Ji2020}, with $P2 = -0.63 \times \log(\lniiopt/\ha) + 0.78 \times \log(\lsiiopt/\ha)$ as a function of $P1 = 0.63 \times \log(\lniiopt/\ha) + 0.51 \times \log(\lsiiopt/\ha) + 0.59 \times \log(\loiiiopt/\hb)$. This diagram provides a good separation between the AGN and SF models at low gas densities. However, increasing the density shifts both AGN and SF models towards lower $P1$ and $P2$ values, causing the AGN models to overlap with the SF region.
Interestingly, both the narrow-line-selected and broad-line-selected high-redshift AGN candidates tend to lie within the SF region of this diagram. This may indicate that these sources have higher gas densities than typically assumed for local AGN. However, several other effects could also contribute to this discrepancy, including contamination by SF emission, enhanced nitrogen abundances, or a density-bounded NLR (see Section~\ref{sec:discumodeluncertainties}).

\subsubsection{Robust diagnostics of AGN emission}\label{sec:TypeIIdiaggood}

\begin{figure*}
\begin{centering}
\includegraphics[width=1\linewidth]{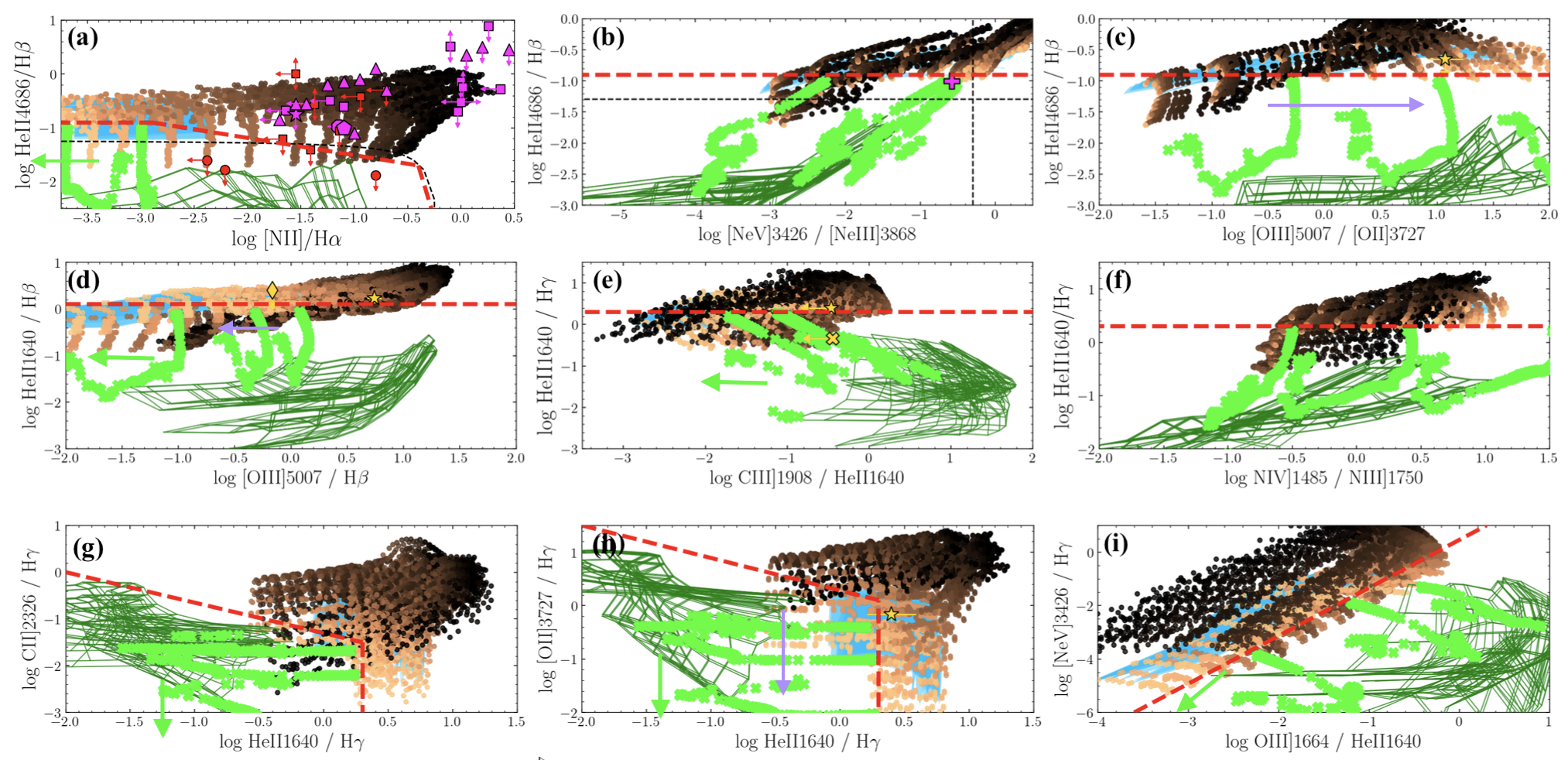}
\includegraphics[width=1\linewidth]{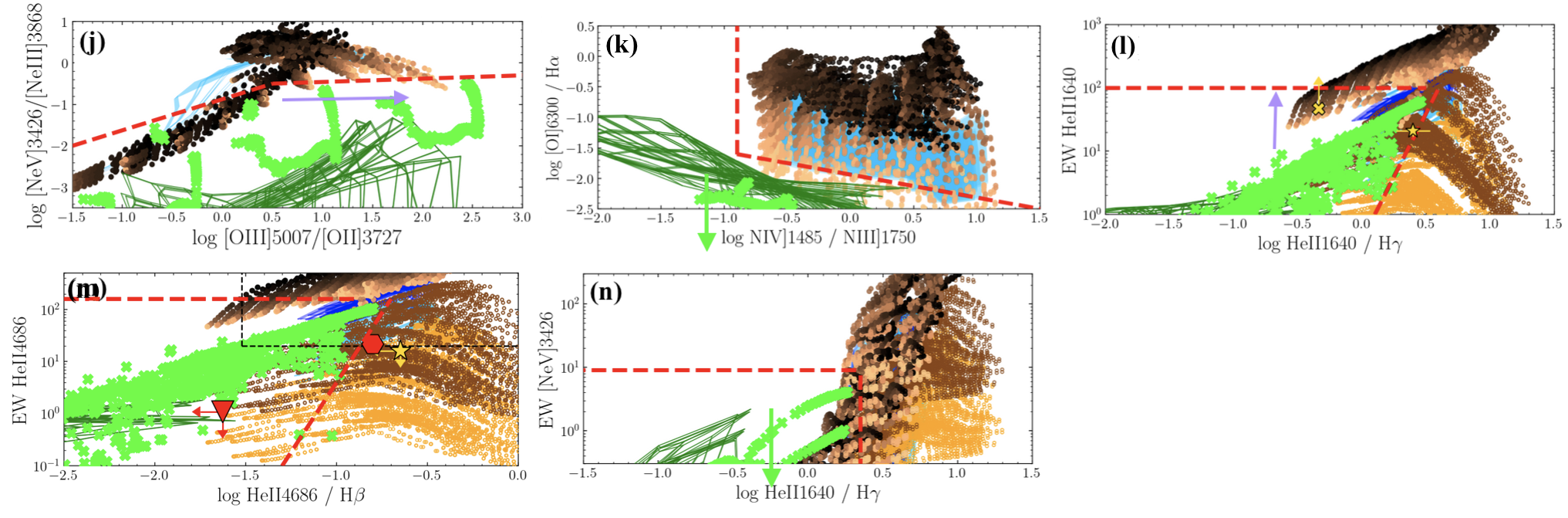}
\includegraphics[width=0.3\linewidth]{fig/NLR_diag_leg_mod.png}
\includegraphics[width=0.5\linewidth]{fig/NLR_diag_leg_obs.png}
\par\end{centering}
\caption{Diagrams in which most of the NLR models are separated from SF-dominated regions (including Pop.~III). Symbols and colors as in Fig.~\ref{fig:NLRdiagbad}. The red lines indicates the proposed demarcations between regions dominated by AGN emission and regions which are degenerate with emission from star-forming regions (Pop I and II and Pop III).
(a) \heiiopt/\hb\ versus \niiopt/\ha.
(b) \heiiopt/\hb\ versus \nev/\neiiiopt.
(c) \heiiopt/\hb\ versus \oiiiopt/\oiiopt.
(d) \heii/\hb\ versus \oiiiopt/\hb.
(e) \heii/\hg\ versus \ciii/\heii.
(f) \heii/\hg\ versus \nivb/\niii.
(g) \cii/\hg\ versus \heii/\hg.
(h) \oiiopt/\hg\ versus \heii/\hg.
(i) \nevopt/\hg\ versus \oiii/\heii.
(j) \nevopt/\neiiiopt\ versus \oiiiopt/\oiiopt.
(k) \oiopt/\ha\ versus \nivb/\niii.
(l) \heii-EW versus \heii/\hg.
(m) \heiiopt-EW versus \heiiopt/\hb.
(n) \nev-EW versus \heii/\hg.
Dashed lines indicate the demarcation criteria separating SF and AGN: \citet{Shirazi2012} in panels (a) and (b), and \citet{Cleri2023} in panel (b) (as used in \citet{Chisholm2024}). In panel (m), the dashed line shows the \citet{Nakajima2022b} criterion used to identify Pop~III emission. The purple arrow shows the direction of high density models.}
\label{fig:NLRdiaggood}
\end{figure*}

\begin{table*}
    \centering
    \caption{Demarcation criteria to select AGN dominated region from line ratio diagrams presented in Fig.~\ref{fig:NLRdiaggood} and Fig.~\ref{fig:NLRdiagdegnh}. (*) the demarcation line for these diagrams are valid only in the low density regime (see Fig.~\ref{fig:NLRdiagdegnh} and text for detail).}
    \label{tab:diagNLRgood}
    \begin{tabular}{lcc}
        \hline
        Diagram & AGN Criteria  \\
        \hline
        log \heiiopt/\hb\ versus log \niiopt/\ha & $x \leq -2.9$ : $y \geq -0.9$\\
        & $-2.9 \leq x \leq -0.4$ : $y \geq -0.32 x -1.83$\\
        & $x \geq -0.4$ : $y \geq -6.5 x -4.3$ \\
        \hline
        log \heiiopt/\hb\ versus log \nevopt/\neiiiopt\ & $y \geq -0.9$ \\
        \hline 
        log \heiiopt/\hb\ versus log \oiiiopt/\oiiopt\ & $y \geq -0.9$ \\
        \hline 
        log \heii/\hb\ versus log \oiiiopt/\hb\ & $y \geq 0.1$ \\
        \hline 
        log \heii/\hg\ versus log \ciii/\heii\ & $y \geq 0.3$ \\
        \hline 
        log \heii/\hg\ versus log \nivb/\niii\ & $y \geq 0.3$ \\
        \hline 
        log \cii/\hg\ versus log \heii/\hg\ & $x \leq 0.3$ : $y \geq -0.65x -1.3$ \\
         & $x \geq 0.3$ \\
         \hline
         log \oiiopt/\hg\ versus log \heii/\hg\ & $x \leq 0.3$ : $y \geq -0.61x +0.28$ \\
         & $x \geq 0.3$ \\ 
         \hline
        log \nevopt/\hg\ versus log \oiii/\heii\ & $y \geq 1.79x +0.46$ \\
         \hline
         log \nevopt/\neiiiopt\ versus log \oiiiopt/\oiiopt\ & $x \leq 0.5$ : $y \geq 0.75x -0.88$ \\
         & $x \geq 0.5$ : $y \geq 0.08x -0.54$\\    
         \hline
         log \neiiiopt/\oiiopt\ versus log \oiiiopt/\oiiopt\ (*) & $x \leq 0$ : $y \geq 0.45x -0.93$ \\
         & $x \geq 0$ : $y \geq 0.95x -0.93$\\     
         \hline
         log \oiopt/\ha\ versus log \nivb/\niii\ & $x \geq -0.9$ : $y \geq -0.38x -1.94$ \\  
         \hline
         log \oiiiopt/\oiiopt\ versus log \oiopt/\oiiiopt\ (*) & $x \leq -1.75$ : $y \geq -3.11x -4.84$ \\
         & $x \geq -1.75$ : $y \geq -0.77x -0.75$\\     
         \hline
         $P2$ versus $P1$ (*) & $x \leq -0.65$ : $y \geq -0.08x +0.4$ \\
         & $x \geq -0.65$ : $y \geq -1.09x -0.26$\\ 
         \hline 
         EW-\heii\ versus log \heii/\hg\ & $y \geq 100$ \\
         & or $\log y \leq 4.0x -0.4$ \\
         \hline 
         EW-\heiiopt\ versus log \heiiopt/\hb\ & $y \geq 160$ \\
         & or $\log y \leq 5.33x +5.93$ \\
         \hline 
         EW-\nevopt\ versus log \heii/\hg\ & $y \geq 9$ \\
         & or $x \geq 0.3$ \\
         \hline 
    \end{tabular}
\end{table*}

In Fig.~\ref{fig:NLRdiaggood}, we present a set of diagnostic diagrams in which most AGN models occupy regions distinct from both the standard SF and Pop~III models. We also propose new demarcation lines to define regions that exclude the majority of non-AGN emission. These demarcation criteria are presented in Table~\ref{tab:diagNLRgood}.
Importantly, these diagrams are much less sensitive to the gas density, making them more reliable diagnostics when no independent density measurement is available.

Panels (a), (b), and (c) show \heiiopt/\hb\ as a function of \lniiopt/\ha, \lnev/\lneiiiopt, and \loiiiopt/\loiiopt, respectively. These diagrams provide a good separation between AGN and SF models, as the AGN models produce stronger \heiiopt/\hb\ ratios, even relative to the Pop~III models.
While the AGN models also tend to exhibit slightly higher \lniiopt/\ha\ and \lnev/\lneiiiopt\ ratios than the SF models, the primary distinction in these diagrams comes from the elevated \heiiopt/\hb\ ratios of the AGN models. As a result, \heiiopt/\hb\ can in principle be combined with a wide range of other line ratios to distinguish AGN emission.
Many of the narrow-line-selected AGN candidates lie above the \citet{Shirazi2012} demarcation criterion in panel (a), whereas some of the broad-line-selected sources fall below it. This could indicate either contamination from SF emission or an additional process suppressing \lheii\ emission in high-redshift AGN \citep[see also][]{Tozzi2023}, as further discussed in Section~\ref{sec:discussion}.
Panels (d), (e), and (f) show very similar behaviour. They present \heii/\hb\ as a function of \loiiiopt/\hb, and \heii/\hg\ as a function of \lciii/\heii\ and \lniv/\lniii, respectively. These three diagnostics provide a good separation between AGN emission and both standard SF and Pop~III emission, owing to the stronger \heii\ emission relative to the hydrogen Balmer lines in the AGN models.

Panels (g) and (h) show \lcii/\hg\ and \loiiopt/\hg\ as a function of \heii/\hg, respectively. The combination of elevated \heii/\hg\ ratios with a ratio of low-ionisation, collisionally excited line to hydrogen Balmer line shifts the AGN models toward the upper-right region of the diagrams relative to the SF models, again providing an efficient diagnostic for identifying AGN emission.

In panel (i), we show \lnev/\hg\ as a function of \loiii/\lheii. This diagram separates AGN from SF emission because the harder ionising radiation field of AGN produces stronger high-ionisation lines, in particular \lnev\ and \lheii.

In panels (j), we show \lnev/\lneiiiopt\ as a function of \loiiiopt/\loiiopt. The AGN models produce stronger \lnev\ emission than the SF models, particularly at fixed \loiiiopt/\loiiopt. 

We show \loiopt/\ha\ as a function of \lniv/\lniii\ in panel (k). This diagram also provides a good separation between AGN and SF emission, owing to the strong \loiopt\ emission produced in the AGN models by the extended partially ionized regions created by penetrating X-ray radiation. We caution however that \loiopt\ emission might be reduced if the NLR is density-bounded (see Section~\ref{sec:discumodeluncertainties}).

In panels (l), (m), and (n), we show \heii-EW\ as a function of \heii/\hg, \heiiopt-EW\ as a function of \heiiopt/\hb, and \lnev-EW\ as a function of \heii/\hg, respectively. Owing to their hard ionising radiation fields, the AGN models populate the upper-right regions of these diagrams, characterized by strong high-ionisation emission lines.
We note, however, that the LRD source from \citet{Brazzini2026} falls in the composite / SF region of the \heiiopt-EW\ as a function of \heiiopt/\hb\ diagram. We discuss this further in Section~\ref{sec:discuhighz}.
As noted by \citet{Rusta2026,Maiolino2026}, increasing the electron density shifts the Pop~III models toward higher \heii-EW, due to the reduced contribution from the two-photon continuum.
We also note some differences with the Pop~III region defined by \citet{Nakajima2022b} in panel (m). This mainly arises from two effects. First, their prescription for computing AGN EWs only considers the type~I AGN case, in which the accretion disc contributes directly to the continuum emission. Here, we consider both this configuration and the type~II AGN case, where the continuum is dominated by nebular emission alone. Second, the SEDs adopted in this work have lower UV/optical continuum levels at fixed ionising radiation, resulting in stronger predicted EWs. As a consequence, part of the AGN model grid overlaps with the Pop~III region defined by \citet{Nakajima2022b}.

In all of the diagnostic diagrams presented above, the emission from the Pop~III candidates of \citet{Wang2024}, \citet{Nakajima2025}, and \citet{Maiolino2026} are broadly consistent with a combination of Pop~III and standard SF emission, but are also compatible with pure NLR emission or a mixture of NLR and SF emission.
From the diagnostic diagrams alone, a combination of NLR and SF emission often appears to provide a good match to the observations. However, it is difficult to determine whether the same combination of NLR and SF models can simultaneously reproduce all of the observed line ratios, or whether different diagrams favour different solutions. A Bayesian fitting approach applied directly to the full set of emission lines (e.g. using \beagle) would allow one to test whether a self-consistent NLR+SF solution exists for each source, or whether more exotic solutions, such as Pop~III emission are required.

We caution that the diagnostics explored above are sensitive to several modelling uncertainties, which we further discuss in Section~\ref{sec:discumodeluncertainties} for both the AGN and SF models. These include the shape of the ionising radiation field, multiple density components, elemental abundances, and the possible contribution from fast radiative shocks.

\section{Discussion}\label{sec:discussion}

The models presented in this work provide a unified framework for interpreting the broad- and narrow-line emission of high-redshift AGN, but their predictions inevitably depend on assumptions regarding the ionising radiation field and the physical conditions of the emitting gas. In this section, we first assess the main modelling uncertainties and their impact on the predicted emission-line properties (Section~\ref{sec:discumodeluncertainties}). We then explore whether plausible alternative assumptions compared to our fiducial models can help explain some of the unusual spectral properties observed in faint high-redshift AGN, in particular their weak high-ionisation emission lines (Section~\ref{sec:discuhighz}).

\subsection{Modelling uncertainties}\label{sec:discumodeluncertainties}

Our fiducial grids necessarily adopt simplified prescriptions for several physical ingredients that remain poorly constrained, particularly for the massive black holes now being uncovered at high redshift. These assumptions affect both the shape of the ionising radiation field and the response of the photoionized gas, and may therefore introduce systematic uncertainties in the predicted line luminosities, line ratios, and equivalent widths. Here, we examine the sensitivity of our results to the adopted metal abundance and dust prescriptions, microturbulence and dissipative heating, gas column density and the escape of ionising photons, gas density and cloud distributions, and the shape of the incident ionising spectrum. We also consider uncertainties in the stellar photoionisation models and the possible contribution of fast radiative shocks. Our aim is not to exhaustively explore the full parameter space, but rather to identify which assumptions can significantly affect the diagnostics and interpretations developed in the previous sections, and which leave our main conclusions largely unchanged.

\subsubsection{Metal and Dust prescription}

In this work, we have only presented SF and NLR models with $\xid = 0.3$. The impact of this parameter has been explored in detail by, e.g., \citet{Gutkin2016,Feltre2016,Plat2019}; it is not well constrain from observations and is strongly degenerate with other parameters. In summary, increasing \xid\ decreases the abundance of metals in the gas phase, which reduces the strength of metal lines for refractory elements, particularly at low metallicity. At high metallicity, however, the reduction of gas-phase coolants and the additional heating provided by dust increase the electron temperature, enhancing collisionally excited lines. In addition, the absorption of ionising photons near the Lyman limit by dust tends to suppress H recombination lines relative to collisionally excited lines and He~II recombination lines.
These effects may slightly shift the predicted line ratios. Yet, they are generally smaller than those induced by the other parameters explored in this work and should not significantly affect any of our main conclusions, in particular about the line-ratio diagnostics presented in Section~\ref{sec:TypeIIdiag}.

Another factor influencing the line emission is the adopted abundance pattern, in particular the \CO\ and \NO\ ratios. In this work, we have considered models with $\CO=\COsol$ and $\CO=0.52~\COsol$, which already span a large fraction of the observed scatter in \CO\ abundances. More extreme values would affect carbon lines and weaken some of the proposed diagnostic diagrams. In particular, the \lciv/\lheii\ and \lciii/\lheii\ ratios depend strongly on \CO: very low \CO\ values can shift SF models toward the AGN region, while very high \CO\ values can move AGN models toward the SF region. It is therefore useful to constrain the \CO\ ratio before applying these diagnostics, or to complement them with line ratios involving \lheii\ and H recombination lines, which primarily trace the hardness of the ionising radiation field and are much less affected by the abundance pattern. In this context, Bayesian fitting tools such as \beagle\ are particularly valuable, as they can simultaneously fit multiple emission lines and infer both the AGN contribution and the \CO\ ratio (if carbon and oxygen lines are included).

In our models, we assume a fixed relation between \NO\ and \logoh\ (see Section~\ref{sec:method}). However, recent \JWST\ observations have revealed that some high-redshift galaxies exhibit elevated \NO\ ratios at low metallicity \citep[e.g.,][]{Bunker2023,Senchyna2023,Marques2024,Topping2024,Topping2025,Morel2025,Napolitano}. To first order, the fluxes of nitrogen lines scale approximately linearly with the nitrogen abundance. Therefore, adopting a value representative of the extreme N-emitters observed at high redshift, at $\logoh=7.5$, increasing $\log(\NO)$ from the fiducial value ($\sim -1.5$) to $-0.5$ would increase the nitrogen-line fluxes, and hence their detectability with \JWST, by approximately the same amount, i.e. by about 1 dex.
In addition, elevated nitrogen abundances would modify some of the diagnostic diagrams presented in Section~\ref{sec:TypeIIdiag}. At high redshift, nitrogen enrichment appears to be associated primarily with the high-ionisation zone, and would therefore preferentially enhance \lniv\ and \lniii\ emission. However, an elevated \NO\ abundance in the low-ionisation zone would strengthen \lniiopt\ emission and thus shift the location of sources in the BPT diagram. In particular, SF models with enhanced \NO\ could move into the AGN region of the diagram.
The $P2$ versus $P1$ diagram would also be affected, as enhanced \lniiopt\ emission would shift AGN models toward higher $P1$ and lower $P2$ values. As a result, some low-metallicity AGN models could move into the SF region of the diagram. Again a fit of multiple emission lines at once, could help constrain both AGN fraction and \NO\ abundance.

\subsubsection{Microturbulence and dissipative heating}

Microturbulence enhances resonant lines through resonance scattering and continuum pumping, while dissipative heating increases the gas temperature and thus strengthens some collisionally excited lines \citep[e.g.,][]{Kraemer1999,Kraemer2007,Kraemer2012,Bottorff2002}. These effects may affect both the NLR and BLR. However, \citet{Kraemer1999} noted that these effects are too small to explain the very bright narrow \lnv\ relative to \lniv\ observed in some AGN. 
Following \citet{Mignoli2019}, we include microturbulence and dissipative heating in the NLR models, as they found that these processes improve the reproduction of the \lnv\ emission observed in type~II AGN at redshifts $\sim 1$--3. 

For the BLR, we include microturbulence in both the fiducial multi-cloud and single-cloud Balmer-break models, considering $\vturb=0$ and $250\ \mathrm{km~s^{-1}}$, without dissipative heating. We additionally explore the impact of dissipative heating following the prescription of \citet{Bottorff2002}, assuming a dissipation length scale equal to the BLR cloud size.
The impact of dissipative heating depends on the gas and ionisation conditions of each cloud ($\Phi$, \nh), with some combinations showing very little impact. However, we find that, in some cases, this additional heating term enhances the hydrogen Balmer lines relative to the \lheii\ lines, as it primarily strengthens low-ionisation and collisionally excited lines \citep[e.g.,][]{Bottorff2002}.

In addition, microturbulence may play a role in the context of LRDs, as it affects Balmer absorption and the shape of the Balmer break \citep[e.g.,][]{Ji2025}. We defer a detailed analysis of LRD spectra, including variations in the microturbulent velocity, to future work.

\subsubsection{Gas column density and escape of ionising photons} 

In the NLR and SF \hii\ region models, the calculations are stopped when the electron fraction drops below 1 per cent, corresponding to ionisation-bounded nebulae. If the hydrogen column density is smaller than the ionisation-bounded limit, the clouds become density bounded and a fraction of the ionising radiation can escape.
This has important consequences for the emission-line spectrum, as the outer regions of the cloud, where low-ionisation species are produced, are truncated. As a result, ratios involving high- to low-ionisation species are enhanced \citep[e.g.,][]{Giammanco2005,Pellegrini2012,Zackrisson2013,Jaskot2013,Nicholls2014,Nakajima2014,Stasinska2015,Jaskot2016,Alexandroff2015,Izotov2018b,DAgostino2019,Plat2019}.
The degeneracy between AGN signatures and density-bounded \hii\ regions has been discussed by \citet{Plat2019}. In particular, although both scenarios can produce elevated ratios of \lheii\ to lower-ionisation lines, such as \lciii\ or \hb, density-bounded \hii\ regions also exhibit significantly reduced EWs and very weak low-ionisation lines, leading to low values of \lniiopt/\ha\ and \loiiopt/\loiiiopt. These characteristics generally allow density-bounded \hii\ regions to be distinguished from NLR emission.

NLR clouds themselves may also be density bounded. In this case, the effect on the line ratios is qualitatively similar to that in \hii\ regions, namely an increase in the ratios of high- to low-ionisation species and a reduction in the overall line fluxes, particularly for lines originating in the low-ionisation zones.
This could influence some diagnostics presented in Section~\ref{sec:TypeIIdiag}, in particular those including \loiopt, in which the NLR models would then move in the SF region.

BLR models are parameterized differently from the NLR and \hii\ region models. They are stopped when either a fixed hydrogen column density is reached or the electron fraction drops below 1 per cent, whichever occurs first (see Section~\ref{sec:method_BLR}). In the fiducial grid, a column density of $\log(N_{\rm H}/{\rm cm}^{-2}) = 23$ is adopted, such that some clouds are density bounded while others are ionisation bounded.
The grid of Balmer break BLR models includes both $\log(N_{\rm H}/{\rm cm}^{-2}) = 23$ and 25. The impact of such large column density is discussed in Section~\ref{sec:TypeI}. In brief, increasing the column density extends the partially ionized region, thus enhancing the emission of lines such as \ha\ and \hb. At the same time, larger column densities increase the optical depth of some lines, reducing their emergent fluxes and strengthening the Balmer break. The relative importance of these competing effects depends on the hydrogen density, inner radius, and temperature of the clouds.
On the other hand, adopting a lower column density of $\log(N_{\rm H}/{\rm cm}^{-2}) \leq 20$, strongly reduces \heii/\hb\ ratio, indicating that this ratio is sensitive not only to the relative rates of ionising photons but also to the properties of the gas.

\subsubsection{Hydrogen density and multicloud models}

Throughout this paper, we adopt constant hydrogen densities of $\nh = 10^{3}\rm cm^{-3}$ for the NLR models and $\nh = 10^{2}\rm cm^{-3}$ for the \hii\ region models.
In addition, since gas densities as high as $\gtrsim 10^{5}$--$10^{6}\rm cm^{-3}$ have been measured in some high redshift galaxies \citep[e.g.,][]{Senchyna2023,Topping2024,Topping2025}, we discuss SF and NLR models with such high density in Section~\ref{sec:TypeIIdiag}.

Measurements of densities of \hii\ regions from \loiiopt\ lines at $z \sim 0$ give values around $10^{2}$~cm$^{-3}$, while the densities measured using \lciii\ or \larivopt\ are higher in the order of $\gtrsim 10^{4}$~cm$^{-3}$. This suggests that the density is higher in the central parts of the \hii\ regions than in the outer parts \citep[e.g.,][]{Osterbrock2006} and decreasing density gradients, as well as density inhomogeneities have been measured in some \hii\ regions \citep[][]{Castaneda1992,Franco2000,Binette2002,Phillips2007,Rubin2011,Mendez2026}.
The presence of multiple density components could shift some of the predicted line ratios. For example, combining a high-density, high-ionisation-parameter component with a low-density, low-ionisation-parameter component could enhance \oiiioptc\ while maintaining a lower \oiiiopt/\oiiopt\ ratio than predicted by a single \hii\ region model, potentially shifting SF models into the AGN region of the \citet{Mazzolari2024} diagram.

The aim of this paper is to provide general diagnostics of AGN emission, and we therefore restrict ourselves to constant-density models. However, when fitting individual sources, more sophisticated models including multiple density components or radial density gradients, for both the SF and NLR emission, may be more appropriate.

Regarding the BLR models, they are composed of multiple clouds spanning a range of densities and radii. As discussed in Sections~\ref{sec:validation_BLR} and \ref{sec:TypeI}, the density has a strong impact on the line ratios (see, e.g., Fig.~\ref{fig:BLRsinglemulti}) and on the strength of the Balmer break.
The radial distribution of the clouds also plays an important role. Clouds located at larger radii have lower ionisation parameters and tend to possess more extended partially ionized regions, while clouds closer to the BH experience higher ionisation parameters, are more likely to be density bounded, and develop larger optical depths in some lines. Consequently, the adopted radial distribution of the BLR clouds has a strong impact on the line ratios and the strength of the Balmer break.
In the fiducial grid, we assume cloud distributions proportional to $1/\nh$ and $1/r$. In Section~\ref{sec:discuhighz}, we explore how alternative density and radial distributions can lead to lower \lheii/\hb\ ratios, potentially explaining some of the \JWST\ observations of high-redshift AGN.

\subsubsection{Shape of the ionising radiation}

As described in Section~\ref{sec:method}, we adopt fixed angles of $45^{\circ}$ between the accretion disc and the NLR clouds, and $22.5^{\circ}$ for the BLR clouds. To quantify the uncertainties associated with the angular distribution of the clouds around the accretion disc, we compute SEDs using the \relqso\ module for viewing angles corresponding to $\cos i = 0.1$ and 0.9, for a model with $\log (\Mbh/\Msun)=7$ and $\log \Fedd=-0.5$.
The accretion disc produces a harder and more luminous radiation field when viewed nearly face-on ($\cos i = 0.9$) than edge-on ($\cos i = 0.1$). In particular, the bolometric luminosity is higher by 0.48 dex, the rate of H-ionising photons increases by 0.7 dex, while the ratio $Q_{\rm HeII}/Q_{\rm H}$ decreases by 0.17 dex. This variation in the hardness of the ionising spectrum is relatively modest and is therefore not expected to have a major impact on the predicted line ratios (at least in the context of these \relqso\ models).
However, \citet{Madau2025} suggested that geometrically thick, non-advective discs in the super-Eddington regime may produce highly anisotropic radiation fields, which could selectively suppress high-ionisation lines, such as \lheii, relative to the Balmer lines. Similarly, \citet{Pacucci2024} found that intrinsically weak X-ray emission, coupled with viewing angle effects, in super-Eddington accretion flows could explain the X-ray weakness observed in some high-redshift AGN \citep[e.g.,][]{Maiolino2025}.

Using \agnslim\ module with X-ray weak SED in the super-Eddington regime, \citet{Lambrides2024} found that this can lead to weaker UV high-ionisation lines. 
Motivated by these results, we compute an additional SED using the \agnslim\ module, adopting the best-fitting parameters inferred by \citet{Jin2023sed} for a weak-line Seyfert galaxy. We then compare this SED to our fiducial \agnslim\ model with the same BH mass, $2\times10^{7}\ \Msun$, and Eddington ratio, $\log \Fedd = 1.31$.
Using these SEDs, we compute NLR models with $Z=0.001$ and $\log U=-2.5$. We find that the change in SED has little effect on the diagnostic diagrams presented in Fig.~\ref{fig:NLRdiaggood}. At fixed metallicity and ionisation parameter, the X-ray-weak SED increases \heiiopt/\hb\ and \heii/\hb\ by $\sim 0.1$ dex relative to the fiducial SED (and are similar to ratios predicted by sub-Eddington model with $\log \Fedd = -1$). Our results probably differ from those of \citet{Lambrides2024}, because that study assumes a lower ionisation parameter to model the X-ray weak SED. We also verify that the effect is similar for the LRD-type models. 
In addition, we also compute NLR models using the X-ray weak SED from \citet{Pacucci2024}, for a BH mass of $10^7$ and Eddington ratio of 2.4, and we find negligible impact on these line ratios compared to our fiducial adopted SED.
Therefore, the weak \heii\ emission observed in some high-redshift AGN is not necessarily a consequence of X-ray weak SED.
  
Nevertheless, all the accretion-disc models considered above, and more generally those used throughout this work (such as the \relqso\ module), are primarily calibrated against low-redshift observations of quasars \citep[][]{Kubota2018}. The properties of accretion discs at high redshift may differ  from these calibrations, resulting in potentially different SEDs.
In particular, the SEDs at low BH masses and/or in the super-Eddington regime remain highly uncertain. For example, using the accretion-disc SEDs of \citet{Pezzulli2017}, \citet{Ji2025} found that super-Eddington models predict slightly weaker \lheii/\hb\ ratios than sub-Eddington models. 
Photon trapping is expected in the super-Eddington regime \citep[][]{Abramowicz1988}, and could result in a softer ionising radiation, as is assumed in \citet{Pezzulli2017} models. 
However, in contrast, the slim-disc models of \citet{Kubota2019} predict enhanced emission from the innermost regions of the disc, such that $Q_{\rm HeII}/Q_{\rm H}$ does not necessarily drop in the super-Eddington regime. 
Alternative accretion-disc models, incorporating geometrically thick discs, instead predict anisotropic radiation fields that can shadow the hardest ionising photons along some line of sights \citep[][]{Madau2025}.
Overall, the ionising SEDs of high-redshift, super-Eddington AGN remain poorly constrained \citep[but see][]{Ighina2025}. If they are intrinsically softer than assumed in this work, producing fewer He$^+$-ionising photons relative to H-ionising photons, they could naturally explain the weaker \heii\ emission observed in some high-redshift AGN.

The accretion disc models presented in this paper assume a BH spin of zero as a fiducial choice. The spin distribution of accreting SMBHs remains uncertain and depends on their accretion and merger histories, as well as on the presence of jets \citep[e.g.,][]{Ricarte2023,Ricarte2025}. In particular, these studies suggest that jet-driven spin-down can substantially reduce BH spins at highly super-Eddington accretion rates.
For a model with $\log (M_{\rm BH}/M_\odot)=7$, increasing the BH spin from 0 to 0.998 produces relatively modest changes in the ionising SED. At $\log \dot{m}=-0.5$, using the \relqso\ module, the bolometric luminosity decreases by 0.10 dex as a result of relativistic transfer, while the rate of H-ionising photons decreases by 0.27 dex and the ratio $Q_{\rm HeII}/Q_{\rm H}$ increases by 0.25 dex. In the super-Eddington case, $\log \dot{m}=1$,  modelled  with \agnslim, the bolometric luminosity decreases by only 0.04 dex, while $Q_{\rm H}$ decreases by 0.26 dex and $Q_{\rm HeII}/Q_{\rm H}$ increases by 0.29 dex. Overall, the change in ionising spectral hardness is modest, suggesting that the effect of BH spin on \lheii/\hb\ is likely smaller than the variations driven by the other model parameters considered here.

Emission lines from SF \hii\ regions are also sensitive to the choice of input stellar population spectra used. In particular, many of the line-ratio diagrams presented in Sec.~\ref{sec:TypeIIdiag} rely on the contrast between the strong \lheii\ emission predicted by NLR models and the weaker emission produced by SF models. The inclusion of updated binary stellar population models could, however, enhance the predicted \lheii\ emission from SF regions, bringing them somewhat closer to the demarcation lines separating SF and AGN models, although the effect is expected to remain modest \citep[e.g.,][]{Stanway2019,Lecroq2024}.

Another parameter that could affect the predicted line ratios is $\alpha$-enhancement, which is expected to be more common in high-redshift star-forming galaxies, given the delay in enrichment from Type Ia supernovae (which dominate Fe production) relative to the rapid enrichment from core-collapse supernovae \citep[which dominate the production of O and other $\alpha$-elements, e.g.,][]{Steidel2016,Topping2020a,Topping2020b,Cullen2021}. While this is not expected to significantly impact the predicted \lheii\ emission, it can increase \loiiiopt/\hb\ and \lniiopt/\ha\ in the SF models, due to a harder ionising radiation at fixed metallicity. This effect contributes to the offset of high-redshift galaxies relative to local galaxies in the BPT diagram \citep[e.g.,][]{Steidel2016,Strom2017,Strom2018,Shapley2019,Topping2020b,Scharre2026}. However, it is generally not strong enough to move the SF models completely into the AGN region of the diagram, especially at low metallicity, suggesting that this effect is unlikely to substantially affect our diagnostics.

\subsubsection{Shock contribution}

High-velocity shocks may contribute to the emission-line spectrum and are strongly degenerate with AGN emission, as they also produce strong high-ionisation lines, such as \lheii\ and \lnev\ \citep[e.g.,][]{Allen2008,Stasinska2015,Jaskot2016,Alarie2019,Plat2019}.
To scale the shock luminosity self-consistently with the stellar population, we assume that all the mechanical energy released by supernovae and stellar winds is converted into fast radiative shocks, thus providing an upper limit on the shock contribution linked to star formation. We estimate the mechanical energy using the predictions of \starburst\ \citep{Leitherer1999} for a constant star-formation history, and normalize the shock models of \citet{Alarie2019} following the prescription described above. Before $\sim 3$ Myr, the \lheii\ luminosity produced by shocks is more than an order of magnitude lower than that emitted by the \hii\ regions associated with the same stellar population. After $\sim 10$ Myr, many supernovae begin to explode, and by $\sim 40$ Myr, the \lheii\ luminosity from shocks becomes comparable to that produced by the \hii\ regions.
Therefore, for shocks to make a significant contribution to \lheii\ emission, either a different prescription for the mechanical energy input or a different star-formation history must be adopted. For example, a galaxy experiencing intense star formation a few tens of Myr ago, but with little or no ongoing star formation, could have an emission-line spectrum dominated by shocks rather than by \hii\ regions.
 
\subsection{Interpretation of faint high-redshift AGN}\label{sec:discuhighz}

Having assessed the main modelling uncertainties in Section~\ref{sec:discumodeluncertainties}, we now turn to their implications for the interpretation of faint high-redshift AGN. In particular, we explore whether plausible differences compared to the fiducial BLR structure and ionising radiation field can account for the unusually weak high-ionisation emission observed in some of these high-redshift JWST AGN.

\begin{figure*}
\begin{centering}
\includegraphics[width=1\linewidth]{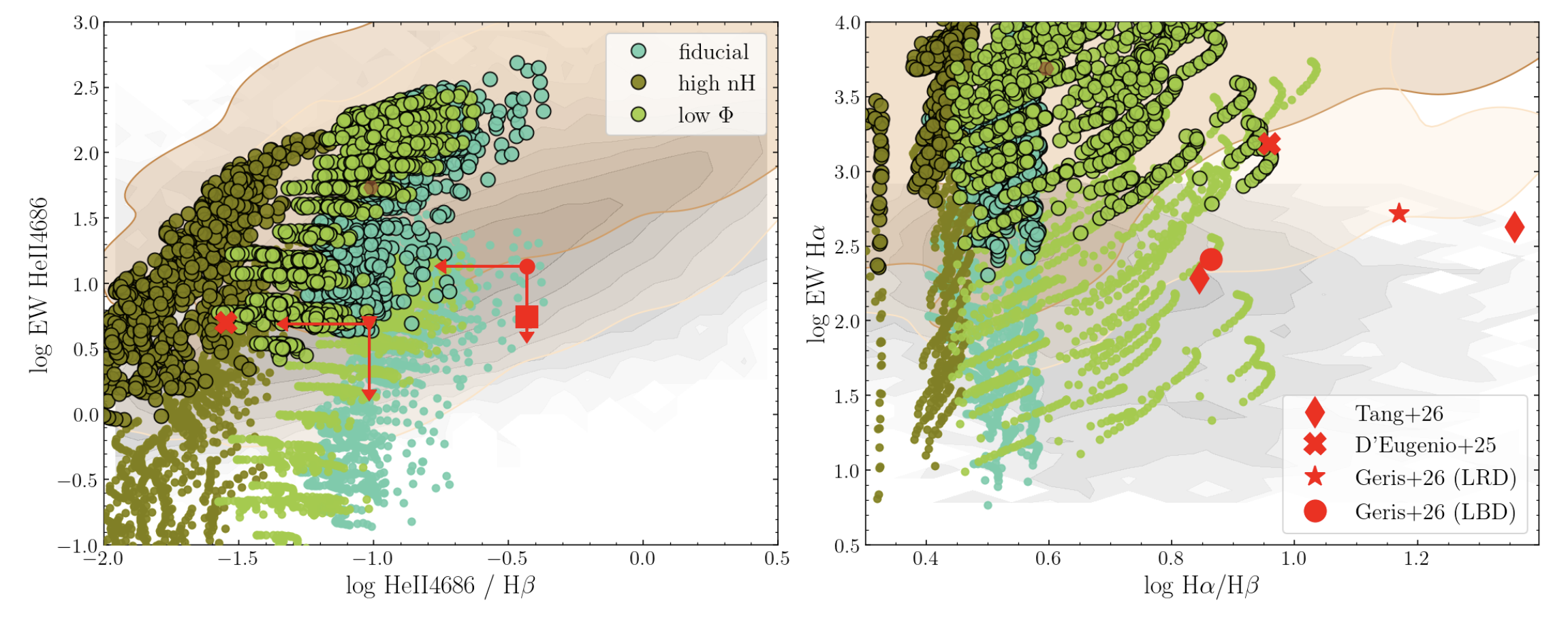}
\par\end{centering}
\caption{Comparison of BLR models adopting different cloud density and radial distributions. The fiducial models, with cloud distribution function proportional to $1/\nh$ and $1/R$, are shown in turquoise, while models with a constant density distribution are shown in khaki and those with a constant radial distribution in light green. Models are shown for BLR covering factors of 1 (large circles with black outlines) and 0.1 (small circles). The distribution of Balmer-break models is shown in dark brown (covering factor of 1) and light brown (covering factor of 0.1). The models are compared with \JWST\ observations of high-redshift candidate type~I AGN and LRDs from \citet{Wang2025,DEugenio2025,Zucchi2025,Brazzini2026,Tang2026} and stack LRD and LBD from \citet{Geris2026}. The position of local AGN from \citet{Liu2019} and \citet{Wu2022} is shown in grey contours. Left : \heiiopt-EW\ as a function of log \heiiopt/\hb, right : \ha-EW as function of log \ha/\hb.}
\label{fig:TypeIhighzvar} 
\end{figure*}

As discussed in Section~\ref{sec:TypeIhighz}, the fiducial BLR models struggle to reproduce the extremely low broad \heiiopt/\hb\ ratios observed in some high-redshift sources. Several explanations have been proposed in the literature, including a softer ionising radiation field illuminating the gas (see Section~\ref{sec:TypeIhighz}).
Here, we explore how varying the adopted BLR cloud density and radial distributions affects these line ratios. In particular, we compute one grid with a density distribution function proportional to 1 instead of the fiducial $1/\nh$, and another with a radial distribution function proportional to 1 instead of the fiducial $1/R$.

The results are shown in Fig.~\ref{fig:TypeIhighzvar}, where we plot \heiiopt-EW\ as a function of \heiiopt/\hb\ (left panel). The models are compared with \JWST\ observations of candidate type~I AGN and LRDs from \citet{Wang2025,Zucchi2025,Brazzini2026}. 
In both alternative distributions, the predicted \heiiopt/\hb\ ratios are shifted toward lower values. A constant density distribution function gives relatively more weight to high-density clouds, while a constant radial distribution function favours clouds at large radii and hence lower incident ionising fluxes. Both changes favour clouds with lower \heiiopt/\hb.
Many uncertainties remain regarding the physical origin of the weak \heiiopt\ emission, including the shape of the incident radiation field, which may be intrinsically softer in the super-Eddington regime, the relative contributions of stellar and AGN ionising radiation to the broad-line emission, and possible obscuration between the accretion disc and the BLR clouds. Nevertheless, our results show that the density and radial distributions of the BLR clouds can also significantly affect the observed line ratios.
In particular, it is plausible that, during the early accretion episodes of high-redshift AGN, the physical properties and structure of the BLR differ from those observed at low redshift, for example through a larger contribution from dense clouds. Moreover, if multiple clouds are present along the same line of sight, clouds located farther from the ionising source may be shielded by those closer to the accretion disc and therefore experience a lower incident ionising flux. 
This effect could qualitatively resemble that of clouds located at larger radii, leading to lower ionisation parameters and potentially weaker \heiiopt/\hb\ ratio, although filtering by the inner clouds could also modify the shape of the incident spectrum.
As explored in Section~\ref{sec:discumodeluncertainties}, lower gas column densities and, for some density and ionisation conditions, dissipative heating can also reduce the \heiiopt/\hb\ ratio. A combination of clouds spanning a range of physical conditions could therefore contribute to the weak \heiiopt/\hb\ ratios observed in some AGN.

In the right panel of Fig.~\ref{fig:TypeIhighzvar}, we show \ha-EW\ as a function of \ha/\hb, together with observations from \citet{Wang2025,DEugenio2025,Tang2026} and LRD and LBD stacks from \citet{Geris2026}. The constant radial weighting shifts the models toward higher \ha\ EWs and larger \ha/\hb\ ratios. For a fixed gas column density, clouds with lower ionisation parameters tend to have a more extended partially ionized region, where collisional excitation and radiative transfer effects may enhance \ha\ emission.

While the observed Balmer break and Balmer absorption require large columns of hydrogen in n=2 state along the line of sight, thereby constraining the physical conditions of the BLR clouds (hydrogen density, temperature, ionising flux, and column density), only a few single-cloud models can reproduce the very low \heiiopt/\hb\ ratios observed when adopting the fiducial accretion-disc SED used in this work (see Section~\ref{sec:TypeIhighz}). A more realistic, inhomogeneous BLR composed of clouds spanning a broad range of densities, radii, and column densities could potentially reproduce both the strong Balmer absorption and the weak \heiiopt\ emission. In this context, the strengths of the Balmer break, absorption features, and emission lines could depend on whether dense clouds intercept the line of sight between the accretion disc and the observer, such that viewing-angle effects may also play an important role. We defer a quantitative fit of LRDs and high-redshift type~I AGN using these models to future work.

Weak narrow \lheii\ emission has also been observed in several high-redshift AGN candidates, particularly in some type~I sources (see Section~\ref{sec:TypeIIdiag}). This is the case, for example, for the LRD GN28074 \citep{Brazzini2026} and the stacked type~I AGN from \citet{Juodvabalis2026}.
 A straightforward explanation for the weakness of the narrow high-ionisation lines could be contamination by SF \hii\ region emission. Indeed, spatial offsets between the blue and the red component have been observed in many LRDs, suggesting a non-negligible contribution from the host galaxy \citep[e.g.,][]{Baggen2026,Barger2026}. However, several sources with weak narrow \lheii\ still occupy the AGN region of some diagnostic diagrams. For example, RUBIES EGS 49140 exhibits elevated \oiiioptc/\hg\ and \loiopt/\ha\ ratios \citep{DEugenio2025}, and similar properties are found for the LRD/LBD stack of \citet{Geris2026}, consistent with a contribution from AGN photoionisation to the narrow-line emission \citep[see also][]{Geris2026}. 
Using a Bayesian fitting analysis combining NLR and \hii-region emission could test whether such a composite model can simultaneously reproduce the weak \lheii\ and the strong \oiiioptc\ and \loiopt\ emission. 
In addition, \citet{DEugenio2025} argued that RUBIES EGS 49140 likely contains both a high-density component ($\nh \gtrsim 10^{6}\ \rm cm^{-3}$) and a lower-density component ($\nh \approx 420\ \rm cm^{-3}$), inferred from the \oiid\ doublet. Such a combination of density components could shift SF models toward the AGN region of the \oiiioptc/\hg\ versus \oiiiopt/\oiiopt\ diagram proposed by \citet{Mazzolari2024} (see Section~\ref{sec:discumodeluncertainties}). Thus, a combination of SF contamination and multiple density components could explain some of the observed line ratios.
Nevertheless, the high \loiopt/\ha\ ratios observed in some sources point toward the presence of an extended, high-temperature partially ionized region. 
One possible explanation is that high-column-density BLR clouds with an ionisation structure that produces appreciable He+ opacity filter the accretion-disc spectrum. Such a screen can preferentially attenuate photons around the He+ edge, while more energetic X-rays penetrate farther, maintaining a warm, partially ionized region \citep[e.g.][]{Binette2003,Kraemer1999b} and providing additional heating to the surrounding ISM, potentially enhancing \loiopt\ and, through the increased electron temperature, \oiiioptc\ emission, while maintaining weak narrow \lheii.

More generally, while the weakness of both the broad and narrow high-ionisation lines may indicate a softer ionising spectrum, our models show that a softer SED is not necessarily required to produce weak \heiiopt/\hb, as variations in the geometry and density distribution of the gas surrounding the BH can also significantly reduce this ratio.
For example, the super-Eddington accretion-disc models of \citet{Madau2026} predict log(\heiiopt/\hb) ratios as low as $-1.16$. 
While the anisotropic ionising radiation field predicted by these models could contribute to the observed BLR properties, its implications for the NLR additionally depend on the geometry and covering factor of the more extended gas.
If the ionising spectrum is intrinsically softer, its effect could also combine with variations in the BLR cloud distributions explored here to produce even lower \heiiopt/\hb ratios.
A combination of these effects could explain the diversity of emission-line properties seen in high-redshift AGN, including the detection of narrow \lheii\ in some LRDs and LBDs \citep[e.g.,][]{Brazzini2026,Geris2026} and of other high-ionisation lines, such as \lnv, in some sources \citep[e.g.,][]{Tang2025}, while these lines remain undetected in others.

\section{Conclusion}\label{sec:conclusion}

We have presented a new suite of photoionisation models designed to interpret the broad- and narrow-line emission of AGN over a wide range of black-hole masses and accretion rates, including the super-Eddington regime. The ionising spectra are computed self-consistently as a function of black-hole mass and Eddington ratio using \relqso\ and \agnslim\ models, and are used as input to the 1D photoionisation code \cloudy\ to construct BLR and NLR models. 
For the BLR, we construct a broad grid of individual-cloud models spanning a wide range of gas densities, incident ionising fluxes, and column densities. We use this grid to construct fiducial multi-cloud LOC models, which we validate against observations of low-redshift type~I AGN, and to identify single-cloud models capable of producing the strong Balmer breaks observed in some LRDs. We further explore variations in the cloud distributions of the multi-cloud models to investigate the weak high-ionisation lines observed in some high-redshift AGN candidates.
For the NLR, we extend previous photoionisation grids from \citet{Feltre2016} by explicitly accounting for variations in the accretion-disc SED with black-hole mass and Eddington ratio.

Our main results can be summarized as follows.

\begin{itemize}

\item The fiducial BLR and NLR models broadly reproduce the emission-line properties of low-redshift AGN. In particular, the multi-cloud BLR models reproduce the main locus of observed broad-line ratios and equivalent widths, while the NLR models recover the location of local AGN in the classical optical diagnostic diagrams. Consistent with previous studies, the comparison of multi-cloud BLR models with single-cloud BLR calculations supports the view that models spanning a distribution of cloud densities and radii generally provide a better match to observed broad-line properties than individual cloud models.

\item Broad Balmer lines provide the most promising route to identifying accreting intermediate-mass black holes with JWST. At $z=6$, assuming a BLR covering factor of 0.4 and a 20-h NIRSpec observation, \ha\ and \hb\ may be detectable for $M_{\rm BH}\gtrsim10^{5.7} {\rm M_\odot}$ at the Eddington limit, while \lheii\ generally requires $M_{\rm BH}\gtrsim10^{6.2}{\rm M_\odot}$. For strongly super-Eddington accretion, broad \ha\ may become detectable down to $M_{\rm BH}\sim10^{4.5} {\rm M_\odot}$ in the fiducial models, and to still lower masses for some dense Balmer-break clouds. These detectability limits should be regarded as optimistic, as broad--narrow line decomposition and dust attenuation can make detection more challenging. Gravitational lensing further extends the accessible mass range. 

\item Narrow \ha\ and \oiiiopt\ are similarly among the most detectable tracers of low-mass accreting black holes. For a black hole accreting at the Eddington limit, these lines become detectable at approximately $M_{\rm BH}\sim10^{5.5} {\rm M_\odot}$ for the adopted NLR covering factor of 0.1, while extreme super-Eddington models can reach substantially lower masses. High-ionisation lines such as \lheii, \lnev, and \lnv, although particularly valuable for identifying AGN activity, generally require larger black-hole masses and/or accretion rates close to or above the Eddington limit.

\item We derive theoretical bolometric corrections for both broad and narrow emission lines. Hydrogen and helium recombination lines provide the most robust tracers of accretion luminosity because they depend less strongly on metallicity than collisionally excited metal lines, which show substantially larger variations with metallicity and ionisation parameter. Nevertheless, the recombination-line corrections still depend on black-hole mass, Eddington ratio and, in the NLR, ionisation parameter. Simple, universal line-to-accretion-luminosity conversions can therefore introduce significant systematic uncertainties, particularly when local calibrations are applied to high-redshift AGN.

\item Classical diagnostic diagrams can fail to identify low-metallicity AGN and accreting black-hole seeds. NLR-model locations in these diagrams are governed primarily by metallicity and ionisation parameter rather than directly by black-hole mass, and several commonly used diagnostics preferentially select AGN with $Z\gtrsim0.004$--$0.008$, depending on the diagnostic, while largely missing lower-metallicity systems. In addition to this metallicity bias, some diagnostic diagrams are sensitive to gas density, which can cause SF and AGN models to cross boundaries calibrated at lower densities, while others can remain degenerate with Pop.~III stellar photoionisation. We identify alternative combinations of UV and optical line ratios that provide improved separation between AGN and both Pop.~II and Pop.~III stellar photoionisation and are less sensitive to the gas-density variations explored here. However, even these more robust diagnostics cannot distinguish an accreting low-metallicity black-hole seed from a more massive low-metallicity AGN. Narrow-line ratios alone are therefore unlikely to provide robust constraints on black-hole mass and Eddington ratio, or to uniquely identify high-redshift black-hole seeds in the intermediate-mass regime.

\item The unusual spectra of faint high-redshift type~I AGN and LRDs need not arise solely from changes in the intrinsic accretion-disc spectrum.
In our models, lower black-hole masses can contribute to larger \ha\ equivalent widths, which may be further enhanced by higher BLR covering factors. Dense, high-column-density clouds can also produce large \ha\ equivalent widths, elevated intrinsic Balmer decrements, or strong Balmer breaks through collisional excitation and radiative transfer effects, with the relative strength of these signatures depending on the cloud physical conditions.
The very weak broad \lheii\ emission observed in some sources remains difficult to reproduce with the fiducial multi-cloud BLR model, but can be reduced by changing the weighting of BLR clouds toward higher densities or larger radii. Weak narrow \lheii\ emission is also observed in some of these systems. 
Changes in BLR properties could also affect the NLR emission by filtering the ionising radiation from the accretion disc before it reaches the more extended gas. Although we do not explicitly model this filtering here, it could contribute to the weakness of narrow high-ionisation lines. The observed diversity of high-ionisation line strengths among faint high-redshift broad-line sources may therefore reflect a combination of variations in the intrinsic SED and in the structure and physical conditions of the line-emitting gas.

\end{itemize}

Looking ahead, these models are designed to be incorporated into \textsc{beagle}, enabling the simultaneous interpretation of both narrow- and broad-line AGN emission within a Bayesian framework. For the NLR, this will allow the ionising spectrum to depend explicitly on black-hole mass and Eddington ratio, while the inclusion of the BLR will provide, for the first time within this framework, a self-consistent treatment of broad-line and AGN continuum emission in type I AGN, alongside the narrow-line and stellar components. This will make it possible to infer AGN and host-galaxy properties jointly from observed spectra, building on the multi-line fitting philosophy already adopted by \textsc{beagle}. 

A remaining challenge is the large dimensionality of the model space, which spans parameters such as metallicity, ionisation parameter, black-hole mass, and Eddington ratio. To reduce this freedom and provide more physically motivated priors, we are also developing a connection between these photoionisation models and modern cosmological simulations such as IllustrisTNG \citep{Pillepich2018} and COLIBRE \citep{Schaye2026}, in the spirit of the approach of \citet{Hirschmann2023}, but extending it to include broad-line emission (Hirschmann et al. in prep.). This will allow us to move beyond uniformly sampled model grids and instead predict the distributions of AGN emission-line properties expected from physically motivated galaxy and black-hole populations at a given cosmic epoch. Such a framework will ultimately enable a more direct comparison between JWST observations and theoretical predictions for the co-evolution of galaxies and accreting black holes across cosmic time.



\section*{Acknowledgements}
AP and MH acknowledge funding from the Swiss National Science Foundation (SNF) via a PRIMA Grant PR00P2 193577 “From cosmic dawn to high noon: the role of black holes for young galaxies”.



\bibliographystyle{mnras}
\bibliography{bibliography.bib}

@ARTICLE{Abazajian2009,
       author = {{Abazajian}, Kevork N. and {Adelman-McCarthy}, Jennifer K. and {Ag{\"u}eros}, Marcel A. and {Allam}, Sahar S. and {Allende Prieto}, Carlos and {An}, Deokkeun and {Anderson}, Kurt S.~J. and {Anderson}, Scott F. and {Annis}, James and {Bahcall}, Neta A. and {Bailer-Jones}, C.~A.~L. and {Barentine}, J.~C. and {Bassett}, Bruce A. and {Becker}, Andrew C. and {Beers}, Timothy C. and {Bell}, Eric F. and {Belokurov}, Vasily and {Berlind}, Andreas A. and {Berman}, Eileen F. and {Bernardi}, Mariangela and {Bickerton}, Steven J. and {Bizyaev}, Dmitry and {Blakeslee}, John P. and {Blanton}, Michael R. and {Bochanski}, John J. and {Boroski}, William N. and {Brewington}, Howard J. and {Brinchmann}, Jarle and {Brinkmann}, J. and {Brunner}, Robert J. and {Budav{\'a}ri}, Tam{\'a}s and {Carey}, Larry N. and {Carliles}, Samuel and {Carr}, Michael A. and {Castander}, Francisco J. and {Cinabro}, David and {Connolly}, A.~J. and {Csabai}, Istv{\'a}n and {Cunha}, Carlos E. and {Czarapata}, Paul C. and {Davenport}, James R.~A. and {de Haas}, Ernst and {Dilday}, Ben and {Doi}, Mamoru and {Eisenstein}, Daniel J. and {Evans}, Michael L. and {Evans}, N.~W. and {Fan}, Xiaohui and {Friedman}, Scott D. and {Frieman}, Joshua A. and {Fukugita}, Masataka and {G{\"a}nsicke}, Boris T. and {Gates}, Evalyn and {Gillespie}, Bruce and {Gilmore}, G. and {Gonzalez}, Belinda and {Gonzalez}, Carlos F. and {Grebel}, Eva K. and {Gunn}, James E. and {Gy{\"o}ry}, Zsuzsanna and {Hall}, Patrick B. and {Harding}, Paul and {Harris}, Frederick H. and {Harvanek}, Michael and {Hawley}, Suzanne L. and {Hayes}, Jeffrey J.~E. and {Heckman}, Timothy M. and {Hendry}, John S. and {Hennessy}, Gregory S. and {Hindsley}, Robert B. and {Hoblitt}, J. and {Hogan}, Craig J. and {Hogg}, David W. and {Holtzman}, Jon A. and {Hyde}, Joseph B. and {Ichikawa}, Shin-ichi and {Ichikawa}, Takashi and {Im}, Myungshin and {Ivezi{\'c}}, {\v{Z}}eljko and {Jester}, Sebastian and {Jiang}, Linhua and {Johnson}, Jennifer A. and {Jorgensen}, Anders M. and {Juri{\'c}}, Mario and {Kent}, Stephen M. and {Kessler}, R. and {Kleinman}, S.~J. and {Knapp}, G.~R. and {Konishi}, Kohki and {Kron}, Richard G. and {Krzesinski}, Jurek and {Kuropatkin}, Nikolay and {Lampeitl}, Hubert and {Lebedeva}, Svetlana and {Lee}, Myung Gyoon and {Lee}, Young Sun and {French Leger}, R. and {L{\'e}pine}, S{\'e}bastien and {Li}, Nolan and {Lima}, Marcos and {Lin}, Huan and {Long}, Daniel C. and {Loomis}, Craig P. and {Loveday}, Jon and {Lupton}, Robert H. and {Magnier}, Eugene and {Malanushenko}, Olena and {Malanushenko}, Viktor and {Mandelbaum}, Rachel and {Margon}, Bruce and {Marriner}, John P. and {Mart{\'\i}nez-Delgado}, David and {Matsubara}, Takahiko and {McGehee}, Peregrine M. and {McKay}, Timothy A. and {Meiksin}, Avery and {Morrison}, Heather L. and {Mullally}, Fergal and {Munn}, Jeffrey A. and {Murphy}, Tara and {Nash}, Thomas and {Nebot}, Ada and {Neilsen}, Eric H., Jr. and {Newberg}, Heidi Jo and {Newman}, Peter R. and {Nichol}, Robert C. and {Nicinski}, Tom and {Nieto-Santisteban}, Maria and {Nitta}, Atsuko and {Okamura}, Sadanori and {Oravetz}, Daniel J. and {Ostriker}, Jeremiah P. and {Owen}, Russell and {Padmanabhan}, Nikhil and {Pan}, Kaike and {Park}, Changbom and {Pauls}, George and {Peoples}, John, Jr. and {Percival}, Will J. and {Pier}, Jeffrey R. and {Pope}, Adrian C. and {Pourbaix}, Dimitri and {Price}, Paul A. and {Purger}, Norbert and {Quinn}, Thomas and {Raddick}, M. Jordan and {Re Fiorentin}, Paola and {Richards}, Gordon T. and {Richmond}, Michael W. and {Riess}, Adam G. and {Rix}, Hans-Walter and {Rockosi}, Constance M. and {Sako}, Masao and {Schlegel}, David J. and {Schneider}, Donald P. and {Scholz}, Ralf-Dieter and {Schreiber}, Matthias R. and {Schwope}, Axel D. and {Seljak}, Uro{\v{s}} and {Sesar}, Branimir and {Sheldon}, Erin and {Shimasaku}, Kazu and {Sibley}, Valena C. and {Simmons}, A.~E. and {Sivarani}, Thirupathi and {Allyn Smith}, J. and {Smith}, Martin C. and {Smol{\v{c}}i{\'c}}, Vernesa and {Snedden}, Stephanie A. and {Stebbins}, Albert and {Steinmetz}, Matthias and {Stoughton}, Chris and {Strauss}, Michael A. and {SubbaRao}, Mark and {Suto}, Yasushi and {Szalay}, Alexander S. and {Szapudi}, Istv{\'a}n and {Szkody}, Paula and {Tanaka}, Masayuki and {Tegmark}, Max and {Teodoro}, Luis F.~A. and {Thakar}, Aniruddha R. and {Tremonti}, Christy A. and {Tucker}, Douglas L. and {Uomoto}, Alan and {Vanden Berk}, Daniel E. and {Vandenberg}, Jan and {Vidrih}, S. and {Vogeley}, Michael S. and {Voges}, Wolfgang and {Vogt}, Nicole P. and {Wadadekar}, Yogesh and {Watters}, Shannon and {Weinberg}, David H. and {West}, Andrew A. and {White}, Simon D.~M. and {Wilhite}, Brian C. and {Wonders}, Alainna C. and {Yanny}, Brian and {Yocum}, D.~R. and {York}, Donald G. and {Zehavi}, Idit and {Zibetti}, Stefano and {Zucker}, Daniel B.},
        title = "{The Seventh Data Release of the Sloan Digital Sky Survey}",
      journal = {\apjs},
         year = 2009,
        month = jun,
       volume = {182},
       number = {2},
        pages = {543-558},
          doi = {10.1088/0067-0049/182/2/543},
archivePrefix = {arXiv},
       eprint = {0812.0649},
 primaryClass = {astro-ph},
       adsurl = {https://ui.adsabs.harvard.edu/abs/2009ApJS..182..543A}
}

@ARTICLE{Abramowicz1988,
       author = {{Abramowicz}, M.~A. and {Czerny}, B. and {Lasota}, J.~P. and {Szuszkiewicz}, E.},
        title = "{Slim Accretion Disks}",
      journal = {\apj},
         year = 1988,
        month = sep,
       volume = {332},
        pages = {646},
          doi = {10.1086/166683},
       adsurl = {https://ui.adsabs.harvard.edu/abs/1988ApJ...332..646A}
}

@ARTICLE{Alarie2019,
       author = {{Alarie}, A. and {Morisset}, C.},
        title = "{Extensive Online Shock Model Database}",
      journal = {\rmxaa},
         year = 2019,
        month = oct,
       volume = {55},
        pages = {377-394},
          doi = {10.22201/ia.01851101p.2019.55.02.21},
archivePrefix = {arXiv},
       eprint = {1908.08579},
 primaryClass = {astro-ph.GA},
       adsurl = {https://ui.adsabs.harvard.edu/abs/2019RMxAA..55..377A}
}

@ARTICLE{Alexandroff2015,
       author = {{Alexandroff}, Rachael M. and {Heckman}, Timothy M. and {Borthakur}, Sanchayeeta and {Overzier}, Roderik and {Leitherer}, Claus},
        title = "{Indirect Evidence for Escaping Ionizing Photons in Local Lyman Break Galaxy Analogs}",
      journal = {\apj},
         year = 2015,
        month = sep,
       volume = {810},
       number = {2},
          eid = {104},
        pages = {104},
          doi = {10.1088/0004-637X/810/2/104},
archivePrefix = {arXiv},
       eprint = {1504.02446},
 primaryClass = {astro-ph.GA},
       adsurl = {https://ui.adsabs.harvard.edu/abs/2015ApJ...810..104A}
}

@ARTICLE{Allen2008,
       author = {{Allen}, Mark G. and {Groves}, Brent A. and {Dopita}, Michael A. and {Sutherland}, Ralph S. and {Kewley}, Lisa J.},
        title = "{The MAPPINGS III Library of Fast Radiative Shock Models}",
      journal = {\apjs},
         year = 2008,
        month = sep,
       volume = {178},
       number = {1},
        pages = {20-55},
          doi = {10.1086/589652},
archivePrefix = {arXiv},
       eprint = {0805.0204},
 primaryClass = {astro-ph},
       adsurl = {https://ui.adsabs.harvard.edu/abs/2008ApJS..178...20A}
}

@ARTICLE{Ananna2024,
       author = {{Ananna}, Tonima Tasnim and {Bogd{\'a}n}, {\'A}kos and {Kov{\'a}cs}, Orsolya E. and {Natarajan}, Priyamvada and {Hickox}, Ryan C.},
        title = "{X-Ray View of Little Red Dots: Do They Host Supermassive Black Holes?}",
      journal = {\apjl},
         year = 2024,
        month = jul,
       volume = {969},
       number = {1},
          eid = {L18},
        pages = {L18},
          doi = {10.3847/2041-8213/ad5669},
archivePrefix = {arXiv},
       eprint = {2404.19010},
 primaryClass = {astro-ph.GA},
       adsurl = {https://ui.adsabs.harvard.edu/abs/2024ApJ...969L..18A}
}

@ARTICLE{Arevalo2025,
       author = {{Arevalo-Gonzalez}, Flor and {Braun}, Titanilla and {Trussler}, James and {Conselice}, Christopher J. and {Harvey}, Thomas and {Adams}, Nathan and {Austin}, Duncan and {Li}, Qiong and {Juod{\v{z}}balis}, Ignas and {Nakajima}, Kimihiko},
        title = "{New methods of identifying AGN in the early Universe using spectroscopy and photometry in the JWST era}",
      journal = {\mnras},
         year = 2025,
        month = dec,
       volume = {544},
       number = {3},
        pages = {2737-2757},
          doi = {10.1093/mnras/staf1836},
archivePrefix = {arXiv},
       eprint = {2501.09585},
 primaryClass = {astro-ph.GA},
       adsurl = {https://ui.adsabs.harvard.edu/abs/2025MNRAS.544.2737A}
}

@ARTICLE{Asada2026,
       author = {{Asada}, Yoshihisa and {Inayoshi}, Kohei and {Fei}, Qinyue and {Fujimoto}, Seiji and {Willott}, Chris},
        title = "{Origins of the UV continuum and Balmer emission lines in Little Red Dots: observational validation of dense gas envelope models enshrouding the AGN}",
      journal = {arXiv e-prints},
         year = 2026,
        month = jan,
          eid = {arXiv:2601.10573},
        pages = {arXiv:2601.10573},
          doi = {10.48550/arXiv.2601.10573},
archivePrefix = {arXiv},
       eprint = {2601.10573},
 primaryClass = {astro-ph.GA},
       adsurl = {https://ui.adsabs.harvard.edu/abs/2026arXiv260110573A}
}

@ARTICLE{Atek2025,
       author = {{Atek}, Hakim and {Chisholm}, John and {Kokorev}, Vasily and {Endsley}, Ryan and {Pan}, Richard and {Furtak}, Lukas and {Chemerynska}, Iryna and {Richard}, Johan and {Claeyssens}, Ad{\'e}la{\"\i}de and {Oesch}, Pascal and {Fujimoto}, Seiji and {Naidu}, Rohan and {Korber}, Damien and {Schaerer}, Daniel and {Blaizot}, Jeremy and {Rosdahl}, Joki and {Adamo}, Angela and {Asada}, Yoshihisa and {Basu}, Arghyadeep and {Beauchesne}, Benjamin and {Berg}, Danielle and {Bezanson}, Rachel and {Bouwens}, Rychard and {Brammer}, Gabriel and {Dessauges-Zavadsky}, Miroslava and {Ellien}, Ama{\"e}l and {Ezziati}, Meriam and {Fei}, Qinyue and {Goovaerts}, Ilias and {Heurtier}, Sylvain and {Hsiao}, Tiger Yu-Yang and {Jecmen}, Michelle and {Khullar}, Gourav and {Kneib}, Jean-Paul and {Labb{\'e}}, Ivo and {Leclercq}, Floriane and {Marques-Chaves}, Rui and {Mason}, Charlotte and {McQuinn}, Kristen B.~W. and {Mu{\~n}oz}, Julian B. and {Natarajan}, Priyamvada and {Saldana-Lopez}, Alberto and {Stephenson}, Mabel G. and {Trebitsch}, Maxime and {Volonteri}, Marta and {Weibel}, Andrea and {Zitrin}, Adi},
        title = "{JWST's GLIMPSE: an overview of the deepest probe of early galaxy formation and cosmic reionization}",
      journal = {arXiv e-prints},
         year = 2025,
        month = nov,
          eid = {arXiv:2511.07542},
        pages = {arXiv:2511.07542},
          doi = {10.48550/arXiv.2511.07542},
archivePrefix = {arXiv},
       eprint = {2511.07542},
 primaryClass = {astro-ph.GA},
       adsurl = {https://ui.adsabs.harvard.edu/abs/2025arXiv251107542A}
}

@ARTICLE{Azadi2025,
       author = {{Azadi}, Mojegan and {Wilkes}, Belinda and {Kuraszkiewicz}, Joanna and {Willner}, Steven. P. and {Ashby}, Matthew L.~N.},
        title = "{A Bolometric Luminosity Correction Recipe for AGN at Any Epoch}",
      journal = {arXiv e-prints},
         year = 2025,
        month = sep,
          eid = {arXiv:2509.19666},
        pages = {arXiv:2509.19666},
          doi = {10.48550/arXiv.2509.19666},
archivePrefix = {arXiv},
       eprint = {2509.19666},
 primaryClass = {astro-ph.GA},
       adsurl = {https://ui.adsabs.harvard.edu/abs/2025arXiv250919666A}
}

@ARTICLE{Backhaus2022,
       author = {{Backhaus}, Bren E. and {Trump}, Jonathan R. and {Cleri}, Nikko J. and {Simons}, Raymond and {Momcheva}, Ivelina and {Papovich}, Casey and {Estrada-Carpenter}, Vicente and {Finkelstein}, Steven L. and {Matharu}, Jasleen and {Ji}, Zhiyuan and {Weiner}, Benjamin and {Giavalisco}, Mauro and {Jung}, Intae},
        title = "{CLEAR: Emission-line Ratios at Cosmic High Noon}",
      journal = {\apj},
         year = 2022,
        month = feb,
       volume = {926},
       number = {2},
          eid = {161},
        pages = {161},
          doi = {10.3847/1538-4357/ac3919},
archivePrefix = {arXiv},
       eprint = {2109.08147},
 primaryClass = {astro-ph.GA},
       adsurl = {https://ui.adsabs.harvard.edu/abs/2022ApJ...926..161B}
}

@ARTICLE{Baggen2024,
       author = {{Baggen}, Josephine F.~W. and {van Dokkum}, Pieter and {Brammer}, Gabriel and {de Graaff}, Anna and {Franx}, Marijn and {Greene}, Jenny and {Labb{\'e}}, Ivo and {Leja}, Joel and {Maseda}, Michael V. and {Nelson}, Erica J. and {Rix}, Hans-Walter and {Wang}, Bingjie and {Weibel}, Andrea},
        title = "{The Small Sizes and High Implied Densities of ``Little Red Dots'' with Balmer Breaks Could Explain Their Broad Emission Lines without an Active Galactic Nucleus}",
      journal = {\apjl},
         year = 2024,
        month = dec,
       volume = {977},
       number = {1},
          eid = {L13},
        pages = {L13},
          doi = {10.3847/2041-8213/ad90b8},
archivePrefix = {arXiv},
       eprint = {2408.07745},
 primaryClass = {astro-ph.GA},
       adsurl = {https://ui.adsabs.harvard.edu/abs/2024ApJ...977L..13B}
}

@ARTICLE{Baggen2026,
       author = {{Baggen}, Josephine F.~W. and {Scoggins}, Matthew T. and {van Dokkum}, Pieter and {Haiman}, Zolt{\'a}n and {Torralba}, Alberto and {Matthee}, Jorryt},
        title = "{Connecting the Dots: UV-bright Companions of Little Red Dots as Lyman─Werner Sources Enabling Direct-collapse Black Hole Formation}",
      journal = {\apjl},
         year = 2026,
        month = may,
       volume = {1002},
       number = {1},
          eid = {L4},
        pages = {L4},
          doi = {10.3847/2041-8213/ae58a5},
archivePrefix = {arXiv},
       eprint = {2602.02702},
 primaryClass = {astro-ph.GA},
       adsurl = {https://ui.adsabs.harvard.edu/abs/2026ApJ..1002L...4B}
}

@ARTICLE{Baldwin1977,
       author = {{Baldwin}, Jack A.},
        title = "{Luminosity Indicators in the Spectra of Quasi-Stellar Objects}",
      journal = {\apj},
         year = 1977,
        month = jun,
       volume = {214},
        pages = {679-684},
          doi = {10.1086/155294},
       adsurl = {https://ui.adsabs.harvard.edu/abs/1977ApJ...214..679B}
}

@ARTICLE{Baldwin81,
       author = {{Baldwin}, J.~A. and {Phillips}, M.~M. and {Terlevich}, R.},
        title = "{Classification parameters for the emission-line spectra of extragalactic objects.}",
      journal = {\pasp},
         year = 1981,
        month = feb,
       volume = {93},
        pages = {5-19},
          doi = {10.1086/130766},
       adsurl = {https://ui.adsabs.harvard.edu/abs/1981PASP...93....5B}
}

@ARTICLE{Baldwin1995,
       author = {{Baldwin}, Jack and {Ferland}, Gary and {Korista}, Kirk and {Verner}, Dima},
        title = "{Locally Optimally Emitting Clouds and the Origin of Quasar Emission Lines}",
      journal = {\apjl},
         year = 1995,
        month = dec,
       volume = {455},
        pages = {L119},
          doi = {10.1086/309827},
archivePrefix = {arXiv},
       eprint = {astro-ph/9510080},
 primaryClass = {astro-ph},
       adsurl = {https://ui.adsabs.harvard.edu/abs/1995ApJ...455L.119B}
}

@INPROCEEDINGS{Baldwin1997,
       author = {{Baldwin}, J.~A.},
        title = "{Broad Emission Lines in Active Galactic Nuclei}",
    booktitle = {IAU Colloquium 159: Emission Lines in Active Galaxies: New Methods and Techniques},
         year = 1997,
       editor = {{Peterson}, Bradley M. and {Cheng}, Fu-Zhen and {Wilson}, Andrew S.},
       series = {Astronomical Society of the Pacific Conference Series},
       volume = {113},
        month = jan,
        pages = {80},
       adsurl = {https://ui.adsabs.harvard.edu/abs/1997ASPC..113...80B}
}

@ARTICLE{Barro2026,
       author = {{Barro}, Guillermo and {P{\'e}rez-Gonz{\'a}lez}, Pablo G. and {Kocevski}, Dale and {Trump}, Jonathan R. and {Dickinson}, Mark and {Arrabal Haro}, Pablo and {Brooks}, Madisyn and {Donnan}, Callum T. and {Dunlop}, James S. and {Finkelstein}, Steven L. and {Franco}, Maximilien and {Gandolfi}, Giovanni and {Giavalisco}, Mauro and {Grogin}, Norman A. and {Hirschmann}, Michaela and {Kartaltepe}, Jeyhan S. and {Koekemoer}, Anton M. and {Larson}, Rebecca L. and {Leung}, Gene C.~K. and {Lucas}, Ray A. and {McGrath}, Elizabeth J. and {Papovich}, Casey and {P{\'e}rez-D{\'\i}az}, Borja and {Somerville}, Rachel S. and {Taylor}, Elizabeth and {Taylor}, Anthony J. and {Tripodi}, Roberta and {Yung}, L.~Y. Aaron and {Wang}, Xin},
        title = "{From ``The Cliff'' to ``Virgil'': Mapping the Spectral Diversity of Little Red Dots with JWST/NIRSpec}",
      journal = {\apj},
         year = 2026,
        month = may,
       volume = {1003},
       number = {1},
          eid = {96},
        pages = {96},
          doi = {10.3847/1538-4357/ae5d2c},
archivePrefix = {arXiv},
       eprint = {2512.15853},
 primaryClass = {astro-ph.GA},
       adsurl = {https://ui.adsabs.harvard.edu/abs/2026ApJ..1003...96B}
}

@ARTICLE{Baskin2005,
       author = {{Baskin}, Alexei and {Laor}, Ari},
        title = "{What controls the [OIII]{\ensuremath{\lambda}}5007 line strength in active galactic nuclei?}",
      journal = {\mnras},
         year = 2005,
        month = apr,
       volume = {358},
       number = {3},
        pages = {1043-1054},
          doi = {10.1111/j.1365-2966.2005.08841.x},
archivePrefix = {arXiv},
       eprint = {astro-ph/0501436},
 primaryClass = {astro-ph},
       adsurl = {https://ui.adsabs.harvard.edu/abs/2005MNRAS.358.1043B}
}

@ARTICLE{Barger2026,
       author = {{Barger}, A.~J. and {Cowie}, L.~L.},
        title = "{Double Dots: Compact Pairs Mark Little Red Dots and High-Redshift Broad-line AGNs}",
      journal = {arXiv e-prints},
         year = 2026,
        month = may,
          eid = {arXiv:2605.27903},
        pages = {arXiv:2605.27903},
          doi = {10.48550/arXiv.2605.27903},
archivePrefix = {arXiv},
       eprint = {2605.27903},
 primaryClass = {astro-ph.GA},
       adsurl = {https://ui.adsabs.harvard.edu/abs/2026arXiv260527903B}
}

@ARTICLE{Begelman2006,
       author = {{Begelman}, Mitchell C. and {Volonteri}, Marta and {Rees}, Martin J.},
        title = "{Formation of supermassive black holes by direct collapse in pre-galactic haloes}",
      journal = {\mnras},
         year = 2006,
        month = jul,
       volume = {370},
       number = {1},
        pages = {289-298},
          doi = {10.1111/j.1365-2966.2006.10467.x},
archivePrefix = {arXiv},
       eprint = {astro-ph/0602363},
 primaryClass = {astro-ph},
       adsurl = {https://ui.adsabs.harvard.edu/abs/2006MNRAS.370..289B}
}

@ARTICLE{Begelman2008,
       author = {{Begelman}, Mitchell C. and {Rossi}, Elena M. and {Armitage}, Philip J.},
        title = "{Quasi-stars: accreting black holes inside massive envelopes}",
      journal = {\mnras},
         year = 2008,
        month = jul,
       volume = {387},
       number = {4},
        pages = {1649-1659},
          doi = {10.1111/j.1365-2966.2008.13344.x},
archivePrefix = {arXiv},
       eprint = {0711.4078},
 primaryClass = {astro-ph},
       adsurl = {https://ui.adsabs.harvard.edu/abs/2008MNRAS.387.1649B}
}

@ARTICLE{Bentz2009,
       author = {{Bentz}, Misty C. and {Peterson}, Bradley M. and {Netzer}, Hagai and {Pogge}, Richard W. and {Vestergaard}, Marianne},
        title = "{The Radius-Luminosity Relationship for Active Galactic Nuclei: The Effect of Host-Galaxy Starlight on Luminosity Measurements. II. The Full Sample of Reverberation-Mapped AGNs}",
      journal = {\apj},
         year = 2009,
        month = may,
       volume = {697},
       number = {1},
        pages = {160-181},
          doi = {10.1088/0004-637X/697/1/160},
archivePrefix = {arXiv},
       eprint = {0812.2283},
 primaryClass = {astro-ph},
       adsurl = {https://ui.adsabs.harvard.edu/abs/2009ApJ...697..160B}
}

@ARTICLE{Bentz2013,
       author = {{Bentz}, Misty C. and {Denney}, Kelly D. and {Grier}, Catherine J. and {Barth}, Aaron J. and {Peterson}, Bradley M. and {Vestergaard}, Marianne and {Bennert}, Vardha N. and {Canalizo}, Gabriela and {De Rosa}, Gisella and {Filippenko}, Alexei V. and {Gates}, Elinor L. and {Greene}, Jenny E. and {Li}, Weidong and {Malkan}, Matthew A. and {Pogge}, Richard W. and {Stern}, Daniel and {Treu}, Tommaso and {Woo}, Jong-Hak},
        title = "{The Low-luminosity End of the Radius-Luminosity Relationship for Active Galactic Nuclei}",
      journal = {\apj},
         year = 2013,
        month = apr,
       volume = {767},
       number = {2},
          eid = {149},
        pages = {149},
          doi = {10.1088/0004-637X/767/2/149},
archivePrefix = {arXiv},
       eprint = {1303.1742},
 primaryClass = {astro-ph.CO},
       adsurl = {https://ui.adsabs.harvard.edu/abs/2013ApJ...767..149B}
}

@ARTICLE{Binette2002,
       author = {{Binette}, L. and {Gonz{\'a}lez-G{\'o}mez}, D.~I. and {Mayya}, Y.~D.},
        title = "{Density Gradients and Internal Dust in the Orion Nebula}",
      journal = {\rmxaa},
         year = 2002,
        month = oct,
       volume = {38},
        pages = {279-288},
archivePrefix = {arXiv},
       eprint = {astro-ph/0210646},
 primaryClass = {astro-ph},
       adsurl = {https://ui.adsabs.harvard.edu/abs/2002RMxAA..38..279B}
}

@ARTICLE{Binette2003,
       author = {{Binette}, L. and {Groves}, B. and {Villar-Mart{\'\i}n}, M. and {Fosbury}, R.~A.~E. and {Axon}, D.~J.},
        title = "{High-z nebulae: Ionization by stars or by an obscured QSO?}",
      journal = {\aap},
         year = 2003,
        month = jul,
       volume = {405},
        pages = {975-980},
          doi = {10.1051/0004-6361:20030718},
       adsurl = {https://ui.adsabs.harvard.edu/abs/2003A&A...405..975B}
}

@ARTICLE{Bogdan2024,
       author = {{Bogd{\'a}n}, {\'A}kos and {Goulding}, Andy D. and {Natarajan}, Priyamvada and {Kov{\'a}cs}, Orsolya E. and {Tremblay}, Grant R. and {Chadayammuri}, Urmila and {Volonteri}, Marta and {Kraft}, Ralph P. and {Forman}, William R. and {Jones}, Christine and {Churazov}, Eugene and {Zhuravleva}, Irina},
        title = "{Evidence for heavy-seed origin of early supermassive black holes from a z ≍ 10 X-ray quasar}",
      journal = {Nature Astronomy},
         year = 2024,
        month = jan,
       volume = {8},
       number = {1},
        pages = {126-133},
          doi = {10.1038/s41550-023-02111-9},
archivePrefix = {arXiv},
       eprint = {2305.15458},
 primaryClass = {astro-ph.GA},
       adsurl = {https://ui.adsabs.harvard.edu/abs/2024NatAs...8..126B}
}

@ARTICLE{Bottorff2000,
       author = {{Bottorff}, Mark and {Ferland}, Gary and {Baldwin}, Jack and {Korista}, Kirk},
        title = "{Observational Constraints on the Internal Velocity Field of Quasar Emission-Line Clouds}",
      journal = {\apj},
         year = 2000,
        month = oct,
       volume = {542},
       number = {2},
        pages = {644-654},
          doi = {10.1086/317051},
       adsurl = {https://ui.adsabs.harvard.edu/abs/2000ApJ...542..644B}
}

@ARTICLE{Brazzini2026,
       author = {{Brazzini}, M. and {D'Eugenio}, F. and {Maiolino}, R. and {Lyu}, J. and {DeCoursey}, C. and {{\"U}bler}, H. and {Ji}, X. and {Juod{\v{z}}balis}, I. and {Scholtz}, J. and {Jones}, G.~C. and {Hainline}, K. and {Dalla Bont{\`a}}, E. and {{\'e}rez-Gonz{\'a}lez}, P.~G. P and {Geris}, S. and {Harshan}, A. and {Feruglio}, C. and {Bischetti}, M. and {Mazzolari}, G. and {Rieke}, G. and {Alberts}, S. and {Trefoloni}, B. and {Carniani}, S. and {Parlanti}, E. and {Marconi}, A. and {Risaliti}, G. and {Ramos Almeida}, C. and {Rinaldi}, P. and {Perna}, M. and {Zamora}, S. and {Lamperti}, I. and {Venturi}, G. and {Cresci}, G. and {Bunker}, Andrew J. and {Ivey}, L.~R.},
        title = "{The Little Blue and Red Dots Rosetta Stones: Non-Gaussian broad lines, hot dust, and X-ray weakness}",
      journal = {arXiv e-prints},
         year = 2026,
        month = jan,
          eid = {arXiv:2601.22214},
        pages = {arXiv:2601.22214},
          doi = {10.48550/arXiv.2601.22214},
archivePrefix = {arXiv},
       eprint = {2601.22214},
 primaryClass = {astro-ph.GA},
       adsurl = {https://ui.adsabs.harvard.edu/abs/2026arXiv260122214B}
}

@ARTICLE{DallaBonta2025,
       author = {{Dalla Bont{\`a}}, E. and {Peterson}, B.~M. and {Grier}, C.~J. and {Berton}, M. and {Brandt}, W.~N. and {Ciroi}, S. and {Corsini}, E.~M. and {Dalla Barba}, B. and {Davies}, R. and {Dehghanian}, M. and {Edelson}, R. and {Foschini}, L. and {Gasparri}, D. and {Ho}, L.~C. and {Horne}, K. and {Iodice}, E. and {Morelli}, L. and {Pizzella}, A. and {Portaluri}, E. and {Shen}, Y. and {Schneider}, D.~P. and {Vestergaard}, M.},
        title = "{Estimating masses of supermassive black holes in active galactic nuclei from the H{\ensuremath{\alpha}} emission line}",
      journal = {\aap},
         year = 2025,
        month = apr,
       volume = {696},
          eid = {A48},
        pages = {A48},
          doi = {10.1051/0004-6361/202452746},
archivePrefix = {arXiv},
       eprint = {2410.21387},
 primaryClass = {astro-ph.GA},
       adsurl = {https://ui.adsabs.harvard.edu/abs/2025A&A...696A..48D}
}

@ARTICLE{Bottorff2002,
       author = {{Bottorff}, Mark and {Ferland}, Gary},
        title = "{Dissipative Heating and Quasar Emission Lines}",
      journal = {\apj},
         year = 2002,
        month = apr,
       volume = {568},
       number = {2},
        pages = {581-591},
          doi = {10.1086/339058},
       adsurl = {https://ui.adsabs.harvard.edu/abs/2002ApJ...568..581B}
}

@ARTICLE{Bruzual2003,
       author = {{Bruzual}, G. and {Charlot}, S.},
        title = "{Stellar population synthesis at the resolution of 2003}",
      journal = {\mnras},
         year = 2003,
        month = oct,
       volume = {344},
       number = {4},
        pages = {1000-1028},
          doi = {10.1046/j.1365-8711.2003.06897.x},
archivePrefix = {arXiv},
       eprint = {astro-ph/0309134},
 primaryClass = {astro-ph},
       adsurl = {https://ui.adsabs.harvard.edu/abs/2003MNRAS.344.1000B}
}

@ARTICLE{Bromm2003,
       author = {{Bromm}, Volker and {Loeb}, Abraham},
        title = "{Formation of the First Supermassive Black Holes}",
      journal = {\apj},
         year = 2003,
        month = oct,
       volume = {596},
       number = {1},
        pages = {34-46},
          doi = {10.1086/377529},
archivePrefix = {arXiv},
       eprint = {astro-ph/0212400},
 primaryClass = {astro-ph},
       adsurl = {https://ui.adsabs.harvard.edu/abs/2003ApJ...596...34B}
}

@ARTICLE{Bunker2023,
       author = {{Bunker}, Andrew J. and {Saxena}, Aayush and {Cameron}, Alex J. and {Willott}, Chris J. and {Curtis-Lake}, Emma and {Jakobsen}, Peter and {Carniani}, Stefano and {Smit}, Renske and {Maiolino}, Roberto and {Witstok}, Joris and {Curti}, Mirko and {D'Eugenio}, Francesco and {Jones}, Gareth C. and {Ferruit}, Pierre and {Arribas}, Santiago and {Charlot}, Stephane and {Chevallard}, Jacopo and {Giardino}, Giovanna and {de Graaff}, Anna and {Looser}, Tobias J. and {L{\"u}tzgendorf}, Nora and {Maseda}, Michael V. and {Rawle}, Tim and {Rix}, Hans-Walter and {Del Pino}, Bruno Rodr{\'\i}guez and {Alberts}, Stacey and {Egami}, Eiichi and {Eisenstein}, Daniel J. and {Endsley}, Ryan and {Hainline}, Kevin and {Hausen}, Ryan and {Johnson}, Benjamin D. and {Rieke}, George and {Rieke}, Marcia and {Robertson}, Brant E. and {Shivaei}, Irene and {Stark}, Daniel P. and {Sun}, Fengwu and {Tacchella}, Sandro and {Tang}, Mengtao and {Williams}, Christina C. and {Willmer}, Christopher N.~A. and {Baker}, William M. and {Baum}, Stefi and {Bhatawdekar}, Rachana and {Bowler}, Rebecca and {Boyett}, Kristan and {Chen}, Zuyi and {Circosta}, Chiara and {Helton}, Jakob M. and {Ji}, Zhiyuan and {Kumari}, Nimisha and {Lyu}, Jianwei and {Nelson}, Erica and {Parlanti}, Eleonora and {Perna}, Michele and {Sandles}, Lester and {Scholtz}, Jan and {Suess}, Katherine A. and {Topping}, Michael W. and {{\"U}bler}, Hannah and {Wallace}, Imaan E.~B. and {Whitler}, Lily},
        title = "{JADES NIRSpec Spectroscopy of GN-z11: Lyman-{\ensuremath{\alpha}} emission and possible enhanced nitrogen abundance in a z = 10.60 luminous galaxy}",
      journal = {\aap},
         year = 2023,
        month = sep,
       volume = {677},
          eid = {A88},
        pages = {A88},
          doi = {10.1051/0004-6361/202346159},
archivePrefix = {arXiv},
       eprint = {2302.07256},
 primaryClass = {astro-ph.GA},
       adsurl = {https://ui.adsabs.harvard.edu/abs/2023A&A...677A..88B}
}

@ARTICLE{Chang2026,
       author = {{Chang}, Seok-Jun and {Gronke}, Max and {Matthee}, Jorryt and {Mason}, Charlotte},
        title = "{Impact of resonance, Raman, and Thomson scattering on hydrogen line formation in Little Red Dots}",
      journal = {\mnras},
         year = 2026,
        month = feb,
       volume = {545},
       number = {4},
          eid = {staf2131},
        pages = {staf2131},
          doi = {10.1093/mnras/staf2131},
archivePrefix = {arXiv},
       eprint = {2508.08768},
 primaryClass = {astro-ph.GA},
       adsurl = {https://ui.adsabs.harvard.edu/abs/2026MNRAS.545f2131C}
}

@ARTICLE{Castaneda1992,
       author = {{Castaneda}, H.~O. and {Vilchez}, J.~M. and {Copetti}, M.~V.~F.},
        title = "{Density studies on giant extragalactic HII regions.}",
      journal = {\aap},
         year = 1992,
        month = jul,
       volume = {260},
        pages = {370-380},
       adsurl = {https://ui.adsabs.harvard.edu/abs/1992A&A...260..370C}
}

@ARTICLE{Castellano2026,
       author = {{Castellano}, M. and {Napolitano}, L. and {Moreschini}, B. and {Calabr{\`o}}, A. and {Christensen}, L. and {Llerena}, M. and {Bakx}, T.~J.~L.~C. and {Belfiore}, F. and {Bevacqua}, D. and {Dickinson}, M. and {Fontana}, A. and {Gandolfi}, G. and {Gasparetto}, T. and {Marconi}, A. and {Mascia}, S. and {Merlin}, E. and {Morishita}, T. and {Nanayakkara}, T. and {Paris}, D. and {Pentericci}, L. and {P{\'e}rez-D{\'\i}az}, B. and {Roberts-Borsani}, G. and {Ruiz}, S. Rojas and {Santini}, P. and {Treu}, T. and {Vanzella}, E. and {Vulcani}, B. and {Wang}, X. and {Yoon}, I. and {Zavala}, J.},
        title = "{Investigating ionising sources and the complex interstellar medium of GHZ2 at z = 12.3}",
      journal = {The Open Journal of Astrophysics},
         year = 2026,
        month = apr,
       volume = {9},
        pages = {60281},
          doi = {10.33232/001c.160281},
archivePrefix = {arXiv},
       eprint = {2512.08490},
 primaryClass = {astro-ph.GA},
       adsurl = {https://ui.adsabs.harvard.edu/abs/2026OJAp....960281C}
}

@ARTICLE{Chavez2025,
       author = {{Chavez Ortiz}, Oscar A. and {Finkelstein}, Steven L. and {Plat}, Adele and {Silcock}, Maddie and {Lake}, Emma Curtis and {Taylor}, Anthony and {Gupta}, Ansh R. and {Napolitano}, Lorenzo and {Castellano}, Marco and {Bromm}, Volker and {Mitsuhashi}, Ikki and {Charlot}, Stephane and {Fontana}, Adriano and {Zavala}, Jorge A. and {Chevallard}, Jacopo and {Burgarella}, Denis and {Hirschmann}, Michaela and {Bakx}, Tom and {Vidal-Garcia}, Alba and {Calabro}, Antonello and {Feltre}, Anna},
        title = "{Significant Evidence of an AGN Contribution in GHZ2 at z = 12.34}",
      journal = {arXiv e-prints},
         year = 2025,
        month = nov,
          eid = {arXiv:2511.03035},
        pages = {arXiv:2511.03035},
          doi = {10.48550/arXiv.2511.03035},
archivePrefix = {arXiv},
       eprint = {2511.03035},
 primaryClass = {astro-ph.GA},
       adsurl = {https://ui.adsabs.harvard.edu/abs/2025arXiv251103035C}
}

@ARTICLE{Chevallard2016,
       author = {{Chevallard}, Jacopo and {Charlot}, St{\'e}phane},
        title = "{Modelling and interpreting spectral energy distributions of galaxies with BEAGLE}",
      journal = {\mnras},
         year = 2016,
        month = oct,
       volume = {462},
       number = {2},
        pages = {1415-1443},
          doi = {10.1093/mnras/stw1756},
archivePrefix = {arXiv},
       eprint = {1603.03037},
 primaryClass = {astro-ph.GA},
       adsurl = {https://ui.adsabs.harvard.edu/abs/2016MNRAS.462.1415C}
}

@ARTICLE{Chisholm2024,
       author = {{Chisholm}, J. and {Berg}, D.~A. and {Endsley}, R. and {Gazagnes}, S. and {Richardson}, C.~T. and {Lambrides}, E. and {Greene}, J. and {Finkelstein}, S. and {Flury}, S. and {Guseva}, N.~G. and {Henry}, A. and {Hutchison}, T.~A. and {Izotov}, Y.~I. and {Marques-Chaves}, R. and {Oesch}, P. and {Papovich}, C. and {Saldana-Lopez}, A. and {Schaerer}, D. and {Stephenson}, M.~G.},
        title = "{[Ne v] emission from a faint epoch of reionization-era galaxy: evidence for a narrow-line intermediate-mass black hole}",
      journal = {\mnras},
         year = 2024,
        month = nov,
       volume = {534},
       number = {3},
        pages = {2633-2652},
          doi = {10.1093/mnras/stae2199},
archivePrefix = {arXiv},
       eprint = {2402.18643},
 primaryClass = {astro-ph.GA},
       adsurl = {https://ui.adsabs.harvard.edu/abs/2024MNRAS.534.2633C}
}

@ARTICLE{Chisholm2026,
       author = {{Chisholm}, John and {Berg}, Danielle A. and {Boylan-Kolchin}, Michael and {de Graaff}, Anna and {Furtak}, Lukas J. and {Kokorev}, Vasily and {Matthee}, Jorryt and {Mu{\~n}oz}, Julian B. and {Naidu}, Rohan P. and {Sander}, Andreas A.~C.},
        title = "{Little Red Dots as Globular Clusters in Formation}",
      journal = {\apjl},
         year = 2026,
        month = jun,
       volume = {1004},
       number = {1},
          eid = {L4},
        pages = {L4},
          doi = {10.3847/2041-8213/ae6dae},
archivePrefix = {arXiv},
       eprint = {2602.15935},
 primaryClass = {astro-ph.GA},
       adsurl = {https://ui.adsabs.harvard.edu/abs/2026ApJ..1004L...4C}
}

@ARTICLE{Clavel1991,
       author = {{Clavel}, J. and {Reichert}, G.~A. and {Alloin}, D. and {Crenshaw}, D.~M. and {Kriss}, G. and {Krolik}, J.~H. and {Malkan}, M.~A. and {Netzer}, H. and {Peterson}, B.~M. and {Wamsteker}, W. and {Altamore}, A. and {Baribaud}, T. and {Barr}, P. and {Beck}, S. and {Binette}, L. and {Bromage}, G.~E. and {Brosch}, N. and {Diaz}, A.~I. and {Filippenko}, A.~V. and {Fricke}, K. and {Gaskell}, C.~M. and {Giommi}, P. and {Glass}, I.~S. and {Gondhalekar}, P. and {Hackney}, R.~L. and {Halpern}, J.~P. and {Hutter}, D.~J. and {Joersaeter}, S. and {Kinney}, A.~L. and {Kollatschny}, W. and {Koratkar}, A. and {Korista}, K.~T. and {Laor}, A. and {Lasota}, J.-P. and {Leibowitz}, E. and {Maoz}, D. and {Martin}, P.~G. and {Mazeh}, T. and {Meurs}, E.~J.~A. and {Nair}, A.~D. and {O'Brien}, P. and {Pelat}, D. and {Perez}, E. and {Perola}, G.~C. and {Ptak}, R.~L. and {Rodriguez-Pascual}, P. and {Rosenblatt}, E.~I. and {Sadun}, A.~C. and {Santos-Lleo}, M. and {Shaw}, R.~A. and {Smith}, P.~S. and {Stirpe}, G.~M. and {Stoner}, R. and {Sun}, W.~H. and {Ulrich}, M.-H. and {van Groningen}, E. and {Zheng}, W.},
        title = "{Steps toward Determination of the Size and Structure of the Broad-Line Region in Active Galactic Nuclei. I. an 8 Month Campaign of Monitoring NGC 5548 with IUE}",
      journal = {\apj},
         year = 1991,
        month = jan,
       volume = {366},
        pages = {64},
          doi = {10.1086/169540},
       adsurl = {https://ui.adsabs.harvard.edu/abs/1991ApJ...366...64C}
}

@ARTICLE{Clegg1987,
       author = {{Clegg}, R.~E.~S.},
        title = "{Collisional effects in He I lines and helium abundances in planetary nebulae.}",
      journal = {\mnras},
         year = 1987,
        month = nov,
       volume = {229},
        pages = {31P-39},
          doi = {10.1093/mnras/229.1.31P},
       adsurl = {https://ui.adsabs.harvard.edu/abs/1987MNRAS.229P..31C}
}

@ARTICLE{Cleri2023,
       author = {{Cleri}, Nikko J. and {Olivier}, Grace M. and {Hutchison}, Taylor A. and {Papovich}, Casey and {Trump}, Jonathan R. and {Amor{\'\i}n}, Ricardo O. and {Backhaus}, Bren E. and {Berg}, Danielle A. and {Fern{\'a}ndez}, Vital and {Finkelstein}, Steven L. and {Fujimoto}, Seiji and {Hirschmann}, Michaela and {Kartaltepe}, Jeyhan S. and {Kocevski}, Dale D. and {Simons}, Raymond C. and {Wilkins}, Stephen M. and {Yung}, L.~Y. Aaron},
        title = "{Using [Ne V]/[Ne III] to Understand the Nature of Extreme-ionization Galaxies}",
      journal = {\apj},
         year = 2023,
        month = aug,
       volume = {953},
       number = {1},
          eid = {10},
        pages = {10},
          doi = {10.3847/1538-4357/acde55},
archivePrefix = {arXiv},
       eprint = {2301.07745},
 primaryClass = {astro-ph.GA},
       adsurl = {https://ui.adsabs.harvard.edu/abs/2023ApJ...953...10C}
}

@ARTICLE{CLOUDY2023,
       author = {{Chatzikos}, M. and {Bianchi}, S. and {Camilloni}, F. and {Chakraborty}, P. and {Gunasekera}, C.~M. and {Guzm{\'a}n}, F. and {Milby}, J.~S. and {Sarkar}, A. and {Shaw}, G. and {van Hoof}, P.~A.~M. and {Ferland}, G.~J.},
        title = "{The 2023 Release of Cloudy}",
      journal = {\rmxaa},
         year = 2023,
        month = oct,
       volume = {59},
        pages = {327-343},
          doi = {10.22201/ia.01851101p.2023.59.02.12},
archivePrefix = {arXiv},
       eprint = {2308.06396},
 primaryClass = {astro-ph.GA},
       adsurl = {https://ui.adsabs.harvard.edu/abs/2023RMxAA..59..327C}
}

@ARTICLE{Cullen2021,
       author = {{Cullen}, F. and {Shapley}, A.~E. and {McLure}, R.~J. and {Dunlop}, J.~S. and {Sanders}, R.~L. and {Topping}, M.~W. and {Reddy}, N.~A. and {Amor{\'\i}n}, R. and {Begley}, R. and {Bolzonella}, M. and {Calabr{\`o}}, A. and {Carnall}, A.~C. and {Castellano}, M. and {Cimatti}, A. and {Cirasuolo}, M. and {Cresci}, G. and {Fontana}, A. and {Fontanot}, F. and {Garilli}, B. and {Guaita}, L. and {Hamadouche}, M. and {Hathi}, N.~P. and {Mannucci}, F. and {McLeod}, D.~J. and {Pentericci}, L. and {Saxena}, A. and {Talia}, M. and {Zamorani}, G.},
        title = "{The NIRVANDELS Survey: a robust detection of {\ensuremath{\alpha}}-enhancement in star-forming galaxies at z ≃ 3.4}",
      journal = {\mnras},
         year = 2021,
        month = jul,
       volume = {505},
       number = {1},
        pages = {903-920},
          doi = {10.1093/mnras/stab1340},
archivePrefix = {arXiv},
       eprint = {2103.06300},
 primaryClass = {astro-ph.GA},
       adsurl = {https://ui.adsabs.harvard.edu/abs/2021MNRAS.505..903C}
}

@ARTICLE{DAgostino2019,
       author = {{D'Agostino}, Joshua J. and {Kewley}, Lisa J. and {Groves}, Brent and {Byler}, Nell and {Sutherland}, Ralph S. and {Nicholls}, David and {Leitherer}, Claus and {Stanway}, Elizabeth R.},
        title = "{Comparison of Theoretical Starburst Photoionization Models for Optical Diagnostics}",
      journal = {\apj},
         year = 2019,
        month = jun,
       volume = {878},
       number = {1},
          eid = {2},
        pages = {2},
          doi = {10.3847/1538-4357/ab1d5e},
archivePrefix = {arXiv},
       eprint = {1905.09528},
 primaryClass = {astro-ph.GA},
       adsurl = {https://ui.adsabs.harvard.edu/abs/2019ApJ...878....2D}
}

@ARTICLE{Davis2007,
       author = {{Davis}, Shane W. and {Woo}, Jong-Hak and {Blaes}, Omer M.},
        title = "{The UV Continuum of Quasars: Models and SDSS Spectral Slopes}",
      journal = {\apj},
         year = 2007,
        month = oct,
       volume = {668},
       number = {2},
        pages = {682-698},
          doi = {10.1086/521393},
archivePrefix = {arXiv},
       eprint = {0707.1456},
 primaryClass = {astro-ph},
       adsurl = {https://ui.adsabs.harvard.edu/abs/2007ApJ...668..682D}
}

@ARTICLE{deGraaff2025a,
       author = {{de Graaff}, Anna and {Hviding}, Raphael E. and {Naidu}, Rohan P. and {Greene}, Jenny E. and {Miller}, Tim B. and {Leja}, Joel and {Matthee}, Jorryt and {Brammer}, Gabriel and {Katz}, Harley and {Bezanson}, Rachel and {Boogaard}, Leindert A. and {Bose}, Sownak and {Chisholm}, John and {Cleri}, Nikko J. and {Dayal}, Pratika and {Feldmann}, Robert and {Fudamoto}, Yoshinobu and {Fujimoto}, Seiji and {Furtak}, Lukas J. and {Glazebrook}, Karl and {Gottumukkala}, Rashmi and {Heintz}, Kasper E. and {Kokorev}, Vasily and {Labbe}, Ivo and {Maseda}, Michael V. and {McConachie}, Ian and {Nanayakkara}, Themiya and {Nelson}, Erica and {Nowaczyk}, Przemys{\l}aw and {Oesch}, Pascal A. and {Rix}, Hans-Walter and {Setton}, David J. and {Torralba}, Alberto and {Walter}, Fabian and {Wang}, Bingjie and {Weibel}, Andrea and {van der Wel}, Arjen},
        title = "{Little Red Dots host Black Hole Stars: A unified family of gas-reddened AGN revealed by JWST/NIRSpec spectroscopy}",
      journal = {arXiv e-prints},
         year = 2025,
        month = nov,
          eid = {arXiv:2511.21820},
        pages = {arXiv:2511.21820},
          doi = {10.48550/arXiv.2511.21820},
archivePrefix = {arXiv},
       eprint = {2511.21820},
 primaryClass = {astro-ph.GA},
       adsurl = {https://ui.adsabs.harvard.edu/abs/2025arXiv251121820D}
}

@ARTICLE{deGraaff2025b,
       author = {{de Graaff}, Anna and {Rix}, Hans-Walter and {Naidu}, Rohan P. and {Labb{\'e}}, Ivo and {Wang}, Bingjie and {Leja}, Joel and {Matthee}, Jorryt and {Katz}, Harley and {Greene}, Jenny E. and {Hviding}, Raphael E. and {Baggen}, Josephine and {Bezanson}, Rachel and {Boogaard}, Leindert A. and {Brammer}, Gabriel and {Dayal}, Pratika and {van Dokkum}, Pieter and {Goulding}, Andy D. and {Hirschmann}, Michaela and {Maseda}, Michael V. and {McConachie}, Ian and {Miller}, Tim B. and {Nelson}, Erica and {Oesch}, Pascal A. and {Setton}, David J. and {Shivaei}, Irene and {Weibel}, Andrea and {Whitaker}, Katherine E. and {Williams}, Christina C.},
        title = "{A remarkable ruby: Absorption in dense gas, rather than evolved stars, drives the extreme Balmer break of a little red dot at z = 3.5}",
      journal = {\aap},
         year = 2025,
        month = sep,
       volume = {701},
          eid = {A168},
        pages = {A168},
          doi = {10.1051/0004-6361/202554681},
archivePrefix = {arXiv},
       eprint = {2503.16600},
 primaryClass = {astro-ph.GA},
       adsurl = {https://ui.adsabs.harvard.edu/abs/2025A&A...701A.168D}
}

@ARTICLE{DEugenio2025,
       author = {{D'Eugenio}, Francesco and {Nelson}, Erica and {Ji}, Xihan and {Baggen}, Josephine and {Greene}, Jenny and {Labb{\'e}}, Ivo and {Pezzulli}, Gabriele and {Brown}, Vanessa and {Maiolino}, Roberto and {Matthee}, Jorryt and {Terlevich}, Elena and {Terlevich}, Roberto and {Torralba}, Alberto and {Carniani}, Stefano},
        title = "{Irony at z=6.68: a bright AGN with forbidden Fe emission and multi-component Balmer absorption}",
      journal = {arXiv e-prints},
         year = 2025,
        month = sep,
          eid = {arXiv:2510.00101},
        pages = {arXiv:2510.00101},
          doi = {10.48550/arXiv.2510.00101},
archivePrefix = {arXiv},
       eprint = {2510.00101},
 primaryClass = {astro-ph.GA},
       adsurl = {https://ui.adsabs.harvard.edu/abs/2025arXiv251000101D}
}

@ARTICLE{DEugenio2026,
       author = {{D'Eugenio}, Francesco and {Maiolino}, Roberto and {Perna}, Michele and {{\"U}bler}, Hannah and {Ji}, Xihan and {McClymont}, William and {Koudmani}, Sophie and {Sijacki}, Debora and {Juod{\v{z}}balis}, Ignas and {Scholtz}, Jan and {Bennett}, Jake S. and {Bunker}, Andrew J. and {Carniani}, Stefano and {Charlot}, St{\'e}phane and {Cresci}, Giovanni and {Curtis-Lake}, Emma and {Dalla Bont{\`a}}, Elena and {Inayoshi}, Kohei and {Jones}, Gareth C. and {Lyu}, Jianwei and {Marconi}, Alessandro and {Mazzolari}, Giovanni and {Nelson}, Erica J. and {Parlanti}, Eleonora and {Robertson}, Brant E. and {Schneider}, Raffaella and {Simmonds}, Charlotte and {Tacchella}, Sandro and {Venturi}, Giacomo and {Willott}, Chris and {Witstok}, Joris and {Witten}, Callum},
        title = "{BlackTHUNDER strikes twice: Balmer-line absorption in an overmassive Little Red Dot at z = 7.04}",
      journal = {\mnras},
         year = 2026,
        month = feb,
          doi = {10.1093/mnras/stag401},
       adsurl = {https://ui.adsabs.harvard.edu/abs/2026MNRAS.tmp..360D}
}

@ARTICLE{Devecchi2009,
       author = {{Devecchi}, B. and {Volonteri}, M.},
        title = "{Formation of the First Nuclear Clusters and Massive Black Holes at High Redshift}",
      journal = {\apj},
         year = 2009,
        month = mar,
       volume = {694},
       number = {1},
        pages = {302-313},
          doi = {10.1088/0004-637X/694/1/302},
archivePrefix = {arXiv},
       eprint = {0810.1057},
 primaryClass = {astro-ph},
       adsurl = {https://ui.adsabs.harvard.edu/abs/2009ApJ...694..302D}
}

@ARTICLE{Dong2008,
       author = {{Dong}, Xiaobo and {Wang}, Tinggui and {Wang}, Jianguo and {Yuan}, Weimin and {Zhou}, Hongyan and {Dai}, Haifeng and {Zhang}, Kai},
        title = "{Broad-line Balmer decrements in blue active galactic nuclei}",
      journal = {\mnras},
         year = 2008,
        month = jan,
       volume = {383},
       number = {2},
        pages = {581-592},
          doi = {10.1111/j.1365-2966.2007.12560.x},
archivePrefix = {arXiv},
       eprint = {0710.1458},
 primaryClass = {astro-ph},
       adsurl = {https://ui.adsabs.harvard.edu/abs/2008MNRAS.383..581D}
}

@ARTICLE{Dopita2002,
       author = {{Dopita}, Michael A. and {Groves}, Brent A. and {Sutherland}, Ralph S. and {Binette}, Luc and {Cecil}, Gerald},
        title = "{Are the Narrow-Line Regions in Active Galaxies Dusty and Radiation Pressure Dominated?}",
      journal = {\apj},
         year = 2002,
        month = jun,
       volume = {572},
       number = {2},
        pages = {753-761},
          doi = {10.1086/340429},
archivePrefix = {arXiv},
       eprint = {astro-ph/0203360},
 primaryClass = {astro-ph},
       adsurl = {https://ui.adsabs.harvard.edu/abs/2002ApJ...572..753D}
}

@ARTICLE{Duras2020,
       author = {{Duras}, F. and {Bongiorno}, A. and {Ricci}, F. and {Piconcelli}, E. and {Shankar}, F. and {Lusso}, E. and {Bianchi}, S. and {Fiore}, F. and {Maiolino}, R. and {Marconi}, A. and {Onori}, F. and {Sani}, E. and {Schneider}, R. and {Vignali}, C. and {La Franca}, F.},
        title = "{Universal bolometric corrections for active galactic nuclei over seven luminosity decades}",
      journal = {\aap},
         year = 2020,
        month = apr,
       volume = {636},
          eid = {A73},
        pages = {A73},
          doi = {10.1051/0004-6361/201936817},
archivePrefix = {arXiv},
       eprint = {2001.09984},
 primaryClass = {astro-ph.GA},
       adsurl = {https://ui.adsabs.harvard.edu/abs/2020A&A...636A..73D}
}

@ARTICLE{Feltre2016,
       author = {{Feltre}, A. and {Charlot}, S. and {Gutkin}, J.},
        title = "{Nuclear activity versus star formation: emission-line diagnostics at ultraviolet and optical wavelengths}",
      journal = {\mnras},
         year = 2016,
        month = mar,
       volume = {456},
       number = {3},
        pages = {3354-3374},
          doi = {10.1093/mnras/stv2794},
archivePrefix = {arXiv},
       eprint = {1511.08217},
 primaryClass = {astro-ph.GA},
       adsurl = {https://ui.adsabs.harvard.edu/abs/2016MNRAS.456.3354F}
}

@ARTICLE{Ferrarese2000,
       author = {{Ferrarese}, Laura and {Merritt}, David},
        title = "{A Fundamental Relation between Supermassive Black Holes and Their Host Galaxies}",
      journal = {\apjl},
         year = 2000,
        month = aug,
       volume = {539},
       number = {1},
        pages = {L9-L12},
          doi = {10.1086/312838},
archivePrefix = {arXiv},
       eprint = {astro-ph/0006053},
 primaryClass = {astro-ph},
       adsurl = {https://ui.adsabs.harvard.edu/abs/2000ApJ...539L...9F}
}

@ARTICLE{Floris2024,
       author = {{Floris}, A. and {Marziani}, P. and {Panda}, S. and {Sniegowska}, M. and {D'Onofrio}, M. and {Deconto-Machado}, A. and {del Olmo}, A. and {Czerny}, B.},
        title = "{Chemical abundances along the quasar main sequence}",
      journal = {\aap},
         year = 2024,
        month = sep,
       volume = {689},
          eid = {A321},
        pages = {A321},
          doi = {10.1051/0004-6361/202450458},
archivePrefix = {arXiv},
       eprint = {2405.04456},
 primaryClass = {astro-ph.GA},
       adsurl = {https://ui.adsabs.harvard.edu/abs/2024A&A...689A.321F}
}

@ARTICLE{Franco2000,
       author = {{Franco}, Jos{\'e} and {Kurtz}, Stan and {Hofner}, Peter and {Testi}, Leonardo and {Garc{\'\i}a-Segura}, Guillermo and {Martos}, Marco},
        title = "{The Density Structure of Highly Compact H II Regions}",
      journal = {\apjl},
         year = 2000,
        month = oct,
       volume = {542},
       number = {2},
        pages = {L143-L146},
          doi = {10.1086/312938},
       adsurl = {https://ui.adsabs.harvard.edu/abs/2000ApJ...542L.143F}
}

@ARTICLE{Fragione2023,
       author = {{Fragione}, Giacomo and {Pacucci}, Fabio},
        title = "{Constraining the Properties of Black Hole Seeds from the Farthest Quasars}",
      journal = {\apjl},
         year = 2023,
        month = dec,
       volume = {958},
       number = {2},
          eid = {L24},
        pages = {L24},
          doi = {10.3847/2041-8213/ad09e5},
archivePrefix = {arXiv},
       eprint = {2308.14986},
 primaryClass = {astro-ph.GA},
       adsurl = {https://ui.adsabs.harvard.edu/abs/2023ApJ...958L..24F}
}

@ARTICLE{Fujimoto2024,
       author = {{Fujimoto}, Seiji and {Wang}, Bingjie and {Weaver}, John R. and {Kokorev}, Vasily and {Atek}, Hakim and {Bezanson}, Rachel and {Labbe}, Ivo and {Brammer}, Gabriel and {Greene}, Jenny E. and {Chemerynska}, Iryna and {Dayal}, Pratika and {de Graaff}, Anna and {Furtak}, Lukas J. and {Oesch}, Pascal A. and {Setton}, David J. and {Price}, Sedona H. and {Miller}, Tim B. and {Williams}, Christina C. and {Whitaker}, Katherine E. and {Zitrin}, Adi and {Cutler}, Sam E. and {Leja}, Joel and {Pan}, Richard and {Coe}, Dan and {van Dokkum}, Pieter and {Feldmann}, Robert and {Fudamoto}, Yoshinobu and {Goulding}, Andy D. and {Khullar}, Gourav and {Marchesini}, Danilo and {Maseda}, Michael and {Nanayakkara}, Themiya and {Nelson}, Erica J. and {Smit}, Renske and {Stefanon}, Mauro and {Weibel}, Andrea},
        title = "{UNCOVER: A NIRSpec Census of Lensed Galaxies at z = 8.50─13.08 Probing a High-AGN Fraction and Ionized Bubbles in the Shadow}",
      journal = {\apj},
         year = 2024,
        month = dec,
       volume = {977},
       number = {2},
          eid = {250},
        pages = {250},
          doi = {10.3847/1538-4357/ad9027},
archivePrefix = {arXiv},
       eprint = {2308.11609},
 primaryClass = {astro-ph.GA},
       adsurl = {https://ui.adsabs.harvard.edu/abs/2024ApJ...977..250F}
}

@ARTICLE{Furtak2023,
       author = {{Furtak}, Lukas J. and {Zitrin}, Adi and {Plat}, Ad{\`e}le and {Fujimoto}, Seiji and {Wang}, Bingjie and {Nelson}, Erica J. and {Labb{\'e}}, Ivo and {Bezanson}, Rachel and {Brammer}, Gabriel B. and {van Dokkum}, Pieter and {Endsley}, Ryan and {Glazebrook}, Karl and {Greene}, Jenny E. and {Leja}, Joel and {Price}, Sedona H. and {Smit}, Renske and {Stark}, Daniel P. and {Weaver}, John R. and {Whitaker}, Katherine E. and {Atek}, Hakim and {Chevallard}, Jacopo and {Curtis-Lake}, Emma and {Dayal}, Pratika and {Feltre}, Anna and {Franx}, Marijn and {Fudamoto}, Yoshinobu and {Marchesini}, Danilo and {Mowla}, Lamiya A. and {Pan}, Richard and {Suess}, Katherine A. and {Vidal-Garc{\'\i}a}, Alba and {Williams}, Christina C.},
        title = "{JWST UNCOVER: Extremely Red and Compact Object at z $_{phot}$ ≃ 7.6 Triply Imaged by A2744}",
      journal = {\apj},
         year = 2023,
        month = aug,
       volume = {952},
       number = {2},
          eid = {142},
        pages = {142},
          doi = {10.3847/1538-4357/acdc9d},
archivePrefix = {arXiv},
       eprint = {2212.10531},
 primaryClass = {astro-ph.GA},
       adsurl = {https://ui.adsabs.harvard.edu/abs/2023ApJ...952..142F}
}

@ARTICLE{Furtak2024,
       author = {{Furtak}, Lukas J. and {Labb{\'e}}, Ivo and {Zitrin}, Adi and {Greene}, Jenny E. and {Dayal}, Pratika and {Chemerynska}, Iryna and {Kokorev}, Vasily and {Miller}, Tim B. and {Goulding}, Andy D. and {de Graaff}, Anna and {Bezanson}, Rachel and {Brammer}, Gabriel B. and {Cutler}, Sam E. and {Leja}, Joel and {Pan}, Richard and {Price}, Sedona H. and {Wang}, Bingjie and {Weaver}, John R. and {Whitaker}, Katherine E. and {Atek}, Hakim and {Bogd{\'a}n}, {\'A}kos and {Charlot}, St{\'e}phane and {Curtis-Lake}, Emma and {van Dokkum}, Pieter and {Endsley}, Ryan and {Feldmann}, Robert and {Fudamoto}, Yoshinobu and {Fujimoto}, Seiji and {Glazebrook}, Karl and {Juneau}, St{\'e}phanie and {Marchesini}, Danilo and {Maseda}, Micheal V. and {Nelson}, Erica and {Oesch}, Pascal A. and {Plat}, Ad{\`e}le and {Setton}, David J. and {Stark}, Daniel P. and {Williams}, Christina C.},
        title = "{A high black-hole-to-host mass ratio in a lensed AGN in the early Universe}",
      journal = {\nat},
         year = 2024,
        month = apr,
       volume = {628},
       number = {8006},
        pages = {57-61},
          doi = {10.1038/s41586-024-07184-8},
archivePrefix = {arXiv},
       eprint = {2308.05735},
 primaryClass = {astro-ph.GA},
       adsurl = {https://ui.adsabs.harvard.edu/abs/2024Natur.628...57F}
}

@ARTICLE{Gardner2006,
       author = {{Gardner}, Jonathan P. and {Mather}, John C. and {Clampin}, Mark and {Doyon}, Rene and {Greenhouse}, Matthew A. and {Hammel}, Heidi B. and {Hutchings}, John B. and {Jakobsen}, Peter and {Lilly}, Simon J. and {Long}, Knox S. and {Lunine}, Jonathan I. and {McCaughrean}, Mark J. and {Mountain}, Matt and {Nella}, John and {Rieke}, George H. and {Rieke}, Marcia J. and {Rix}, Hans-Walter and {Smith}, Eric P. and {Sonneborn}, George and {Stiavelli}, Massimo and {Stockman}, H.~S. and {Windhorst}, Rogier A. and {Wright}, Gillian S.},
        title = "{The James Webb Space Telescope}",
      journal = {\ssr},
         year = 2006,
        month = apr,
       volume = {123},
       number = {4},
        pages = {485-606},
          doi = {10.1007/s11214-006-8315-7},
archivePrefix = {arXiv},
       eprint = {astro-ph/0606175},
 primaryClass = {astro-ph},
       adsurl = {https://ui.adsabs.harvard.edu/abs/2006SSRv..123..485G}
}

@ARTICLE{Gardner2023,
       author = {{Gardner}, Jonathan P. and {Mather}, John C. and {Abbott}, Randy and {Abell}, James S. and {Abernathy}, Mark and {Abney}, Faith E. and {Abraham}, John G. and {Abraham}, Roberto and {Abul-Huda}, Yasin M. and {Acton}, Scott and {Adams}, Cynthia K. and {Adams}, Evan and {Adler}, David S. and {Adriaensen}, Maarten and {Aguilar}, Jonathan Albert and {Ahmed}, Mansoor and {Ahmed}, Nasif S. and {Ahmed}, Tanjira and {Albat}, R{\"u}deger and {Albert}, Lo{\"\i}c and {Alberts}, Stacey and {Aldridge}, David and {Allen}, Mary Marsha and {Allen}, Shaune S. and {Altenburg}, Martin and {Altunc}, Serhat and {Alvarez}, Jose Lorenzo and {{\'A}lvarez-M{\'a}rquez}, Javier and {Alves de Oliveira}, Catarina and {Ambrose}, Leslie L. and {Anandakrishnan}, Satya M. and {Andersen}, Gregory C. and {Anderson}, Harry James and {Anderson}, Jay and {Anderson}, Kristen and {Anderson}, Sara M. and {Aprea}, Julio and {Archer}, Benita J. and {Arenberg}, Jonathan W. and {Argyriou}, Ioannis and {Arribas}, Santiago and {Artigau}, {\'E}tienne and {Arvai}, Amanda Rose and {Atcheson}, Paul and {Atkinson}, Charles B. and {Averbukh}, Jesse and {Aymergen}, Cagatay and {Bacinski}, John J. and {Baggett}, Wayne E. and {Bagnasco}, Giorgio and {Baker}, Lynn L. and {Balzano}, Vicki Ann and {Banks}, Kimberly A. and {Baran}, David A. and {Barker}, Elizabeth A. and {Barrett}, Larry K. and {Barringer}, Bruce O. and {Barto}, Allison and {Bast}, William and {Baudoz}, Pierre and {Baum}, Stefi and {Beatty}, Thomas G. and {Beaulieu}, Mathilde and {Bechtold}, Kathryn and {Beck}, Tracy and {Beddard}, Megan M. and {Beichman}, Charles and {Bellagama}, Larry and {Bely}, Pierre and {Berger}, Timothy W. and {Bergeron}, Louis E. and {Bernier}, Antoine-Darveau and {Bertch}, Maria D. and {Beskow}, Charlotte and {Betz}, Laura E. and {Biagetti}, Carl P. and {Birkmann}, Stephan and {Bjorklund}, Kurt F. and {Blackwood}, James D. and {Blazek}, Ronald Paul and {Blossfeld}, Stephen and {Bluth}, Marcel and {Boccaletti}, Anthony and {Boegner}, Jr., Martin E. and {Bohlin}, Ralph C. and {Boia}, John Joseph and {B{\"o}ker}, Torsten and {Bonaventura}, N. and {Bond}, Nicholas A. and {Bosley}, Kari Ann and {Boucarut}, Rene A. and {Bouchet}, Patrice and {Bouwman}, Jeroen and {Bower}, Gary and {Bowers}, Ariel S. and {Bowers}, Charles W. and {Boyce}, Leslye A. and {Boyer}, Christine T. and {Boyer}, Martha L. and {Boyer}, Michael and {Boyer}, Robert and {Bradley}, Larry D. and {Brady}, Gregory R. and {Brandl}, Bernhard R. and {Brannen}, Judith L. and {Breda}, David and {Bremmer}, Harold G. and {Brennan}, David and {Bresnahan}, Pamela A. and {Bright}, Stacey N. and {Broiles}, Brian J. and {Bromenschenkel}, Asa and {Brooks}, Brian H. and {Brooks}, Keira J. and {Brown}, Bob and {Brown}, Bruce and {Brown}, Thomas M. and {Bruce}, Barry W. and {Bryson}, Jonathan G. and {Bujanda}, Edwin D. and {Bullock}, Blake M. and {Bunker}, A.~J. and {Bureo}, Rafael and {Burt}, Irving J. and {Bush}, James Aaron and {Bushouse}, Howard A. and {Bussman}, Marie C. and {Cabaud}, Olivier and {Cale}, Steven and {Calhoon}, Charles D. and {Calvani}, Humberto and {Canipe}, Alicia M. and {Caputo}, Francis M. and {Cara}, Mihai and {Carey}, Larkin and {Case}, Michael Eli and {Cesari}, Thaddeus and {Cetorelli}, Lee D. and {Chance}, Don R. and {Chandler}, Lynn and {Chaney}, Dave and {Chapman}, George N. and {Charlot}, S. and {Chayer}, Pierre and {Cheezum}, Jeffrey I. and {Chen}, Bin and {Chen}, Christine H. and {Cherinka}, Brian and {Chichester}, Sarah C. and {Chilton}, Zachary S. and {Chittiraibalan}, Dharini and {Clampin}, Mark and {Clark}, Charles R. and {Clark}, Kerry W. and {Clark}, Stephanie M. and {Claybrooks}, Edward E. and {Cleveland}, Keith A. and {Cohen}, Andrew L. and {Cohen}, Lester M. and {Col{\'o}n}, Knicole D. and {Coleman}, Benee L. and {Colina}, Luis and {Comber}, Brian J. and {Comeau}, Thomas M. and {Comer}, Thomas and {Conde Reis}, Alain and {Connolly}, Dennis C. and {Conroy}, Kyle E. and {Contos}, Adam R. and {Contreras}, James and {Cook}, Neil J. and {Cooper}, James L. and {Cooper}, Rachel Aviva and {Correia}, Michael F. and {Correnti}, Matteo and {Cossou}, Christophe and {Costanza}, Brian F. and {Coulais}, Alain and {Cox}, Colin R. and {Coyle}, Ray T. and {Cracraft}, Misty M. and {Crew}, Keith A. and {Curtis}, Gary J. and {Cusveller}, Bianca and {Da Costa Maciel}, Cleyciane and {Dailey}, Christopher T. and {Daugeron}, Fr{\'e}d{\'e}ric and {Davidson}, Greg S. and {Davies}, James E. and {Davis}, Katherine Anne and {Davis}, Michael S. and {Day}, Ratna and {de Chambure}, Daniel and {de Jong}, Pauline and {De Marchi}, Guido and {Dean}, Bruce H. and {Decker}, John E. and {Delisa}, Amy S. and {Dell}, Lawrence C. and {Dellagatta}, Gail},
        title = "{The James Webb Space Telescope Mission}",
      journal = {\pasp},
         year = 2023,
        month = jun,
       volume = {135},
       number = {1048},
          eid = {068001},
        pages = {068001},
          doi = {10.1088/1538-3873/acd1b5},
archivePrefix = {arXiv},
       eprint = {2304.04869},
 primaryClass = {astro-ph.IM},
       adsurl = {https://ui.adsabs.harvard.edu/abs/2023PASP..135f8001G}
}

@ARTICLE{Gentile2026,
       author = {{Gentile}, Fabrizio and {Giavalisco}, Mauro and {Daddi}, Emanuele and {Elbaz}, David and {Billand}, Jean-Baptiste and {Franco}, Maximilen and {Magnelli}, Benjamin and {Barro}, Guillermo and {Cheng}, Yingjie and {Cleri}, Nikko J. and {Davis}, Kelcey and {Delvecchio}, Ivan and {Dickinson}, Mark and {Finkelstein}, Steven L. and {Gandolfi}, Giovanni and {Hirschmann}, Michaela and {Hu}, Weida and {Kocevski}, Dale and {Koekemoer}, Anton M. and {Lucas}, Ray and {Mascia}, Sara and {Napolitano}, Lorenzo and {Papovich}, Casey and {P{\'e}rez-D{\'\i}az}, Borja and {Perez-Gonzalez}, Pablo and {Trump}, Jonathan R. and {Wang}, Xin and {Yung}, L.~Y. Aaron},
        title = "{The quasi-star model for Little Red Dots: potential and challenges}",
      journal = {arXiv e-prints},
         year = 2026,
        month = jun,
          eid = {arXiv:2606.06575},
        pages = {arXiv:2606.06575},
          doi = {10.48550/arXiv.2606.06575},
archivePrefix = {arXiv},
       eprint = {2606.06575},
 primaryClass = {astro-ph.GA},
       adsurl = {https://ui.adsabs.harvard.edu/abs/2026arXiv260606575G}
}

@ARTICLE{Geris2026,
       author = {{Geris}, Sophia and {Maiolino}, Roberto and {Ji}, Xihan and {Risaliti}, Guido and {Lanzuisi}, Giorgio and {D'Eugenio}, Francesco and {Isobe}, Yuki and {Jones}, Gareth and {Harshan}, Anishya and {Brazzini}, Matilde and {Juod{\v{z}}balis}, Ignas and {Scholtz}, Jan and {Rinaldi}, Pierluigi and {{\"U}bler}, Hannah and {Baker}, William and {Bunker}, Andrew J. and {Brusa}, Marcella and {Carniani}, Stefano and {Charlot}, Stephane and {Curti}, Mirko and {Comastri}, Andrea and {Lake}, Emma Curtis and {Gilli}, Roberto and {Hainline}, Kevin and {Madau}, Piero and {Marchesi}, Stefano and {Mazzolari}, Giovanni and {Napolitano}, Lorenzo and {Parlanti}, Eleonora and {Pentericci}, Laura and {Ramos Almeida}, Cristina and {Robertson}, Brant and {Silcock}, Maddie S. and {Tripodi}, Roberta and {Venturi}, Giacomo and {Vignali}, Cristian and {Vito}, Fabio and {Zhu}, Yongda},
        title = "{Little Red and Blue Dots: AGN-excited narrow lines, Lyman-$α$ emission, and resemblance to standard quasars}",
      journal = {arXiv e-prints},
         year = 2026,
        month = jun,
          eid = {arXiv:2606.21614},
        pages = {arXiv:2606.21614},
          doi = {10.48550/arXiv.2606.21614},
archivePrefix = {arXiv},
       eprint = {2606.21614},
 primaryClass = {astro-ph.GA},
       adsurl = {https://ui.adsabs.harvard.edu/abs/2026arXiv260621614G}
}

@ARTICLE{Giammanco2005,
       author = {{Giammanco}, C. and {Beckman}, J.~E. and {Cedr{\'e}s}, B.},
        title = "{Effects of photon escape on diagnostic diagrams for H II regions}",
      journal = {\aap},
         year = 2005,
        month = aug,
       volume = {438},
       number = {2},
        pages = {599-610},
          doi = {10.1051/0004-6361:20042268},
archivePrefix = {arXiv},
       eprint = {astro-ph/0504234},
 primaryClass = {astro-ph},
       adsurl = {https://ui.adsabs.harvard.edu/abs/2005A&A...438..599G}
}

@ARTICLE{Goulding2023,
       author = {{Goulding}, Andy D. and {Greene}, Jenny E. and {Setton}, David J. and {Labbe}, Ivo and {Bezanson}, Rachel and {Miller}, Tim B. and {Atek}, Hakim and {Bogd{\'a}n}, {\'A}kos and {Brammer}, Gabriel and {Chemerynska}, Iryna and {Cutler}, Sam E. and {Dayal}, Pratika and {Fudamoto}, Yoshinobu and {Fujimoto}, Seiji and {Furtak}, Lukas J. and {Kokorev}, Vasily and {Khullar}, Gourav and {Leja}, Joel and {Marchesini}, Danilo and {Natarajan}, Priyamvada and {Nelson}, Erica and {Oesch}, Pascal A. and {Pan}, Richard and {Papovich}, Casey and {Price}, Sedona H. and {van Dokkum}, Pieter and {Wang}, Bingjie and {Weaver}, John R. and {Whitaker}, Katherine E. and {Zitrin}, Adi},
        title = "{UNCOVER: The Growth of the First Massive Black Holes from JWST/NIRSpec-Spectroscopic Redshift Confirmation of an X-Ray Luminous AGN at z = 10.1}",
      journal = {\apjl},
         year = 2023,
        month = sep,
       volume = {955},
       number = {1},
          eid = {L24},
        pages = {L24},
          doi = {10.3847/2041-8213/acf7c5},
archivePrefix = {arXiv},
       eprint = {2308.02750},
 primaryClass = {astro-ph.GA},
       adsurl = {https://ui.adsabs.harvard.edu/abs/2023ApJ...955L..24G}
}

@ARTICLE{Greene2005,
       author = {{Greene}, Jenny E. and {Ho}, Luis C.},
        title = "{Estimating Black Hole Masses in Active Galaxies Using the H{\ensuremath{\alpha}} Emission Line}",
      journal = {\apj},
         year = 2005,
        month = sep,
       volume = {630},
       number = {1},
        pages = {122-129},
          doi = {10.1086/431897},
archivePrefix = {arXiv},
       eprint = {astro-ph/0508335},
 primaryClass = {astro-ph},
       adsurl = {https://ui.adsabs.harvard.edu/abs/2005ApJ...630..122G}
}

@ARTICLE{Greene2024,
       author = {{Greene}, Jenny E. and {Labbe}, Ivo and {Goulding}, Andy D. and {Furtak}, Lukas J. and {Chemerynska}, Iryna and {Kokorev}, Vasily and {Dayal}, Pratika and {Volonteri}, Marta and {Williams}, Christina C. and {Wang}, Bingjie and {Setton}, David J. and {Burgasser}, Adam J. and {Bezanson}, Rachel and {Atek}, Hakim and {Brammer}, Gabriel and {Cutler}, Sam E. and {Feldmann}, Robert and {Fujimoto}, Seiji and {Glazebrook}, Karl and {de Graaff}, Anna and {Khullar}, Gourav and {Leja}, Joel and {Marchesini}, Danilo and {Maseda}, Michael V. and {Matthee}, Jorryt and {Miller}, Tim B. and {Naidu}, Rohan P. and {Nanayakkara}, Themiya and {Oesch}, Pascal A. and {Pan}, Richard and {Papovich}, Casey and {Price}, Sedona H. and {van Dokkum}, Pieter and {Weaver}, John R. and {Whitaker}, Katherine E. and {Zitrin}, Adi},
        title = "{UNCOVER Spectroscopy Confirms the Surprising Ubiquity of Active Galactic Nuclei in Red Sources at z > 5}",
      journal = {\apj},
         year = 2024,
        month = mar,
       volume = {964},
       number = {1},
          eid = {39},
        pages = {39},
          doi = {10.3847/1538-4357/ad1e5f},
archivePrefix = {arXiv},
       eprint = {2309.05714},
 primaryClass = {astro-ph.GA},
       adsurl = {https://ui.adsabs.harvard.edu/abs/2024ApJ...964...39G}
}

@ARTICLE{Greene2026,
       author = {{Greene}, Jenny E. and {Setton}, David J. and {Furtak}, Lukas J. and {Naidu}, Rohan P. and {Volonteri}, Marta and {Dayal}, Pratika and {Labbe}, Ivo and {van Dokkum}, Pieter and {Bezanson}, Rachel and {Brammer}, Gabriel and {Cutler}, Sam E. and {Glazebrook}, Karl and {de Graaff}, Anna and {Hirschmann}, Michaela and {Hviding}, Raphael E. and {Kokorev}, Vasily and {Leja}, Joel and {Liu}, Hanpu and {Ma}, Yilun and {Matthee}, Jorryt and {Nanayakkara}, Themiya and {Oesch}, Pascal A. and {Pan}, Richard and {Price}, Sedona H. and {Spilker}, Justin S. and {Wang}, Bingjie and {Weaver}, John R. and {Whitaker}, Katherine E. and {Williams}, Christina C. and {Zitrin}, Adi},
        title = "{What You See Is What You Get: Empirically Measured Bolometric Luminosities of Little Red Dots}",
      journal = {\apj},
         year = 2026,
        month = jan,
       volume = {996},
       number = {2},
          eid = {129},
        pages = {129},
          doi = {10.3847/1538-4357/ae1836},
archivePrefix = {arXiv},
       eprint = {2509.05434},
 primaryClass = {astro-ph.GA},
       adsurl = {https://ui.adsabs.harvard.edu/abs/2026ApJ...996..129G}
}

@ARTICLE{Grier2017,
       author = {{Grier}, C.~J. and {Trump}, J.~R. and {Shen}, Yue and {Horne}, Keith and {Kinemuchi}, Karen and {McGreer}, Ian D. and {Starkey}, D.~A. and {Brandt}, W.~N. and {Hall}, P.~B. and {Kochanek}, C.~S. and {Chen}, Yuguang and {Denney}, K.~D. and {Greene}, Jenny E. and {Ho}, L.~C. and {Homayouni}, Y. and {I-Hsiu Li}, Jennifer and {Pei}, Liuyi and {Peterson}, B.~M. and {Petitjean}, P. and {Schneider}, D.~P. and {Sun}, Mouyuan and {AlSayyad}, Yusura and {Bizyaev}, Dmitry and {Brinkmann}, Jonathan and {Brownstein}, Joel R. and {Bundy}, Kevin and {Dawson}, K.~S. and {Eftekharzadeh}, Sarah and {Fernandez-Trincado}, J.~G. and {Gao}, Yang and {Hutchinson}, Timothy A. and {Jia}, Siyao and {Jiang}, Linhua and {Oravetz}, Daniel and {Pan}, Kaike and {Paris}, Isabelle and {Ponder}, Kara A. and {Peters}, Christina and {Rogerson}, Jesse and {Simmons}, Audrey and {Smith}, Robyn and {Wang}, Ran},
        title = "{The Sloan Digital Sky Survey Reverberation Mapping Project: H{\ensuremath{\alpha}} and H{\ensuremath{\beta}} Reverberation Measurements from First-year Spectroscopy and Photometry}",
      journal = {\apj},
         year = 2017,
        month = dec,
       volume = {851},
       number = {1},
          eid = {21},
        pages = {21},
          doi = {10.3847/1538-4357/aa98dc},
archivePrefix = {arXiv},
       eprint = {1711.03114},
 primaryClass = {astro-ph.GA},
       adsurl = {https://ui.adsabs.harvard.edu/abs/2017ApJ...851...21G}
}

@ARTICLE{Grier2017b,
       author = {{Grier}, C.~J. and {Pancoast}, A. and {Barth}, A.~J. and {Fausnaugh}, M.~M. and {Brewer}, B.~J. and {Treu}, T. and {Peterson}, B.~M.},
        title = "{The Structure of the Broad-line Region in Active Galactic Nuclei. II. Dynamical Modeling of Data From the AGN10 Reverberation Mapping Campaign}",
      journal = {\apj},
         year = 2017,
        month = nov,
       volume = {849},
       number = {2},
          eid = {146},
        pages = {146},
          doi = {10.3847/1538-4357/aa901b},
archivePrefix = {arXiv},
       eprint = {1705.02346},
 primaryClass = {astro-ph.GA},
       adsurl = {https://ui.adsabs.harvard.edu/abs/2017ApJ...849..146G}
}

@ARTICLE{Groves2004,
       author = {{Groves}, Brent A. and {Dopita}, Michael A. and {Sutherland}, Ralph S.},
        title = "{Dusty, Radiation Pressure-Dominated Photoionization. I. Model Description, Structure, and Grids}",
      journal = {\apjs},
         year = 2004,
        month = jul,
       volume = {153},
       number = {1},
        pages = {9-73},
          doi = {10.1086/421113},
archivePrefix = {arXiv},
       eprint = {astro-ph/0404175},
 primaryClass = {astro-ph},
       adsurl = {https://ui.adsabs.harvard.edu/abs/2004ApJS..153....9G}
}

@ARTICLE{Gupta2026,
       author = {{Gupta}, Ansh R. and {Taylor}, Anthony and {Curtis-Lake}, Emma and {Silcock}, Maddie and {Ch{\'a}vez Ortiz}, {\'O}scar A. and {Finkelstein}, Steven L. and {Akins}, Hollis B. and {Backhaus}, Bren E. and {Barro}, Guillermo and {Bisigello}, Laura and {Brooks}, Madisyn and {Casey}, Caitlin M. and {Charlot}, Stephane and {Chevallard}, Jacopo and {Feltre}, Anna and {Gandolfi}, Giovanni and {Giavalisco}, Mauro and {Grogin}, Norman A. and {Hirschmann}, Michaela and {Hsiao}, Tiger Yu-Yang and {Jeon}, Junehyoung and {Jogee}, Shardha and {Kartaltepe}, Jeyhan S. and {Kocevski}, Dale D. and {Koekemoer}, Anton M. and {Kokorev}, Vasily and {Leung}, Gene C.~K. and {Lucas}, Ray A. and {Pacucci}, Fabio and {Pirzkal}, Nor and {Plat}, Adele and {Somerville}, Rachel S. and {Trump}, Jonathan R. and {Vidal-Garc{\'\i}a}, Alba and {Wang}, Xin and {Yung}, L.~Y. Aaron},
        title = "{A Rapid Evolution in the Observed Mbh/M* Relation at z > 3 Revealed via Spectro-photometric SED-Modeling}",
      journal = {arXiv e-prints},
         year = 2026,
        month = may,
          eid = {arXiv:2605.30414},
        pages = {arXiv:2605.30414},
          doi = {10.48550/arXiv.2605.30414},
archivePrefix = {arXiv},
       eprint = {2605.30414},
 primaryClass = {astro-ph.GA},
       adsurl = {https://ui.adsabs.harvard.edu/abs/2026arXiv260530414G}
}

@ARTICLE{Gutkin2016,
       author = {{Gutkin}, Julia and {Charlot}, St{\'e}phane and {Bruzual}, Gustavo},
        title = "{Modelling the nebular emission from primeval to present-day star-forming galaxies}",
      journal = {\mnras},
         year = 2016,
        month = oct,
       volume = {462},
       number = {2},
        pages = {1757-1774},
          doi = {10.1093/mnras/stw1716},
archivePrefix = {arXiv},
       eprint = {1607.06086},
 primaryClass = {astro-ph.GA},
       adsurl = {https://ui.adsabs.harvard.edu/abs/2016MNRAS.462.1757G}
}

@ARTICLE{Hagen2023,
       author = {{Hagen}, Scott and {Done}, Chris},
        title = "{Estimating black hole spin from AGN SED fitting: the impact of general-relativistic ray tracing}",
      journal = {\mnras},
         year = 2023,
        month = nov,
       volume = {525},
       number = {3},
        pages = {3455-3467},
          doi = {10.1093/mnras/stad2499},
archivePrefix = {arXiv},
       eprint = {2304.01253},
 primaryClass = {astro-ph.HE},
       adsurl = {https://ui.adsabs.harvard.edu/abs/2023MNRAS.525.3455H}
}

@ARTICLE{Hamann1993,
       author = {{Hamann}, Fred and {Ferland}, Gary},
        title = "{The Chemical Evolution of QSOs and the Implications for Cosmology and Galaxy Formation}",
      journal = {\apj},
         year = 1993,
        month = nov,
       volume = {418},
        pages = {11},
          doi = {10.1086/173366},
       adsurl = {https://ui.adsabs.harvard.edu/abs/1993ApJ...418...11H}
}

@ARTICLE{Hamann1996,
       author = {{Hamann}, Fred and {Korista}, Kirk T.},
        title = "{On the Scattering Contributions to N V lambda 1240 and C IV lambda 1549 in QSOs}",
      journal = {\apj},
         year = 1996,
        month = jun,
       volume = {464},
        pages = {158},
          doi = {10.1086/177307},
       adsurl = {https://ui.adsabs.harvard.edu/abs/1996ApJ...464..158H}
}

@article{Hamann2002,
doi = {10.1086/324289},
url = {https://dx.doi.org/10.1086/324289},
year = {2002},
month = {jan},
publisher = {},
volume = {564},
number = {2},
pages = {592},
author = {Fred Hamann and K. T. Korista and G. J. Ferland and Craig Warner and Jack Baldwin},
title = {Metallicities and Abundance Ratios from Quasar Broad Emission Lines},
journal = {The Astrophysical Journal}
}

@ARTICLE{Harikane2023,
       author = {{Harikane}, Yuichi and {Zhang}, Yechi and {Nakajima}, Kimihiko and {Ouchi}, Masami and {Isobe}, Yuki and {Ono}, Yoshiaki and {Hatano}, Shun and {Xu}, Yi and {Umeda}, Hiroya},
        title = "{A JWST/NIRSpec First Census of Broad-line AGNs at z = 4-7: Detection of 10 Faint AGNs with M $_{BH}$ {}10$^{6}$-{}10$^{8}$ M $_{☉}$ and Their Host Galaxy Properties}",
      journal = {\apj},
         year = 2023,
        month = dec,
       volume = {959},
       number = {1},
          eid = {39},
        pages = {39},
          doi = {10.3847/1538-4357/ad029e},
archivePrefix = {arXiv},
       eprint = {2303.11946},
 primaryClass = {astro-ph.GA},
       adsurl = {https://ui.adsabs.harvard.edu/abs/2023ApJ...959...39H}
}

@ARTICLE{Hirschmann2019,
       author = {{Hirschmann}, Michaela and {Charlot}, Stephane and {Feltre}, Anna and {Naab}, Thorsten and {Somerville}, Rachel S. and {Choi}, Ena},
        title = "{Synthetic nebular emission from massive galaxies - II. Ultraviolet-line diagnostics of dominant ionizing sources}",
      journal = {\mnras},
         year = 2019,
        month = jul,
       volume = {487},
       number = {1},
        pages = {333-353},
          doi = {10.1093/mnras/stz1256},
archivePrefix = {arXiv},
       eprint = {1811.07909},
 primaryClass = {astro-ph.GA},
       adsurl = {https://ui.adsabs.harvard.edu/abs/2019MNRAS.487..333H}
}

@ARTICLE{Hirschmann2023,
       author = {{Hirschmann}, Michaela and {Charlot}, Stephane and {Feltre}, Anna and {Curtis-Lake}, Emma and {Somerville}, Rachel S. and {Chevallard}, Jacopo and {Choi}, Ena and {Nelson}, Dylan and {Morisset}, Christophe and {Plat}, Adele and {Vidal-Garcia}, Alba},
        title = "{Emission-line properties of IllustrisTNG galaxies: from local diagnostic diagrams to high-redshift predictions for JWST}",
      journal = {\mnras},
         year = 2023,
        month = dec,
       volume = {526},
       number = {3},
        pages = {3610-3636},
          doi = {10.1093/mnras/stad2955},
archivePrefix = {arXiv},
       eprint = {2212.02522},
 primaryClass = {astro-ph.GA},
       adsurl = {https://ui.adsabs.harvard.edu/abs/2023MNRAS.526.3610H}
}

@ARTICLE{Horne2021,
       author = {{Horne}, Keith and {De Rosa}, G. and {Peterson}, B.~M. and {Barth}, A.~J. and {Ely}, J. and {Fausnaugh}, M.~M. and {Kriss}, G.~A. and {Pei}, L. and {Bentz}, M.~C. and {Cackett}, E.~M. and {Edelson}, R. and {Eracleous}, M. and {Goad}, M.~R. and {Grier}, C.~J. and {Kaastra}, J. and {Kochanek}, C.~S. and {Krongold}, Y. and {Mathur}, S. and {Netzer}, H. and {Proga}, D. and {Tejos}, N. and {Vestergaard}, M. and {Villforth}, C. and {Adams}, S.~M. and {Anderson}, M.~D. and {Ar{\'e}valo}, P. and {Beatty}, T.~G. and {Bennert}, V.~N. and {Bigley}, A. and {Bisogni}, S. and {Borman}, G.~A. and {Boroson}, T.~A. and {Bottorff}, M.~C. and {Brandt}, W.~N. and {Breeveld}, A.~A. and {Brotherton}, M. and {Brown}, J.~E. and {Brown}, J.~S. and {Canalizo}, G. and {Carini}, M.~T. and {Clubb}, K.~I. and {Comerford}, J.~M. and {Corsini}, E.~M. and {Crenshaw}, D.~M. and {Croft}, S. and {Croxall}, K.~V. and {Dalla Bont{\`a}}, E. and {Deason}, A.~J. and {Dehghanian}, M. and {De Lorenzo-C{\'a}ceres}, A. and {Denney}, K.~D. and {Dietrich}, M. and {Done}, C. and {Efimova}, N.~V. and {Evans}, P.~A. and {Ferland}, G.~J. and {Filippenko}, A.~V. and {Flatland}, K. and {Fox}, O.~D. and {Gardner}, E. and {Gates}, E.~L. and {Gehrels}, N. and {Geier}, S. and {Gelbord}, J.~M. and {Gonzalez}, L. and {Gorjian}, V. and {Greene}, J.~E. and {Grupe}, D. and {Gupta}, A. and {Hall}, P.~B. and {Henderson}, C.~B. and {Hicks}, S. and {Holmbeck}, E. and {Holoien}, T.~W.-S. and {Hutchison}, T. and {Im}, M. and {Jensen}, J.~J. and {Johnson}, C.~A. and {Joner}, M.~D. and {Jones}, J. and {Kaspi}, S. and {Kelly}, P.~L. and {Kennea}, J.~A. and {Kim}, M. and {Kim}, S. and {Kim}, S.~C. and {King}, A. and {Klimanov}, S.~A. and {Korista}, K.~T. and {Lau}, M.~W. and {Lee}, J.~C. and {Leonard}, D.~C. and {Li}, Miao and {Lira}, P. and {Lochhaas}, C. and {Ma}, Zhiyuan and {MacInnis}, F. and {Malkan}, M.~A. and {Manne-Nicholas}, E.~R. and {Mauerhan}, J.~C. and {McGurk}, R. and {McHardy}, I.~M. and {Montuori}, C. and {Morelli}, L. and {Mosquera}, A. and {Mudd}, D. and {M{\"u}ller-S{\'a}nchez}, F. and {Nazarov}, S.~V. and {Norris}, R.~P. and {Nousek}, J.~A. and {Nguyen}, M.~L. and {Ochner}, P. and {Okhmat}, D.~N. and {Pancoast}, A. and {Papadakis}, I. and {Parks}, J.~R. and {Penny}, M.~T. and {Pizzella}, A. and {Pogge}, R.~W. and {Poleski}, R. and {Pott}, J.-U. and {Rafter}, S.~E. and {Rix}, H.-W. and {Runnoe}, J. and {Saylor}, D.~A. and {Schimoia}, J.~S. and {Schn{\"u}lle}, K. and {Scott}, B. and {Sergeev}, S.~G. and {Shappee}, B.~J. and {Shivvers}, I. and {Siegel}, M. and {Simonian}, G.~V. and {Siviero}, A. and {Skielboe}, A. and {Somers}, G. and {Spencer}, M. and {Starkey}, D. and {Stevens}, D.~J. and {Sung}, H.-I. and {Tayar}, J. and {Treu}, T. and {Turner}, C.~S. and {Uttley}, P. and {Van Saders}, J. and {Vican}, L. and {Villanueva}, Jr., S. and {Weiss}, Y. and {Woo}, J.-H. and {Yan}, H. and {Young}, S. and {Yuk}, H. and {Zheng}, W. and {Zhu}, W. and {Zu}, Y.},
        title = "{Space Telescope and Optical Reverberation Mapping Project. IX. Velocity-Delay Maps for Broad Emission Lines in NGC 5548}",
      journal = {\apj},
         year = 2021,
        month = feb,
       volume = {907},
       number = {2},
          eid = {76},
        pages = {76},
          doi = {10.3847/1538-4357/abce60},
archivePrefix = {arXiv},
       eprint = {2003.01448},
 primaryClass = {astro-ph.GA},
       adsurl = {https://ui.adsabs.harvard.edu/abs/2021ApJ...907...76H}
}

@ARTICLE{Huang2023,
       author = {{Huang}, Jiamu and {Lin}, Douglas N.~C. and {Shields}, Gregory},
        title = "{Metal enrichment due to embedded stars in AGN discs}",
      journal = {\mnras},
         year = 2023,
        month = nov,
       volume = {525},
       number = {4},
        pages = {5702-5718},
          doi = {10.1093/mnras/stad2642},
archivePrefix = {arXiv},
       eprint = {2308.15761},
 primaryClass = {astro-ph.GA},
       adsurl = {https://ui.adsabs.harvard.edu/abs/2023MNRAS.525.5702H}
}

@ARTICLE{Ighina2025,
       author = {{Ighina}, Luca and {Caccianiga}, Alessandro and {Connor}, Thomas and {Moretti}, Alberto and {Pacucci}, Fabio and {Reynolds}, Cormac and {Afonso}, Jos{\'e} and {Arsioli}, Bruno and {Belladitta}, Silvia and {Broderick}, Jess W. and {Dallacasa}, Daniele and {Della Ceca}, Roberto and {Haardt}, Francesco and {Lambrides}, Erini and {Leung}, James K. and {Lupi}, Alessandro and {Matute}, Israel and {Rigamonti}, Fabio and {Severgnini}, Paola and {Seymour}, Nick and {Tavecchio}, Fabrizio and {Vignali}, Cristian},
        title = "{X-Ray Investigation of Possible Super-Eddington Accretion in a Radio-loud Quasar at z = 6.13}",
      journal = {\apjl},
         year = 2025,
        month = sep,
       volume = {990},
       number = {2},
          eid = {L56},
        pages = {L56},
          doi = {10.3847/2041-8213/aded0a},
archivePrefix = {arXiv},
       eprint = {2509.04559},
 primaryClass = {astro-ph.GA},
       adsurl = {https://ui.adsabs.harvard.edu/abs/2025ApJ...990L..56I}
}

@ARTICLE{Ilic2012,
       author = {{Ili{\'c}}, D. and {Popovi{\'c}}, L. {\v{C}}. and {La Mura}, G. and {Ciroi}, S. and {Rafanelli}, P.},
        title = "{The analysis of the broad hydrogen Balmer line ratios: Possible implications for the physical properties of the broad line region of AGNs}",
      journal = {\aap},
         year = 2012,
        month = jul,
       volume = {543},
          eid = {A142},
        pages = {A142},
          doi = {10.1051/0004-6361/201219299},
archivePrefix = {arXiv},
       eprint = {1205.3950},
 primaryClass = {astro-ph.CO},
       adsurl = {https://ui.adsabs.harvard.edu/abs/2012A&A...543A.142I}
}

@ARTICLE{Inayoshi2025,
       author = {{Inayoshi}, Kohei and {Maiolino}, Roberto},
        title = "{Extremely Dense Gas around Little Red Dots and High-redshift Active Galactic Nuclei: A Nonstellar Origin of the Balmer Break and Absorption Features}",
      journal = {\apjl},
         year = 2025,
        month = feb,
       volume = {980},
       number = {2},
          eid = {L27},
        pages = {L27},
          doi = {10.3847/2041-8213/adaebd},
archivePrefix = {arXiv},
       eprint = {2409.07805},
 primaryClass = {astro-ph.GA},
       adsurl = {https://ui.adsabs.harvard.edu/abs/2025ApJ...980L..27I}
}

@ARTICLE{Inayoshi2026,
       author = {{Inayoshi}, Kohei and {Murase}, Kohta and {Kashiyama}, Kazumi},
        title = "{Spectral Uniformity of Little Red Dots: A Natural Outcome of Coevolving Seed Black Holes and Nascent Starbursts}",
      journal = {\apj},
         year = 2026,
        month = mar,
       volume = {1000},
       number = {1},
          eid = {90},
        pages = {90},
          doi = {10.3847/1538-4357/ae42ce},
archivePrefix = {arXiv},
       eprint = {2509.19422},
 primaryClass = {astro-ph.GA},
       adsurl = {https://ui.adsabs.harvard.edu/abs/2026ApJ..1000...90I}
}

@ARTICLE{Izotov1997,
       author = {{Izotov}, Yuri I. and {Thuan}, Trinh X. and {Lipovetsky}, Valentin A.},
        title = "{The Primordial Helium Abundance: Systematic Effects and a New Determination}",
      journal = {\apjs},
         year = 1997,
        month = jan,
       volume = {108},
       number = {1},
        pages = {1-39},
          doi = {10.1086/312956},
       adsurl = {https://ui.adsabs.harvard.edu/abs/1997ApJS..108....1I}
}

@ARTICLE{Izotov2018b,
       author = {{Izotov}, Y.~I. and {Worseck}, G. and {Schaerer}, D. and {Guseva}, N.~G. and {Thuan}, T.~X. and {Fricke}, Verhamme, A. and {Orlitov{\'a}}, I.},
        title = "{Low-redshift Lyman continuum leaking galaxies with high [O III]/[O II] ratios}",
      journal = {\mnras},
         year = 2018,
        month = aug,
       volume = {478},
       number = {4},
        pages = {4851-4865},
          doi = {10.1093/mnras/sty1378},
archivePrefix = {arXiv},
       eprint = {1805.09865},
 primaryClass = {astro-ph.GA},
       adsurl = {https://ui.adsabs.harvard.edu/abs/2018MNRAS.478.4851I}
}

@ARTICLE{Jaskot2013,
       author = {{Jaskot}, A.~E. and {Oey}, M.~S.},
        title = "{The Origin and Optical Depth of Ionizing Radiation in the ``Green Pea'' Galaxies}",
      journal = {\apj},
         year = 2013,
        month = apr,
       volume = {766},
       number = {2},
          eid = {91},
        pages = {91},
          doi = {10.1088/0004-637X/766/2/91},
archivePrefix = {arXiv},
       eprint = {1301.0530},
 primaryClass = {astro-ph.CO},
       adsurl = {https://ui.adsabs.harvard.edu/abs/2013ApJ...766...91J}
}

@ARTICLE{Jaskot2016,
       author = {{Jaskot}, A.~E. and {Ravindranath}, S.},
        title = "{Photoionization Models for the Semi-forbidden C III] 1909 Emission in Star-forming Galaxies}",
      journal = {\apj},
         year = 2016,
        month = dec,
       volume = {833},
       number = {2},
          eid = {136},
        pages = {136},
          doi = {10.3847/1538-4357/833/2/136},
archivePrefix = {arXiv},
       eprint = {1610.03778},
 primaryClass = {astro-ph.GA},
       adsurl = {https://ui.adsabs.harvard.edu/abs/2016ApJ...833..136J}
}

@ARTICLE{Ji2020,
       author = {{Ji}, Xihan and {Yan}, Renbin},
        title = "{Constraining photoionization models with a reprojected optical diagnostic diagram}",
      journal = {\mnras},
         year = 2020,
        month = dec,
       volume = {499},
       number = {4},
        pages = {5749-5764},
          doi = {10.1093/mnras/staa3259},
archivePrefix = {arXiv},
       eprint = {2007.09159},
 primaryClass = {astro-ph.GA},
       adsurl = {https://ui.adsabs.harvard.edu/abs/2020MNRAS.499.5749J}
}

@ARTICLE{Ji2025,
       author = {{Ji}, Xihan and {Maiolino}, Roberto and {{\"U}bler}, Hannah and {Scholtz}, Jan and {D'Eugenio}, Francesco and {Sun}, Fengwu and {Perna}, Michele and {Turner}, Hannah and {Carniani}, Stefano and {Arribas}, Santiago and {Bennett}, Jake S. and {Bunker}, Andrew and {Charlot}, St{\'e}phane and {Cresci}, Giovanni and {Curti}, Mirko and {Egami}, Eiichi and {Fabian}, Andy and {Inayoshi}, Kohei and {Isobe}, Yuki and {Jones}, Gareth and {Juod{\v{z}}balis}, Ignas and {Kumari}, Nimisha and {Lyu}, Jianwei and {Mazzolari}, Giovanni and {Parlanti}, Eleonora and {Robertson}, Brant and {Rodr{\'\i}guez Del Pino}, Bruno and {Schneider}, Raffaella and {Sijacki}, Debora and {Tacchella}, Sandro and {Trinca}, Alessandro and {Valiante}, Rosa and {Venturi}, Giacomo and {Volonteri}, Marta and {Willott}, Chris and {Witten}, Callum and {Witstok}, Joris},
        title = "{BlackTHUNDER ─ A non-stellar Balmer break in a black hole-dominated little red dot at z = 7.04}",
      journal = {\mnras},
         year = 2025,
        month = dec,
       volume = {544},
       number = {4},
        pages = {3900-3935},
          doi = {10.1093/mnras/staf1867},
archivePrefix = {arXiv},
       eprint = {2501.13082},
 primaryClass = {astro-ph.GA},
       adsurl = {https://ui.adsabs.harvard.edu/abs/2025MNRAS.544.3900J}
}

@ARTICLE{Jin2023sed,
       author = {{Jin}, Chichuan and {Done}, Chris and {Ward}, Martin and {Panessa}, Francesca and {Liu}, Bo and {Liu}, He-Yang},
        title = "{The extreme super-eddington NLS1 RX J0134.2-4258 - II. A weak-line Seyfert linking to the weak-line quasar}",
      journal = {\mnras},
         year = 2023,
        month = feb,
       volume = {518},
       number = {4},
        pages = {6065-6082},
          doi = {10.1093/mnras/stac3513},
archivePrefix = {arXiv},
       eprint = {2208.06581},
 primaryClass = {astro-ph.HE},
       adsurl = {https://ui.adsabs.harvard.edu/abs/2023MNRAS.518.6065J}
}

@ARTICLE{Juodvabalis2024,
       author = {{Juod{\v{z}}balis}, Ignas and {Ji}, Xihan and {Maiolino}, Roberto and {D'Eugenio}, Francesco and {Scholtz}, Jan and {Risaliti}, Guido and {Fabian}, Andrew C. and {Mazzolari}, Giovanni and {Gilli}, Roberto and {Prandoni}, Isabella and {Arribas}, Santiago and {Bunker}, Andrew J. and {Carniani}, Stefano and {Charlot}, St{\'e}phane and {Curtis-Lake}, Emma and {de Graaff}, Anna and {Hainline}, Kevin and {Parlanti}, Eleonora and {Perna}, Michele and {P{\'e}rez-Gonz{\'a}lez}, Pablo G. and {Robertson}, Brant and {Tacchella}, Sandro and {{\"U}bler}, Hannah and {Williams}, Christina C. and {Willott}, Chris and {Witstok}, Joris},
        title = "{JADES - the Rosetta stone of JWST-discovered AGN: deciphering the intriguing nature of early AGN}",
      journal = {\mnras},
         year = 2024,
        month = nov,
       volume = {535},
       number = {1},
        pages = {853-873},
          doi = {10.1093/mnras/stae2367},
archivePrefix = {arXiv},
       eprint = {2407.08643},
 primaryClass = {astro-ph.GA},
       adsurl = {https://ui.adsabs.harvard.edu/abs/2024MNRAS.535..853J}
}

@ARTICLE{Juodvabalis2026,
       author = {{Juod{\v{z}}balis}, Ignas and {Maiolino}, Roberto and {Baker}, William M. and {Lake}, Emma Curtis and {Scholtz}, Jan and {D'Eugenio}, Francesco and {Trefoloni}, Bartolomeo and {Isobe}, Yuki and {Tacchella}, Sandro and {Bunker}, Andrew J. and {Carniani}, Stefano and {Charlot}, St{\'e}phane and {Jones}, Gareth C. and {Parlanti}, Eleonora and {Perna}, Michele and {Rinaldi}, Pierluigi and {Robertson}, Brant and {{\"U}bler}, Hannah and {Venturi}, Giacomo and {Willott}, Chris},
        title = "{JADES: comprehensive census of broad-line AGN from Reionization to Cosmic Noon revealed by JWST}",
      journal = {\mnras},
         year = 2026,
        month = jan,
          doi = {10.1093/mnras/stag086},
archivePrefix = {arXiv},
       eprint = {2504.03551},
 primaryClass = {astro-ph.GA},
       adsurl = {https://ui.adsabs.harvard.edu/abs/2026MNRAS.tmp...94J}
}

@ARTICLE{Kauffmann2003,
       author = {{Kauffmann}, Guinevere and {Heckman}, Timothy M. and {Tremonti}, Christy and {Brinchmann}, Jarle and {Charlot}, St{\'e}phane and {White}, Simon D.~M. and {Ridgway}, Susan E. and {Brinkmann}, Jon and {Fukugita}, Masataka and {Hall}, Patrick B. and {Ivezi{\'c}}, {\v{Z}}eljko and {Richards}, Gordon T. and {Schneider}, Donald P.},
        title = "{The host galaxies of active galactic nuclei}",
      journal = {\mnras},
         year = 2003,
        month = dec,
       volume = {346},
       number = {4},
        pages = {1055-1077},
          doi = {10.1111/j.1365-2966.2003.07154.x},
archivePrefix = {arXiv},
       eprint = {astro-ph/0304239},
 primaryClass = {astro-ph},
       adsurl = {https://ui.adsabs.harvard.edu/abs/2003MNRAS.346.1055K}
}

@ARTICLE{Kewley2001,
       author = {{Kewley}, L.~J. and {Dopita}, M.~A. and {Sutherland}, R.~S. and {Heisler}, C.~A. and {Trevena}, J.},
        title = "{Theoretical Modeling of Starburst Galaxies}",
      journal = {\apj},
         year = 2001,
        month = jul,
       volume = {556},
       number = {1},
        pages = {121-140},
          doi = {10.1086/321545},
archivePrefix = {arXiv},
       eprint = {astro-ph/0106324},
 primaryClass = {astro-ph},
       adsurl = {https://ui.adsabs.harvard.edu/abs/2001ApJ...556..121K}
}

@ARTICLE{Kocevski2023,
       author = {{Kocevski}, Dale D. and {Onoue}, Masafusa and {Inayoshi}, Kohei and {Trump}, Jonathan R. and {Arrabal Haro}, Pablo and {Grazian}, Andrea and {Dickinson}, Mark and {Finkelstein}, Steven L. and {Kartaltepe}, Jeyhan S. and {Hirschmann}, Michaela and {Aird}, James and {Holwerda}, Benne W. and {Fujimoto}, Seiji and {Juneau}, St{\'e}phanie and {Amor{\'\i}n}, Ricardo O. and {Backhaus}, Bren E. and {Bagley}, Micaela B. and {Barro}, Guillermo and {Bell}, Eric F. and {Bisigello}, Laura and {Calabr{\`o}}, Antonello and {Cleri}, Nikko J. and {Cooper}, M.~C. and {Ding}, Xuheng and {Grogin}, Norman A. and {Ho}, Luis C. and {Hutchison}, Taylor A. and {Inoue}, Akio K. and {Jiang}, Linhua and {Jones}, Brenda and {Koekemoer}, Anton M. and {Li}, Wenxiu and {Li}, Zhengrong and {McGrath}, Elizabeth J. and {Molina}, Juan and {Papovich}, Casey and {P{\'e}rez-Gonz{\'a}lez}, Pablo G. and {Pirzkal}, Nor and {Wilkins}, Stephen M. and {Yang}, Guang and {Yung}, L.~Y. Aaron},
        title = "{Hidden Little Monsters: Spectroscopic Identification of Low-mass, Broad-line AGNs at z > 5 with CEERS}",
      journal = {\apjl},
         year = 2023,
        month = sep,
       volume = {954},
       number = {1},
          eid = {L4},
        pages = {L4},
          doi = {10.3847/2041-8213/ace5a0},
archivePrefix = {arXiv},
       eprint = {2302.00012},
 primaryClass = {astro-ph.GA},
       adsurl = {https://ui.adsabs.harvard.edu/abs/2023ApJ...954L...4K}
}

@ARTICLE{Kocevski2025,
       author = {{Kocevski}, Dale D. and {Finkelstein}, Steven L. and {Barro}, Guillermo and {Taylor}, Anthony J. and {Calabr{\`o}}, Antonello and {Laloux}, Brivael and {Buchner}, Johannes and {Trump}, Jonathan R. and {Leung}, Gene C.~K. and {Yang}, Guang and {Dickinson}, Mark and {P{\'e}rez-Gonz{\'a}lez}, Pablo G. and {Pacucci}, Fabio and {Inayoshi}, Kohei and {Somerville}, Rachel S. and {McGrath}, Elizabeth J. and {Akins}, Hollis B. and {Bagley}, Micaela B. and {Bowler}, Rebecca A.~A. and {Bisigello}, Laura and {Carnall}, Adam and {Casey}, Caitlin M. and {Cheng}, Yingjie and {Cleri}, Nikko J. and {Costantin}, Luca and {Cullen}, Fergus and {Davis}, Kelcey and {Donnan}, Callum T. and {Dunlop}, James S. and {Ellis}, Richard S. and {Ferguson}, Henry C. and {Fujimoto}, Seiji and {Fontana}, Adriano and {Giavalisco}, Mauro and {Grazian}, Andrea and {Grogin}, Norman A. and {Hathi}, Nimish P. and {Hirschmann}, Michaela and {Huertas-Company}, Marc and {Holwerda}, Benne W. and {Illingworth}, Garth and {Juneau}, St{\'e}phanie and {Kartaltepe}, Jeyhan S. and {Koekemoer}, Anton M. and {Li}, Wenxiu and {Lucas}, Ray A. and {Magee}, Dan and {Mason}, Charlotte and {McLeod}, Derek J. and {McLure}, Ross J. and {Napolitano}, Lorenzo and {Papovich}, Casey and {Pirzkal}, Nor and {Rodighiero}, Giulia and {Santini}, Paola and {Wilkins}, Stephen M. and {Yung}, L.~Y. Aaron},
        title = "{The Rise of Faint, Red Active Galactic Nuclei at z > 4: A Sample of Little Red Dots in the JWST Extragalactic Legacy Fields}",
      journal = {\apj},
         year = 2025,
        month = jun,
       volume = {986},
       number = {2},
          eid = {126},
        pages = {126},
          doi = {10.3847/1538-4357/adbc7d},
archivePrefix = {arXiv},
       eprint = {2404.03576},
 primaryClass = {astro-ph.GA},
       adsurl = {https://ui.adsabs.harvard.edu/abs/2025ApJ...986..126K}
}

@ARTICLE{Kokorev2023,
       author = {{Kokorev}, Vasily and {Fujimoto}, Seiji and {Labbe}, Ivo and {Greene}, Jenny E. and {Bezanson}, Rachel and {Dayal}, Pratika and {Nelson}, Erica J. and {Atek}, Hakim and {Brammer}, Gabriel and {Caputi}, Karina I. and {Chemerynska}, Iryna and {Cutler}, Sam E. and {Feldmann}, Robert and {Fudamoto}, Yoshinobu and {Furtak}, Lukas J. and {Goulding}, Andy D. and {de Graaff}, Anna and {Leja}, Joel and {Marchesini}, Danilo and {Miller}, Tim B. and {Nanayakkara}, Themiya and {Oesch}, Pascal A. and {Pan}, Richard and {Price}, Sedona H. and {Setton}, David J. and {Smit}, Renske and {Stefanon}, Mauro and {Wang}, Bingjie and {Weaver}, John R. and {Whitaker}, Katherine E. and {Williams}, Christina C. and {Zitrin}, Adi},
        title = "{UNCOVER: A NIRSpec Identification of a Broad-line AGN at z = 8.50}",
      journal = {\apjl},
         year = 2023,
        month = nov,
       volume = {957},
       number = {1},
          eid = {L7},
        pages = {L7},
          doi = {10.3847/2041-8213/ad037a},
archivePrefix = {arXiv},
       eprint = {2308.11610},
 primaryClass = {astro-ph.GA},
       adsurl = {https://ui.adsabs.harvard.edu/abs/2023ApJ...957L...7K}
}

@ARTICLE{Kokorev2024,
       author = {{Kokorev}, Vasily and {Caputi}, Karina I. and {Greene}, Jenny E. and {Dayal}, Pratika and {Trebitsch}, Maxime and {Cutler}, Sam E. and {Fujimoto}, Seiji and {Labb{\'e}}, Ivo and {Miller}, Tim B. and {Iani}, Edoardo and {Navarro-Carrera}, Rafael and {Rinaldi}, Pierluigi},
        title = "{A Census of Photometrically Selected Little Red Dots at 4 < z < 9 in JWST Blank Fields}",
      journal = {\apj},
         year = 2024,
        month = jun,
       volume = {968},
       number = {1},
          eid = {38},
        pages = {38},
          doi = {10.3847/1538-4357/ad4265},
archivePrefix = {arXiv},
       eprint = {2401.09981},
 primaryClass = {astro-ph.GA},
       adsurl = {https://ui.adsabs.harvard.edu/abs/2024ApJ...968...38K}
}

@ARTICLE{Kokorev2025,
       author = {{Kokorev}, Vasily and {Chisholm}, John and {Naidu}, Rohan P. and {Fujimoto}, Seiji and {Atek}, Hakim and {Brammer}, Gabriel and {Finkelstein}, Steven L. and {Akins}, Hollis B. and {Berg}, Danielle A. and {Furtak}, Lukas J. and {Fei}, Qinyue and {Hsiao}, Tiger Yu-Yang and {Labb{\'e}}, Ivo and {Matthee}, Jorryt and {Mu{\~n}oz}, Julian B. and {Oesch}, Pascal A. and {Pan}, Richard and {Rinaldi}, Pierluigi and {Saldana-Lopez}, Alberto and {Schaerer}, Daniel and {Volonteri}, Marta and {Zitrin}, Adi},
        title = "{The Deepest GLIMPSE of a Dense Gas Cocoon Enshrouding a Little Red Dot}",
      journal = {arXiv e-prints},
         year = 2025,
        month = nov,
          eid = {arXiv:2511.07515},
        pages = {arXiv:2511.07515},
          doi = {10.48550/arXiv.2511.07515},
archivePrefix = {arXiv},
       eprint = {2511.07515},
 primaryClass = {astro-ph.GA},
       adsurl = {https://ui.adsabs.harvard.edu/abs/2025arXiv251107515K}
}

@ARTICLE{Korista1997,
       author = {{Korista}, Kirk and {Baldwin}, Jack and {Ferland}, Gary and {Verner}, Dima},
        title = "{An Atlas of Computed Equivalent Widths of Quasar Broad Emission Lines}",
      journal = {\apjs},
         year = 1997,
        month = jan,
       volume = {108},
       number = {2},
        pages = {401-415},
          doi = {10.1086/312966},
archivePrefix = {arXiv},
       eprint = {astro-ph/9611220},
 primaryClass = {astro-ph},
       adsurl = {https://ui.adsabs.harvard.edu/abs/1997ApJS..108..401K}
}

@ARTICLE{Korista2000,
       author = {{Korista}, Kirk T. and {Goad}, Michael R.},
        title = "{Locally Optimally Emitting Clouds and the Variable Broad Emission Line Spectrum of NGC 5548}",
      journal = {\apj},
         year = 2000,
        month = jun,
       volume = {536},
       number = {1},
        pages = {284-298},
          doi = {10.1086/308930},
archivePrefix = {arXiv},
       eprint = {astro-ph/0001399},
 primaryClass = {astro-ph},
       adsurl = {https://ui.adsabs.harvard.edu/abs/2000ApJ...536..284K}
}

@ARTICLE{Korista2004,
       author = {{Korista}, Kirk T. and {Goad}, Michael R.},
        title = "{What the Optical Recombination Lines Can Tell Us about the Broad-Line Regions of Active Galactic Nuclei}",
      journal = {\apj},
         year = 2004,
        month = may,
       volume = {606},
       number = {2},
        pages = {749-762},
          doi = {10.1086/383193},
archivePrefix = {arXiv},
       eprint = {astro-ph/0402506},
 primaryClass = {astro-ph},
       adsurl = {https://ui.adsabs.harvard.edu/abs/2004ApJ...606..749K}
}

@ARTICLE{Kormendy2013,
       author = {{Kormendy}, John and {Ho}, Luis C.},
        title = "{Coevolution (Or Not) of Supermassive Black Holes and Host Galaxies}",
      journal = {\araa},
         year = 2013,
        month = aug,
       volume = {51},
       number = {1},
        pages = {511-653},
          doi = {10.1146/annurev-astro-082708-101811},
archivePrefix = {arXiv},
       eprint = {1304.7762},
 primaryClass = {astro-ph.CO},
       adsurl = {https://ui.adsabs.harvard.edu/abs/2013ARA&A..51..511K}
}

@ARTICLE{Kraemer1999b,
       author = {{Kraemer}, S.~B. and {Turner}, T.~J. and {Crenshaw}, D.~M. and {George}, I.~M.},
        title = "{The Effect of Intrinsic Ultraviolet Absorbers on the Ionizing Continuum and Narrow Emission Line Ratios in Seyfert Galaxies}",
      journal = {\apj},
         year = 1999,
        month = jul,
       volume = {519},
       number = {1},
        pages = {69-79},
          doi = {10.1086/307352},
archivePrefix = {arXiv},
       eprint = {astro-ph/9902161},
 primaryClass = {astro-ph},
       adsurl = {https://ui.adsabs.harvard.edu/abs/1999ApJ...519...69K}
}

@INPROCEEDINGS{Kraemer1999,
       author = {{Kraemer}, S.~B. and {Crenshaw}, D.~M.},
        title = "{Resolved Spectroscopy of the Narrow-Line Region in NGC 1068. II. Physical Conditions Near the NGC 1068 ``Hot-Spot''}",
    booktitle = {American Astronomical Society Meeting Abstracts},
         year = 1999,
       series = {American Astronomical Society Meeting Abstracts},
       volume = {195},
        month = dec,
          eid = {115.05},
        pages = {115.05},
       adsurl = {https://ui.adsabs.harvard.edu/abs/1999AAS...19511505K}
}

@ARTICLE{Kraemer2007,
       author = {{Kraemer}, S.~B. and {Bottorff}, M.~C. and {Crenshaw}, D.~M.},
        title = "{On the Effects of Dissipative Turbulence on the Narrow Emission-Line Ratios in Seyfert Galaxies}",
      journal = {\apj},
         year = 2007,
        month = oct,
       volume = {668},
       number = {2},
        pages = {730-737},
          doi = {10.1086/521272},
archivePrefix = {arXiv},
       eprint = {0706.4317},
 primaryClass = {astro-ph},
       adsurl = {https://ui.adsabs.harvard.edu/abs/2007ApJ...668..730K}
}

@INPROCEEDINGS{Kraemer2012,
       author = {{Kraemer}, S.~B. and {Crenshaw}, D.~M. and {Bottorff}, M.~C. and {Turner}, T.~J. and {Miller}, L.},
        title = "{Exploring Micro-Turbulence in Emission and Absorption in AGN}",
    booktitle = {AGN Winds in Charleston},
         year = 2012,
       editor = {{Chartas}, G. and {Hamann}, F. and {Leighly}, K.~M.},
       series = {Astronomical Society of the Pacific Conference Series},
       volume = {460},
        month = aug,
        pages = {57},
       adsurl = {https://ui.adsabs.harvard.edu/abs/2012ASPC..460...57K}
}

@ARTICLE{Kubota2018,
       author = {{Kubota}, Aya and {Done}, Chris},
        title = "{A physical model of the broad-band continuum of AGN and its implications for the UV/X relation and optical variability}",
      journal = {\mnras},
         year = 2018,
        month = oct,
       volume = {480},
       number = {1},
        pages = {1247-1262},
          doi = {10.1093/mnras/sty1890},
archivePrefix = {arXiv},
       eprint = {1804.00171},
 primaryClass = {astro-ph.HE},
       adsurl = {https://ui.adsabs.harvard.edu/abs/2018MNRAS.480.1247K}
}

@ARTICLE{Kubota2019,
       author = {{Kubota}, Aya and {Done}, Chris},
        title = "{Modelling the spectral energy distribution of super-Eddington quasars}",
      journal = {\mnras},
         year = 2019,
        month = oct,
       volume = {489},
       number = {1},
        pages = {524-533},
          doi = {10.1093/mnras/stz2140},
archivePrefix = {arXiv},
       eprint = {1905.02920},
 primaryClass = {astro-ph.GA},
       adsurl = {https://ui.adsabs.harvard.edu/abs/2019MNRAS.489..524K}
}

@ARTICLE{Labbe2023,
       author = {{Labb{\'e}}, Ivo and {van Dokkum}, Pieter and {Nelson}, Erica and {Bezanson}, Rachel and {Suess}, Katherine A. and {Leja}, Joel and {Brammer}, Gabriel and {Whitaker}, Katherine and {Mathews}, Elijah and {Stefanon}, Mauro and {Wang}, Bingjie},
        title = "{A population of red candidate massive galaxies  600 Myr after the Big Bang}",
      journal = {\nat},
         year = 2023,
        month = apr,
       volume = {616},
       number = {7956},
        pages = {266-269},
          doi = {10.1038/s41586-023-05786-2},
archivePrefix = {arXiv},
       eprint = {2207.12446},
 primaryClass = {astro-ph.GA},
       adsurl = {https://ui.adsabs.harvard.edu/abs/2023Natur.616..266L}
}

@ARTICLE{Lamareille2010,
       author = {{Lamareille}, F.},
        title = "{Spectral classification of emission-line galaxies from the Sloan Digital Sky Survey. I. An improved classification for high-redshift galaxies}",
      journal = {\aap},
         year = 2010,
        month = jan,
       volume = {509},
          eid = {A53},
        pages = {A53},
          doi = {10.1051/0004-6361/200913168},
archivePrefix = {arXiv},
       eprint = {0910.4814},
 primaryClass = {astro-ph.CO},
       adsurl = {https://ui.adsabs.harvard.edu/abs/2010A&A...509A..53L}
}

@ARTICLE{Lambrides2024,
       author = {{Lambrides}, Erini and {Garofali}, Kristen and {Larson}, Rebecca and {Ptak}, Andrew and {Chiaberge}, Marco and {Long}, Arianna S. and {Hutchison}, Taylor A. and {Norman}, Colin and {McKinney}, Jed and {Akins}, Hollis B. and {Berg}, Danielle A. and {Chisholm}, John and {Civano}, Francesca and {Cloonan}, Aidan P. and {Endsley}, Ryan and {Faisst}, Andreas L. and {Gilli}, Roberto and {Gillman}, Steven and {Hirschmann}, Michaela and {Kartaltepe}, Jeyhan S. and {Kocevski}, Dale D. and {Kokorev}, Vasily and {Pacucci}, Fabio and {Richardson}, Chris T. and {Stiavelli}, Massimo and {Whalen}, Kelly E.},
        title = "{The Case for Super-Eddington Accretion: Connecting Weak X-ray and UV Line Emission in JWST Broad-Line AGN During the First Gyr of Cosmic Time}",
      journal = {arXiv e-prints},
         year = 2024,
        month = sep,
          eid = {arXiv:2409.13047},
        pages = {arXiv:2409.13047},
          doi = {10.48550/arXiv.2409.13047},
archivePrefix = {arXiv},
       eprint = {2409.13047},
 primaryClass = {astro-ph.HE},
       adsurl = {https://ui.adsabs.harvard.edu/abs/2024arXiv240913047L}
}

@ARTICLE{Lecroq2024,
       author = {{Lecroq}, Marie and {Charlot}, St{\'e}phane and {Bressan}, Alessandro and {Bruzual}, Gustavo and {Costa}, Guglielmo and {Iorio}, Giuliano and {Spera}, Mario and {Mapelli}, Michela and {Chen}, Yang and {Chevallard}, Jacopo and {Dall'Amico}, Marco},
        title = "{Nebular emission from young stellar populations including binary stars}",
      journal = {\mnras},
         year = 2024,
        month = jan,
       volume = {527},
       number = {3},
        pages = {9480-9504},
          doi = {10.1093/mnras/stad3838},
archivePrefix = {arXiv},
       eprint = {2312.08432},
 primaryClass = {astro-ph.GA},
       adsurl = {https://ui.adsabs.harvard.edu/abs/2024MNRAS.527.9480L}
}

@ARTICLE{Lecroq2025,
       author = {{Lecroq}, Marie and {Charlot}, St{\'e}phane and {Bressan}, Alessandro and {Bruzual}, Gustavo and {Costa}, Guglielmo and {Iorio}, Giuliano and {Mapelli}, Michela and {Santoliquido}, Filippo and {Shepherd}, Kendall and {Spera}, Mario},
        title = "{A new prescription for the spectral properties of population III stellar populations}",
      journal = {\aap},
         year = 2025,
        month = mar,
       volume = {695},
          eid = {A17},
        pages = {A17},
          doi = {10.1051/0004-6361/202452463},
archivePrefix = {arXiv},
       eprint = {2502.14028},
 primaryClass = {astro-ph.GA},
       adsurl = {https://ui.adsabs.harvard.edu/abs/2025A&A...695A..17L}
}

@ARTICLE{Leitherer1999,
       author = {{Leitherer}, Claus and {Schaerer}, Daniel and {Goldader}, Jeffrey D. and {Delgado}, Rosa M. Gonz{\'a}lez and {Robert}, Carmelle and {Kune}, Denis Foo and {de Mello}, Du{\'\i}lia F. and {Devost}, Daniel and {Heckman}, Timothy M.},
        title = "{Starburst99: Synthesis Models for Galaxies with Active Star Formation}",
      journal = {\apjs},
         year = 1999,
        month = jul,
       volume = {123},
       number = {1},
        pages = {3-40},
          doi = {10.1086/313233},
archivePrefix = {arXiv},
       eprint = {astro-ph/9902334},
 primaryClass = {astro-ph},
       adsurl = {https://ui.adsabs.harvard.edu/abs/1999ApJS..123....3L}
}

@ARTICLE{Liu2019,
       author = {{Liu}, He-Yang and {Liu}, Wen-Juan and {Dong}, Xiao-Bo and {Zhou}, Hongyan and {Wang}, Tinggui and {Lu}, Honglin and {Yuan}, Weimin},
        title = "{A Comprehensive and Uniform Sample of Broad-line Active Galactic Nuclei from the SDSS DR7}",
      journal = {\apjs},
         year = 2019,
        month = aug,
       volume = {243},
       number = {2},
          eid = {21},
        pages = {21},
          doi = {10.3847/1538-4365/ab298b},
archivePrefix = {arXiv},
       eprint = {1906.05597},
 primaryClass = {astro-ph.GA},
       adsurl = {https://ui.adsabs.harvard.edu/abs/2019ApJS..243...21L}
}

@ARTICLE{Liu2026,
       author = {{Liu}, Hanpu and {Jiang}, Yan-Fei and {Quataert}, Eliot and {Greene}, Jenny E. and {Ma}, Yilun and {Lin}, Xiaojing},
        title = "{Synthetic Spectral Library of Optically Thick Atmospheres for Little Red Dots}",
      journal = {arXiv e-prints},
         year = 2026,
        month = mar,
          eid = {arXiv:2603.02317},
        pages = {arXiv:2603.02317},
          doi = {10.48550/arXiv.2603.02317},
archivePrefix = {arXiv},
       eprint = {2603.02317},
 primaryClass = {astro-ph.GA},
       adsurl = {https://ui.adsabs.harvard.edu/abs/2026arXiv260302317L}
}

@ARTICLE{Lodato2006,
       author = {{Lodato}, Giuseppe and {Natarajan}, Priyamvada},
        title = "{Supermassive black hole formation during the assembly of pre-galactic discs}",
      journal = {\mnras},
         year = 2006,
        month = oct,
       volume = {371},
       number = {4},
        pages = {1813-1823},
          doi = {10.1111/j.1365-2966.2006.10801.x},
archivePrefix = {arXiv},
       eprint = {astro-ph/0606159},
 primaryClass = {astro-ph},
       adsurl = {https://ui.adsabs.harvard.edu/abs/2006MNRAS.371.1813L}
}

@ARTICLE{Loeb1994,
       author = {{Loeb}, Abraham and {Rasio}, Frederic A.},
        title = "{Collapse of Primordial Gas Clouds and the Formation of Quasar Black Holes}",
      journal = {\apj},
         year = 1994,
        month = sep,
       volume = {432},
        pages = {52},
          doi = {10.1086/174548},
archivePrefix = {arXiv},
       eprint = {astro-ph/9401026},
 primaryClass = {astro-ph},
       adsurl = {https://ui.adsabs.harvard.edu/abs/1994ApJ...432...52L}
}

@ARTICLE{Madau2001,
       author = {{Madau}, Piero and {Rees}, Martin J.},
        title = "{Massive Black Holes as Population III Remnants}",
      journal = {\apjl},
         year = 2001,
        month = apr,
       volume = {551},
       number = {1},
        pages = {L27-L30},
          doi = {10.1086/319848},
archivePrefix = {arXiv},
       eprint = {astro-ph/0101223},
 primaryClass = {astro-ph},
       adsurl = {https://ui.adsabs.harvard.edu/abs/2001ApJ...551L..27M}
}

@ARTICLE{Madau2024,
       author = {{Madau}, Piero and {Haardt}, Francesco},
        title = "{X-Ray Weak Active Galactic Nuclei from Super-Eddington Accretion onto Infant Black Holes}",
      journal = {\apjl},
         year = 2024,
        month = dec,
       volume = {976},
       number = {2},
          eid = {L24},
        pages = {L24},
          doi = {10.3847/2041-8213/ad90e1},
archivePrefix = {arXiv},
       eprint = {2410.00417},
 primaryClass = {astro-ph.GA},
       adsurl = {https://ui.adsabs.harvard.edu/abs/2024ApJ...976L..24M}
}

@ARTICLE{Madau2025,
       author = {{Madau}, Piero},
        title = "{Chasing the Light: Shadowing, Collimation, and the Super-Eddington Growth of Infant Black Holes in JWST-Discovered AGNs}",
      journal = {arXiv e-prints},
         year = 2025,
        month = jan,
          eid = {arXiv:2501.09854},
        pages = {arXiv:2501.09854},
          doi = {10.48550/arXiv.2501.09854},
archivePrefix = {arXiv},
       eprint = {2501.09854},
 primaryClass = {astro-ph.HE},
       adsurl = {https://ui.adsabs.harvard.edu/abs/2025arXiv250109854M}
}

@ARTICLE{Madau2026,
       author = {{Madau}, Piero and {Maiolino}, Roberto},
        title = "{Little Red Dots as Obscured Little Blue Dots: A Super-Eddington Unification Model}",
      journal = {arXiv e-prints},
         year = 2026,
        month = feb,
          eid = {arXiv:2602.22386},
        pages = {arXiv:2602.22386},
          doi = {10.48550/arXiv.2602.22386},
archivePrefix = {arXiv},
       eprint = {2602.22386},
 primaryClass = {astro-ph.GA},
       adsurl = {https://ui.adsabs.harvard.edu/abs/2026arXiv260222386M}
}

@ARTICLE{Madau2026b,
       author = {{Madau}, Piero and {Maiolino}, Roberto and {Scholtz}, Jan and {D'Eugenio}, Francesco},
        title = "{Wings of little dots: Exponential broad lines from a stratified BLR}",
      journal = {arXiv e-prints},
         year = 2026,
        month = apr,
          eid = {arXiv:2604.04216},
        pages = {arXiv:2604.04216},
          doi = {10.48550/arXiv.2604.04216},
archivePrefix = {arXiv},
       eprint = {2604.04216},
 primaryClass = {astro-ph.GA},
       adsurl = {https://ui.adsabs.harvard.edu/abs/2026arXiv260404216M}
}

@ARTICLE{Magorrian1998,
       author = {{Magorrian}, John and {Tremaine}, Scott and {Richstone}, Douglas and {Bender}, Ralf and {Bower}, Gary and {Dressler}, Alan and {Faber}, S.~M. and {Gebhardt}, Karl and {Green}, Richard and {Grillmair}, Carl and {Kormendy}, John and {Lauer}, Tod},
        title = "{The Demography of Massive Dark Objects in Galaxy Centers}",
      journal = {\aj},
         year = 1998,
        month = jun,
       volume = {115},
       number = {6},
        pages = {2285-2305},
          doi = {10.1086/300353},
archivePrefix = {arXiv},
       eprint = {astro-ph/9708072},
 primaryClass = {astro-ph},
       adsurl = {https://ui.adsabs.harvard.edu/abs/1998AJ....115.2285M}
}

@ARTICLE{Maiolino2024,
       author = {{Maiolino}, Roberto and {Scholtz}, Jan and {Curtis-Lake}, Emma and {Carniani}, Stefano and {Baker}, William and {de Graaff}, Anna and {Tacchella}, Sandro and {{\"U}bler}, Hannah and {D'Eugenio}, Francesco and {Witstok}, Joris and {Curti}, Mirko and {Arribas}, Santiago and {Bunker}, Andrew J. and {Charlot}, St{\'e}phane and {Chevallard}, Jacopo and {Eisenstein}, Daniel J. and {Egami}, Eiichi and {Ji}, Zhiyuan and {Jones}, Gareth C. and {Lyu}, Jianwei and {Rawle}, Tim and {Robertson}, Brant and {Rujopakarn}, Wiphu and {Perna}, Michele and {Sun}, Fengwu and {Venturi}, Giacomo and {Williams}, Christina C. and {Willott}, Chris},
        title = "{JADES: The diverse population of infant black holes at 4 < z < 11: Merging, tiny, poor, but mighty}",
      journal = {\aap},
         year = 2024,
        month = nov,
       volume = {691},
          eid = {A145},
        pages = {A145},
          doi = {10.1051/0004-6361/202347640},
archivePrefix = {arXiv},
       eprint = {2308.01230},
 primaryClass = {astro-ph.GA},
       adsurl = {https://ui.adsabs.harvard.edu/abs/2024A&A...691A.145M}
}

@ARTICLE{Maiolino2025,
       author = {{Maiolino}, Roberto and {Risaliti}, Guido and {Signorini}, Matilde and {Trefoloni}, Bartolomeo and {Juod{\v{z}}balis}, Ignas and {Scholtz}, Jan and {{\"U}bler}, Hannah and {D'Eugenio}, Francesco and {Carniani}, Stefano and {Fabian}, Andy and {Ji}, Xihan and {Mazzolari}, Giovanni and {Bertola}, Elena and {Brusa}, Marcella and {Bunker}, Andrew J. and {Charlot}, Stephane and {Comastri}, Andrea and {Cresci}, Giovanni and {DeCoursey}, Christa Noel and {Egami}, Eiichi and {Fiore}, Fabrizio and {Gilli}, Roberto and {Perna}, Michele and {Tacchella}, Sandro and {Venturi}, Giacomo},
        title = "{JWST meets Chandra: a large population of Compton thick, feedback-free, and intrinsically X-ray weak AGN, with a sprinkle of SNe}",
      journal = {\mnras},
         year = 2025,
        month = apr,
       volume = {538},
       number = {3},
        pages = {1921-1943},
          doi = {10.1093/mnras/staf359},
archivePrefix = {arXiv},
       eprint = {2405.00504},
 primaryClass = {astro-ph.GA},
       adsurl = {https://ui.adsabs.harvard.edu/abs/2025MNRAS.538.1921M}
}

@ARTICLE{Maiolino2026,
       author = {{Maiolino}, Roberto and {{\"U}bler}, Hannah and {Perna}, Michele and {Witstok}, Joris and {Jones}, Gareth C. and {Perez-Gonzalez}, Pablo G. and {Nakajima}, Kimihiko and {Rusta}, Elka and {Salvadori}, Stefania and {Tacchella}, Sandro and {Madau}, Piero and {Trussler}, James A.~A. and {D'Eugenio}, Francesco and {Ji}, Xihan and {Scholtz}, Jan and {Carniani}, Stefano and {Isobe}, Yuki and {Katz}, Harley and {Arribas}, Santiago and {Baker}, William M. and {B{\"o}ker}, Torsten and {Bromm}, Volker and {Bunker}, Andrew J. and {Charlot}, Stephane and {Chevallard}, Jacopo and {Curti}, Mirko and {Curtis-Lake}, Emma and {Eisenstein}, Daniel and {Egami}, Eiichi and {Ferrara}, Andrea and {Graziani}, Luca and {Hainline}, Kevin and {Helton}, Jakob M. and {Ivey}, Lucy and {Jonson}, Benjamin and {Koller}, Maria and {Kumari}, Nimisha and {Marconi}, Alessandro and {Mazzolari}, Giovanni and {Laporte}, Nicolas and {Parlanti}, Eleonora and {Pascalau}, Robert and {Pentericci}, Laura and {Rinaldi}, Pierluigi and {Robertson}, Brant and {Rodr{\'\i}guez Del Pino}, Bruno and {Schneider}, Raffaella and {Venditti}, Alessandra and {Venturi}, Giacomo and {Willmer}, Christopher N.~A. and {Witten}, Callum and {Zamora}, Sandra},
        title = "{The search for Population III: Confirmation of a HeII emitter with no metal lines at z=10.6}",
      journal = {arXiv e-prints},
         year = 2026,
        month = mar,
          eid = {arXiv:2603.20362},
        pages = {arXiv:2603.20362},
          doi = {10.48550/arXiv.2603.20362},
archivePrefix = {arXiv},
       eprint = {2603.20362},
 primaryClass = {astro-ph.GA},
       adsurl = {https://ui.adsabs.harvard.edu/abs/2026arXiv260320362M}
}

@ARTICLE{Marconi2004,
       author = {{Marconi}, A. and {Risaliti}, G. and {Gilli}, R. and {Hunt}, L.~K. and {Maiolino}, R. and {Salvati}, M.},
        title = "{Local supermassive black holes, relics of active galactic nuclei and the X-ray background}",
      journal = {\mnras},
         year = 2004,
        month = jun,
       volume = {351},
       number = {1},
        pages = {169-185},
          doi = {10.1111/j.1365-2966.2004.07765.x},
archivePrefix = {arXiv},
       eprint = {astro-ph/0311619},
 primaryClass = {astro-ph},
       adsurl = {https://ui.adsabs.harvard.edu/abs/2004MNRAS.351..169M}
}

@ARTICLE{Marques2024,
       author = {{Marques-Chaves}, R. and {Schaerer}, D. and {Kuruvanthodi}, A. and {Korber}, D. and {Prantzos}, N. and {Charbonnel}, C. and {Weibel}, A. and {Izotov}, Y.~I. and {Messa}, M. and {Brammer}, G. and {Dessauges-Zavadsky}, M. and {Oesch}, P.},
        title = "{Extreme N-emitters at high redshift: Possible signatures of supermassive stars and globular cluster or black hole formation in action}",
      journal = {\aap},
         year = 2024,
        month = jan,
       volume = {681},
          eid = {A30},
        pages = {A30},
          doi = {10.1051/0004-6361/202347411},
archivePrefix = {arXiv},
       eprint = {2307.04234},
 primaryClass = {astro-ph.GA},
       adsurl = {https://ui.adsabs.harvard.edu/abs/2024A&A...681A..30M}
}

@ARTICLE{Marziani2015,
       author = {{Marziani}, P. and {Sulentic}, J.~W. and {Negrete}, C.~A. and {Dultzin}, D. and {Del Olmo}, A. and {Mart{\'\i}nez Carballo}, M.~A. and {Zwitter}, T. and {Bachev}, R.},
        title = "{UV spectral diagnostics for low redshift quasars: estimating physical conditions and radius of the broad line region}",
      journal = {\apss},
         year = 2015,
        month = apr,
       volume = {356},
       number = {2},
        pages = {339-346},
          doi = {10.1007/s10509-014-2136-z},
archivePrefix = {arXiv},
       eprint = {1410.3146},
 primaryClass = {astro-ph.GA},
       adsurl = {https://ui.adsabs.harvard.edu/abs/2015Ap&SS.356..339M}
}

@ARTICLE{Matthee2026,
       author = {{Matthee}, Jorryt and {Torralba}, Alberto and {Pezzulli}, Gabriele and {Naidu}, Rohan P. and {Chisholm}, John and {Mascia}, Sara and {Greene}, Jenny E. and {Ishikawa}, Yuzo and {Gronke}, Max and {Wuyts}, Stijn and {Bordoloi}, Rongmon and {Brammer}, Gabriel and {Chang}, Seok-Jun and {Eilers}, Anna-Christina and {de Graaff}, Anna and {Hviding}, Raphael E. and {Iani}, Edoardo and {Illingworth}, Garth and {Kashino}, Daichi and {Labbe}, Ivo and {Ma}, Yilun and {Maseda}, Michael V. and {Meyer}, Romain and {Nelson}, Erica and {Oesch}, Pascal and {Xiao}, Mengyuan},
        title = "{The Engine and its Flows: Little Red Dot spectra are shaped by the column densities of their gas envelopes}",
      journal = {arXiv e-prints},
         year = 2026,
        month = mar,
          eid = {arXiv:2603.17667},
        pages = {arXiv:2603.17667},
          doi = {10.48550/arXiv.2603.17667},
archivePrefix = {arXiv},
       eprint = {2603.17667},
 primaryClass = {astro-ph.GA},
       adsurl = {https://ui.adsabs.harvard.edu/abs/2026arXiv260317667M}
}

@ARTICLE{Matthee2024,
       author = {{Matthee}, Jorryt and {Naidu}, Rohan P. and {Brammer}, Gabriel and {Chisholm}, John and {Eilers}, Anna-Christina and {Goulding}, Andy and {Greene}, Jenny and {Kashino}, Daichi and {Labbe}, Ivo and {Lilly}, Simon J. and {Mackenzie}, Ruari and {Oesch}, Pascal A. and {Weibel}, Andrea and {Wuyts}, Stijn and {Xiao}, Mengyuan and {Bordoloi}, Rongmon and {Bouwens}, Rychard and {van Dokkum}, Pieter and {Illingworth}, Garth and {Kramarenko}, Ivan and {Maseda}, Michael V. and {Mason}, Charlotte and {Meyer}, Romain A. and {Nelson}, Erica J. and {Reddy}, Naveen A. and {Shivaei}, Irene and {Simcoe}, Robert A. and {Yue}, Minghao},
        title = "{Little Red Dots: An Abundant Population of Faint Active Galactic Nuclei at z {\ensuremath{\sim}} 5 Revealed by the EIGER and FRESCO JWST Surveys}",
      journal = {\apj},
         year = 2024,
        month = mar,
       volume = {963},
       number = {2},
          eid = {129},
        pages = {129},
          doi = {10.3847/1538-4357/ad2345},
archivePrefix = {arXiv},
       eprint = {2306.05448},
 primaryClass = {astro-ph.GA},
       adsurl = {https://ui.adsabs.harvard.edu/abs/2024ApJ...963..129M}
}

@ARTICLE{Mazzolari2024,
       author = {{Mazzolari}, Giovanni and {{\"U}bler}, Hannah and {Maiolino}, Roberto and {Ji}, Xihan and {Nakajima}, Kimihiko and {Feltre}, Anna and {Scholtz}, Jan and {D'Eugenio}, Francesco and {Curti}, Mirko and {Mignoli}, Marco and {Marconi}, Alessandro},
        title = "{New AGN diagnostic diagrams based on the [OIII]{\ensuremath{\lambda}}4363 auroral line}",
      journal = {\aap},
         year = 2024,
        month = nov,
       volume = {691},
          eid = {A345},
        pages = {A345},
          doi = {10.1051/0004-6361/202450407},
archivePrefix = {arXiv},
       eprint = {2404.10811},
 primaryClass = {astro-ph.GA},
       adsurl = {https://ui.adsabs.harvard.edu/abs/2024A&A...691A.345M}
}

@ARTICLE{Mazzolari2025,
       author = {{Mazzolari}, Giovanni and {Scholtz}, Jan and {Maiolino}, Roberto and {Gilli}, Roberto and {Traina}, Alberto and {L{\'o}pez}, Ivan E. and {{\"U}bler}, Hannah and {Trefoloni}, Bartolomeo and {D'Eugenio}, Francesco and {Ji}, Xihan and {Mignoli}, Marco and {Vito}, Fabio and {Vignali}, Cristian and {Brusa}, Marcella},
        title = "{Narrow-line AGN selection in CEERS: Spectroscopic selection, physical properties, and X-ray and radio analysis}",
      journal = {\aap},
         year = 2025,
        month = aug,
       volume = {700},
          eid = {A12},
        pages = {A12},
          doi = {10.1051/0004-6361/202451860},
archivePrefix = {arXiv},
       eprint = {2408.15615},
 primaryClass = {astro-ph.GA},
       adsurl = {https://ui.adsabs.harvard.edu/abs/2025A&A...700A..12M}
}

@ARTICLE{McKaig2024,
       author = {{McKaig}, Jeffrey D. and {Satyapal}, Shobita and {Laor}, Ari and {Abel}, Nicholas P. and {Doan}, Sara M. and {Ricci}, Claudio and {Cann}, Jenna M.},
        title = "{Why Are Optical Coronal Lines Faint in Active Galactic Nuclei?}",
      journal = {\apj},
         year = 2024,
        month = nov,
       volume = {976},
       number = {1},
          eid = {130},
        pages = {130},
          doi = {10.3847/1538-4357/ad7a79},
archivePrefix = {arXiv},
       eprint = {2408.15229},
 primaryClass = {astro-ph.GA},
       adsurl = {https://ui.adsabs.harvard.edu/abs/2024ApJ...976..130M}
}

@ARTICLE{Mendez2026,
       author = {{M{\'e}ndez-Delgado}, J. Eduardo and {Morisset}, Christophe and {Henney}, William J. and {Drory}, Niv and {Egorov}, Oleg V. and {S{\'a}nchez}, Sebasti{\'a}n F. and {Kollmeier}, Juna A. and {Kreckel}, Kathryn and {Blanc}, Guillermo and {Stasi{\'n}ska}, Gra{\.z}yna and {Johnston}, Evelyn J. and {Ibarra-Medel}, H{\'e}ctor J. and {Mej{\'\i}a-Narv{\'a}ez}, Alfredo J. and {Esteban}, C{\'e}sar and {Garc{\'\i}a-Rojas}, Jorge and {Singh}, Amrita and {Katkov}, Ivan Yu. and {Skillman}, Evan D. and {Orozco-Duarte}, Rogelio and {Zinchenko}, Igor A. and {Lugo-Aranda}, Alejandra Z. and {Wofford}, Aida and {Glover}, Simon C.~O. and {Egorova}, Evgeniya and {Zerme{\~n}o}, Rodolfo de J. and {Casta{\~n}eda-Carlos}, Lesly C. and {Liang}, Fu-Heng and {Sattler}, Natascha and {Cruz-Gonz{\'a}lez}, Irene and {Brownstein}, Joel R. and {Hilder}, Thomas and {Schneider}, Donald P.},
        title = "{There is no single density: star-forming regions and galaxies hold more dense ionized gas than long assumed}",
      journal = {arXiv e-prints},
         year = 2026,
        month = jul,
          eid = {arXiv:2607.07973},
        pages = {arXiv:2607.07973},
archivePrefix = {arXiv},
       eprint = {2607.07973},
 primaryClass = {astro-ph.GA},
       adsurl = {https://ui.adsabs.harvard.edu/abs/2026arXiv260707973M}
}

@ARTICLE{Mignoli2019,
       author = {{Mignoli}, M. and {Feltre}, A. and {Bongiorno}, A. and {Calura}, F. and {Gilli}, R. and {Vignali}, C. and {Zamorani}, G. and {Lilly}, S.~J. and {Le F{\`e}vre}, O. and {Bardelli}, S. and {Bolzonella}, M. and {Bordoloi}, R. and {Le Brun}, V. and {Caputi}, K.~I. and {Cimatti}, A. and {Diener}, C. and {Garilli}, B. and {Koekemoer}, A.~M. and {Maier}, C. and {Mainieri}, V. and {Peng}, Y. and {P{\'e}rez Montero}, E. and {Silverman}, J.~D. and {Zucca}, E.},
        title = "{Obscured AGN at 1.5 < z < 3.0 from the zCOSMOS-deep Survey . I. Properties of the emitting gas in the narrow-line region}",
      journal = {\aap},
         year = 2019,
        month = jun,
       volume = {626},
          eid = {A9},
        pages = {A9},
          doi = {10.1051/0004-6361/201935062},
archivePrefix = {arXiv},
       eprint = {1903.11085},
 primaryClass = {astro-ph.GA},
       adsurl = {https://ui.adsabs.harvard.edu/abs/2019A&A...626A...9M}
}

@ARTICLE{Morel2025,
       author = {{Morel}, I. and {Schaerer}, D. and {Marques-Chaves}, R. and {Prantzos}, N. and {Charbonnel}, C. and {Brammer}, G. and {Xiao}, M. and {Dessauges-Zavadsky}, M.},
        title = "{Discovery of new N-emitters over a wide redshift range}",
      journal = {arXiv e-prints},
         year = 2025,
        month = nov,
          eid = {arXiv:2511.20484},
        pages = {arXiv:2511.20484},
          doi = {10.48550/arXiv.2511.20484},
archivePrefix = {arXiv},
       eprint = {2511.20484},
 primaryClass = {astro-ph.GA},
       adsurl = {https://ui.adsabs.harvard.edu/abs/2025arXiv251120484M}
}

@ARTICLE{Naidu2025,
       author = {{Naidu}, Rohan P. and {Matthee}, Jorryt and {Katz}, Harley and {de Graaff}, Anna and {Oesch}, Pascal and {Smith}, Aaron and {Greene}, Jenny E. and {Brammer}, Gabriel and {Weibel}, Andrea and {Hviding}, Raphael and {Chisholm}, John and {Labb\textbackslash'e}, Ivo and {Simcoe}, Robert A. and {Witten}, Callum and {Atek}, Hakim and {Baggen}, Josephine F.~W. and {Belli}, Sirio and {Bezanson}, Rachel and {Boogaard}, Leindert A. and {Bose}, Sownak and {Covelo-Paz}, Alba and {Dayal}, Pratika and {Fudamoto}, Yoshinobu and {Furtak}, Lukas J. and {Giovinazzo}, Emma and {Goulding}, Andy and {Gronke}, Max and {Heintz}, Kasper E. and {Hirschmann}, Michaela and {Illingworth}, Garth and {Inoue}, Akio K. and {Johnson}, Benjamin D. and {Leja}, Joel and {Leonova}, Ecaterina and {McConachie}, Ian and {Maseda}, Michael V. and {Natarajan}, Priyamvada and {Nelson}, Erica and {Setton}, David J. and {Shivaei}, Irene and {Sobral}, David and {Stefanon}, Mauro and {Tacchella}, Sandro and {Toft}, Sune and {Torralba}, Alberto and {van Dokkum}, Pieter and {van der Wel}, Arjen and {Volonteri}, Marta and {Walter}, Fabian and {Wang}, Bingjie and {Watson}, Darach},
        title = "{A ``Black Hole Star'' Reveals the Remarkable Gas-Enshrouded Hearts of the Little Red Dots}",
      journal = {arXiv e-prints},
         year = 2025,
        month = mar,
          eid = {arXiv:2503.16596},
        pages = {arXiv:2503.16596},
          doi = {10.48550/arXiv.2503.16596},
archivePrefix = {arXiv},
       eprint = {2503.16596},
 primaryClass = {astro-ph.GA},
       adsurl = {https://ui.adsabs.harvard.edu/abs/2025arXiv250316596N}
}

@ARTICLE{Nagao2006,
       author = {{Nagao}, T. and {Marconi}, A. and {Maiolino}, R.},
        title = "{The evolution of the broad-line region among SDSS quasars}",
      journal = {\aap},
         year = 2006,
        month = feb,
       volume = {447},
       number = {1},
        pages = {157-172},
          doi = {10.1051/0004-6361:20054024},
archivePrefix = {arXiv},
       eprint = {astro-ph/0510385},
 primaryClass = {astro-ph},
       adsurl = {https://ui.adsabs.harvard.edu/abs/2006A&A...447..157N}
}

@ARTICLE{Nakajima2014,
       author = {{Nakajima}, Kimihiko and {Ouchi}, Masami},
        title = "{Ionization state of inter-stellar medium in galaxies: evolution, SFR-M$_{*}$-Z dependence, and ionizing photon escape}",
      journal = {\mnras},
         year = 2014,
        month = jul,
       volume = {442},
       number = {1},
        pages = {900-916},
          doi = {10.1093/mnras/stu902},
archivePrefix = {arXiv},
       eprint = {1309.0207},
 primaryClass = {astro-ph.CO},
       adsurl = {https://ui.adsabs.harvard.edu/abs/2014MNRAS.442..900N}
}

@ARTICLE{Nakajima2018,
       author = {{Nakajima}, K. and {Schaerer}, D. and {Le F{\`e}vre}, O. and {Amor{\'\i}n}, R. and {Talia}, M. and {Lemaux}, B.~C. and {Tasca}, L.~A.~M. and {Vanzella}, E. and {Zamorani}, G. and {Bardelli}, S. and {Grazian}, A. and {Guaita}, L. and {Hathi}, N.~P. and {Pentericci}, L. and {Zucca}, E.},
        title = "{The VIMOS Ultra Deep Survey: Nature, ISM properties, and ionizing spectra of CIII]{\ensuremath{\lambda}}1909 emitters at z = 2-4}",
      journal = {\aap},
         year = 2018,
        month = may,
       volume = {612},
          eid = {A94},
        pages = {A94},
          doi = {10.1051/0004-6361/201731935},
archivePrefix = {arXiv},
       eprint = {1709.03990},
 primaryClass = {astro-ph.GA},
       adsurl = {https://ui.adsabs.harvard.edu/abs/2018A&A...612A..94N}
}

@ARTICLE{Nakajima2022b,
       author = {{Nakajima}, K. and {Maiolino}, R.},
        title = "{Diagnostics for PopIII galaxies and direct collapse black holes in the early universe}",
      journal = {\mnras},
         year = 2022,
        month = jul,
       volume = {513},
       number = {4},
        pages = {5134-5147},
          doi = {10.1093/mnras/stac1242},
archivePrefix = {arXiv},
       eprint = {2204.11870},
 primaryClass = {astro-ph.GA},
       adsurl = {https://ui.adsabs.harvard.edu/abs/2022MNRAS.513.5134N}
}

@ARTICLE{Nakajima2025,
       author = {{Nakajima}, Kimihiko and {Ouchi}, Masami and {Harikane}, Yuichi and {Vanzella}, Eros and {Ono}, Yoshiaki and {Isobe}, Yuki and {Nishigaki}, Moka and {Tsujimoto}, Takuji and {Nakamura}, Fumitaka and {Xu}, Yi and {Umeda}, Hiroya and {Zhang}, Yechi},
        title = "{An Ultra-Faint, Chemically Primitive Galaxy Forming in the Reionization Era}",
      journal = {arXiv e-prints},
         year = 2025,
        month = jun,
          eid = {arXiv:2506.11846},
        pages = {arXiv:2506.11846},
          doi = {10.48550/arXiv.2506.11846},
archivePrefix = {arXiv},
       eprint = {2506.11846},
 primaryClass = {astro-ph.GA},
       adsurl = {https://ui.adsabs.harvard.edu/abs/2025arXiv250611846N}
}

@ARTICLE{Nandal2026,
       author = {{Nandal}, Devesh and {Loeb}, Abraham},
        title = "{Supermassive Stars Match the Spectral Signatures of JWST's Little Red Dots}",
      journal = {\apj},
         year = 2026,
        month = feb,
       volume = {998},
       number = {1},
          eid = {124},
        pages = {124},
          doi = {10.3847/1538-4357/ae32f3},
archivePrefix = {arXiv},
       eprint = {2507.12618},
 primaryClass = {astro-ph.GA},
       adsurl = {https://ui.adsabs.harvard.edu/abs/2026ApJ...998..124N}
}

@ARTICLE{Napolitano,
       author = {{Napolitano}, Lorenzo and {Castellano}, Marco and {Pentericci}, Laura and {Vignali}, Cristian and {Gilli}, Roberto and {Fontana}, Adriano and {Santini}, Paola and {Treu}, Tommaso and {Calabr{\`o}}, Antonello and {Llerena}, Mario and {Piconcelli}, Enrico and {Zappacosta}, Luca and {Mascia}, Sara and {Tripodi}, Roberta and {Arrabal Haro}, Pablo and {Bergamini}, Pietro and {Bakx}, Tom J.~L.~C. and {Dickinson}, Mark and {Glazebrook}, Karl and {Henry}, Alaina and {Leethochawalit}, Nicha and {Mazzolari}, Giovanni and {Merlin}, Emiliano and {Morishita}, Takahiro and {Nanayakkara}, Themiya and {Paris}, Diego and {Puccetti}, Simonetta and {Roberts-Borsani}, Guido and {Rojas Ruiz}, Sofia and {Rosati}, Piero and {Vanzella}, Eros and {Vito}, Fabio and {Vulcani}, Benedetta and {Wang}, Xin and {Yoon}, Ilsang and {Zavala}, Jorge A.},
        title = "{The Dual Nature of GHZ9: Coexisting Active Galactic Nuclei and Star Formation Activity in a Remote X-Ray Source at z = 10.145}",
      journal = {\apj},
         year = 2025,
        month = aug,
       volume = {989},
       number = {1},
          eid = {75},
        pages = {75},
          doi = {10.3847/1538-4357/ade706},
archivePrefix = {arXiv},
       eprint = {2410.18763},
 primaryClass = {astro-ph.GA},
       adsurl = {https://ui.adsabs.harvard.edu/abs/2025ApJ...989...75N}
}

@ARTICLE{Netzer1993,
       author = {{Netzer}, Hagai and {Laor}, Ari},
        title = "{Dust in the Narrow-Line Region of Active Galactic Nuclei}",
      journal = {\apjl},
         year = 1993,
        month = feb,
       volume = {404},
        pages = {L51},
          doi = {10.1086/186741},
       adsurl = {https://ui.adsabs.harvard.edu/abs/1993ApJ...404L..51N}
}

@ARTICLE{Netzer2007,
       author = {{Netzer}, Hagai and {Trakhtenbrot}, Benny},
        title = "{Cosmic Evolution of Mass Accretion Rate and Metallicity in Active Galactic Nuclei}",
      journal = {\apj},
         year = 2007,
        month = jan,
       volume = {654},
       number = {2},
        pages = {754-763},
          doi = {10.1086/509650},
archivePrefix = {arXiv},
       eprint = {astro-ph/0607654},
 primaryClass = {astro-ph},
       adsurl = {https://ui.adsabs.harvard.edu/abs/2007ApJ...654..754N}
}

@ARTICLE{Netzer2009,
       author = {{Netzer}, Hagai},
        title = "{Accretion and star formation rates in low-redshift type II active galactic nuclei}",
      journal = {\mnras},
         year = 2009,
        month = nov,
       volume = {399},
       number = {4},
        pages = {1907-1920},
          doi = {10.1111/j.1365-2966.2009.15434.x},
archivePrefix = {arXiv},
       eprint = {0907.3575},
 primaryClass = {astro-ph.GA},
       adsurl = {https://ui.adsabs.harvard.edu/abs/2009MNRAS.399.1907N}
}

@ARTICLE{Netzer2019,
       author = {{Netzer}, Hagai},
        title = "{Bolometric correction factors for active galactic nuclei}",
      journal = {\mnras},
         year = 2019,
        month = oct,
       volume = {488},
       number = {4},
        pages = {5185-5191},
          doi = {10.1093/mnras/stz2016},
archivePrefix = {arXiv},
       eprint = {1907.09534},
 primaryClass = {astro-ph.GA},
       adsurl = {https://ui.adsabs.harvard.edu/abs/2019MNRAS.488.5185N}
}

@ARTICLE{Nicholls2014,
       author = {{Nicholls}, David C. and {Dopita}, Michael A. and {Sutherland}, Ralph S. and {Jerjen}, Helmut and {Kewley}, Lisa J.},
        title = "{Metal-poor Dwarf Galaxies in the SIGRID Galaxy Sample. II. The Electron Temperature-Abundance Calibration and the Parameters that Affect it}",
      journal = {\apj},
         year = 2014,
        month = jul,
       volume = {790},
       number = {1},
          eid = {75},
        pages = {75},
          doi = {10.1088/0004-637X/790/1/75},
archivePrefix = {arXiv},
       eprint = {1405.7170},
 primaryClass = {astro-ph.GA},
       adsurl = {https://ui.adsabs.harvard.edu/abs/2014ApJ...790...75N}
}

@ARTICLE{Nikopoulos2026,
       author = {{Nikopoulos}, G.~P. and {Watson}, D. and {Sneppen}, A. and {Rusakov}, V. and {Heintz}, K.~E. and {Witstok}, J. and {Brammer}, G.},
        title = "{Evidence of violation of case B recombination in little red dots}",
      journal = {\aap},
         year = 2026,
        month = jun,
       volume = {710},
          eid = {A136},
        pages = {A136},
          doi = {10.1051/0004-6361/202557604},
archivePrefix = {arXiv},
       eprint = {2510.06362},
 primaryClass = {astro-ph.GA},
       adsurl = {https://ui.adsabs.harvard.edu/abs/2026A&A...710A.136N}
}

@INPROCEEDINGS{Novikov1973,
       author = {{Novikov}, I.~D. and {Thorne}, K.~S.},
        title = "{Astrophysics of black holes.}",
    booktitle = {Black Holes (Les Astres Occlus)},
         year = 1973,
       editor = {{Dewitt}, C. and {Dewitt}, B.~S.},
        month = jan,
        pages = {343-450},
       adsurl = {https://ui.adsabs.harvard.edu/abs/1973blho.conf..343N}
}

@ARTICLE{Onoue2023,
       author = {{Onoue}, Masafusa and {Inayoshi}, Kohei and {Ding}, Xuheng and {Li}, Wenxiu and {Li}, Zhengrong and {Molina}, Juan and {Inoue}, Akio K. and {Jiang}, Linhua and {Ho}, Luis C.},
        title = "{A Candidate for the Least-massive Black Hole in the First 1.1 Billion Years of the Universe}",
      journal = {\apjl},
         year = 2023,
        month = jan,
       volume = {942},
       number = {1},
          eid = {L17},
        pages = {L17},
          doi = {10.3847/2041-8213/aca9d3},
archivePrefix = {arXiv},
       eprint = {2209.07325},
 primaryClass = {astro-ph.GA},
       adsurl = {https://ui.adsabs.harvard.edu/abs/2023ApJ...942L..17O}
}

@BOOK{Osterbrock2006,
       author = {{Osterbrock}, Donald E. and {Ferland}, Gary J.},
        title = "{Astrophysics of gaseous nebulae and active galactic nuclei}",
         year = 2006,
       adsurl = {https://ui.adsabs.harvard.edu/abs/2006agna.book.....O}
}

@ARTICLE{Pacucci2020,
       author = {{Pacucci}, Fabio and {Loeb}, Abraham},
        title = "{Separating Accretion and Mergers in the Cosmic Growth of Black Holes with X-Ray and Gravitational-wave Observations}",
      journal = {\apj},
         year = 2020,
        month = jun,
       volume = {895},
       number = {2},
          eid = {95},
        pages = {95},
          doi = {10.3847/1538-4357/ab886e},
archivePrefix = {arXiv},
       eprint = {2004.07246},
 primaryClass = {astro-ph.GA},
       adsurl = {https://ui.adsabs.harvard.edu/abs/2020ApJ...895...95P}
}

@ARTICLE{Pacucci2022,
       author = {{Pacucci}, Fabio and {Loeb}, Abraham},
        title = "{The search for the farthest quasar: consequences for black hole growth and seed models}",
      journal = {\mnras},
         year = 2022,
        month = jan,
       volume = {509},
       number = {2},
        pages = {1885-1891},
          doi = {10.1093/mnras/stab3071},
archivePrefix = {arXiv},
       eprint = {2110.10176},
 primaryClass = {astro-ph.GA},
       adsurl = {https://ui.adsabs.harvard.edu/abs/2022MNRAS.509.1885P}
}

@ARTICLE{Pacucci2023,
       author = {{Pacucci}, Fabio and {Nguyen}, Bao and {Carniani}, Stefano and {Maiolino}, Roberto and {Fan}, Xiaohui},
        title = "{JWST CEERS and JADES Active Galaxies at z = 4-7 Violate the Local M $_{{\ensuremath{\bullet}}}$-M $_{{\ensuremath{\star}}}$ Relation at >3{\ensuremath{\sigma}}: Implications for Low-mass Black Holes and Seeding Models}",
      journal = {\apjl},
         year = 2023,
        month = nov,
       volume = {957},
       number = {1},
          eid = {L3},
        pages = {L3},
          doi = {10.3847/2041-8213/ad0158},
archivePrefix = {arXiv},
       eprint = {2308.12331},
 primaryClass = {astro-ph.GA},
       adsurl = {https://ui.adsabs.harvard.edu/abs/2023ApJ...957L...3P}
}

@ARTICLE{Pacucci2024,
       author = {{Pacucci}, Fabio and {Narayan}, Ramesh},
        title = "{Mildly Super-Eddington Accretion onto Slowly Spinning Black Holes Explains the X-Ray Weakness of the Little Red Dots}",
      journal = {\apj},
         year = 2024,
        month = nov,
       volume = {976},
       number = {1},
          eid = {96},
        pages = {96},
          doi = {10.3847/1538-4357/ad84f7},
archivePrefix = {arXiv},
       eprint = {2407.15915},
 primaryClass = {astro-ph.HE},
       adsurl = {https://ui.adsabs.harvard.edu/abs/2024ApJ...976...96P}
}

@ARTICLE{Pacucci2026,
       author = {{Pacucci}, Fabio and {Ferrara}, Andrea and {Kocevski}, Dale D.},
        title = "{The Little Red Dots Are Direct Collapse Black Holes}",
      journal = {arXiv e-prints},
         year = 2026,
        month = jan,
          eid = {arXiv:2601.14368},
        pages = {arXiv:2601.14368},
          doi = {10.48550/arXiv.2601.14368},
archivePrefix = {arXiv},
       eprint = {2601.14368},
 primaryClass = {astro-ph.GA},
       adsurl = {https://ui.adsabs.harvard.edu/abs/2026arXiv260114368P}
}

@ARTICLE{Panda2019,
       author = {{Panda}, Swayamtrupta and {Czerny}, Bo{\.z}ena and {Done}, Chris and {Kubota}, Aya},
        title = "{CLOUDY View of the Warm Corona}",
      journal = {\apj},
         year = 2019,
        month = apr,
       volume = {875},
       number = {2},
          eid = {133},
        pages = {133},
          doi = {10.3847/1538-4357/ab11cb},
archivePrefix = {arXiv},
       eprint = {1901.02962},
 primaryClass = {astro-ph.HE},
       adsurl = {https://ui.adsabs.harvard.edu/abs/2019ApJ...875..133P}
}

@ARTICLE{Pellegrini2012,
       author = {{Pellegrini}, E.~W. and {Oey}, M.~S. and {Winkler}, P.~F. and {Points}, S.~D. and {Smith}, R.~C. and {Jaskot}, A.~E. and {Zastrow}, J.},
        title = "{The Optical Depth of H II Regions in the Magellanic Clouds}",
      journal = {\apj},
         year = 2012,
        month = aug,
       volume = {755},
       number = {1},
          eid = {40},
        pages = {40},
          doi = {10.1088/0004-637X/755/1/40},
archivePrefix = {arXiv},
       eprint = {1202.3334},
 primaryClass = {astro-ph.CO},
       adsurl = {https://ui.adsabs.harvard.edu/abs/2012ApJ...755...40P}
}

@ARTICLE{PerezGonzalez2026,
       author = {{P{\'e}rez-Gonz{\'a}lez}, Pablo G. and {Barro}, Guillermo and {Carniani}, Stefano and {D'Eugenio}, Francesco and {Rieke}, George H. and {Tripodi}, Roberta and {Bunker}, Andrew J. and {Ji}, Xihan and {Marques-Chaves}, Rui and {Schaerer}, Daniel and {Venturi}, Giacomo and {Ar{\'e}valo-Gonz{\'a}lez}, Flor and {Arribas}, Santiago and {Rinaldi}, Pierluigi and {Rodr{\'\i}guez Del Pino}, Bruno and {Witstok}, Joris and {Bhatawdekar}, Rachana and {Boogaard}, Leindert A. and {Charlot}, Stephane and {Chevallard}, Jacopo and {Costantin}, Luca and {Curti}, Mirko and {Curtis-Lake}, Emma and {Daddi}, Emanuele and {Davis}, Kelcey and {Dickinson}, Mark and {Donnan}, Callum T. and {Donnan}, Fergus R. and {Dunlop}, James S. and {Eisenstein}, Daniel J. and {Ferguson}, Henry C. and {Fern{\'a}ndez Aranda}, Rom{\'a}n and {Finkelstein}, Steven L. and {Fujimoto}, Seiji and {Gandolfi}, Giovanni and {Giavalisco}, Mauro and {Grogin}, Norman A. and {Hamed}, Mahmoud and {Hirschmann}, Michaela and {Kartaltepe}, Jeyhan S. and {Kocevski}, Dale D. and {Koekemoer}, Anton M. and {Leung}, Gene C.~K. and {Lofaro}, Cristina M. and {Lucas}, Ray A. and {McLeod}, Derek J. and {Melinder}, Jens and {{\"O}stlin}, Goran and {Papovich}, Casey and {Pentericci}, Laura and {P{\'e}rez-D{\'\i}az}, Borja and {Rieke}, Marcia and {Scholtz}, Jan and {Somerville}, Rachel S. and {Stanton}, Thomas M. and {Stevenson}, Struan D. and {Shivaei}, Irene and {Tacchella}, Sandro and {Trump}, Jonathan R. and {{\"U}bler}, Hannah and {Wang}, Xin and {Williams}, Christina C. and {Willmer}, Christopher N.~A. and {Yung}, L.~Y. Aaron and {Zhu}, Yongda},
        title = "{Little Red Dots: One Photometric Tag Concealing Diverse Spectroscopic Flavors of Massive Star Formation and Black Hole Activity}",
      journal = {arXiv e-prints},
         year = 2026,
        month = feb,
          eid = {arXiv:2602.20247},
        pages = {arXiv:2602.20247},
          doi = {10.48550/arXiv.2602.20247},
archivePrefix = {arXiv},
       eprint = {2602.20247},
 primaryClass = {astro-ph.GA},
       adsurl = {https://ui.adsabs.harvard.edu/abs/2026arXiv260220247P}
}

@INCOLLECTION{Peterson2006,
       author = {{Peterson}, B.~M.},
        title = "{The Broad-Line Region in Active Galactic Nuclei}",
    booktitle = {Physics of Active Galactic Nuclei at all Scales},
         year = 2006,
       editor = {{Alloin}, Danielle},
       volume = {693},
        pages = {77},
          doi = {10.1007/3-540-34621-X_3},
       adsurl = {https://ui.adsabs.harvard.edu/abs/2006LNP...693...77P}
}

@ARTICLE{Pezzulli2017,
       author = {{Pezzulli}, Edwige and {Valiante}, Rosa and {Orofino}, Maria C. and {Schneider}, Raffaella and {Gallerani}, Simona and {Sbarrato}, Tullia},
        title = "{Faint progenitors of luminous z {\ensuremath{\sim}} 6 quasars: Why do not we see them?}",
      journal = {\mnras},
         year = 2017,
        month = apr,
       volume = {466},
       number = {2},
        pages = {2131-2142},
          doi = {10.1093/mnras/stw3243},
archivePrefix = {arXiv},
       eprint = {1612.04188},
 primaryClass = {astro-ph.GA},
       adsurl = {https://ui.adsabs.harvard.edu/abs/2017MNRAS.466.2131P}
}

@ARTICLE{Phillips2007,
       author = {{Phillips}, J.~P.},
        title = "{Density gradients in Galactic ultra-compact HII regions}",
      journal = {\mnras},
         year = 2007,
        month = sep,
       volume = {380},
       number = {1},
        pages = {369-373},
          doi = {10.1111/j.1365-2966.2007.12078.x},
       adsurl = {https://ui.adsabs.harvard.edu/abs/2007MNRAS.380..369P}
}

@ARTICLE{Pillepich2018,
       author = {{Pillepich}, Annalisa and {Springel}, Volker and {Nelson}, Dylan and {Genel}, Shy and {Naiman}, Jill and {Pakmor}, R{\"u}diger and {Hernquist}, Lars and {Torrey}, Paul and {Vogelsberger}, Mark and {Weinberger}, Rainer and {Marinacci}, Federico},
        title = "{Simulating galaxy formation with the IllustrisTNG model}",
      journal = {\mnras},
         year = 2018,
        month = jan,
       volume = {473},
       number = {3},
        pages = {4077-4106},
          doi = {10.1093/mnras/stx2656},
archivePrefix = {arXiv},
       eprint = {1703.02970},
 primaryClass = {astro-ph.GA},
       adsurl = {https://ui.adsabs.harvard.edu/abs/2018MNRAS.473.4077P}
}

@ARTICLE{Pirzkal2024,
       author = {{Pirzkal}, Nor and {Rothberg}, Barry and {Papovich}, Casey and {Shen}, Lu and {Leung}, Gene C.~K. and {Bagley}, Micaela B. and {Finkelstein}, Steven L. and {Vanderhoof}, Brittany N. and {Lotz}, Jennifer M. and {Koekemoer}, Anton M. and {Hathi}, Nimish P. and {Cheng}, Yingjie and {Cleri}, Nikko J. and {Grogin}, Norman A. and {Yung}, L.~Y. Aaron and {Dickinson}, Mark and {Ferguson}, Henry C. and {Gardner}, Jonathan P. and {Jung}, Intae and {Kartaltepe}, Jeyhan S. and {Ryan}, Russell and {Simons}, Raymond C. and {Ravindranath}, Swara and {Berg}, Danielle A. and {Backhaus}, Bren E. and {Casey}, Caitlin M. and {Castellano}, Marco and {Ch{\'a}vez Ortiz}, {\'O}scar A. and {Chworowsky}, Katherine and {Cox}, Isabella G. and {Dav{\'e}}, Romeel and {Davis}, Kelcey and {Estrada-Carpenter}, Vicente and {Fontana}, Adriano and {Fujimoto}, Seiji and {Giavalisco}, Mauro and {Grazian}, Andrea and {Hutchison}, Taylor A. and {Jaskot}, Anne E. and {Kewley}, Lisa J. and {Kirkpatrick}, Allison and {Kocevski}, Dale D. and {Larson}, Rebecca L. and {Matharu}, Jasleen and {Natarajan}, Priyamvada and {Pentericci}, Laura and {P{\'e}rez-Gonz{\'a}lez}, Pablo G. and {Snyder}, Gregory F. and {Somerville}, Rachel S. and {Trump}, Jonathan R. and {Wilkins}, Stephen M.},
        title = "{The Next Generation Deep Extragalactic Exploratory Public Near-infrared Slitless Survey Epoch 1 (NGDEEP-NISS1): Extragalactic Star-formation and Active Galactic Nuclei at 0.5 < z < 3.6}",
      journal = {\apj},
         year = 2024,
        month = jul,
       volume = {969},
       number = {2},
          eid = {90},
        pages = {90},
          doi = {10.3847/1538-4357/ad429c},
archivePrefix = {arXiv},
       eprint = {2312.09972},
 primaryClass = {astro-ph.GA},
       adsurl = {https://ui.adsabs.harvard.edu/abs/2024ApJ...969...90P}
}

@ARTICLE{Plat2019,
       author = {{Plat}, A. and {Charlot}, S. and {Bruzual}, G. and {Feltre}, A. and {Vidal-Garc{\'\i}a}, A. and {Morisset}, C. and {Chevallard}, J. and {Todt}, H.},
        title = "{Constraints on the production and escape of ionizing radiation from the emission-line spectra of metal-poor star-forming galaxies}",
      journal = {\mnras},
         year = 2019,
        month = nov,
       volume = {490},
       number = {1},
        pages = {978-1009},
          doi = {10.1093/mnras/stz2616},
archivePrefix = {arXiv},
       eprint = {1909.07386},
 primaryClass = {astro-ph.GA},
       adsurl = {https://ui.adsabs.harvard.edu/abs/2019MNRAS.490..978P}
}

@INPROCEEDINGS{Pontoppidan2016,
       author = {{Pontoppidan}, Klaus M. and {Pickering}, Timothy E. and {Laidler}, Victoria G. and {Gilbert}, Karoline and {Sontag}, Christopher D. and {Slocum}, Christine and {Sienkiewicz}, Mark J. and {Hanley}, Christopher and {Earl}, Nicholas M. and {Pueyo}, Laurent and {Ravindranath}, Swara and {Karakla}, Diane M. and {Robberto}, Massimo and {Noriega-Crespo}, Alberto and {Barker}, Elizabeth A.},
        title = "{Pandeia: a multi-mission exposure time calculator for JWST and WFIRST}",
    booktitle = {Observatory Operations: Strategies, Processes, and Systems VI},
         year = 2016,
       editor = {{Peck}, Alison B. and {Seaman}, Robert L. and {Benn}, Chris R.},
       series = {Society of Photo-Optical Instrumentation Engineers (SPIE) Conference Series},
       volume = {9910},
        month = jul,
          eid = {991016},
        pages = {991016},
          doi = {10.1117/12.2231768},
archivePrefix = {arXiv},
       eprint = {1707.02202},
 primaryClass = {astro-ph.IM},
       adsurl = {https://ui.adsabs.harvard.edu/abs/2016SPIE.9910E..16P}
}

@ARTICLE{Popovic2003,
       author = {{Popovi{\'c}}, L. {\v{C}}.},
        title = "{Balmer Lines as Diagnostics of Physical Conditions in Active Galactic Nuclei Broad Emission Line Regions}",
      journal = {\apj},
         year = 2003,
        month = dec,
       volume = {599},
       number = {1},
        pages = {140-146},
          doi = {10.1086/376401},
archivePrefix = {arXiv},
       eprint = {astro-ph/0304390},
 primaryClass = {astro-ph},
       adsurl = {https://ui.adsabs.harvard.edu/abs/2003ApJ...599..140P}
}

@ARTICLE{Portegies2002,
       author = {{Portegies Zwart}, Simon F. and {McMillan}, Stephen L.~W.},
        title = "{The Runaway Growth of Intermediate-Mass Black Holes in Dense Star Clusters}",
      journal = {\apj},
         year = 2002,
        month = sep,
       volume = {576},
       number = {2},
        pages = {899-907},
          doi = {10.1086/341798},
archivePrefix = {arXiv},
       eprint = {astro-ph/0201055},
 primaryClass = {astro-ph},
       adsurl = {https://ui.adsabs.harvard.edu/abs/2002ApJ...576..899P}
}

@ARTICLE{Portegies2004,
       author = {{Portegies Zwart}, Simon F. and {Baumgardt}, Holger and {Hut}, Piet and {Makino}, Junichiro and {McMillan}, Stephen L.~W.},
        title = "{Formation of massive black holes through runaway collisions in dense young star clusters}",
      journal = {\nat},
         year = 2004,
        month = apr,
       volume = {428},
       number = {6984},
        pages = {724-726},
          doi = {10.1038/nature02448},
archivePrefix = {arXiv},
       eprint = {astro-ph/0402622},
 primaryClass = {astro-ph},
       adsurl = {https://ui.adsabs.harvard.edu/abs/2004Natur.428..724P}
}

@ARTICLE{Radovich1994,
       author = {{Radovich}, M. and {Rafanelli}, P.},
        title = "{Multi-density models of the Broad Line Region in Active Galactic Nuclei: A comparative study with single-cloud models}",
      journal = {Astronomische Nachrichten},
         year = 1994,
        month = jun,
       volume = {315},
       number = {4},
        pages = {265-277},
          doi = {10.1002/asna.2103150402},
       adsurl = {https://ui.adsabs.harvard.edu/abs/1994AN....315..265R}
}

@ARTICLE{Raiter2010,
       author = {{Raiter}, A. and {Schaerer}, D. and {Fosbury}, R.~A.~E.},
        title = "{Predicted UV properties of very metal-poor starburst galaxies}",
      journal = {\aap},
         year = 2010,
        month = nov,
       volume = {523},
          eid = {A64},
        pages = {A64},
          doi = {10.1051/0004-6361/201015236},
archivePrefix = {arXiv},
       eprint = {1008.2114},
 primaryClass = {astro-ph.CO},
       adsurl = {https://ui.adsabs.harvard.edu/abs/2010A&A...523A..64R}
}

@ARTICLE{Rees1984,
       author = {{Rees}, Martin J.},
        title = "{Black Hole Models for Active Galactic Nuclei}",
      journal = {\araa},
         year = 1984,
        month = jan,
       volume = {22},
        pages = {471-506},
          doi = {10.1146/annurev.aa.22.090184.002351},
       adsurl = {https://ui.adsabs.harvard.edu/abs/1984ARA&A..22..471R}
}

@ARTICLE{Ren2025,
       author = {{Ren}, Wenke and {Silverman}, John D. and {Faisst}, Andreas L. and {Fujimoto}, Seiji and {Yan}, Lin and {Liu}, Zhaoxuan and {Tsujita}, Akiyoshi and {Aravena}, Manuel and {Davies}, Rebecca L. and {De Looze}, Ilse and {Dessauges-Zavadsky}, Miroslava and {Herrera-Camus}, Rodrigo and {Ibar}, Edo and {Jones}, Gareth C. and {Kartaltepe}, Jeyhan S. and {Koekemoer}, Anton M. and {Lin}, Yu-Heng and {Mitsuhashi}, Ikki and {Molina}, Juan and {Nanni}, Ambra and {Relano}, Monica and {Romano}, Michael and {Sanders}, David B. and {Solimano}, Manuel and {Veraldi}, Enrico and {Villanueva}, Vicente and {Wang}, Wuji and {Zamorani}, Giovanni},
        title = "{The ALPINE─CRISTAL─JWST survey: revealing less massive black holes in high-redshift galaxies}",
      journal = {\mnras},
         year = 2025,
        month = nov,
       volume = {544},
       number = {1},
        pages = {211-233},
          doi = {10.1093/mnras/staf1709},
archivePrefix = {arXiv},
       eprint = {2509.02027},
 primaryClass = {astro-ph.GA},
       adsurl = {https://ui.adsabs.harvard.edu/abs/2025MNRAS.544..211R}
}

@ARTICLE{Ricarte2023,
       author = {{Ricarte}, Angelo and {Narayan}, Ramesh and {Curd}, Brandon},
        title = "{Recipes for Jet Feedback and Spin Evolution of Black Holes with Strongly Magnetized Super-Eddington Accretion Disks}",
      journal = {\apjl},
         year = 2023,
        month = sep,
       volume = {954},
       number = {1},
          eid = {L22},
        pages = {L22},
          doi = {10.3847/2041-8213/aceda5},
archivePrefix = {arXiv},
       eprint = {2307.04621},
 primaryClass = {astro-ph.HE},
       adsurl = {https://ui.adsabs.harvard.edu/abs/2023ApJ...954L..22R}
}

@ARTICLE{Ricarte2025,
       author = {{Ricarte}, Angelo and {Natarajan}, Priyamvada and {Narayan}, Ramesh and {Palumbo}, Daniel C.~M.},
        title = "{Multimessenger Probes of Supermassive Black Hole Spin Evolution}",
      journal = {\apj},
         year = 2025,
        month = feb,
       volume = {980},
       number = {1},
          eid = {136},
        pages = {136},
          doi = {10.3847/1538-4357/ad9ea9},
archivePrefix = {arXiv},
       eprint = {2410.07477},
 primaryClass = {astro-ph.HE},
       adsurl = {https://ui.adsabs.harvard.edu/abs/2025ApJ...980..136R}
}

@ARTICLE{Richards2006,
       author = {{Richards}, Gordon T. and {Lacy}, Mark and {Storrie-Lombardi}, Lisa J. and {Hall}, Patrick B. and {Gallagher}, S.~C. and {Hines}, Dean C. and {Fan}, Xiaohui and {Papovich}, Casey and {Vanden Berk}, Daniel E. and {Trammell}, George B. and {Schneider}, Donald P. and {Vestergaard}, Marianne and {York}, Donald G. and {Jester}, Sebastian and {Anderson}, Scott F. and {Budav{\'a}ri}, Tam{\'a}s and {Szalay}, Alexander S.},
        title = "{Spectral Energy Distributions and Multiwavelength Selection of Type 1 Quasars}",
      journal = {\apjs},
         year = 2006,
        month = oct,
       volume = {166},
       number = {2},
        pages = {470-497},
          doi = {10.1086/506525},
archivePrefix = {arXiv},
       eprint = {astro-ph/0601558},
 primaryClass = {astro-ph},
       adsurl = {https://ui.adsabs.harvard.edu/abs/2006ApJS..166..470R}
}

@ARTICLE{Rubin2011,
       author = {{Rubin}, Robert H. and {Simpson}, Janet P. and {O'Dell}, C.~R. and {McNabb}, Ian A. and {Colgan}, Sean W.~J. and {Zhuge}, Scott Y. and {Ferland}, Gary J. and {Hidalgo}, Sergio A.},
        title = "{Spitzer reveals what is behind Orion's Bar}",
      journal = {\mnras},
         year = 2011,
        month = jan,
       volume = {410},
       number = {2},
        pages = {1320-1348},
          doi = {10.1111/j.1365-2966.2010.17522.x},
archivePrefix = {arXiv},
       eprint = {1008.2736},
 primaryClass = {astro-ph.GA},
       adsurl = {https://ui.adsabs.harvard.edu/abs/2011MNRAS.410.1320R}
}

@ARTICLE{Ruff2012,
       author = {{Ruff}, Andrea J. and {Floyd}, David J.~E. and {Webster}, Rachel L. and {Korista}, Kirk T. and {Landt}, Hermine},
        title = "{New Constraints on the Quasar Broad Emission Line Region}",
      journal = {\apj},
         year = 2012,
        month = jul,
       volume = {754},
       number = {1},
          eid = {18},
        pages = {18},
          doi = {10.1088/0004-637X/754/1/18},
archivePrefix = {arXiv},
       eprint = {1206.0721},
 primaryClass = {astro-ph.GA},
       adsurl = {https://ui.adsabs.harvard.edu/abs/2012ApJ...754...18R}
}

@ARTICLE{Rusakov2026,
       author = {{Rusakov}, V. and {Watson}, D. and {Nikopoulos}, G.~P. and {Brammer}, G. and {Gottumukkala}, R. and {Harvey}, T. and {Heintz}, K.~E. and {Damgaard}, R. and {Sim}, S.~A. and {Sneppen}, A. and {Vijayan}, A.~P. and {Adams}, N. and {Austin}, D. and {Conselice}, C.~J. and {Goolsby}, C.~M. and {Toft}, S. and {Witstok}, J.},
        title = "{Little red dots as young supermassive black holes in dense ionized cocoons}",
      journal = {\nat},
         year = 2026,
        month = jan,
       volume = {649},
       number = {8097},
        pages = {574-579},
          doi = {10.1038/s41586-025-09900-4},
archivePrefix = {arXiv},
       eprint = {2503.16595},
 primaryClass = {astro-ph.GA},
       adsurl = {https://ui.adsabs.harvard.edu/abs/2026Natur.649..574R}
}

@ARTICLE{Rusta2025,
       author = {{Rusta}, Elka and {Salvadori}, Stefania and {Gelli}, Viola and {Schaerer}, Daniel and {Marconi}, Alessandro and {Koutsouridou}, Ioanna and {Carniani}, Stefano},
        title = "{Metal-polluted Population III Galaxies and How to Find Them}",
      journal = {\apjl},
         year = 2025,
        month = aug,
       volume = {989},
       number = {2},
          eid = {L32},
        pages = {L32},
          doi = {10.3847/2041-8213/adf4e3},
archivePrefix = {arXiv},
       eprint = {2506.17400},
 primaryClass = {astro-ph.GA},
       adsurl = {https://ui.adsabs.harvard.edu/abs/2025ApJ...989L..32R}
}

@ARTICLE{Rusta2026,
       author = {{Rusta}, Elka and {Salvadori}, Stefania and {Maiolino}, Roberto and {Gelli}, Viola and {Koutsouridou}, Ioanna and {Carniani}, Stefano and {{\"U}bler}, Hannah and {Marconi}, Alessandro and {Schaerer}, Daniel},
        title = "{The Pristine He II Emitter near GN-z11: Constraining the Mass Distribution of the First Stars}",
      journal = {\apjl},
         year = 2026,
        month = may,
       volume = {1003},
       number = {1},
          eid = {L14},
        pages = {L14},
          doi = {10.3847/2041-8213/ae64e1},
archivePrefix = {arXiv},
       eprint = {2603.20363},
 primaryClass = {astro-ph.GA},
       adsurl = {https://ui.adsabs.harvard.edu/abs/2026ApJ..1003L..14R}
}

@ARTICLE{Sanders1989,
       author = {{Sanders}, D.~B. and {Phinney}, E.~S. and {Neugebauer}, G. and {Soifer}, B.~T. and {Matthews}, K.},
        title = "{Continuum Energy Distributions of Quasars: Shapes and Origins}",
      journal = {\apj},
         year = 1989,
        month = dec,
       volume = {347},
        pages = {29},
          doi = {10.1086/168094},
       adsurl = {https://ui.adsabs.harvard.edu/abs/1989ApJ...347...29S}
}

@ARTICLE{Sarkar2021,
       author = {{Sarkar}, A. and {Ferland}, G.~J. and {Chatzikos}, M. and {Guzm{\'a}n}, F. and {van Hoof}, P.~A.~M. and {Smyth}, R.~T. and {Ramsbottom}, C.~A. and {Keenan}, F.~P. and {Ballance}, C.~P.},
        title = "{Improved Fe II Emission-line Models for AGNs Using New Atomic Data Sets}",
      journal = {\apj},
         year = 2021,
        month = jan,
       volume = {907},
       number = {1},
          eid = {12},
        pages = {12},
          doi = {10.3847/1538-4357/abcaa6},
archivePrefix = {arXiv},
       eprint = {2011.09007},
 primaryClass = {astro-ph.GA},
       adsurl = {https://ui.adsabs.harvard.edu/abs/2021ApJ...907...12S}
}

@ARTICLE{Scarlata2024,
       author = {{Scarlata}, C. and {Hayes}, M. and {Panagia}, N. and {Mehta}, V. and {Haardt}, F. and {Bagley}, M.},
        title = "{On the universal validity of Case B recombination theory}",
      journal = {arXiv e-prints},
         year = 2024,
        month = apr,
          eid = {arXiv:2404.09015},
        pages = {arXiv:2404.09015},
          doi = {10.48550/arXiv.2404.09015},
archivePrefix = {arXiv},
       eprint = {2404.09015},
 primaryClass = {astro-ph.GA},
       adsurl = {https://ui.adsabs.harvard.edu/abs/2024arXiv240409015S}
}

@ARTICLE{Scharre2026,
       author = {{Scharr{\'e}}, Lucie and {Hirschmann}, Michaela and {Plat}, Ad{\`e}le and {Charlot}, Stephane and {Somerville}, Rachel S. and {Curtis-Lake}, Emma and {De Lucia}, Gabriella and {Dessauges-Zavadsky}, Miroslava and {Feltre}, Anna and {Farcy}, Marion and {Lah{\'e}n}, Natalia and {Vijayan}, Aswin P. and {Wilkins}, Stephen M.},
        title = "{Origins of Extreme Emission-Line Ratios in z > 3 Galaxies: Insights from the Lumen Model}",
      journal = {arXiv e-prints},
         year = 2026,
        month = may,
          eid = {arXiv:2605.06769},
        pages = {arXiv:2605.06769},
          doi = {10.48550/arXiv.2605.06769},
archivePrefix = {arXiv},
       eprint = {2605.06769},
 primaryClass = {astro-ph.GA},
       adsurl = {https://ui.adsabs.harvard.edu/abs/2026arXiv260506769S}
}

@ARTICLE{Scholtz2025,
       author = {{Scholtz}, Jan and {Maiolino}, Roberto and {D'Eugenio}, Francesco and {Curtis-Lake}, Emma and {Carniani}, Stefano and {Charlot}, Stephane and {Curti}, Mirko and {Silcock}, Maddie S. and {Arribas}, Santiago and {Baker}, William and {Bhatawdekar}, Rachana and {Boyett}, Kristan and {Bunker}, Andrew J. and {Chevallard}, Jacopo and {Circosta}, Chiara and {Eisenstein}, Daniel J. and {Hainline}, Kevin and {Hausen}, Ryan and {Ji}, Xihan and {Ji}, Zhiyuan and {Johnson}, Benjamin D. and {Kumari}, Nimisha and {Looser}, Tobias J. and {Lyu}, Jianwei and {Maseda}, Michael V. and {Parlanti}, Eleonora and {Perna}, Michele and {Rieke}, Marcia and {Robertson}, Brant and {Del Pino}, Bruno Rodr{\'\i}guez and {Sun}, Fengwu and {Tacchella}, Sandro and {{\"U}bler}, Hannah and {Venturi}, Giacomo and {Williams}, Christina C. and {Willmer}, Christopher N.~A. and {Willott}, Chris and {Witstok}, Joris},
        title = "{JADES: A large population of obscured, narrow-line active galactic nuclei at high redshift}",
      journal = {\aap},
         year = 2025,
        month = may,
       volume = {697},
          eid = {A175},
        pages = {A175},
          doi = {10.1051/0004-6361/202348804},
archivePrefix = {arXiv},
       eprint = {2311.18731},
 primaryClass = {astro-ph.GA},
       adsurl = {https://ui.adsabs.harvard.edu/abs/2025A&A...697A.175S}
}

@ARTICLE{Senchyna2023,
       author = {{Senchyna}, Peter and {Plat}, Adele and {Stark}, Daniel P. and {Rudie}, Gwen C.},
        title = "{GN-z11 in context: possible signatures of globular cluster precursors at redshift 10}",
      journal = {arXiv e-prints},
         year = 2023,
        month = mar,
          eid = {arXiv:2303.04179},
        pages = {arXiv:2303.04179},
          doi = {10.48550/arXiv.2303.04179},
archivePrefix = {arXiv},
       eprint = {2303.04179},
 primaryClass = {astro-ph.GA},
       adsurl = {https://ui.adsabs.harvard.edu/abs/2023arXiv230304179S}
}

@ARTICLE{Shakura1973,
       author = {{Shakura}, N.~I. and {Sunyaev}, R.~A.},
        title = "{Black holes in binary systems. Observational appearance.}",
      journal = {\aap},
         year = 1973,
        month = jan,
       volume = {24},
        pages = {337-355},
       adsurl = {https://ui.adsabs.harvard.edu/abs/1973A&A....24..337S}
}

@ARTICLE{Shapley2019,
       author = {{Shapley}, Alice E. and {Sanders}, Ryan L. and {Shao}, Peng and {Reddy}, Naveen A. and {Kriek}, Mariska and {Coil}, Alison L. and {Mobasher}, Bahram and {Siana}, Brian and {Shivaei}, Irene and {Freeman}, William R. and {Azadi}, Mojegan and {Price}, Sedona H. and {Leung}, Gene C.~K. and {Fetherolf}, Tara and {de Groot}, Laura and {Zick}, Tom and {Fornasini}, Francesca M. and {Barro}, Guillermo},
        title = "{The MOSDEF Survey: Sulfur Emission-line Ratios Provide New Insights into Evolving Interstellar Medium Conditions at High Redshift}",
      journal = {\apjl},
         year = 2019,
        month = aug,
       volume = {881},
       number = {2},
          eid = {L35},
        pages = {L35},
          doi = {10.3847/2041-8213/ab385a},
archivePrefix = {arXiv},
       eprint = {1907.07189},
 primaryClass = {astro-ph.GA},
       adsurl = {https://ui.adsabs.harvard.edu/abs/2019ApJ...881L..35S}
}

@ARTICLE{Schaye2026,
       author = {{Schaye}, Joop and {Chaikin}, Evgenii and {Schaller}, Matthieu and {Ploeckinger}, Sylvia and {Hu{\v{s}}ko}, Filip and {McGibbon}, Robert J. and {Trayford}, James W. and {Ben{\'\i}tez-Llambay}, Alejandro and {Correa}, Camila and {Frenk}, Carlos S. and {Richings}, Alexander J. and {Forouhar Moreno}, Victor J. and {Bah{\'e}}, Yannick M. and {Borrow}, Josh and {Durrant}, Anna and {Gebek}, Andrea and {Helly}, John C. and {Jenkins}, Adrian and {Lacey}, Cedric G. and {Ludlow}, Aaron and {Nobels}, Folkert S.~J.},
        title = "{The COLIBRE project: cosmological hydrodynamical simulations of galaxy formation and evolution}",
      journal = {\mnras},
         year = 2026,
        month = may,
       volume = {548},
       number = {1},
          eid = {stag375},
        pages = {stag375},
          doi = {10.1093/mnras/stag375},
archivePrefix = {arXiv},
       eprint = {2508.21126},
 primaryClass = {astro-ph.GA},
       adsurl = {https://ui.adsabs.harvard.edu/abs/2026MNRAS.548ag375S}
}

@ARTICLE{Shemmer2002,
       author = {{Shemmer}, Ohad and {Netzer}, Hagai},
        title = "{Is There a Metallicity-Luminosity Relationship in Active Galactic Nuclei? The Case of Narrow-Line Seyfert 1 Galaxies}",
      journal = {\apjl},
         year = 2002,
        month = mar,
       volume = {567},
       number = {1},
        pages = {L19-L22},
          doi = {10.1086/339797},
archivePrefix = {arXiv},
       eprint = {astro-ph/0201437},
 primaryClass = {astro-ph},
       adsurl = {https://ui.adsabs.harvard.edu/abs/2002ApJ...567L..19S}
}

@ARTICLE{Shen2024,
       author = {{Shen}, Yue and {Grier}, Catherine J. and {Horne}, Keith and {Stone}, Zachary and {Li}, Jennifer I. and {Yang}, Qian and {Homayouni}, Yasaman and {Trump}, Jonathan R. and {Anderson}, Scott F. and {Brandt}, W.~N. and {Hall}, Patrick B. and {Ho}, Luis C. and {Jiang}, Linhua and {Petitjean}, Patrick and {Schneider}, Donald P. and {Tao}, Charling and {Donnan}, Fergus. R. and {AlSayyad}, Yusra and {Bershady}, Matthew A. and {Blanton}, Michael R. and {Bizyaev}, Dmitry and {Bundy}, Kevin and {Chen}, Yuguang and {Davis}, Megan C. and {Dawson}, Kyle and {Fan}, Xiaohui and {Greene}, Jenny E. and {Gr{\"o}ller}, Hannes and {Guo}, Yucheng and {Ibarra-Medel}, H{\'e}ctor and {Jiang}, Yuanzhe and {Keenan}, Ryan P. and {Kollmeier}, Juna A. and {Lejoly}, Cassandra and {Li}, Zefeng and {de la Macorra}, Axel and {Moe}, Maxwell and {Nie}, Jundan and {Rossi}, Graziano and {Smith}, Paul S. and {Tee}, Wei Leong and {Weijmans}, Anne-Marie and {Xu}, Jiachuan and {Yue}, Minghao and {Zhou}, Xu and {Zhou}, Zhimin and {Zou}, Hu},
        title = "{The Sloan Digital Sky Survey Reverberation Mapping Project: Key Results}",
      journal = {\apjs},
         year = 2024,
        month = jun,
       volume = {272},
       number = {2},
          eid = {26},
        pages = {26},
          doi = {10.3847/1538-4365/ad3936},
archivePrefix = {arXiv},
       eprint = {2305.01014},
 primaryClass = {astro-ph.GA},
       adsurl = {https://ui.adsabs.harvard.edu/abs/2024ApJS..272...26S}
}

@ARTICLE{Shirazi2012,
       author = {{Shirazi}, Maryam and {Brinchmann}, Jarle},
        title = "{Strongly star forming galaxies in the local Universe with nebular He II{\ensuremath{\lambda}}4686 emission}",
      journal = {\mnras},
         year = 2012,
        month = apr,
       volume = {421},
       number = {2},
        pages = {1043-1063},
          doi = {10.1111/j.1365-2966.2012.20439.x},
archivePrefix = {arXiv},
       eprint = {1201.1290},
 primaryClass = {astro-ph.CO},
       adsurl = {https://ui.adsabs.harvard.edu/abs/2012MNRAS.421.1043S}
}

@ARTICLE{Schneider2010,
       author = {{Schneider}, Donald P. and {Richards}, Gordon T. and {Hall}, Patrick B. and {Strauss}, Michael A. and {Anderson}, Scott F. and {Boroson}, Todd A. and {Ross}, Nicholas P. and {Shen}, Yue and {Brandt}, W.~N. and {Fan}, Xiaohui and {Inada}, Naohisa and {Jester}, Sebastian and {Knapp}, G.~R. and {Krawczyk}, Coleman M. and {Thakar}, Anirudda R. and {Vanden Berk}, Daniel E. and {Voges}, Wolfgang and {Yanny}, Brian and {York}, Donald G. and {Bahcall}, Neta A. and {Bizyaev}, Dmitry and {Blanton}, Michael R. and {Brewington}, Howard and {Brinkmann}, J. and {Eisenstein}, Daniel and {Frieman}, Joshua A. and {Fukugita}, Masataka and {Gray}, Jim and {Gunn}, James E. and {Hibon}, Pascale and {Ivezi{\'c}}, {\v{Z}}eljko and {Kent}, Stephen M. and {Kron}, Richard G. and {Lee}, Myung Gyoon and {Lupton}, Robert H. and {Malanushenko}, Elena and {Malanushenko}, Viktor and {Oravetz}, Dan and {Pan}, K. and {Pier}, Jeffrey R. and {Price}, III, Ted N. and {Saxe}, David H. and {Schlegel}, David J. and {Simmons}, Audry and {Snedden}, Stephanie A. and {SubbaRao}, Mark U. and {Szalay}, Alexander S. and {Weinberg}, David H.},
        title = "{The Sloan Digital Sky Survey Quasar Catalog. V. Seventh Data Release}",
      journal = {\aj},
         year = 2010,
        month = jun,
       volume = {139},
       number = {6},
          eid = {2360},
        pages = {2360},
          doi = {10.1088/0004-6256/139/6/2360},
archivePrefix = {arXiv},
       eprint = {1004.1167},
 primaryClass = {astro-ph.CO},
       adsurl = {https://ui.adsabs.harvard.edu/abs/2010AJ....139.2360S}
}

@ARTICLE{Silcock2025,
       author = {{Silcock}, M.~S. and {Curtis-Lake}, E. and {Smith}, D.~J.~B. and {Wallace}, I.~E.~B. and {Vidal-Garc{\'\i}a}, A. and {Plat}, A. and {Hirschmann}, M. and {Feltre}, A. and {Chevallard}, J. and {Charlot}, S. and {Carniani}, S. and {Bunker}, A.~J.},
        title = "{Characterizing the z ≍ 7.66 Type-II AGN candidate SMACS S06355 using BEAGLE-AGN and JWST NIRSpec/NIRCam}",
      journal = {\mnras},
         year = 2025,
        month = aug,
       volume = {541},
       number = {4},
        pages = {3822-3836},
          doi = {10.1093/mnras/staf1087},
archivePrefix = {arXiv},
       eprint = {2410.18193},
 primaryClass = {astro-ph.GA},
       adsurl = {https://ui.adsabs.harvard.edu/abs/2025MNRAS.541.3822S}
}

@ARTICLE{Sneppen2026,
       author = {{Sneppen}, A. and {Watson}, D. and {Matthews}, J.~H. and {Nikopoulos}, G. and {Allen}, N. and {Brammer}, G. and {Damgaard}, R. and {Heintz}, K.~E. and {Knigge}, C. and {Long}, K.~S. and {Rusakov}, V. and {Sim}, S.~A. and {Witstok}, J.},
        title = "{Inside the cocoon: a comprehensive explanation of the spectra of Little Red Dots}",
      journal = {arXiv e-prints},
         year = 2026,
        month = jan,
          eid = {arXiv:2601.18864},
        pages = {arXiv:2601.18864},
          doi = {10.48550/arXiv.2601.18864},
archivePrefix = {arXiv},
       eprint = {2601.18864},
 primaryClass = {astro-ph.GA},
       adsurl = {https://ui.adsabs.harvard.edu/abs/2026arXiv260118864S}
}

@ARTICLE{Sneppen2026b,
       author = {{Sneppen}, Albert and {Matthews}, James H. and {Watson}, Darach and {Cameron}, Alex J. and {Sim}, Stuart A. and {Witstok}, Joris and {Brammer}, Gabriel B. and {Heintz}, Kasper E. and {Nikopoulos}, Georgios},
        title = "{Paschen Jumps in Little Red Dots: Evidence for Nebular Continua}",
      journal = {arXiv e-prints},
         year = 2026,
        month = apr,
          eid = {arXiv:2604.09399},
        pages = {arXiv:2604.09399},
          doi = {10.48550/arXiv.2604.09399},
archivePrefix = {arXiv},
       eprint = {2604.09399},
 primaryClass = {astro-ph.GA},
       adsurl = {https://ui.adsabs.harvard.edu/abs/2026arXiv260409399S}
}

@ARTICLE{Stanway2019,
       author = {{Stanway}, E.~R. and {Eldridge}, J.~J.},
        title = "{Initial mass function variations cannot explain the ionizing spectrum of low metallicity starbursts}",
      journal = {\aap},
         year = 2019,
        month = jan,
       volume = {621},
          eid = {A105},
        pages = {A105},
          doi = {10.1051/0004-6361/201834359},
archivePrefix = {arXiv},
       eprint = {1811.03856},
 primaryClass = {astro-ph.GA},
       adsurl = {https://ui.adsabs.harvard.edu/abs/2019A&A...621A.105S}
}

@ARTICLE{Stasinska2015,
       author = {{Stasi{\'n}ska}, G. and {Izotov}, Yu. and {Morisset}, C. and {Guseva}, N.},
        title = "{Excitation properties of galaxies with the highest [O iii]/[O ii] ratios. No evidence for massive escape of ionizing photons}",
      journal = {\aap},
         year = 2015,
        month = apr,
       volume = {576},
          eid = {A83},
        pages = {A83},
          doi = {10.1051/0004-6361/201425389},
archivePrefix = {arXiv},
       eprint = {1503.00320},
 primaryClass = {astro-ph.GA},
       adsurl = {https://ui.adsabs.harvard.edu/abs/2015A&A...576A..83S}
}

@ARTICLE{Steidel2016,
       author = {{Steidel}, Charles C. and {Strom}, Allison L. and {Pettini}, Max and {Rudie}, Gwen C. and {Reddy}, Naveen A. and {Trainor}, Ryan F.},
        title = "{Reconciling the Stellar and Nebular Spectra of High-redshift Galaxies}",
      journal = {\apj},
         year = 2016,
        month = aug,
       volume = {826},
       number = {2},
          eid = {159},
        pages = {159},
          doi = {10.3847/0004-637X/826/2/159},
archivePrefix = {arXiv},
       eprint = {1605.07186},
 primaryClass = {astro-ph.GA},
       adsurl = {https://ui.adsabs.harvard.edu/abs/2016ApJ...826..159S}
}

@ARTICLE{Stern2012,
       author = {{Stern}, Jonathan and {Laor}, Ari},
        title = "{Type 1 AGN at low z- I. Emission properties}",
      journal = {\mnras},
         year = 2012,
        month = jun,
       volume = {423},
       number = {1},
        pages = {600-631},
          doi = {10.1111/j.1365-2966.2012.20901.x},
archivePrefix = {arXiv},
       eprint = {1203.3158},
 primaryClass = {astro-ph.CO},
       adsurl = {https://ui.adsabs.harvard.edu/abs/2012MNRAS.423..600S}
}

@ARTICLE{Stevenson2026,
       author = {{Stevenson}, Struan D. and {Carnall}, Adam C. and {Leung}, Ho-Hin and {Taylor}, Elizabeth and {Cullen}, Fergus and {Dunlop}, James S. and {McLeod}, Derek J. and {McLure}, Ross J. and {Begley}, Ryan and {Arellano-C{\'o}rdova}, Karla Z. and {Barrufet}, Laia and {Bondestam}, Cecilia and {Donnan}, Callum T. and {Ellis}, Richard S. and {Grogin}, Norman A. and {Koekemoer}, Anton M. and {Liu}, Feng-Yuan and {P{\'e}rez-Gonz{\'a}lez}, Pablo G. and {Rowlands}, Kate and {Sanders}, Ryan L. and {Scholte}, Dirk and {Shapley}, Alice E. and {Skarbinski}, Maya and {Stanton}, Thomas M. and {Wild}, Vivienne},
        title = "{PRIMER and JADES reveal an abundance of massive quiescent galaxies at 2 < z < 5}",
      journal = {\mnras},
         year = 2026,
        month = jan,
       volume = {545},
       number = {3},
          eid = {staf2087},
        pages = {staf2087},
          doi = {10.1093/mnras/staf2087},
archivePrefix = {arXiv},
       eprint = {2509.06913},
 primaryClass = {astro-ph.GA},
       adsurl = {https://ui.adsabs.harvard.edu/abs/2026MNRAS.545f2087S}
}

@ARTICLE{Strom2017,
       author = {{Strom}, Allison L. and {Steidel}, Charles C. and {Rudie}, Gwen C. and {Trainor}, Ryan F. and {Pettini}, Max and {Reddy}, Naveen A.},
        title = "{Nebular Emission Line Ratios in z ≃ 2-3 Star-forming Galaxies with KBSS-MOSFIRE: Exploring the Impact of Ionization, Excitation, and Nitrogen-to-Oxygen Ratio}",
      journal = {\apj},
         year = 2017,
        month = feb,
       volume = {836},
       number = {2},
          eid = {164},
        pages = {164},
          doi = {10.3847/1538-4357/836/2/164},
archivePrefix = {arXiv},
       eprint = {1608.02587},
 primaryClass = {astro-ph.GA},
       adsurl = {https://ui.adsabs.harvard.edu/abs/2017ApJ...836..164S}
}

@ARTICLE{Strom2018,
       author = {{Strom}, Allison L. and {Steidel}, Charles C. and {Rudie}, Gwen C. and {Trainor}, Ryan F. and {Pettini}, Max},
        title = "{Measuring the Physical Conditions in High-redshift Star-forming Galaxies: Insights from KBSS-MOSFIRE}",
      journal = {\apj},
         year = 2018,
        month = dec,
       volume = {868},
       number = {2},
          eid = {117},
        pages = {117},
          doi = {10.3847/1538-4357/aae1a5},
archivePrefix = {arXiv},
       eprint = {1711.08820},
 primaryClass = {astro-ph.GA},
       adsurl = {https://ui.adsabs.harvard.edu/abs/2018ApJ...868..117S}
}

@ARTICLE{Tang2025,
       author = {{Tang}, Mengtao and {Stark}, Daniel P. and {Plat}, Ad{\`e}le and {Feltre}, Anna and {Katz}, Harley and {Senchyna}, Peter and {Mason}, Charlotte A. and {Whitler}, Lily and {Chen}, Zuyi and {Topping}, Michael W.},
        title = "{JWST/NIRSpec Observations of High-ionization Emission Lines in Galaxies at High Redshift}",
      journal = {\apj},
         year = 2025,
        month = oct,
       volume = {991},
       number = {2},
          eid = {217},
        pages = {217},
          doi = {10.3847/1538-4357/adfd57},
archivePrefix = {arXiv},
       eprint = {2505.06359},
 primaryClass = {astro-ph.GA},
       adsurl = {https://ui.adsabs.harvard.edu/abs/2025ApJ...991..217T}
}

@ARTICLE{Tang2026,
       author = {{Tang}, Mengtao and {Stark}, Daniel P. and {Mason}, Charlotte A. and {Chen}, Zuyi and {Katz}, Harley and {Gronke}, Max and {Furtak}, Lukas J. and {Chang}, Seok-Jun and {Matthee}, Jorryt and {Whitler}, Lily and {Zitrin}, Adi and {Endsley}, Ryan and {Gelli}, Viola and {Roychowdhury}, Tamojeet and {Senchyna}, Peter and {Topping}, Michael W. and {Zhang}, Meng},
        title = "{SPURS: Evidence for Clumpy Neutral Envelopes and Ionized IGM Surrounding Little Red Dots in Abell 2744 from Ultra-Deep Rest-UV Spectroscopy}",
      journal = {arXiv e-prints},
         year = 2026,
        month = apr,
          eid = {arXiv:2604.03563},
        pages = {arXiv:2604.03563},
archivePrefix = {arXiv},
       eprint = {2604.03563},
 primaryClass = {astro-ph.GA},
       adsurl = {https://ui.adsabs.harvard.edu/abs/2026arXiv260403563T}
}

@ARTICLE{Taylor2025b,
       author = {{Taylor}, Anthony J. and {Finkelstein}, Steven L. and {Kocevski}, Dale D. and {Jeon}, Junehyoung and {Bromm}, Volker and {Amor{\'\i}n}, Ricardo O. and {Arrabal Haro}, Pablo and {Backhaus}, Bren E. and {Bagley}, Micaela B. and {Banados}, Eduardo and {Bhatawdekar}, Rachana and {Brooks}, Madisyn and {Calabr{\`o}}, Antonello and {Ch{\'a}vez Ortiz}, {\'O}scar A. and {Cheng}, Yingjie and {Cleri}, Nikko J. and {Cole}, Justin W. and {Davis}, Kelcey and {Dickinson}, Mark and {Donnan}, Callum and {Dunlop}, James S. and {Ellis}, Richard S. and {Fern{\'a}ndez}, Vital and {Fontana}, Adriano and {Fujimoto}, Seiji and {Giavalisco}, Mauro and {Grazian}, Andrea and {Guo}, Jingsong and {Hathi}, Nimish P. and {Holwerda}, Benne W. and {Hirschmann}, Michaela and {Inayoshi}, Kohei and {Kartaltepe}, Jeyhan S. and {Khusanova}, Yana and {Koekemoer}, Anton M. and {Kokorev}, Vasily and {Larson}, Rebecca L. and {Leung}, Gene C.~K. and {Lucas}, Ray A. and {McLeod}, Derek J. and {Napolitano}, Lorenzo and {Onoue}, Masafusa and {Pacucci}, Fabio and {Papovich}, Casey and {P{\'e}rez-Gonz{\'a}lez}, Pablo G. and {Pirzkal}, Nor and {Somerville}, Rachel S. and {Trump}, Jonathan R. and {Wilkins}, Stephen M. and {Yung}, L.~Y. Aaron and {Zhang}, Haowen},
        title = "{Broad-line AGNs at 3.5 < z < 6: The Black Hole Mass Function and a Connection with Little Red Dots}",
      journal = {\apj},
         year = 2025,
        month = jun,
       volume = {986},
       number = {2},
          eid = {165},
        pages = {165},
          doi = {10.3847/1538-4357/add15b},
archivePrefix = {arXiv},
       eprint = {2409.06772},
 primaryClass = {astro-ph.GA},
       adsurl = {https://ui.adsabs.harvard.edu/abs/2025ApJ...986..165T}
}

@ARTICLE{Taylor2025,
       author = {{Taylor}, Anthony J. and {Kokorev}, Vasily and {Kocevski}, Dale D. and {Akins}, Hollis B. and {Cullen}, Fergus and {Dickinson}, Mark and {Finkelstein}, Steven L. and {Arrabal Haro}, Pablo and {Bromm}, Volker and {Giavalisco}, Mauro and {Inayoshi}, Kohei and {Juneau}, St{\'e}phanie and {Leung}, Gene C.~K. and {P{\'e}rez-Gonz{\'a}lez}, Pablo G. and {Somerville}, Rachel S. and {Trump}, Jonathan R. and {Amor{\'\i}n}, Ricardo O. and {Barro}, Guillermo and {Burgarella}, Denis and {Brooks}, Madisyn and {Carnall}, Adam C. and {Casey}, Caitlin M. and {Cheng}, Yingjie and {Chisholm}, John and {Chworowsky}, Katherine and {Davis}, Kelcey and {Donnan}, Callum T. and {Dunlop}, James S. and {Ellis}, Richard S. and {Fern{\'a}ndez}, Vital and {Fujimoto}, Seiji and {Grogin}, Norman A. and {Gupta}, Ansh R. and {Hathi}, Nimish P. and {Jung}, Intae and {Hirschmann}, Michaela and {Kartaltepe}, Jeyhan S. and {Koekemoer}, Anton M. and {Larson}, Rebecca L. and {Leung}, Ho-Hin and {Llerena}, Mario and {Lucas}, Ray A. and {McLeod}, Derek J. and {McLure}, Ross and {Napolitano}, Lorenzo and {Papovich}, Casey and {Stanton}, Thomas M. and {Tripodi}, Roberta and {Wang}, Xin and {Wilkins}, Stephen M. and {Yung}, L.~Y. Aaron and {Zavala}, Jorge A.},
        title = "{CAPERS-LRD-z9: A Gas-enshrouded Little Red Dot Hosting a Broad-line Active Galactic Nucleus at z = 9.288}",
      journal = {\apjl},
         year = 2025,
        month = aug,
       volume = {989},
       number = {1},
          eid = {L7},
        pages = {L7},
          doi = {10.3847/2041-8213/ade789},
archivePrefix = {arXiv},
       eprint = {2505.04609},
 primaryClass = {astro-ph.GA},
       adsurl = {https://ui.adsabs.harvard.edu/abs/2025ApJ...989L...7T}
}

@ARTICLE{Telles2014,
       author = {{Telles}, Eduardo and {Thuan}, Trinh X. and {Izotov}, Yuri I. and {Carrasco}, Eleazar R.},
        title = "{A Gemini/GMOS study of the physical conditions and kinematics of the blue compact dwarf galaxy Mrk 996}",
      journal = {\aap},
         year = 2014,
        month = jan,
       volume = {561},
          eid = {A64},
        pages = {A64},
          doi = {10.1051/0004-6361/201219270},
archivePrefix = {arXiv},
       eprint = {1311.2192},
 primaryClass = {astro-ph.GA},
       adsurl = {https://ui.adsabs.harvard.edu/abs/2014A&A...561A..64T}
}

@ARTICLE{Temple2020,
       author = {{Temple}, Matthew J. and {Ferland}, Gary J. and {Rankine}, Amy L. and {Hewett}, Paul C. and {Badnell}, N.~R. and {Ballance}, Connor P. and {Del Zanna}, Giulio and {Dufresne}, Roger P.},
        title = "{Fe III emission in quasars: evidence for a dense turbulent medium}",
      journal = {\mnras},
         year = 2020,
        month = aug,
       volume = {496},
       number = {3},
        pages = {2565-2576},
          doi = {10.1093/mnras/staa1717},
archivePrefix = {arXiv},
       eprint = {2006.08617},
 primaryClass = {astro-ph.GA},
       adsurl = {https://ui.adsabs.harvard.edu/abs/2020MNRAS.496.2565T}
}

@ARTICLE{Temple2021,
       author = {{Temple}, Matthew J. and {Ferland}, Gary J. and {Rankine}, Amy L. and {Chatzikos}, Marios and {Hewett}, Paul C.},
        title = "{High-ionization emission-line ratios from quasar broad-line regions: metallicity or density?}",
      journal = {\mnras},
         year = 2021,
        month = aug,
       volume = {505},
       number = {3},
        pages = {3247-3259},
          doi = {10.1093/mnras/stab1610},
archivePrefix = {arXiv},
       eprint = {2106.01379},
 primaryClass = {astro-ph.GA},
       adsurl = {https://ui.adsabs.harvard.edu/abs/2021MNRAS.505.3247T}
}

@ARTICLE{Temple2023,
       author = {{Temple}, Matthew J. and {Matthews}, James H. and {Hewett}, Paul C. and {Rankine}, Amy L. and {Richards}, Gordon T. and {Banerji}, Manda and {Ferland}, Gary J. and {Knigge}, Christian and {Stepney}, Matthew},
        title = "{Testing AGN outflow and accretion models with C IV and He II emission line demographics in z {\ensuremath{\approx}} 2 quasars}",
      journal = {\mnras},
         year = 2023,
        month = jul,
       volume = {523},
       number = {1},
        pages = {646-666},
          doi = {10.1093/mnras/stad1448},
archivePrefix = {arXiv},
       eprint = {2301.02675},
 primaryClass = {astro-ph.GA},
       adsurl = {https://ui.adsabs.harvard.edu/abs/2023MNRAS.523..646T}
}

@ARTICLE{Thomas2016,
       author = {{Thomas}, A.~D. and {Groves}, B.~A. and {Sutherland}, R.~S. and {Dopita}, M.~A. and {Kewley}, L.~J. and {Jin}, C.},
        title = "{A Physically based Model of the Ionizing Radiation from Active Galaxies for Photoionization Modeling}",
      journal = {\apj},
         year = 2016,
        month = dec,
       volume = {833},
       number = {2},
          eid = {266},
        pages = {266},
          doi = {10.3847/1538-4357/833/2/266},
archivePrefix = {arXiv},
       eprint = {1611.05165},
 primaryClass = {astro-ph.GA},
       adsurl = {https://ui.adsabs.harvard.edu/abs/2016ApJ...833..266T}
}

@ARTICLE{Topping2020a,
       author = {{Topping}, Michael W. and {Shapley}, Alice E. and {Reddy}, Naveen A. and {Sanders}, Ryan L. and {Coil}, Alison L. and {Kriek}, Mariska and {Mobasher}, Bahram and {Siana}, Brian},
        title = "{The MOSDEF-LRIS Survey: The connection between massive stars and ionized gas in individual galaxies at z {\ensuremath{\sim}} 2}",
      journal = {\mnras},
         year = 2020,
        month = dec,
       volume = {499},
       number = {2},
        pages = {1652-1665},
          doi = {10.1093/mnras/staa2941},
archivePrefix = {arXiv},
       eprint = {2008.02282},
 primaryClass = {astro-ph.GA},
       adsurl = {https://ui.adsabs.harvard.edu/abs/2020MNRAS.499.1652T}
}

@ARTICLE{Topping2020b,
       author = {{Topping}, Michael W. and {Shapley}, Alice E. and {Reddy}, Naveen A. and {Sanders}, Ryan L. and {Coil}, Alison L. and {Kriek}, Mariska and {Mobasher}, Bahram and {Siana}, Brian},
        title = "{The MOSDEF-LRIS Survey: the interplay between massive stars and ionized gas in high-redshift star-forming galaxies}",
      journal = {\mnras},
         year = 2020,
        month = jul,
       volume = {495},
       number = {4},
        pages = {4430-4444},
          doi = {10.1093/mnras/staa1410},
archivePrefix = {arXiv},
       eprint = {1912.10243},
 primaryClass = {astro-ph.GA},
       adsurl = {https://ui.adsabs.harvard.edu/abs/2020MNRAS.495.4430T}
}

@ARTICLE{Topping2024,
       author = {{Topping}, Michael W. and {Stark}, Daniel P. and {Senchyna}, Peter and {Plat}, Adele and {Zitrin}, Adi and {Endsley}, Ryan and {Charlot}, St{\'e}phane and {Furtak}, Lukas J. and {Maseda}, Michael V. and {Smit}, Renske and {Mainali}, Ramesh and {Chevallard}, Jacopo and {Molyneux}, Stephen and {Rigby}, Jane R.},
        title = "{Metal-poor star formation at $z>6$ with JWST: new insight into hard radiation fields and nitrogen enrichment on 20 pc scales}",
      journal = {arXiv e-prints},
         year = 2024,
        month = jan,
          eid = {arXiv:2401.08764},
        pages = {arXiv:2401.08764},
          doi = {10.48550/arXiv.2401.08764},
archivePrefix = {arXiv},
       eprint = {2401.08764},
 primaryClass = {astro-ph.GA},
       adsurl = {https://ui.adsabs.harvard.edu/abs/2024arXiv240108764T}
}

@ARTICLE{Topping2025,
       author = {{Topping}, Michael W. and {Stark}, Daniel P. and {Senchyna}, Peter and {Chen}, Zuyi and {Zitrin}, Adi and {Endsley}, Ryan and {Charlot}, St{\'e}phane and {Furtak}, Lukas J. and {Maseda}, Michael V. and {Plat}, Adele and {Smit}, Renske and {Mainali}, Ramesh and {Chevallard}, Jacopo and {Molyneux}, Stephen and {Rigby}, Jane R.},
        title = "{Deep Rest-UV JWST/NIRSpec Spectroscopy of Early Galaxies: The Demographics of C IV and N-emitters in the Reionization Era}",
      journal = {\apj},
         year = 2025,
        month = feb,
       volume = {980},
       number = {2},
          eid = {225},
        pages = {225},
          doi = {10.3847/1538-4357/ada95c},
archivePrefix = {arXiv},
       eprint = {2407.19009},
 primaryClass = {astro-ph.GA},
       adsurl = {https://ui.adsabs.harvard.edu/abs/2025ApJ...980..225T}
}

@ARTICLE{Torralba2026,
       author = {{Torralba}, Alberto and {Matthee}, Jorryt and {Pezzulli}, Gabriele and {Naidu}, Rohan P. and {Ishikawa}, Yuzo and {Brammer}, Gabriel B. and {Chang}, Seok-Jun and {Chisholm}, John and {de Graaff}, Anna and {D'Eugenio}, Francesco and {Di Cesare}, Claudia and {Eilers}, Anna-Christina and {Greene}, Jenny E. and {Gronke}, Max and {Iani}, Edoardo and {Kokorev}, Vasily and {Kotiwale}, Gauri and {Kramarenko}, Ivan and {Ma}, Yilun and {Mascia}, Sara and {Navarrete}, Benjam{\'\i}n and {Nelson}, Erica and {Oesch}, Pascal and {Simcoe}, Robert A. and {Wuyts}, Stijn},
        title = "{The warm outer layer of a little red dot as the source of [Fe II] and collisional Balmer lines with scattering wings}",
      journal = {\aap},
         year = 2026,
        month = feb,
       volume = {707},
          eid = {A75},
        pages = {A75},
          doi = {10.1051/0004-6361/202557537},
archivePrefix = {arXiv},
       eprint = {2510.00103},
 primaryClass = {astro-ph.GA},
       adsurl = {https://ui.adsabs.harvard.edu/abs/2026A&A...707A..75T}
}

@ARTICLE{Tozzi2023,
       author = {{Tozzi}, Giulia and {Maiolino}, Roberto and {Cresci}, Giovanni and {Piotrowska}, Joanna M. and {Belfiore}, Francesco and {Curti}, Mirko and {Mannucci}, Filippo and {Marconi}, Alessandro},
        title = "{Unveiling hidden active nuclei in MaNGA star-forming galaxies with He II {\ensuremath{\lambda}}4686 line emission}",
      journal = {\mnras},
         year = 2023,
        month = may,
       volume = {521},
       number = {1},
        pages = {1264-1276},
          doi = {10.1093/mnras/stad506},
archivePrefix = {arXiv},
       eprint = {2302.04282},
 primaryClass = {astro-ph.GA},
       adsurl = {https://ui.adsabs.harvard.edu/abs/2023MNRAS.521.1264T}
}

@ARTICLE{Vidal2024,
       author = {{Vidal-Garc{\'\i}a}, A. and {Plat}, A. and {Curtis-Lake}, E. and {Feltre}, A. and {Hirschmann}, M. and {Chevallard}, J. and {Charlot}, S.},
        title = "{BEAGLE-AGN I: simultaneous constraints on the properties of gas in star-forming and AGN narrow-line regions in galaxies}",
      journal = {\mnras},
         year = 2024,
        month = jan,
       volume = {527},
       number = {3},
        pages = {7217-7241},
          doi = {10.1093/mnras/stad3252},
archivePrefix = {arXiv},
       eprint = {2211.13648},
 primaryClass = {astro-ph.GA},
       adsurl = {https://ui.adsabs.harvard.edu/abs/2024MNRAS.527.7217V}
}

@ARTICLE{Vanni2024,
       author = {{Vanni}, Irene and {Salvadori}, Stefania and {D'Odorico}, Valentina and {Becker}, George D. and {Cupani}, Guido},
        title = "{Chemical Diagnostics to Unveil Environments Enriched by First Stars}",
      journal = {\apjl},
         year = 2024,
        month = jun,
       volume = {967},
       number = {2},
          eid = {L22},
        pages = {L22},
          doi = {10.3847/2041-8213/ad46fa},
archivePrefix = {arXiv},
       eprint = {2402.18640},
 primaryClass = {astro-ph.GA},
       adsurl = {https://ui.adsabs.harvard.edu/abs/2024ApJ...967L..22V}
}

@ARTICLE{Veilleux1987,
       author = {{Veilleux}, Sylvain and {Osterbrock}, Donald E.},
        title = "{Spectral Classification of Emission-Line Galaxies}",
      journal = {\apjs},
         year = 1987,
        month = feb,
       volume = {63},
        pages = {295},
          doi = {10.1086/191166},
       adsurl = {https://ui.adsabs.harvard.edu/abs/1987ApJS...63..295V}
}

@ARTICLE{Venditti2026,
       author = {{Venditti}, Alessandra and {Schaerer}, Daniel and {Zackrisson}, Erik and {Asada}, Yoshihisa and {Katz}, Harley and {Salvadori}, Stefania and {Vanzella}, Eros and {Mu{\~n}oz}, Julian B. and {Storck}, Anatole and {Bunker}, Andrew J. and {Trinca}, Alessandro and {Scholte}, Dirk and {Pacucci}, Fabio and {P{\'e}rez-Gonz{\'a}lez}, Pablo G. and {Fujimoto}, Seiji and {Charbonnel}, Corinne and {Maiolino}, Roberto and {Ferrara}, Andrea and {Giavalisco}, Mauro and {Schneider}, Raffaella and {Baggen}, Josephine and {Atek}, Hakim and {Bromm}, Volker and {Caputi}, Karina and {Ciesla}, Laure and {Dayal}, Pratika and {Kobayashi}, Chiaki and {Castellano}, Marco and {Santini}, Paola},
        title = "{How can we finally see the first light? Status and perspective in the search for Population III stars}",
      journal = {The Open Journal of Astrophysics},
         year = 2026,
        month = aug,
       volume = {9},
        pages = {67811},
          doi = {10.33232/001c.167811},
archivePrefix = {arXiv},
       eprint = {2607.00167},
 primaryClass = {astro-ph.GA},
       adsurl = {https://ui.adsabs.harvard.edu/abs/2026OJAp....967811V}
}

@ARTICLE{Volonteri2003,
       author = {{Volonteri}, Marta and {Haardt}, Francesco and {Madau}, Piero},
        title = "{The Assembly and Merging History of Supermassive Black Holes in Hierarchical Models of Galaxy Formation}",
      journal = {\apj},
         year = 2003,
        month = jan,
       volume = {582},
       number = {2},
        pages = {559-573},
          doi = {10.1086/344675},
archivePrefix = {arXiv},
       eprint = {astro-ph/0207276},
 primaryClass = {astro-ph},
       adsurl = {https://ui.adsabs.harvard.edu/abs/2003ApJ...582..559V}
}

@ARTICLE{Volonteri2021,
       author = {{Volonteri}, Marta and {Habouzit}, M{\'e}lanie and {Colpi}, Monica},
        title = "{The origins of massive black holes}",
      journal = {Nature Reviews Physics},
         year = 2021,
        month = sep,
       volume = {3},
       number = {11},
        pages = {732-743},
          doi = {10.1038/s42254-021-00364-9},
archivePrefix = {arXiv},
       eprint = {2110.10175},
 primaryClass = {astro-ph.GA},
       adsurl = {https://ui.adsabs.harvard.edu/abs/2021NatRP...3..732V}
}

@ARTICLE{Wang2022,
       author = {{Wang}, Shu and {Jiang}, Linhua and {Shen}, Yue and {Ho}, Luis C. and {Vestergaard}, Marianne and {Ba{\~n}ados}, Eduardo and {Willott}, Chris J. and {Wu}, Jin and {Zou}, Siwei and {Yang}, Jinyi and {Wang}, Feige and {Fan}, Xiaohui and {Wu}, Xue-Bing},
        title = "{Metallicity in Quasar Broad-line Regions at Redshift 6}",
      journal = {\apj},
         year = 2022,
        month = feb,
       volume = {925},
       number = {2},
          eid = {121},
        pages = {121},
          doi = {10.3847/1538-4357/ac3a69},
archivePrefix = {arXiv},
       eprint = {2112.07799},
 primaryClass = {astro-ph.GA},
       adsurl = {https://ui.adsabs.harvard.edu/abs/2022ApJ...925..121W}
}

@ARTICLE{Wang2024,
       author = {{Wang}, Xin and {Cheng}, Cheng and {Ge}, Junqiang and {Meng}, Xiao-Lei and {Daddi}, Emanuele and {Yan}, Haojing and {Ji}, Zhiyuan and {Jin}, Yifei and {Jones}, Tucker and {Malkan}, Matthew A. and {Arrabal Haro}, Pablo and {Brammer}, Gabriel and {Oguri}, Masamune and {Hou}, Meicun and {Zhang}, Shiwu},
        title = "{A Strong He II {\ensuremath{\lambda}}1640 Emitter with an Extremely Blue UV Spectral Slope at z = 8.16: Presence of Population III Stars?}",
      journal = {\apjl},
         year = 2024,
        month = jun,
       volume = {967},
       number = {2},
          eid = {L42},
        pages = {L42},
          doi = {10.3847/2041-8213/ad4ced},
archivePrefix = {arXiv},
       eprint = {2212.04476},
 primaryClass = {astro-ph.GA},
       adsurl = {https://ui.adsabs.harvard.edu/abs/2024ApJ...967L..42W}
}

@ARTICLE{Wang2025,
       author = {{Wang}, Bingjie and {Leja}, Joel and {Katz}, Harley and {Inayoshi}, Kohei and {Cleri}, Nikko J. and {de Graaff}, Anna and {Hviding}, Raphael E. and {van Dokkum}, Pieter and {Greene}, Jenny E. and {Labb{\'e}}, Ivo and {Matthee}, Jorryt and {McConachie}, Ian and {Naidu}, Rohan P. and {Nelson}, Erica J.},
        title = "{The Missing Hard Photons of Little Red Dots: Their Incident Ionizing Spectra Resemble Massive Stars}",
      journal = {arXiv e-prints},
         year = 2025,
        month = aug,
          eid = {arXiv:2508.18358},
        pages = {arXiv:2508.18358},
          doi = {10.48550/arXiv.2508.18358},
archivePrefix = {arXiv},
       eprint = {2508.18358},
 primaryClass = {astro-ph.GA},
       adsurl = {https://ui.adsabs.harvard.edu/abs/2025arXiv250818358W}
}

@ARTICLE{Williams2018,
       author = {{Williams}, Peter R. and {Pancoast}, Anna and {Treu}, Tommaso and {Brewer}, Brendon J. and {Barth}, Aaron J. and {Bennert}, Vardha N. and {Buehler}, Tabitha and {Canalizo}, Gabriela and {Cenko}, S. Bradley and {Clubb}, Kelsey I. and {Cooper}, Michael C. and {Filippenko}, Alexei V. and {Gates}, Elinor and {Hoenig}, Sebastian F. and {Joner}, Michael D. and {Kandrashoff}, Michael T. and {Laney}, Clifton David and {Lazarova}, Mariana S. and {Li}, Weidong and {Malkan}, Matthew A. and {Rex}, Jacob and {Silverman}, Jeffrey M. and {Tollerud}, Erik and {Walsh}, Jonelle L. and {Woo}, Jong-Hak},
        title = "{The Lick AGN Monitoring Project 2011: Dynamical Modeling of the Broad-line Region}",
      journal = {\apj},
         year = 2018,
        month = oct,
       volume = {866},
       number = {2},
          eid = {75},
        pages = {75},
          doi = {10.3847/1538-4357/aae086},
archivePrefix = {arXiv},
       eprint = {1809.05113},
 primaryClass = {astro-ph.GA},
       adsurl = {https://ui.adsabs.harvard.edu/abs/2018ApJ...866...75W}
}

@ARTICLE{Wilkins2025,
       author = {{Wilkins}, Stephen M. and {Vijayan}, Aswin P. and {Hagen}, Scott and {Caruana}, Joseph and {Conselice}, Christopher J. and {Done}, Chris and {Hirschmann}, Michaela and {Irodotou}, Dimitrios and {Lovell}, Christopher C. and {Matthee}, Jorryt and {Plat}, Ad{\`e}le and {Roper}, William J. and {Taylor}, Anthony J.},
        title = "{First Light and Reionization Epoch Simulations (FLARES) -- XVIII: the ionising emissivities and hydrogen recombination line properties of early AGN}",
      journal = {arXiv e-prints},
         year = 2025,
        month = may,
          eid = {arXiv:2505.05257},
        pages = {arXiv:2505.05257},
          doi = {10.48550/arXiv.2505.05257},
archivePrefix = {arXiv},
       eprint = {2505.05257},
 primaryClass = {astro-ph.GA},
       adsurl = {https://ui.adsabs.harvard.edu/abs/2025arXiv250505257W}
}

@PROCEEDINGS{Wiklind2013,
        title = "{The First Galaxies}",
    booktitle = {The First Galaxies},
         year = 2013,
       editor = {{Wiklind}, Tommy and {Mobasher}, Bahram and {Bromm}, Volker},
       series = {Astrophysics and Space Science Library},
       volume = {396},
        month = jan,
          doi = {10.1007/978-3-642-32362-1},
       adsurl = {https://ui.adsabs.harvard.edu/abs/2013ASSL..396.....W}
}

@ARTICLE{Wu2022,
       author = {{Wu}, Qiaoya and {Shen}, Yue},
        title = "{A Catalog of Quasar Properties from Sloan Digital Sky Survey Data Release 16}",
      journal = {\apjs},
         year = 2022,
        month = dec,
       volume = {263},
       number = {2},
          eid = {42},
        pages = {42},
          doi = {10.3847/1538-4365/ac9ead},
archivePrefix = {arXiv},
       eprint = {2209.03987},
 primaryClass = {astro-ph.GA},
       adsurl = {https://ui.adsabs.harvard.edu/abs/2022ApJS..263...42W}
}

@ARTICLE{Yue2024,
       author = {{Yue}, Minghao and {Eilers}, Anna-Christina and {Ananna}, Tonima Tasnim and {Panagiotou}, Christos and {Kara}, Erin and {Miyaji}, Takamitsu},
        title = "{Stacking X-Ray Observations of ``Little Red Dots'': Implications for Their Active Galactic Nucleus Properties}",
      journal = {\apjl},
         year = 2024,
        month = oct,
       volume = {974},
       number = {2},
          eid = {L26},
        pages = {L26},
          doi = {10.3847/2041-8213/ad7eba},
archivePrefix = {arXiv},
       eprint = {2404.13290},
 primaryClass = {astro-ph.GA},
       adsurl = {https://ui.adsabs.harvard.edu/abs/2024ApJ...974L..26Y}
}


\appendix

\section{JWST flux limit for high BH mass}\label{sec:JWSTlimhighBHmass}

\begin{figure*}
\begin{centering}
\includegraphics[width=1\linewidth]{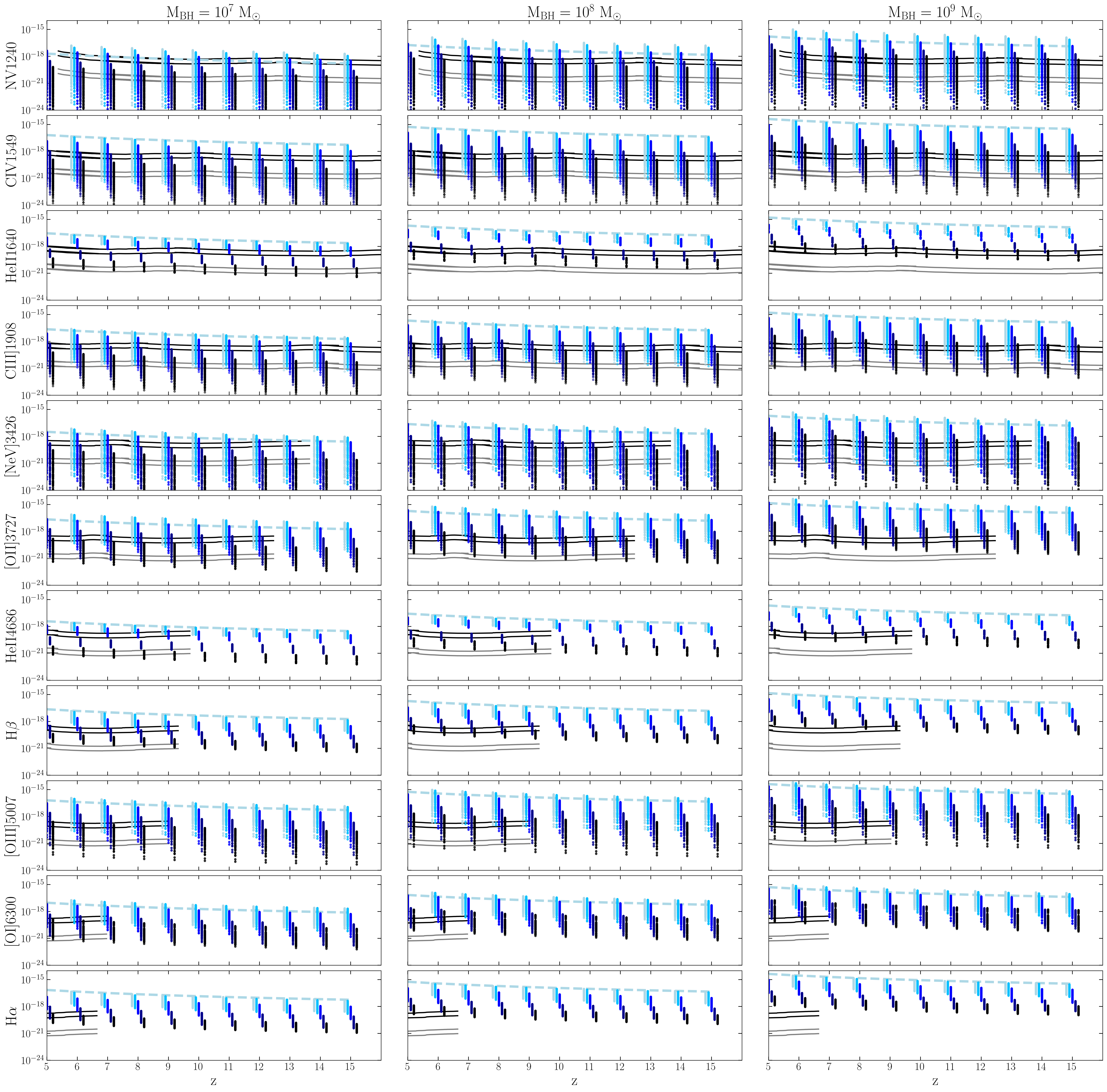}
\includegraphics[width=0.9\linewidth]{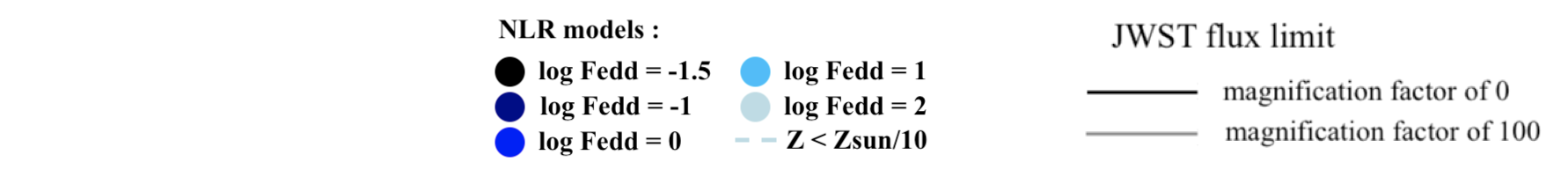}
\par\end{centering}
\caption{Same as Figure~\ref{fig:NLRdetecJWST}, but for BH masses from $10^7$ to $10^9~\Msun$.
}
\label{fig:NLRdetecJWSThighmass}
\end{figure*}


\label{lastpage}
\end{document}